\documentclass[11pt,twoside,dvips]{book}
\usepackage{fancyhdr}
\fancypagestyle{plain}{%
\fancyhf{} 
\fancyfoot[R]{\textsf{ILC-Reference Design Report\hspace{0.5cm}II-\thepage}}}
\usepackage{epsfig}
\usepackage{rotate}
\usepackage[figuresright]{rotating}
\usepackage{lscape}
\usepackage{hyperref}
\usepackage{url}
\usepackage{graphicx}
\usepackage{color}
\usepackage{epstopdf}
\usepackage{epsf}
\usepackage{hhline}
\renewcommand{\arraystretch}{1.25}
\usepackage{trc_style}
\usepackage{ulem}
\usepackage{verbatim}

\usepackage{eso-pic}
\usepackage{type1cm}

\usepackage{axodraw}

\renewcommand{\thepage}{\arabic{chapter}.\arabic{section}-\arabic{page}}

\newcommand{\ee}    {e^+e^-}
\newcommand{\lept}  {\ell^+\ell^-}
\newcommand{\ff}    {f \bar f}
\newcommand{\bb}    {b \bar b}
\newcommand{\rts}   {\sqrt{s}}
\def\Ecm{\ifmmode {\,\mathrm{ E_{cm}}}\else
                   \textrm{E_{cm}}\fi}%
\def\TeV{\ifmmode {\,\mathrm{ Te\kern -0.05em V}}\else
                   \textrm{Te\kern -0.1em V}\fi}%
\def\GeV{\ifmmode {\,\mathrm{ Ge\kern -0.05em V}}\else
                   \textrm{Ge\kern -0.1em V}\fi}%
\def\MeV{\ifmmode {\,\mathrm{ Me\kern -0.05em V}}\else
                   \textrm{Me\kern -0.1em V}\fi}%
\newcommand{\fb}{\,{\rm fb}}

\newcommand{\fbi}{\, \fb^{-1}}

\newcommand{\gvf}  {g_{ V,f}}
\newcommand{\gaf}  {g_{ A,f}}
\newcommand{\gve}  {g_{ V,e}}
\newcommand{\gae}  {g_{ A,e}}
\newcommand{\gvb}  {g_{ V,b}}
\newcommand{\gab}  {g_{ A,b}}

\newcommand{\ppl}  {{\cal P}_{\rm{e}^+}}
\newcommand{\pmi}  {{\cal P}_{\rm{e}^-}}

\newcommand{\pol}  {{\cal P}}

\newcommand{\MZ}{M_{{Z}}}
\newcommand{\MW}{M_{{W}}}

\newcommand{\ALR}{A_{\rm LR}}
\newcommand {\stl}  {\sin^2 \theta_{\rm eff}^l}

\newcommand{\Gll}        {\Gamma_{\ell}}

\newcommand{\Gtot}       {\Gamma_{\mathrm{tot}}}
\newcommand{\Ghad}       {\Gamma_{\mathrm{had}}}
\newcommand{\Rhad}       {R_{\mathrm{had}}}

\def\lsim{\raise0.3ex\hbox{$\;<$\kern-0.75em\raise-1.1ex\hbox{$\sim\;$}}}
\def\gsim{\raise0.3ex\hbox{$\;>$\kern-0.75em\raise-1.1ex\hbox{$\sim\;$}}}
\def\tb{\tan\beta}

\begin{document}
\frontmatter


{\sffamily\bfseries
\begin{titlepage}
\begin{center}
~
 ~ \vskip 4cm

    {\Huge I}{\huge NTERNATIONAL} 
    {\Huge L}{\huge INEAR} 
    {\Huge C}{\huge OLLIDER}
    
  \vskip 1.2cm

    {\Huge R}{\huge EFERENCE}
    {\Huge D}{\huge ESIGN}
    {\Huge R}{\huge EPORT}

  \vskip 1.2cm

\vskip 3cm

{\huge ILC Global Design Effort and} \\
    
  \vskip 0.5cm

{\huge World Wide Study }

  \vskip 3cm

    {\huge AUGUST, 2007}

\end{center}
\end{titlepage}

\newpage\thispagestyle{empty}
~
 ~ \vskip 2cm

{\LARGE Volume 1:~~~EXECUTIVE SUMMARY}
 \vskip 0.5cm
{\Large Editors:} 
 \vskip 0.25cm
{\Large James~Brau, Yasuhiro~Okada, Nicholas~Walker}

  \vskip 1.5cm

{\LARGE Volume 2:~~~PHYSICS AT THE ILC}
 \vskip 0.5cm
{\Large Editors:} 
 \vskip 0.25cm
{\Large Abdelhak~Djouadi, Joseph~Lykken, Klaus~M{\"o}nig} 
 \vskip 0.25cm
{\Large Yasuhiro~Okada, Mark~Oreglia, Satoru~Yamashita}

  \vskip 1.5cm

{\LARGE Volume 3:~~~ACCELERATOR}
 \vskip 0.5cm
{\Large Editors:} 
 \vskip 0.25cm
{\Large Nan~Phinney, Nobukazu~Toge, Nicholas~Walker}

  \vskip 1.5cm

{\LARGE Volume 4:~~~DETECTORS}
 \vskip 0.5cm
{\Large Editors:} 
 \vskip 0.25cm
{\Large Ties~Behnke, Chris~Damerell, John~Jaros, Akiya~Miyamoto}

\newpage\thispagestyle{empty}
~
 ~ \vskip 4cm
{\Huge Volume 2:~~~PHYSICS AT THE ILC}
 \vskip 1cm
{\LARGE Editors:} 
 \vskip 0.5cm
{\LARGE Abdelhak~Djouadi, Joseph~Lykken, Klaus~M{\"o}nig} 
 \vskip 0.4cm
{\LARGE Yasuhiro~Okada, Mark~Oreglia, Satoru~Yamashita}

\newpage\thispagestyle{empty}

}

\cleardoublepage\setcounter{page}{1}
\chapter*{List of Contributors} 

\begin{center}

\begin{center}

Gerald~Aarons$^{203}$,
Toshinori~Abe$^{290}$,
Jason~Abernathy$^{293}$,
Medina~Ablikim$^{87}$,
Halina~Abramowicz$^{216}$,
David~Adey$^{236}$,
Catherine~Adloff$^{128}$,
Chris~Adolphsen$^{203}$,
Konstantin~Afanaciev$^{11,47}$,
Ilya~Agapov$^{192,35}$
% ...

\end{center}

\clearpage

\chapter*{List of Institutions}

\begin{center}

{\sl $^{1}$ Abdus Salam International Centre for Theoretical Physics, Strada Costriera 11, 34014 Trieste, Italy}

{\sl $^{2}$ Academy, RPR, National Institute of Physics and Nuclear Engineering `Horia Hulubei' (IFIN-HH), Str. Atomistilor no. 407, P.O. Box MG-6, R-76900 Bucharest - Magurele, Romania}

{\sl $^{3}$ AGH University of Science and Technology Akademia Gorniczo-Hutnicza im. Stanislawa Staszica w Krakowie al. Mickiewicza 30 PL-30-059 Cracow, Poland}

{\sl $^{4}$ Albert-Ludwigs Universit{\"a}t Freiburg, Physikalisches Institut, Hermann-Herder Str. 3, D-79104 Freiburg, Germany}

{\sl $^{5}$ Aligarh Muslim University, Aligarh, Uttar Pradesh 202002, India}

{\sl $^{6}$ Amberg Engineering AG, Trockenloostr. 21, P.O.Box 27, 8105 Regensdorf-Watt, Switzerland}

{\sl $^{7}$ Angstromquelle Karlsruhe (ANKA), Forschungszentrum Karlsruhe, Hermann-von-Helmholtz-Platz 1, D-76344 Eggenstein-Leopoldshafen, Germany}

{\sl $^{8}$ Argonne National Laboratory (ANL), 9700 S. Cass Avenue, Argonne, IL 60439, USA}

{\sl $^{9}$ Baylor University, Department of Physics, 101 Bagby Avenue, Waco, TX 76706, USA}

{\sl $^{10}$ Beijing University, Department of Physics, Beijing, China 100871}

{\sl $^{11}$ Belarusian State University, National Scientific \& Educational Center, Particle \& HEP Physics, M. Bogdanovich St., 153, 240040 Minsk, Belarus}

{\sl $^{12}$ Benares Hindu University, Benares, Varanasi 221005, India}

{\sl $^{13}$ Bhabha Atomic Research Centre, Trombay, Mumbai 400085, India}

{\sl $^{14}$ Birla Institute of Technology and Science, EEE Dept., Pilani, Rajasthan, India}

{\sl $^{15}$ Bogazici University, Physics Department, 34342 Bebek / Istanbul, 80820 Istanbul, Turkey}

{\sl $^{16}$ Boston University, Department of Physics, 590 Commonwealth Avenue, Boston, MA 02215, USA}

{\sl $^{17}$ Brandenburg University of Technology, Postfach 101344, D-03013 Cottbus, Germany}

{\sl $^{18}$ Brno University of Technology, Anton\'insk\'a; 548/1, CZ 601 90 Brno, Czech Republic}

{\sl $^{19}$ Brookhaven National Laboratory (BNL), P.O.Box 5000, Upton, NY 11973-5000, USA}

{\sl $^{20}$ Brown University, Department of Physics, Box 1843, Providence, RI 02912, USA}

{\sl $^{21}$ Budkar Institute for Nuclear Physics (BINP), 630090 Novosibirsk, Russia}

{\sl $^{22}$ Calcutta University, Department of Physics, 92 A.P.C. Road, Kolkata 700009, India}

{\sl $^{23}$ California Institute of Technology, Physics, Mathematics and Astronomy (PMA), 1200 East California Blvd, Pasadena, CA 91125, USA}

{\sl $^{24}$ Carleton University, Department of Physics, 1125 Colonel By Drive, Ottawa, Ontario, Canada K1S 5B6}

{\sl $^{25}$ Carnegie Mellon University, Department of Physics, Wean Hall 7235, Pittsburgh, PA 15213, USA}

{\sl $^{26}$ CCLRC Daresbury Laboratory, Daresbury, Warrington, Cheshire WA4 4AD, UK }

{\sl $^{27}$ CCLRC Rutherford Appleton Laboratory, Chilton, Didcot, Oxton OX11 0QX, UK }

{\sl $^{28}$ CEA Saclay, DAPNIA, F-91191 Gif-sur-Yvette, France}

{\sl $^{29}$ CEA Saclay, Service de Physique Th{\'e}orique, CEA/DSM/SPhT, F-91191 Gif-sur-Yvette Cedex, France}

{\sl $^{30}$ Center for High Energy Physics (CHEP) / Kyungpook National University, 1370 Sankyuk-dong, Buk-gu, Daegu 702-701, Korea}

{\sl $^{31}$ Center for High Energy Physics (TUHEP), Tsinghua University, Beijing, China 100084}

{\sl $^{32}$ Centre de Physique Theorique, CNRS - Luminy, Universiti d'Aix - Marseille II, Campus of Luminy, Case 907, 13288 Marseille Cedex 9, France}

{\sl $^{33}$ Centro de Investigaciones Energ\'eticas, Medioambientales y Technol\'ogicas, CIEMAT, Avenia Complutense 22, E-28040 Madrid, Spain}

{\sl $^{34}$ Centro Nacional de Microelectr\'onica (CNM), Instituto de Microelectr\'onica de Barcelona (IMB), Campus UAB, 08193 Cerdanyola del Vall\`es (Bellaterra), Barcelona, Spain}

{\sl $^{35}$ CERN, CH-1211 Gen\`eve 23, Switzerland}

{\sl $^{36}$ Charles University, Institute of Particle \& Nuclear Physics, Faculty of Mathematics and Physics, V Holesovickach 2, CZ-18000 Praque 8, Czech Republic}

{\sl $^{37}$ Chonbuk National University, Physics Department, Chonju 561-756, Korea}

{\sl $^{38}$ Cockcroft Institute, Daresbury, Warrington WA4 4AD, UK }

{\sl $^{39}$ College of William and Mary, Department of Physics, Williamsburg, VA, 23187, USA}

{\sl $^{40}$ Colorado State University, Department of Physics, Fort Collins, CO 80523, USA}

{\sl $^{41}$ Columbia University, Department of Physics, New York, NY 10027-6902, USA}

{\sl $^{42}$ Concordia University, Department of Physics, 1455 De Maisonneuve Blvd. West, Montreal, Quebec, Canada H3G 1M8}

{\sl $^{43}$ Cornell University, Laboratory for Elementary-Particle Physics (LEPP), Ithaca, NY 14853, USA}

{\sl $^{44}$ Cukurova University, Department of Physics, Fen-Ed. Fakultesi 01330, Balcali, Turkey}

{\sl $^{45}$ D.~V. Efremov Research Institute, SINTEZ, 196641 St. Petersburg, Russia}

{\sl $^{46}$ Dartmouth College, Department of Physics and Astronomy, 6127 Wilder Laboratory, Hanover, NH 03755, USA}

{\sl $^{47}$ DESY-Hamburg site, Deutsches Elektronen-Synchrotoron in der Helmholtz-Gemeinschaft, Notkestrasse 85, 22607 Hamburg, Germany}

{\sl $^{48}$ DESY-Zeuthen site, Deutsches Elektronen-Synchrotoron in der Helmholtz-Gemeinschaft, Platanenallee 6, D-15738 Zeuthen, Germany}

{\sl $^{49}$ Durham University,  Department of Physics, Ogen Center for Fundamental Physics, South Rd., Durham DH1 3LE, UK}

{\sl $^{50}$ Ecole Polytechnique, Laboratoire Leprince-Ringuet (LLR), Route de Saclay, F-91128 Palaiseau Cedex, France}

{\sl $^{51}$ Ege University, Department of Physics, Faculty of Science, 35100 Izmir, Turkey}

{\sl $^{52}$ Enrico Fermi Institute, University of Chicago, 5640 S. Ellis Avenue, RI-183, Chicago, IL 60637, USA}

{\sl $^{53}$ Ewha Womans University, 11-1 Daehyun-Dong, Seodaemun-Gu, Seoul, 120-750, Korea}

{\sl $^{54}$ Fermi National Accelerator Laboratory (FNAL), P.O.Box 500, Batavia, IL 60510-0500, USA}

{\sl $^{55}$ Fujita Gakuen Health University, Department of Physics, Toyoake, Aichi 470-1192, Japan}

{\sl $^{56}$ Fukui University of Technology, 3-6-1 Gakuen, Fukui-shi, Fukui 910-8505, Japan}

{\sl $^{57}$ Fukui University, Department of Physics, 3-9-1 Bunkyo, Fukui-shi, Fukui 910-8507, Japan}

{\sl $^{58}$ Georg-August-Universit{\"a}t G{\"o}ttingen, II. Physikalisches Institut, Friedrich-Hund-Platz 1, 37077 G{\"o}ttingen, Germany}

{\sl $^{59}$ Global Design Effort}

{\sl $^{60}$ Gomel State University, Department of Physics, Ul. Sovietskaya 104, 246699 Gomel, Belarus}

{\sl $^{61}$ Guangxi University, College of Physics science and Engineering Technology, Nanning, China 530004}

{\sl $^{62}$ Hanoi University of Technology, 1 Dai Co Viet road, Hanoi, Vietnam}

{\sl $^{63}$ Hanson Professional Services, Inc., 1525 S. Sixth St., Springfield, IL 62703, USA}

{\sl $^{64}$ Harish-Chandra Research Institute, Chhatnag Road, Jhusi, Allahabad 211019, India}

{\sl $^{65}$ Helsinki Institute of Physics (HIP), P.O. Box 64, FIN-00014 University of Helsinki, Finland}

{\sl $^{66}$ Henan Normal University, College of Physics and Information Engineering, Xinxiang, China 453007}

{\sl $^{67}$ High Energy Accelerator Research Organization, KEK, 1-1 Oho, Tsukuba, Ibaraki 305-0801, Japan}

{\sl $^{68}$ Hiroshima University, Department of Physics, 1-3-1 Kagamiyama, Higashi-Hiroshima, Hiroshima 739-8526, Japan}

{\sl $^{69}$ Humboldt Universit{\"a}t zu Berlin, Fachbereich Physik, Institut f\"ur Elementarteilchenphysik, Newtonstr. 15, D-12489 Berlin, Germany}

{\sl $^{70}$ Hungarian Academy of Sciences, KFKI Research Institute for Particle and Nuclear Physics, P.O. Box 49, H-1525 Budapest, Hungary}

{\sl $^{71}$ Ibaraki University, College of Technology, Department of Physics, Nakanarusawa 4-12-1, Hitachi, Ibaraki 316-8511, Japan}

{\sl $^{72}$ Imperial College, Blackett Laboratory, Department of Physics, Prince Consort Road, London, SW7 2BW, UK}

{\sl $^{73}$ Indian Association for the Cultivation of Science, Department of Theoretical Physics and Centre for Theoretical Sciences, Kolkata 700032, India}

{\sl $^{74}$ Indian Institute of Science, Centre for High Energy Physics, Bangalore 560012, Karnataka, India}

{\sl $^{75}$ Indian Institute of Technology, Bombay, Powai, Mumbai 400076, India}

{\sl $^{76}$ Indian Institute of Technology, Guwahati, Guwahati, Assam 781039, India}

{\sl $^{77}$ Indian Institute of Technology, Kanpur, Department of Physics,  IIT Post Office, Kanpur 208016, India}

{\sl $^{78}$ Indiana University - Purdue University, Indianapolis, Department of Physics, 402 N. Blackford St., LD 154, Indianapolis, IN 46202, USA}

{\sl $^{79}$ Indiana University, Department of Physics, Swain Hall West 117, 727 E. 3rd St., Bloomington, IN 47405-7105, USA}

{\sl $^{80}$ Institucio Catalana de Recerca i Estudis, ICREA,  Passeig Lluis Companys, 23, Barcelona 08010, Spain}

{\sl $^{81}$ Institut de Physique Nucl\'eaire, F-91406 Orsay, France }

{\sl $^{82}$ Institut f\"ur Theorie Elektromagnetischer Felder (TEMF), Technische Universit\"at Darmstadt, Schlo{\ss}gartenstr. 8, D-64289 Darmstadt, Germany}

{\sl $^{83}$ Institut National de Physique Nucleaire et de Physique des Particules, 3, Rue Michel- Ange, 75794 Paris Cedex 16, France}

{\sl $^{84}$ Institut Pluridisciplinaire Hubert Curien, 23 Rue du Loess - BP28, 67037 Strasbourg Cedex 2, France}

{\sl $^{85}$ Institute for Chemical Research, Kyoto University, Gokasho, Uji, Kyoto 611-0011, Japan}

{\sl $^{86}$ Institute for Cosmic Ray Research, University of Tokyo, 5-1-5 Kashiwa-no-Ha, Kashiwa, Chiba 277-8582, Japan}

{\sl $^{87}$ Institute of High Energy Physics - IHEP, Chinese Academy of Sciences, P.O. Box 918, Beijing, China 100049}

{\sl $^{88}$ Institute of Mathematical Sciences, Taramani, C.I.T. Campus, Chennai 600113, India}

{\sl $^{89}$ Institute of Physics and Electronics, Vietnamese Academy of Science and Technology (VAST), 10 Dao-Tan, Ba-Dinh, Hanoi 10000, Vietnam}

{\sl $^{90}$ Institute of Physics, ASCR, Academy of Science of the Czech Republic, Division of Elementary Particle Physics, Na Slovance 2, CS-18221 Prague 8, Czech Republic}

{\sl $^{91}$ Institute of Physics, Pomorska 149/153, PL-90-236 Lodz, Poland}

{\sl $^{92}$ Institute of Theoretical and Experimetal Physics, B. Cheremushkinskawa, 25, RU-117259, Moscow, Russia}

{\sl $^{93}$ Institute of Theoretical Physics, Chinese Academy of Sciences, P.O.Box 2735, Beijing, China 100080}

{\sl $^{94}$ Instituto de Fisica Corpuscular (IFIC), Centro Mixto CSIC-UVEG, Edificio Investigacion Paterna, Apartado 22085, 46071 Valencia, Spain}

{\sl $^{95}$ Instituto de Fisica de Cantabria, (IFCA, CSIC-UC), Facultad de Ciencias, Avda. Los Castros s/n, 39005 Santander, Spain}

{\sl $^{96}$ Instituto Nazionale di Fisica Nucleare (INFN), Laboratorio LASA, Via Fratelli Cervi 201, 20090 Segrate, Italy}

{\sl $^{97}$ Instituto Nazionale di Fisica Nucleare (INFN), Sezione di Ferrara, via Paradiso 12, I-44100 Ferrara, Italy}

{\sl $^{98}$ Instituto Nazionale di Fisica Nucleare (INFN), Sezione di Firenze, Via G. Sansone 1, I-50019 Sesto Fiorentino (Firenze), Italy}

{\sl $^{99}$ Instituto Nazionale di Fisica Nucleare (INFN), Sezione di Lecce, via Arnesano, I-73100 Lecce, Italy}

{\sl $^{100}$ Instituto Nazionale di Fisica Nucleare (INFN), Sezione di Napoli, Complesso Universit{\'a} di Monte Sant'Angelo,via, I-80126 Naples, Italy}

{\sl $^{101}$ Instituto Nazionale di Fisica Nucleare (INFN), Sezione di Pavia, Via Bassi 6, I-27100 Pavia, Italy}

{\sl $^{102}$ Instituto Nazionale di Fisica Nucleare (INFN), Sezione di Pisa, Edificio C - Polo Fibonacci Largo B. Pontecorvo, 3, I-56127 Pisa, Italy}

{\sl $^{103}$ Instituto Nazionale di Fisica Nucleare (INFN), Sezione di Torino, c/o Universit{\'a}' di Torino facolt{\'a}' di Fisica, via P Giuria 1, 10125 Torino, Italy}

{\sl $^{104}$ Instituto Nazionale di Fisica Nucleare (INFN), Sezione di Trieste, Padriciano 99, I-34012 Trieste (Padriciano), Italy}

{\sl $^{105}$ Inter-University Accelerator Centre, Aruna Asaf Ali Marg, Post Box 10502, New Delhi 110067, India}

{\sl $^{106}$ International Center for Elementary Particle Physics, University of Tokyo, Hongo 7-3-1, Bunkyo District, Tokyo 113-0033, Japan}

{\sl $^{107}$ Iowa State University, Department of Physics, High Energy Physics Group, Ames, IA 50011, USA}

{\sl $^{108}$ Jagiellonian University, Institute of Physics, Ul. Reymonta 4, PL-30-059 Cracow, Poland}

{\sl $^{109}$ Jamia Millia Islamia, Centre for Theoretical Physics, Jamia Nagar, New Delhi 110025, India}

{\sl $^{110}$ Jamia Millia Islamia, Department of Physics, Jamia Nagar, New Delhi 110025, India}

{\sl $^{111}$ Japan Aerospace Exploration Agency, Sagamihara Campus, 3-1-1 Yoshinodai, Sagamihara, Kanagawa 220-8510 , Japan}

{\sl $^{112}$ Japan Atomic Energy Agency, 4-49 Muramatsu, Tokai-mura, Naka-gun, Ibaraki 319-1184, Japan}

{\sl $^{113}$ Johannes Gutenberg Universit{\"a}t Mainz, Institut f{\"u}r Physik, 55099 Mainz, Germany}

{\sl $^{114}$ Johns Hopkins University, Applied Physics Laboratory, 11100 Johns Hopkins RD., Laurel, MD 20723-6099, USA}

{\sl $^{115}$ Joint Institute for Nuclear Research (JINR), Joliot-Curie 6, 141980, Dubna, Moscow Region, Russia}

{\sl $^{116}$ Kansas State University, Department of Physics, 116 Cardwell Hall, Manhattan, KS 66506, USA}

{\sl $^{117}$ KCS Corp., 2-7-25 Muramatsukita, Tokai, Ibaraki 319-1108, Japan}

{\sl $^{118}$ Kharkov Institute of Physics and Technology, National Science Center, 1, Akademicheskaya St., Kharkov, 61108, Ukraine}

{\sl $^{119}$ Kinki University, Department of Physics, 3-4-1 Kowakae, Higashi-Osaka, Osaka 577-8502, Japan}

{\sl $^{120}$ Kobe University, Faculty of Science, 1-1 Rokkodai-cho, Nada-ku, Kobe, Hyogo 657-8501, Japan}

{\sl $^{121}$ Kogakuin University, Department of Physics, Shinjuku Campus, 1-24-2 Nishi-Shinjuku, Shinjuku-ku, Tokyo 163-8677, Japan}

{\sl $^{122}$ Konkuk University, 93-1 Mojin-dong, Kwanglin-gu, Seoul 143-701, Korea}

{\sl $^{123}$ Korea Advanced Institute of Science \& Technology, Department of Physics, 373-1 Kusong-dong, Yusong-gu, Taejon 305-701, Korea}

{\sl $^{124}$ Korea Institute for Advanced Study (KIAS), School of Physics, 207-43 Cheongryangri-dong, Dongdaemun-gu, Seoul 130-012, Korea}

{\sl $^{125}$ Korea University, Department of Physics, Seoul 136-701, Korea}

{\sl $^{126}$ Kyoto University, Department of Physics, Kitashirakawa-Oiwakecho, Sakyo-ku, Kyoto 606-8502, Japan}

{\sl $^{127}$ L.P.T.A., UMR 5207 CNRS-UM2, Universit{\'e} Montpellier II, Case Courrier 070, B{\^a}t. 13, place Eug{\`e}ne Bataillon, 34095 Montpellier Cedex 5, France}

{\sl $^{128}$ Laboratoire d'Annecy-le-Vieux de Physique des Particules (LAPP), Chemin du Bellevue, BP 110, F-74941 Annecy-le-Vieux Cedex, France}

{\sl $^{129}$ Laboratoire d'Annecy-le-Vieux de Physique Theorique (LAPTH), Chemin de Bellevue, BP 110, F-74941 Annecy-le-Vieux Cedex, France}

{\sl $^{130}$ Laboratoire de l'Acc\'el\'erateur Lin\'eaire (LAL), Universit\'e Paris-Sud 11, B\^atiment 200, 91898 Orsay, France}

{\sl $^{131}$ Laboratoire de Physique Corpusculaire de Clermont-Ferrand (LPC), Universit\'e Blaise Pascal, I.N.2.P.3./C.N.R.S., 24 avenue des Landais, 63177 Aubi\`ere Cedex, France}

{\sl $^{132}$ Laboratoire de Physique Subatomique et de Cosmologie (LPSC), Universit\'e Joseph Fourier (Grenoble 1), 53, ave. des Marthyrs, F-38026 Grenoble Cedex, France}

{\sl $^{133}$ Laboratoire de Physique Theorique, Universit\'e de Paris-Sud XI, Batiment 210, F-91405 Orsay Cedex, France}

{\sl $^{134}$ Laboratori Nazionali di Frascati, via E. Fermi, 40, C.P. 13, I-00044 Frascati, Italy}

{\sl $^{135}$ Laboratory of High Energy Physics and Cosmology, Department of Physics, Hanoi National University, 334 Nguyen Trai, Hanoi, Vietnam}

{\sl $^{136}$ Lancaster University, Physics Department, Lancaster LA1 4YB, UK}

{\sl $^{137}$ Lawrence Berkeley National Laboratory (LBNL), 1 Cyclotron Rd, Berkeley, CA 94720, USA}

{\sl $^{138}$ Lawrence Livermore National Laboratory (LLNL), Livermore, CA 94551, USA}

{\sl $^{139}$ Lebedev Physical Institute, Leninsky Prospect 53, RU-117924 Moscow, Russia}

{\sl $^{140}$ Liaoning Normal University, Department of Physics, Dalian, China 116029}

{\sl $^{141}$ Lomonosov Moscow State University, Skobeltsyn Institute of Nuclear Physics (MSU SINP), 1(2), Leninskie gory, GSP-1, Moscow 119991, Russia}

{\sl $^{142}$ Los Alamos National Laboratory (LANL), P.O.Box 1663, Los Alamos, NM 87545, USA}

{\sl $^{143}$ Louisiana Technical University, Department of Physics, Ruston, LA 71272, USA}

{\sl $^{144}$ Ludwig-Maximilians-Universit{\"a}t M{\"u}nchen, Department f{\"u}r Physik, Schellingstr. 4, D-80799 Munich, Germany}

{\sl $^{145}$ Lunds Universitet, Fysiska Institutionen, Avdelningen f{\"o}r Experimentell H{\"o}genergifysik, Box 118, 221 00 Lund, Sweden}

{\sl $^{146}$ Massachusetts Institute of Technology, Laboratory for Nuclear Science \& Center for Theoretical Physics, 77 Massachusetts Ave., NW16, Cambridge, MA 02139, USA}

{\sl $^{147}$ Max-Planck-Institut f{\"u}r Physik (Werner-Heisenberg-Institut), F{\"o}hringer Ring 6, 80805 M{\"u}nchen, Germany}

{\sl $^{148}$ McGill University, Department of Physics, Ernest Rutherford Physics Bldg., 3600 University Ave., Montreal, Quebec, H3A 2T8 Canada}

{\sl $^{149}$ Meiji Gakuin University, Department of Physics, 2-37 Shirokanedai 1-chome, Minato-ku, Tokyo 244-8539, Japan}

{\sl $^{150}$ Michigan State University, Department of Physics and Astronomy, East Lansing, MI 48824, USA}

{\sl $^{151}$ Middle East Technical University, Department of Physics, TR-06531 Ankara, Turkey}

{\sl $^{152}$ Mindanao Polytechnic State College, Lapasan, Cagayan de Oro City 9000, Phillipines}

{\sl $^{153}$ MSU-Iligan Institute of Technology, Department of Physics, Andres Bonifacio Avenue, 9200 Iligan City, Phillipines}

{\sl $^{154}$ Nagasaki Institute of Applied Science, 536 Abamachi, Nagasaki-Shi, Nagasaki 851-0193, Japan}

{\sl $^{155}$ Nagoya University, Fundamental Particle Physics Laboratory, Division of Particle and Astrophysical Sciences, Furo-cho, Chikusa-ku, Nagoya, Aichi 464-8602, Japan}

{\sl $^{156}$ Nanchang University, Department of Physics, Nanchang, China 330031}

{\sl $^{157}$ Nanjing University, Department of Physics, Nanjing, China 210093}

{\sl $^{158}$ Nankai University, Department of Physics, Tianjin, China 300071}

{\sl $^{159}$ National Central University, High Energy Group, Department of Physics, Chung-li, Taiwan 32001}

{\sl $^{160}$ National Institute for Nuclear \& High Energy Physics, PO Box 41882, 1009 DB Amsterdam, Netherlands}

{\sl $^{161}$ National Institute of Radiological Sciences, 4-9-1 Anagawa, Inaga, Chiba 263-8555, Japan}

{\sl $^{162}$ National Synchrotron Radiation Laboratory, University of Science and Technology of china, Hefei, Anhui, China 230029}

{\sl $^{163}$ National Synchrotron Research Center, 101 Hsin-Ann Rd., Hsinchu Science Part, Hsinchu, Taiwan 30076}

{\sl $^{164}$ National Taiwan University, Physics Department, Taipei, Taiwan 106}

{\sl $^{165}$ Niels Bohr Institute (NBI), University of Copenhagen, Blegdamsvej 17, DK-2100 Copenhagen, Denmark}

{\sl $^{166}$ Niigata University, Department of Physics, Ikarashi, Niigata 950-218, Japan}

{\sl $^{167}$ Nikken Sekkai Ltd., 2-18-3 Iidabashi, Chiyoda-Ku, Tokyo 102-8117, Japan}

{\sl $^{168}$ Nippon Dental University, 1-9-20 Fujimi, Chiyoda-Ku, Tokyo 102-8159, Japan}

{\sl $^{169}$ North Asia University, Akita 010-8515, Japan}

{\sl $^{170}$ North Eastern Hill University, Department of Physics, Shillong 793022, India}

{\sl $^{171}$ Northern Illinois University, Department of Physics, DeKalb, Illinois 60115-2825, USA}

{\sl $^{172}$ Northwestern University, Department of Physics and Astronomy, 2145 Sheridan Road., Evanston, IL 60208, USA}

{\sl $^{173}$ Novosibirsk State University (NGU), Department of Physics, Pirogov st. 2, 630090 Novosibirsk, Russia}

{\sl $^{174}$ Obninsk State Technical University for Nuclear Engineering (IATE), Obninsk, Russia}

{\sl $^{175}$ Ochanomizu University, Department of Physics, Faculty of Science, 1-1 Otsuka 2, Bunkyo-ku, Tokyo 112-8610, Japan}

{\sl $^{176}$ Osaka University, Laboratory of Nuclear Studies, 1-1 Machikaneyama, Toyonaka, Osaka 560-0043, Japan}

{\sl $^{177}$ {\"O}sterreichische Akademie der Wissenschaften, Institut f{\"u}r Hochenergiephysik, Nikolsdorfergasse 18, A-1050 Vienna, Austria}

{\sl $^{178}$ Panjab University, Chandigarh 160014, India}

{\sl $^{179}$ Pavel Sukhoi Gomel State Technical University, ICTP Affiliated Centre \& Laboratory for Physical Studies, October Avenue, 48, 246746, Gomel, Belarus}

{\sl $^{180}$ Pavel Sukhoi Gomel State Technical University, Physics Department, October Ave. 48, 246746 Gomel, Belarus}

{\sl $^{181}$ Physical Research Laboratory, Navrangpura, Ahmedabad 380 009, Gujarat, India}

{\sl $^{182}$ Pohang Accelerator Laboratory (PAL), San-31 Hyoja-dong, Nam-gu, Pohang, Gyeongbuk 790-784, Korea}

{\sl $^{183}$ Polish Academy of Sciences (PAS), Institute of Physics, Al. Lotnikow 32/46, PL-02-668 Warsaw, Poland}

{\sl $^{184}$ Primera Engineers Ltd., 100 S Wacker Drive, Suite 700, Chicago, IL 60606, USA}

{\sl $^{185}$ Princeton University, Department of Physics, P.O. Box 708, Princeton, NJ 08542-0708, USA}

{\sl $^{186}$ Purdue University, Department of Physics, West Lafayette, IN 47907, USA}

{\sl $^{187}$ Pusan National University, Department of Physics, Busan 609-735, Korea}

{\sl $^{188}$ R. W. Downing Inc., 6590 W. Box Canyon Dr., Tucson, AZ 85745, USA}

{\sl $^{189}$ Raja Ramanna Center for Advanced Technology, Indore 452013, India}

{\sl $^{190}$ Rheinisch-Westf{\"a}lische Technische Hochschule (RWTH), Physikalisches Institut, Physikzentrum, Sommerfeldstrasse 14, D-52056 Aachen, Germany}

{\sl $^{191}$ RIKEN, 2-1 Hirosawa, Wako, Saitama 351-0198, Japan}

{\sl $^{192}$ Royal Holloway, University of London (RHUL), Department of Physics, Egham, Surrey TW20 0EX, UK }

{\sl $^{193}$ Saga University, Department of Physics, 1 Honjo-machi, Saga-shi, Saga 840-8502, Japan}

{\sl $^{194}$ Saha Institute of Nuclear Physics, 1/AF Bidhan Nagar, Kolkata 700064, India}

{\sl $^{195}$ Salalah College of Technology (SCOT), Engineering Department, Post Box No. 608, Postal Code 211, Salalah, Sultanate of Oman}

{\sl $^{196}$ Saube Co., Hanabatake, Tsukuba, Ibaraki 300-3261, Japan}

{\sl $^{197}$ Seoul National University, San 56-1, Shinrim-dong, Kwanak-gu, Seoul 151-742, Korea}

{\sl $^{198}$ Shandong University, 27 Shanda Nanlu, Jinan, China 250100}

{\sl $^{199}$ Shanghai Institute of Applied Physics, Chinese Academy of Sciences, 2019 Jiaruo Rd., Jiading, Shanghai, China 201800}

{\sl $^{200}$ Shinshu University, 3-1-1, Asahi, Matsumoto, Nagano 390-8621, Japan}

{\sl $^{201}$ Sobolev Institute of Mathematics, Siberian Branch of the Russian Academy of Sciences, 4 Acad. Koptyug Avenue, 630090 Novosibirsk, Russia}

{\sl $^{202}$ Sokendai, The Graduate University for Advanced Studies, Shonan Village, Hayama, Kanagawa 240-0193, Japan}

{\sl $^{203}$ Stanford Linear Accelerator Center (SLAC), 2575 Sand Hill Road, Menlo Park, CA 94025, USA}

{\sl $^{204}$ State University of New York at Binghamton, Department of Physics, PO Box 6016, Binghamton, NY 13902, USA}

{\sl $^{205}$ State University of New York at Buffalo, Department of Physics \& Astronomy, 239 Franczak Hall, Buffalo, NY 14260, USA}

{\sl $^{206}$ State University of New York at Stony Brook, Department of Physics and Astronomy, Stony Brook, NY 11794-3800, USA}

{\sl $^{207}$ Sumitomo Heavy Industries, Ltd., Natsushima-cho, Yokosuka, Kanagawa 237-8555, Japan}

{\sl $^{208}$ Sungkyunkwan University (SKKU), Natural Science Campus 300, Physics Research Division, Chunchun-dong, Jangan-gu, Suwon, Kyunggi-do 440-746, Korea}

{\sl $^{209}$ Swiss Light Source (SLS), Paul Scherrer Institut (PSI), PSI West, CH-5232 Villigen PSI, Switzerland}

{\sl $^{210}$ Syracuse University, Department of Physics, 201 Physics Building, Syracuse, NY 13244-1130, USA}

{\sl $^{211}$ Tata Institute of Fundamental Research, School of Natural Sciences, Homi Bhabha Rd., Mumbai 400005, India}

{\sl $^{212}$ Technical Institute of Physics and Chemistry, Chinese Academy of Sciences, 2 North 1st St., Zhongguancun, Beijing, China 100080}

{\sl $^{213}$ Technical University of Lodz, Department of Microelectronics and Computer Science, al. Politechniki 11, 90-924 Lodz, Poland}

{\sl $^{214}$ Technische Universit{\"a}t Dresden, Institut f{\"u}r Kern- und Teilchenphysik, D-01069 Dresden, Germany}

{\sl $^{215}$ Technische Universit{\"a}t Dresden, Institut f{\"u}r Theoretische Physik,D-01062 Dresden, Germany}

{\sl $^{216}$ Tel-Aviv University, School of Physics and Astronomy, Ramat Aviv, Tel Aviv 69978, Israel}

{\sl $^{217}$ Texas A\&M University, Physics Department, College Station, 77843-4242 TX, USA}

{\sl $^{218}$ Texas Tech University, Department of Physics, Campus Box 41051, Lubbock, TX 79409-1051, USA}

{\sl $^{219}$ The Henryk Niewodniczanski Institute of Nuclear Physics (NINP), High Energy Physics Lab, ul. Radzikowskiego 152, PL-31342 Cracow, Poland}

{\sl $^{220}$ Thomas Jefferson National Accelerator Facility (TJNAF), 12000 Jefferson Avenue, Newport News, VA 23606, USA}

{\sl $^{221}$ Tohoku Gakuin University, Faculty of Technology, 1-13-1 Chuo, Tagajo, Miyagi 985-8537, Japan}

{\sl $^{222}$ Tohoku University, Department of Physics, Aoba District, Sendai, Miyagi 980-8578, Japan}

{\sl $^{223}$ Tokyo Management College, Computer Science Lab, Ichikawa, Chiba 272-0001, Japan}

{\sl $^{224}$ Tokyo University of Agriculture Technology, Department of Applied Physics, Naka-machi, Koganei, Tokyo 183-8488, Japan}

{\sl $^{225}$ Toyama University, Department of Physics, 3190 Gofuku, Toyama-shi 930-8588, Japan}

{\sl $^{226}$ TRIUMF, 4004 Wesbrook Mall, Vancouver, BC V6T 2A3, Canada}

{\sl $^{227}$ Tufts University, Department of Physics and Astronomy, Robinson Hall, Medford, MA 02155, USA}

{\sl $^{228}$ Universidad Aut\`onoma de Madrid (UAM), Facultad de Ciencias C-XI, Departamento de Fisica Teorica, Cantoblanco, Madrid 28049, Spain}

{\sl $^{229}$ Universitat Aut\`onoma de Barcelona, Institut de Fisica d'Altes Energies (IFAE), Campus UAB, Edifici Cn, E-08193 Bellaterra, Barcelona, Spain}

{\sl $^{230}$ University College of London (UCL), High Energy Physics Group, Physics and Astronomy Department, Gower Street, London WC1E 6BT, UK }

{\sl $^{231}$ University College, National University of Ireland (Dublin), Department of Experimental Physics, Science Buildings, Belfield, Dublin 4, Ireland}

{\sl $^{232}$ University de Barcelona, Facultat de F\'isica, Av. Diagonal, 647, Barcelona 08028, Spain}

{\sl $^{233}$ University of Abertay Dundee, Department of Physics, Bell St, Dundee, DD1 1HG, UK}

{\sl $^{234}$ University of Auckland, Department of Physics, Private Bag, Auckland 1, New Zealand}

{\sl $^{235}$ University of Bergen, Institute of Physics, Allegaten 55, N-5007 Bergen, Norway}

{\sl $^{236}$ University of Birmingham, School of Physics and Astronomy, Particle Physics Group, Edgbaston, Birmingham B15 2TT, UK}

{\sl $^{237}$ University of Bristol, H. H. Wills Physics Lab, Tyndall Ave., Bristol BS8 1TL, UK}

{\sl $^{238}$ University of British Columbia, Department of Physics and Astronomy, 6224 Agricultural Rd., Vancouver, BC V6T 1Z1, Canada}

{\sl $^{239}$ University of California Berkeley, Department of Physics, 366 Le Conte Hall, \#7300, Berkeley, CA 94720, USA}

{\sl $^{240}$ University of California Davis, Department of Physics, One Shields Avenue, Davis, CA 95616-8677, USA}

{\sl $^{241}$ University of California Irvine, Department of Physics and Astronomy, High Energy Group, 4129 Frederick Reines Hall, Irvine, CA 92697-4575 USA}

{\sl $^{242}$ University of California Riverside, Department of Physics, Riverside, CA 92521, USA}

{\sl $^{243}$ University of California Santa Barbara, Department of Physics, Broida Hall, Mail Code 9530, Santa Barbara, CA 93106-9530, USA}

{\sl $^{244}$ University of California Santa Cruz, Department of Astronomy and Astrophysics, 1156 High Street, Santa Cruz, CA 05060, USA}

{\sl $^{245}$ University of California Santa Cruz, Institute for Particle Physics, 1156 High Street, Santa Cruz, CA 95064, USA}

{\sl $^{246}$ University of Cambridge, Cavendish Laboratory, J J Thomson Avenue, Cambridge CB3 0HE, UK}

{\sl $^{247}$ University of Colorado at Boulder, Department of Physics, 390 UCB, University of Colorado, Boulder, CO 80309-0390, USA}

{\sl $^{248}$ University of Delhi, Department of Physics and Astrophysics, Delhi 110007, India}

{\sl $^{249}$ University of Delhi, S.G.T.B. Khalsa College, Delhi 110007, India}

{\sl $^{250}$ University of Dundee, Department of Physics, Nethergate, Dundee, DD1 4HN,  Scotland, UK}

{\sl $^{251}$ University of Edinburgh, School of Physics, James Clerk Maxwell Building, The King's Buildings, Mayfield Road, Edinburgh EH9 3JZ, UK}

{\sl $^{252}$ University of Essex, Department of Physics, Wivenhoe Park, Colchester CO4 3SQ, UK}

{\sl $^{253}$ University of Florida, Department of Physics, Gainesville, FL 32611, USA}

{\sl $^{254}$ University of Glasgow, Department of Physics \& Astronomy, University Avenue, Glasgow G12 8QQ, Scotland, UK}

{\sl $^{255}$ University of Hamburg, Physics Department, Institut f{\"u}r Experimentalphysik, Luruper Chaussee 149, 22761 Hamburg, Germany}

{\sl $^{256}$ University of Hawaii, Department of Physics and Astronomy, HEP, 2505 Correa Rd., WAT 232, Honolulu, HI 96822-2219, USA}

{\sl $^{257}$ University of Heidelberg, Kirchhoff Institute of Physics, Albert {\"U}berle Strasse 3-5, DE-69120 Heidelberg, Germany}

{\sl $^{258}$ University of Helsinki, Department of Physical Sciences, P.O. Box 64 (Vaino Auerin katu 11), FIN-00014, Helsinki, Finland}

{\sl $^{259}$ University of Hyogo, School of Science, Kouto 3-2-1, Kamigori, Ako, Hyogo 678-1297,  Japan}

{\sl $^{260}$ University of Illinois at Urbana-Champaign, Department of Phys., High Energy Physics, 441 Loomis Lab. of Physics1110 W. Green St., Urbana, IL 61801-3080, USA}

{\sl $^{261}$ University of Iowa, Department of Physics and Astronomy, 203 Van Allen Hall, Iowa City, IA 52242-1479, USA}

{\sl $^{262}$ University of Kansas, Department of Physics and Astronomy, Malott Hall, 1251 Wescoe Hall Drive, Room 1082, Lawrence, KS 66045-7582, USA}

{\sl $^{263}$ University of Liverpool, Department of Physics, Oliver Lodge Lab, Oxford St., Liverpool L69 7ZE, UK}

{\sl $^{264}$ University of Louisville, Department of Physics, Louisville, KY 40292, USA}

{\sl $^{265}$ University of Manchester, School of Physics and Astronomy, Schuster Lab, Manchester M13 9PL, UK}

{\sl $^{266}$ University of Maryland, Department of Physics and Astronomy, Physics Building (Bldg. 082), College Park, MD 20742, USA}

{\sl $^{267}$ University of Melbourne, School of Physics, Victoria 3010, Australia}

{\sl $^{268}$ University of Michigan, Department of Physics, 500 E. University Ave., Ann Arbor, MI 48109-1120, USA}

{\sl $^{269}$ University of Minnesota, 148 Tate Laboratory Of Physics, 116 Church St. S.E., Minneapolis, MN 55455, USA}

{\sl $^{270}$ University of Mississippi, Department of Physics and Astronomy, 108 Lewis Hall, PO Box 1848, Oxford, Mississippi 38677-1848, USA}

{\sl $^{271}$ University of Montenegro, Faculty of Sciences and Math., Department of Phys., P.O. Box 211, 81001 Podgorica, Serbia and Montenegro}

{\sl $^{272}$ University of New Mexico, New Mexico Center for Particle Physics, Department of Physics and Astronomy, 800 Yale Boulevard N.E., Albuquerque, NM 87131, USA}

{\sl $^{273}$ University of Notre Dame, Department of Physics, 225 Nieuwland Science Hall, Notre Dame, IN 46556, USA}

{\sl $^{274}$ University of Oklahoma, Department of Physics and Astronomy, Norman, OK 73071, USA}

{\sl $^{275}$ University of Oregon, Department of Physics, 1371 E. 13th Ave., Eugene, OR 97403, USA}

{\sl $^{276}$ University of Oxford, Particle Physics Department, Denys Wilkinson Bldg., Keble Road, Oxford OX1 3RH England, UK }

{\sl $^{277}$ University of Patras, Department of Physics, GR-26100 Patras, Greece}

{\sl $^{278}$ University of Pavia, Department of Nuclear and Theoretical Physics, via Bassi 6, I-27100 Pavia, Italy}

{\sl $^{279}$ University of Pennsylvania, Department of Physics and Astronomy, 209 South 33rd Street, Philadelphia, PA 19104-6396, USA}

{\sl $^{280}$ University of Puerto Rico at Mayaguez, Department of Physics, P.O. Box 9016, Mayaguez, 00681-9016 Puerto Rico}

{\sl $^{281}$ University of Regina, Department of Physics, Regina, Saskatchewan, S4S 0A2 Canada}

{\sl $^{282}$ University of Rochester, Department of Physics and Astronomy, Bausch \& Lomb Hall, P.O. Box 270171, 600 Wilson Boulevard, Rochester, NY 14627-0171 USA}

{\sl $^{283}$ University of Science and Technology of China, Department of Modern Physics (DMP), Jin Zhai Road 96, Hefei, China 230026}

{\sl $^{284}$ University of Silesia, Institute of Physics, Ul. Uniwersytecka 4, PL-40007 Katowice, Poland}

{\sl $^{285}$ University of Southampton, School of Physics and Astronomy, Highfield, Southampton S017 1BJ, England, UK}

{\sl $^{286}$ University of Strathclyde, Physics Department, John Anderson Building, 107 Rottenrow, Glasgow, G4 0NG, Scotland, UK}

{\sl $^{287}$ University of Sydney, Falkiner High Energy Physics Group, School of Physics, A28, Sydney, NSW 2006, Australia}

{\sl $^{288}$ University of Texas, Center for Accelerator Science and Technology, Arlington, TX 76019, USA}

{\sl $^{289}$ University of Tokushima, Institute of Theoretical Physics, Tokushima-shi 770-8502, Japan}

{\sl $^{290}$ University of Tokyo, Department of Physics, 7-3-1 Hongo, Bunkyo District, Tokyo 113-0033, Japan}

{\sl $^{291}$ University of Toronto, Department of Physics, 60 St. George St., Toronto M5S 1A7, Ontario, Canada}

{\sl $^{292}$ University of Tsukuba, Institute of Physics, 1-1-1 Ten'nodai, Tsukuba, Ibaraki 305-8571, Japan}

{\sl $^{293}$ University of Victoria, Department of Physics and Astronomy, P.O.Box 3055 Stn Csc, Victoria, BC V8W 3P6, Canada}

{\sl $^{294}$ University of Warsaw, Institute of Physics, Ul. Hoza 69, PL-00 681 Warsaw, Poland}

{\sl $^{295}$ University of Warsaw, Institute of Theoretical Physics, Ul. Hoza 69, PL-00 681 Warsaw, Poland}

{\sl $^{296}$ University of Washington, Department of Physics, PO Box 351560, Seattle, WA 98195-1560, USA}

{\sl $^{297}$ University of Wisconsin, Physics Department, Madison, WI 53706-1390, USA}

{\sl $^{298}$ University of Wuppertal, Gau{\ss}stra{\ss}e 20, D-42119 Wuppertal, Germany}

{\sl $^{299}$ Universit\'e Claude Bernard Lyon-I, Institut de Physique Nucl\'eaire de Lyon (IPNL), 4, rue Enrico Fermi, F-69622 Villeurbanne Cedex, France}

{\sl $^{300}$ Universit\'e de Gen\`eve, Section de Physique, 24, quai E. Ansermet, 1211 Gen\`eve 4, Switzerland}

{\sl $^{301}$ Universit\'e Louis Pasteur (Strasbourg I), UFR de Sciences Physiques, 3-5 Rue de l'Universit\'e, F-67084 Strasbourg Cedex, France}

{\sl $^{302}$ Universit\'e Pierre et Marie Curie (Paris VI-VII) (6-7) (UPMC), Laboratoire de Physique Nucl\'eaire et de Hautes Energies (LPNHE), 4 place Jussieu, Tour 33, Rez de chausse, 75252 Paris Cedex 05, France}

{\sl $^{303}$ Universit{\"a}t Bonn, Physikalisches Institut, Nu{\ss}allee 12, 53115 Bonn, Germany}

{\sl $^{304}$ Universit{\"a}t Karlsruhe, Institut f{\"u}r Physik, Postfach 6980, Kaiserstrasse 12, D-76128 Karlsruhe, Germany}

{\sl $^{305}$ Universit{\"a}t Rostock, Fachbereich Physik, Universit{\"a}tsplatz 3, D-18051 Rostock, Germany}

{\sl $^{306}$ Universit{\"a}t Siegen, Fachbereich f{\"u}r Physik, Emmy Noether Campus, Walter-Flex-Str.3, D-57068 Siegen, Germany}

{\sl $^{307}$ Universit{\`a} de Bergamo, Dipartimento di Fisica, via Salvecchio, 19, I-24100 Bergamo, Italy}

{\sl $^{308}$ Universit{\`a} degli Studi di Roma La Sapienza, Dipartimento di Fisica, Istituto Nazionale di Fisica Nucleare, Piazzale Aldo Moro 2, I-00185 Rome, Italy}

{\sl $^{309}$ Universit{\`a} degli Studi di Trieste, Dipartimento di Fisica, via A. Valerio 2, I-34127 Trieste, Italy}

{\sl $^{310}$ Universit{\`a} degli Studi di ``Roma Tre'', Dipartimento di Fisica ``Edoardo Amaldi'', Istituto Nazionale di Fisica Nucleare, Via della Vasca Navale 84, 00146 Roma, Italy}

{\sl $^{311}$ Universit{\`a} dell'Insubria in Como, Dipartimento di Scienze CC.FF.MM., via Vallegio 11, I-22100 Como, Italy}

{\sl $^{312}$ Universit{\`a} di Pisa, Departimento di Fisica 'Enrico Fermi', Largo Bruno Pontecorvo 3, I-56127 Pisa, Italy}

{\sl $^{313}$ Universit{\`a} di Salento, Dipartimento di Fisica, via Arnesano, C.P. 193, I-73100 Lecce, Italy}

{\sl $^{314}$ Universit{\`a} di Udine, Dipartimento di Fisica, via delle Scienze, 208, I-33100 Udine, Italy}

{\sl $^{315}$ Variable Energy Cyclotron Centre, 1/AF, Bidhan Nagar, Kolkata 700064, India}

{\sl $^{316}$ VINCA Institute of Nuclear Sciences, Laboratory of Physics, PO Box 522, YU-11001 Belgrade, Serbia and Montenegro}

{\sl $^{317}$ Vinh University, 182 Le Duan, Vinh City, Nghe An Province, Vietnam}

{\sl $^{318}$ Virginia Polytechnic Institute and State University, Physics Department, Blacksburg, VA 2406, USA}

{\sl $^{319}$ Visva-Bharati University, Department of Physics, Santiniketan 731235, India}

{\sl $^{320}$ Waseda University, Advanced Research Institute for Science and Engineering, Shinjuku, Tokyo 169-8555, Japan}

{\sl $^{321}$ Wayne State University, Department of Physics, Detroit, MI 48202, USA}

{\sl $^{322}$ Weizmann Institute of Science, Department of Particle Physics, P.O. Box 26, Rehovot 76100, Israel}

{\sl $^{323}$ Yale University, Department of Physics, New Haven, CT 06520, USA}

{\sl $^{324}$ Yonsei University, Department of Physics, 134 Sinchon-dong, Sudaemoon-gu, Seoul 120-749, Korea}

{\sl $^{325}$ Zhejiang University, College of Science, Department of Physics, Hangzhou, China 310027}

{\sl * deceased } 

\end{center}

\end{center}

\chapter*{Acknowledgements} 
We would like to acknowledge the support and guidance of the International Committee on Future Accelerators (ICFA), chaired by A. Wagner of DESY, and the International Linear Collider Steering Committee (ILCSC), chaired by S. Kurokawa of KEK, who established the ILC Global Design Effort, as well as the World Wide Study of the Physics and Detectors.
      
\medskip
We are grateful to the ILC Machine Advisory Committee (MAC), chaired by F. Willeke of DESY and the International ILC Cost Review Committee, chaired by L. Evans of CERN, for their advice on the ILC Reference Design. We also thank the consultants who particpated in the Conventional Facilities Review at CalTech and in the RDR Cost Review at SLAC. 

\medskip 
We would like to thank the directors of the institutions who have hosted ILC meetings: KEK, ANL/FNAL/SLAC/U. Colorado (Snowmass), INFN/Frascati, IIT/Bangalore, TRIUMF/U. British Columbia, U. Valencia, IHEP/Beijing and DESY.

\medskip 
We are grateful for the support of the Funding Agencies for Large Colliders (FALC), chaired by R. Petronzio of INFN, and we thank all of the international, regional and national funding agencies whose generous support has made the ILC 
Reference Design possible.

\medskip 
Each of the GDE regional teams in the Americas, Asia and Europe are grateful for the support of their local scientific societies, industrial forums, advisory committees and reviewers.

\cleardoublepage

\tableofcontents %

\setcounter{secnumdepth}{0}





\mainmatter
\setcounter{secnumdepth}{4}
\setcounter{page}{1} \setcounter{chapter}{0}



\chapter{Introduction}


\section{Questions about the universe}

\begin{itemize}

\item {\it What is the universe? How did it begin?}
\vspace*{-2mm}

\item{\it What are matter and energy? What are space and time?} 
\vspace*{-2mm}
\end{itemize}

Throughout human history, scientific theories and experiments of
increasing power and sophistication have addressed these basic
questions about the universe. The resulting knowledge has revolutionized
our view of the world around us, transforming our society and advancing
our civilization.

Everyday phenomena are governed by universal laws and principles whose
natural realm is at scales of time and distance far removed from our
direct experience. Particle physics is a primary avenue of inquiry into
these most basic workings of the universe. Experiments using particle
accelerators convert matter into energy and back to matter again,
exploiting the insights summarized by the equation $E\!=\!mc^2$. Other
experiments exploit naturally occurring particles, such as neutrinos
from the Sun or cosmic rays striking Earth's atmosphere. Many experiments
use exquisitely sensitive detectors to search for rare phenomena or
exotic particles. Physicists combine astrophysical observations
with results from laboratory experiments, pushing towards a great
intellectual synthesis of the laws of the large with laws of the small.

The triumph of 20th century particle physics was the development  of the
Standard Model and the confirmation of many of its aspects. Experiments
determined the particle constituents of ordinary matter, and identified four
forces that hold matter together and transform it from one form to another.
Particle interactions were found to obey precise laws of relativity and quantum
theory. Remarkable features of quantum physics were observed, including the real
effects of ``virtual'' particles on the visible world.

Building on this success, particle physicists are now able to address
questions that are even more fundamental, and explore some of the
deepest mysteries in science. The scope of these questions is
illustrated by this summary from the 
report {\it Quantum Universe}~\cite{Albrecht:2005np}:
\begin{enumerate}
\item{\it Are there undiscovered principles of nature?}
\vspace*{-2mm}
\item{\it How can we solve the mystery of dark energy?}
\vspace*{-2mm}
\item{\it Are there extra dimensions of space?}
\vspace*{-2mm}
\item{\it Do all the forces become one?}
\vspace*{-2mm}
\item{\it Why are there so many particles?}
\vspace*{-2mm}
\item{\it What is dark matter? How can we make it in the laboratory?}
\vspace*{-2mm}
\item{\it What are neutrinos telling us?}
\vspace*{-2mm}
\item{\it How did the universe begin?}
\vspace*{-2mm}
\item{\it What happened to the antimatter?}
\vspace*{-2mm}
\end{enumerate} 

A worldwide program of particle physics investigations, using multiple
approaches, is already underway to explore this compelling scientific landscape.
As emphasized in many  scientific studies 
\cite{EPP2010:2006,CERNSG:2006,:2003mg,OECD:2002,JLC:2002,Aguilar-Saavedra:2001rg,Abe:2001gc,Abe:2001nq,ECFA-DESY},
the International Linear Collider is expected to play a central role in what is
likely to be an era of revolutionary advances. As already documented in
\cite{Bagger:2006dqu}, discoveries from the ILC could have breakthrough impact
on many of these fundamental questions.

Many of the scientific opportunities for the ILC involve the Higgs particle and
related new phenomena at Terascale energies. The Standard Model boldly
hypothesizes a new form of  Terascale energy, called the Higgs field, that
permeates the entire universe. Elementary particles acquire mass by interacting
with this field. The Higgs field also breaks a fundamental electroweak force
into two forces, the electromagnetic and weak forces, which are observed by
experiments in very different forms.

So far, there is no direct experimental evidence for a Higgs field
or the Higgs particle that should accompany it. Furthermore, quantum
effects of the type already observed in experiments should destabilize
the Higgs boson of the Standard Model, preventing its operation at Terascale
energies. The proposed antidotes for this quantum instability mostly
involve dramatic phenomena at the Terascale: new forces, a new principle
of nature called supersymmetry, or even extra dimensions of space.

Thus for particle physicists the Higgs boson is at the center of a
much broader program of discovery, taking off from a long list
of questions.
Is there really a Higgs boson? If not, what are the mechanisms that give
mass to particles and break the electroweak force? If there is a Higgs boson,
does it differ from the hypothetical Higgs of the Standard Model?
Is there more than one Higgs particle?  
What are the new phenomena that stabilize the Higgs boson at the Terascale?
What properties of Higgs boson inform us about these new phenomena?

Another major opportunity for the ILC is to shed light on the dark
side of the universe. Astrophysical data shows that dark matter
dominates over visible matter, and that almost all of this dark matter
cannot be composed of known particles. This data, combined with the
concordance model of
Big Bang cosmology, suggests that dark matter is comprised of new
particles that interact weakly with ordinary matter and have
Terascale masses. It is truely remarkable that astrophysics and
cosmology, completely independently of the particle physics
considerations reviewed above, point to new phenomena at the
Terascale.

If Terascale dark matter exists, experiments at the ILC should
be able to produce such particles in the laboratory and study
their properties. Another list of  questions will then
beckon. Do these new particles really have the correct properties
to be the dark matter? Do they account for all of the dark matter,
or only part of it? What do their properties tell us about the
evolution of the universe? How is dark matter connected to
new principles or forces of nature? 

A third cluster of scientific opportunities for the ILC focus 
on Einstein's vision of an ultimate unified theory.
Particle physics data already suggests that three of the
fundamental forces originated from a single ``grand'' unified
force in the first instant of the Big Bang. Experiments at the
ILC could test this idea and look for evidence of a related unified
origin of matter involving supersymmetry.
A theoretical framework called string theory
goes beyond grand unification to include gravity, extra spatial
dimensions, and new fundamental entities called superstrings.
Theoretical models to explain the properties of neutrinos,
and account for the mysterious dominance of matter over antimatter,
also posit unification at high energies.
While the realm of unification is almost certainly beyond the
direct reach of experiments, different unification models
predict different patterns of new phenomena at Terascale energies.
ILC experiments could distinguish among these patterns,
effectively providing a telescopic view of ultimate
unification. Combined with future data from astrophysics,
this view should also give insights about our cosmic origins.


\section{The new landscape of particle physics}

During the next few years, experiments at CERN's Large Hadron Collider
will have the first
direct look at Terascale physics. Like the discovery of an uncharted
continent, this exploration of the Terascale will transform forever the
geography of our universe. Equally compelling will be the interplay of
LHC discoveries with other experiments and observations, including
those that can probe the fundamental nature of dark matter, neutrinos and
sources of matter--antimatter asymmetry. Some aspects of the new phenomena
may fit well with existing speculative theoretical frameworks, suggesting
a radical rewriting of the laws of nature. Other aspects may be initially
ambiguous or mystifying, with data raising more questions than it answers.
Particle physics should be entering a new era of intellectual
ferment and revolutionary advance, unparalleled in the past half-century.

No one knows what will be found at the LHC, but the discovery potential
of the LHC experiments is well studied \cite{atlastdr,CMSTDR}.
If there is a Higgs
boson, it is almost certain to be found by the ATLAS and CMS experiments.
Its mass should be measured with an accuracy between 0.1 and
1\%, and at least one of its decay modes should be observed. If the Higgs
particle decays into more than one type of particle, the LHC
experiments should measure the ratio of the Higgs couplings to
those different particles, with an accuracy between about 7 and 30\%.
If there is more than one type of Higgs boson, ATLAS and CMS will have
a reasonably good chance of seeing both the lighter and heavier Higgs
bosons.
In favorable cases, these experiments will have some ability to
discriminate the spin and CP properties of the Higgs particle.

Thus for LHC there are three possible outcomes with respect to the Higgs
particle.
The first is that a Higgs boson has been found, and at first look its
properties seem consistent with the Standard Model. Then the compelling
issue will be whether a more complete and precise experimental analysis
reveals nonstandard properties. This will be especially compelling if
other new phenomena, possibly related to the Higgs sector, have also been discovered.
The second possible outcome is that a Higgs boson is found with gross
features at variance with the Standard Model. This variation could be
something as simple as a Higgs mass of 200 GeV or more, which would
conflict with existing precision data without other new phenomena to
compensate for it. The variation could also come from a large deviation
in the predicted pattern of Higgs decay or the discovery of multiple Higgs
particles. The third possible outcome is that no Higgs boson is
discovered. In this case particle physicists will need either a radical
rethink of the origin of mass, or new experimental tools to uncover a
``hidden'' or ``invisible'' Higgs boson.

For all of these possible outcomes, the ILC will be essential
to move forward on our understanding of the Higgs mechanism and of its relation to
other new fundamental phenomena. This claim is
documented in many detailed studies which are reviewed in this report.

LHC experiments have impressive capabilities to discover new heavy
particles, especially particles which are strongly produced in
proton-proton collisions, or particles seen as resonances in the
production of pairs of fermions or gauge bosons. ATLAS and
CMS could detect a new $Z'$ gauge boson as heavy
as 5 TeV \cite{Cousins},
and the squarks and gluinos of supersymmetry even if they are as
heavy as 2.5 TeV \cite{atlastdr}.
New particles associated with the existence of extra
spatial dimensions could be seen, if  their masses are less than a
few TeV \cite{atlastdr,CMSTDR}.

The discovery of a $Z'$ particle would indicate a new fundamental
force of nature. LHC measurements may discriminate somewhat between
possible origins of the new force, but this potential is limited to
$Z'$ particles lighter than 2.5 TeV in the most optimistic
scenarios, and 1 TeV in others \cite{Cousins}. Through precision
measurements of how the $Z'$ interacts with other particles, the ILC
could determine the properties of this new force, its origins, its
relation to the other forces in a unified framework, and its role in
the earliest moments of the Big Bang.

If supersymmetry is responsible for the existence of the Terascale and
a light Higgs boson, then signals of superpartner particles
should be seen at LHC. Since supersymmetry is an organizing principle
of nature (like relativity), it can be realized in an infinite variety
of ways. Thus a supersymmetry signal will raise two urgent issues.
The first is whether the new heavy particles seen at LHC are
actually superpartners, with the spins and couplings to other particles
predicted by supersymmetry. Some results bearing on this may be
available from LHC, but only ILC can provide an unequivocal answer.
The second issue involves a set of fundamental questions: How does
supersymmetry manifest itself in nature? What mechanism makes it appear
as a ``broken'' symmetry? Is supersymmetry related to unification at
a higher energy scale? How is supersymmetry related to the Higgs mechanism? What
role did supersymmetry play in our cosmic origins?
Definitive answers to these questions will require precise measurements
of the entire roster of superpartner particles as well as the Higgs
particles. To achieve this, physicists will need to extract the best
possible results from the LHC and the ILC in a combined
analysis \cite{Weiglein:2004hn}, supplemented
by signals or constraints from future B physics experiments and other
precision measurements.

Supersymmetry is a good example to illustrate the possibility of an
exciting interplay between different experiments and observations.
Missing energy signatures at the LHC may indicate a weakly
interacting massive particle consistent with the lightest neutralino
of supersymmetry. At the same time, next generation direct or indirect
dark matter searches may see a signal for weakly interacting exotic
particles in our galactic halo. Are these particles neutralinos? If so,
are neutralinos responsible for all of the dark matter, or only part
of it? Does the model for supersymmetry preferred by collider data
predict the observed abundance of dark matter, or do cosmologists need
to change their assumptions about the early history of the universe?
For all of these questions, detailed studies show the central importance
of ILC measurements.

Other new physics models which might be observed at the next
generation colliders could involve extra spatial dimensions or new
strong forces. These are exciting possibilities that can also lead
to confusion, calling for ILC to reveal their true nature. In some
scenarios the new phenomena are effectively hidden from the LHC
detectors, but are revealed as small deviations in couplings
measured at the ILC. In favorable cases the LHC experiments could
uncover strong evidence for the existence of extra dimensions. In
this event the ILC will be essential to explore the size, shape,
origins and impact of this expanded universe.

\section{Running scenarios}
The basic parameters needed for the planned physics program are
detailed in Ref.~\cite{scope} and confirmed by the machine design.
The maximal center of mass energy is designed to be $\rts = 500 \GeV$, with
a possible upgrade to $1 \TeV$, 
where physics runs must be possible for every energy above $\rts = 200
\GeV$ and some luminosity for calibration runs is needed at
$\rts = 91 \GeV$. For mass measurements threshold scans are required
so that it must be possible to change the beam energy fast in small
steps.

The total luminosity is required to be around $500 \fbi$ within the
first four years and about $1000\fbi$ during the first phase of
operation. For the electron beam, polarization with a degree of larger
than $\pm 80\%$ is mandatory. For the positron beam, a polarization of
more than $\pm 50\%$ is useful \cite{gudi} which should be relatively easy to
achieve with the undulator positron source in the present ILC design.
To reduce systematic uncertainties, the polarization direction has to
be switchable on the train by train basis.  Beam energy and
polarization have to be stable and measurable at a level of about
0.1\%.

Contrary to a hadron machine, an $e^+e^-$ collider produces at a given
time events 
at one fixed center of mass energy $\rts$ and, if polarization
should be exploited in the analyses, fixed polarization.
A physics study has
to assume a certain value for the integrated luminosity and polarization mix
which may be in conflict with other studies. To check whether
this feature does not prevent the ILC from doing the many precision
measurements claimed in the individual analyses, in a toy study a
scenario with many new particles has been performed
\cite{grannis_scenario}. This study assumes supersymmetry with all
sleptons, the lightest chargino and the lightest two neutralinos in
the ILC energy range. In addition, the top quark and a light 
Higgs boson are visible. A first run is done at $\rts\!=\!500\GeV$ to get a
first measurement of the particle masses to optimize the threshold
scans. The rest of the time is spent with these scans for precision
measurements. Those analyses that do not require a given beam energy
apart from being above production thresholds are done during the scans.
This applies especially to the precision Higgs measurements.
It has been shown that in such a scenario, a precision close to the one
claimed in the isolated studies can be reached for all relevant
observables.

A representative set of physics scenarios has been studied and in all
cases it has been found that a $\rts\!=\!500\GeV$ collider adds enough to
our physics knowledge to justify the project. However, in all cases, an
upgrade to $\rts\!\sim\! 1 \TeV$ increases significantly the value of the
ILC. In the following chapters, also the case for an upgrade to
$\rts=1\TeV$ after the first phase of ILC running will be presented.


In addition to the standard $e^+e^-$ running at $\rts > 200\GeV$, the ILC offers
some options that can be realized with reasonable modifications if required by
physics.

In the GigaZ mode, the ILC can run with high luminosity and both beams
polarized on the $Z$--boson resonance, producing $10^9$ hadronic $Z$ decays in
less than a year or at the $W$--boson pair production threshold to measure the
$W$ boson mass with high precision \cite{gigaz}. 
This requires only minor modifications to the machine.

With relatively few modifications, both arms can accelerate electrons
resulting in an $e^- e^-$ collider \cite{Heusch:2005hb}.
This mode can especially be useful to measure the selectron mass if it
exists in the ILC energy range.

If the electrons are collided with a very intense laser beam about 1\,mm in
front of the interaction point, a high energy photon beam can be produced with
a similar beam quality as the undisturbed electron beam. Converting only one
or both beams this results in an $e \gamma$ or $\gamma \gamma$ collider
\cite{Ginzburg:1981ik,ggtdr}.  This mode requires a larger crossing angle than
$e^+e^-$ and the installation of a large laser system \cite{Bechtel:2006mr}. The
feasibility of such a laser system has not yet been proven.

In the following, it will be assumed that all options are technically
possible and they will be implemented when they are required by the
ILC and LHC data.

To exploit fully the physics program of ILC will take a long time of possibly
around 20 to 30 years. Possible options will certainly be realized only towards
the end of the program.

\section{Physics and the detectors}

Detectors at the ILC face a very different set of challenges
compared to the current state-of-the-art employed for LEP/SLD and
hadron colliders~\cite{Dawson:2004xz}. While ILC detectors will
enjoy lower rates, less background and lower radiation doses than
those at the LHC, the ILC will be pursuing physics that places
challenging demands on precision measurements and particle tracking
and identification. The reasons for this can be illustrated by
several important physics processes, namely measuring the properties 
of a Higgs boson, identifying strong electroweak  symmetry
breaking, identifying supersymmetric (SUSY) particles and their
properties, and precision electroweak studies. These are just a few examples
taken from benchmark studies for ILC
detectors~\cite{Battaglia:2006bv}.

\begin{figure}[htbp]
\centerline{ \epsfig{file=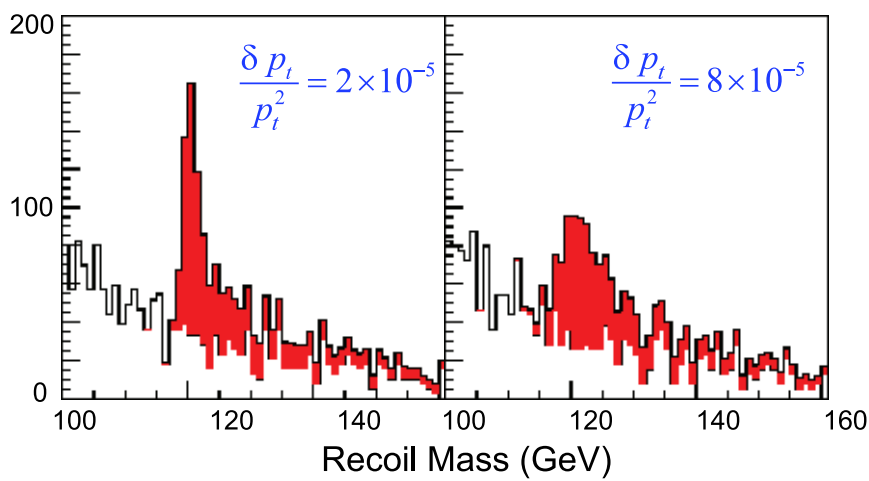,width=4.5in}} 
\vspace*{-4mm}
\caption[Tracking resolution for a Higgs recoiling against dimuons at a 
500 GeV ILC] 
{Histogram of mass recoiling from dimuons at $\sqrt s = 500~\GeV$ for a
Higgs boson mass of 120~\GeV, for two values of the tracking
resolution; from Ref.~\cite{Hewett:2005ec}.} \label{fig:recoil}
\vspace*{-1mm}
\end{figure}

The Higgs boson(s) of the Standard Model (SM), minimal
supersymmetric extension of the SM (MSSM), or extended models will require precision
measurements of their mass and couplings in order to identify the
theory \cite{Strong-review}. The golden measurement channel of Higgs 
production is $\ee\!\rightarrow\! Z H\! \rightarrow\! \lept X$,
with the Higgs mass measured by its recoil from the $Z$ boson. The mass must
be measured to a precision sufficient to cleanly separate the
resonance from backgrounds -- a precision of approximately 50~\MeV\
is usually sufficient. This will require a resolution 
$\delta (1/p)$ better than
$7\times10^{-5}$ GeV$^{-1}$ for a low mass Higgs boson, and that requires
tracking performance an order of magnitude better than that achieved
by LEP/SLD detectors. The need for this performance is illustrated
in Figure~\ref{fig:recoil}, which shows the impact of tracker
resolution on the significance of signal compared to expected
backgrounds. The Higgs mass measurement also requires precise
knowledge of the center of mass energy, and this requires precision
measurement of the luminosity--weighted energy spectrum in order to
measure the beamstrahlung energy loss (more information on this subject
can be found in the top quark chapter).

\begin{figure}[htbp]
\begin{minipage}{7cm}
\vspace*{1cm}
\hspace*{-.8cm}
\epsfig{file=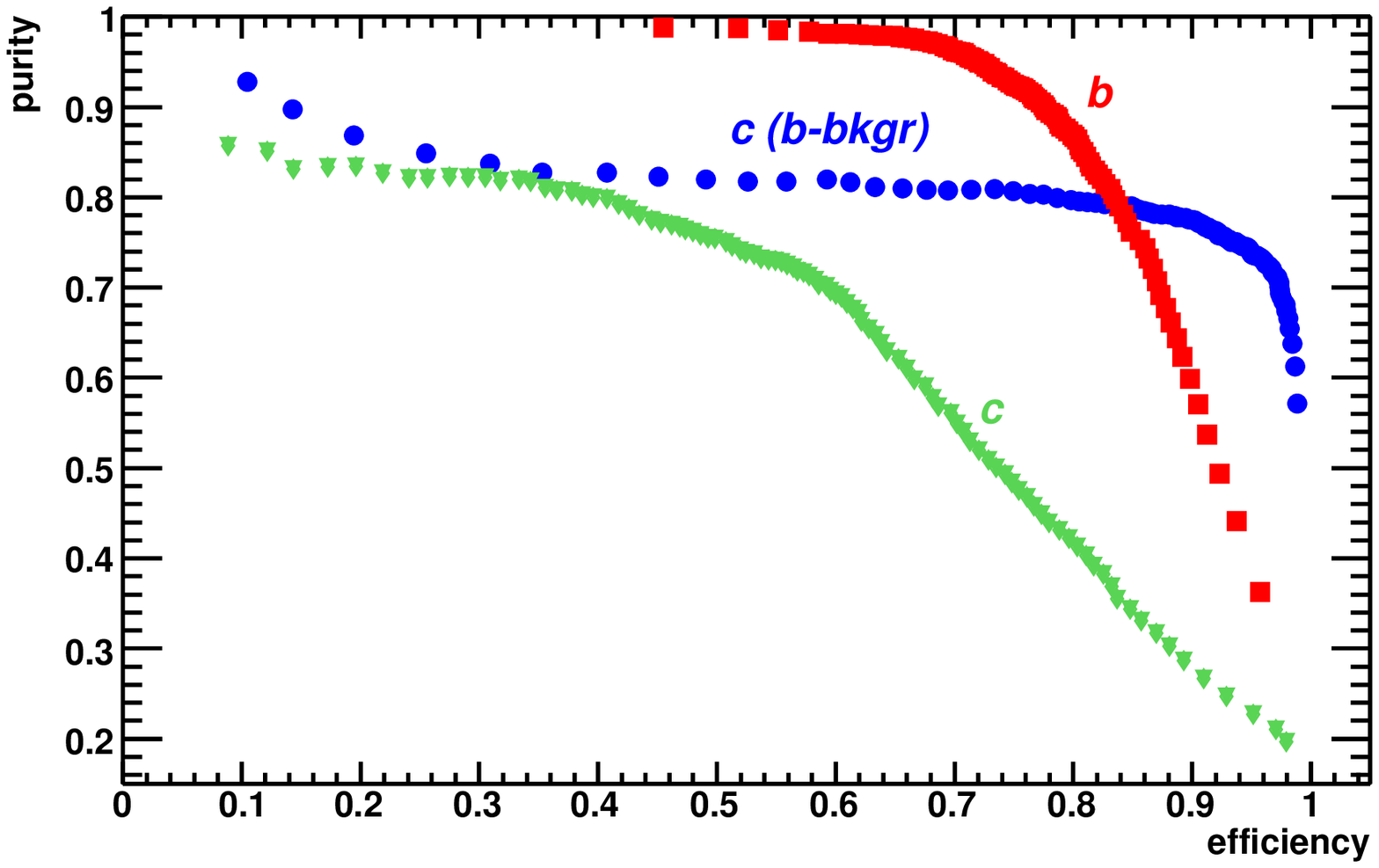,width=9cm,,height=7cm} \hspace*{1cm}
\end{minipage}
\hspace*{1cm}
\begin{minipage}{7cm}
\epsfig{file=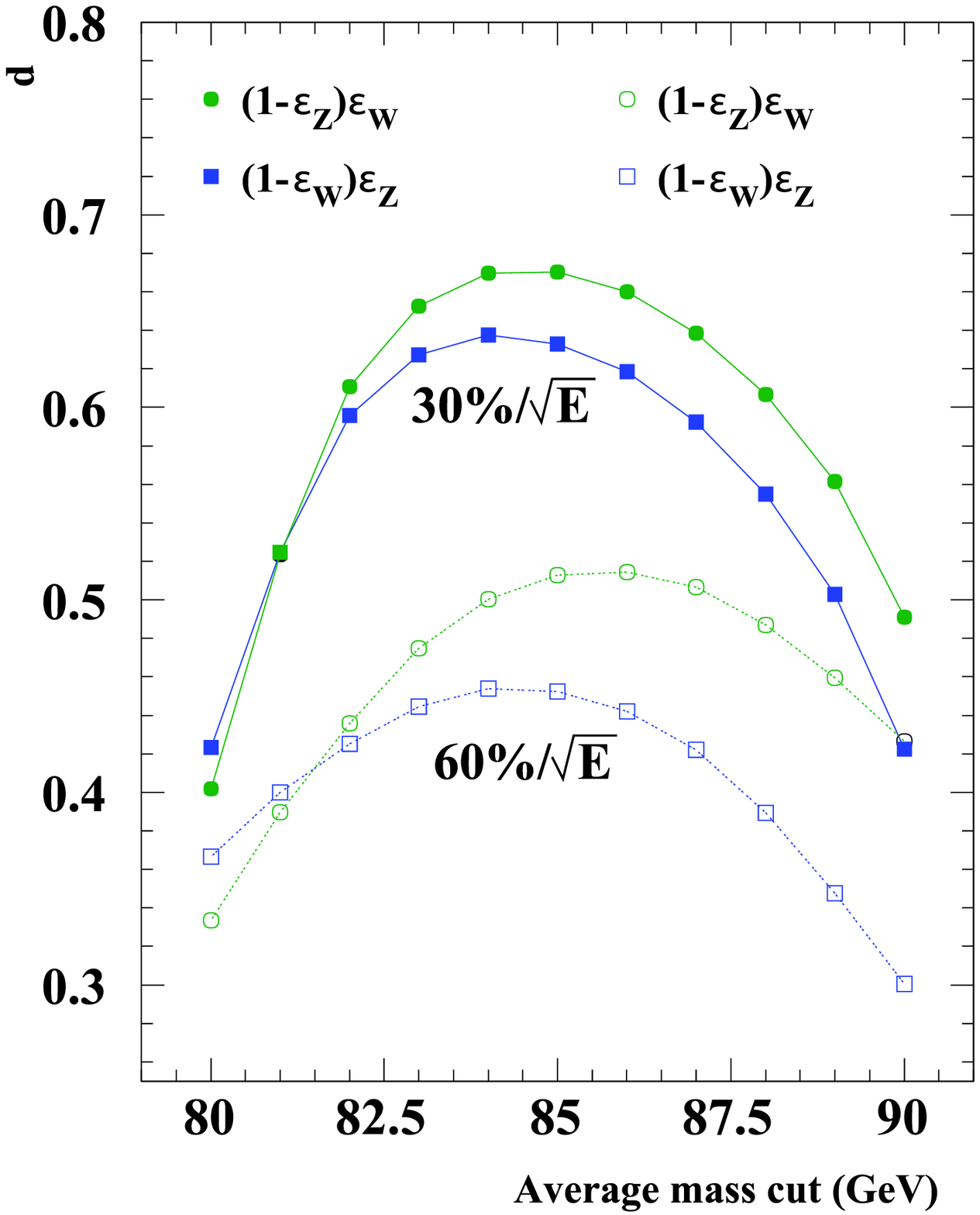,width=7cm}
\end{minipage} 
\vspace*{-4mm}
\caption[Simulated purity/efficiency for b,c tagging and purity for WW/ZZ separation] 
{Left: purity v.s. efficiency for tagging of $b$ and $c$ jets in a simulated VTX
detector described in the text; the points labeled ``$c$ ($b$ bkgr)'' indicate 
the case where only $b$--quark backgrounds are present in the $c$--study; from
Ref.~\cite{Hillert2006}. Right: purity factor $d$ (for ``dilution'')  for  the
process $\ee \rightarrow \nu\bar{\nu} WW/$ $\ee ZZ$ as a function of invariant 
mass cut for two values of the energy resolution; from 
Ref.~\cite{Behnke:2001qq}.
}
\label{fig:btag}
\vspace*{-4mm}
\end{figure}

Because of the important role played by heavy $t,b,c$ quarks  and the
tau lepton in the SM and essentially all new physics models, the ILC
detectors will require excellent vertex detection in a challenging
high rate environment of low energy $e^{+}e^{-}$ pairs. An even
stronger requirement on the vertex detector is imposed by the desire
to measure vertex charge with good efficiency. This is useful for
reducing large combinatoric jet backgrounds and to distinguish $b$
from $\bar b$ for measurement of forward--backward
asymmetries, which are very sensitive to new physics, or for
establishing CP violation. To make the requisite improvements over
the LEP/SLD detectors, the impact parameters will have to be
measured to (5 $\oplus$ 10/p) $\mu$m (momentum $p$ in \GeV), and this
will require putting finely-segmented ($20 \times 20~\mu {\rm m}^{2}$)
silicon arrays within 1.5~cm of the beamline. Figure~\ref{fig:btag} (left)
shows the purity/efficiency obtained with a 5--layer vertex detector
with inner radius 1.5cm, ladder thickness 0.1\% $X_0$ and resolution
$3.5~\mu$m; this study uses a ``fast'' version of the simulation program.

\begin{figure}[htbp]
\centerline{ \epsfig{file=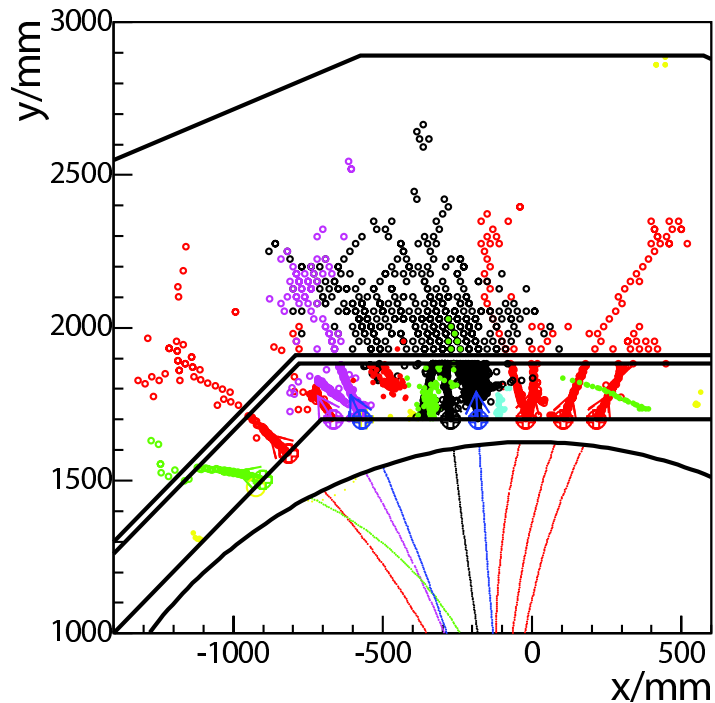,width=4in}} 
\vspace*{-7mm}
\caption[Simulation of a 100 GeV jet using MOKKA for the TESLA TDR detector]
{Simulation of a 100~\GeV\ jet using the MOKKA simulation of the
TESLA TDR detector; colors show tracks-cluster associations using
PandoraPFA; from Ref.~\cite{Thomson:2006dc}.} 
\label{fig:thomson}
\vspace*{-2mm}
\end{figure}

Excellent resolution on jet energy, which is essential for the unambiguous
identification of many decay channels, enhances the impact of
precision measurements, and lowers the integrated luminosity needed
for many measurements. Figure~\ref{fig:btag} (right) demonstrates the
luminosity dependence on jet energy resolution. Distinguishing $WW$
from $ZZ$ production at ILC energies is challenging, but essential for
matching branching fractions to a model, such as identifying strong
electroweak symmetry breaking or supersymmetric  parameters.  The
low ILC backgrounds permit association of tracks and calorimeter
clusters, making possible unprecedented jet energy measurement.
However, to achieve $WW/ZZ$ separation the detectors must measure jet
energy about a factor of two better than the best achieved so far.
The jet energy resolution must be roughly
5~\GeV\ , corresponding to an energy resolution of
30\%/$\sqrt(E_{\rm jet})$ for the 100--150~\GeV\ jets common at higher
center of mass energies. Depending on the quark content, jets of
these energies deposit roughly 65\% of the visible energy in the
form of charged particles, 25\% in the form of photons, and 10\% as
neutral hadrons. In the relatively clean environment of ILC, the
required energy resolution translates into a factor 2 improvement in
hadron calorimeter performance over those currently operating. To
meet such a goal, the method of "particle flow" association of
tracks and calorimeter clusters must be validated.
Figure~\ref{fig:thomson} shows the "particle flow" for a jet in an
ILC detector.

If low energy supersymmetry is indeed realized, one of the more important tasks for the ILC will be
to identify SUSY particle spectra and decay chains, and to establish
if SUSY particles could be some or all of the dark matter. Since the
lightest SUSY particle will not be observable, the detectors must be
extremely hermetic, particularly at extreme polar angles. To achieve
these goals the effect of beam crossing angle, beamstrahlung and
machine backgrounds must be well understood, and development of instrumentation is necessary to measure
the luminosity spectrum and beam polarization.

%
\chapter{Higgs physics}
\label{sec:higgs}
\newcommand{\eei}{e^+e^-}

The search and the study of Higgs bosons is one of the main missions of present
and future high--energy colliders. The observation of these particles is of
major importance for the present understanding of the interactions of the
fundamental particles and the generation of their masses.  In the Standard Model
(SM), the existence of one isodoublet scalar field is required, the neutral
component of which acquires a non--zero vacuum expectation value leading to the
spontaneous breaking of the electroweak symmetry and the generation of the 
gauge boson and fermion masses.  In this picture, one degree of freedom among
the four degrees of freedom of the original isodoublet field is left over,
corresponding to a physical scalar particle, the Higgs boson \cite{Higgs}. The
discovery of this new type of matter particle is  considered as being of
profound importance.  In fact, despite of its numerous successes in explaining
the present data, the SM  is not complete before this particle is
experimentally observed and its fundamental properties studied in detail.
Furthermore, even if we understand that the Higgs field is the source of
particle masses, the origin of electroweak symmetry breaking  itself needs to be
explained and its dynamics to be clarified.  Very little is known about this
symmetry breaking  and important questions include: does the dynamics involve
new strong interactions and/or sizable CP violation, and, if elementary Higgs
particles indeed exist in nature, how many fields are there and in which gauge
representations do they appear. Theoretical realizations span a wide range of
scenarios extending from weak to strong breaking mechanisms.  Examples, on one
side, are  models involving light fundamental Higgs fields, such as the SM
and its supersymmetric extensions which include two--Higgs doublets in the
minimal version and additional singlet fields or higher representations in
extended versions; on the other  side, there are new strong interaction and
extra--dimensional models without a fundamental Higgs field.  Furthermore, the
electroweak symmetry breaking  mechanism might be related to other fundamental
questions of particle physics and cosmology. For instance, the Higgs sector
could play an important role in the annihilation of the new particles that are
responsible of the cosmological dark matter and might shed light  on how the
baryon--antibaryon asymmetry proceeded in the early universe. It might also
explain  how and why the three generations of quarks and leptons are different.

Only detailed investigation of the properties of the Higgs particles will answer
these questions. The ILC is a unique tool in this context and it could play an
extremely important role: high--precision measurements would allow to determine
with a high level of confidence the profile of the Higgs bosons and their 
fundamental properties and would provide a unique opportunity  to establish
experimentally the mechanism that generates the particle masses.

\section{The Higgs sector of the SM and beyond}

\subsection{The Higgs boson in the SM}

The Standard Model makes use of one isodoublet complex scalar field and, 
after spontaneous electroweak symmetry breaking (EWSB), three would--be
Goldstone bosons among the four degrees of freedom are absorbed to build up the
longitudinal components of the $W^\pm,Z$ gauge bosons and generate their masses;
the fermion masses are generated through a Yukawa interaction with the same
scalar field. The remaining degree of freedom corresponds to the unique  Higgs
particle of the model with the $J^{\rm PC}=0^{++}$ assignment of spin, parity
and charge conjugation quantum numbers \cite{Higgs,HHG,Djouadi:2005gi}.  Since
the Higgs couplings to fermions and gauge bosons are related to the  masses of
these particles and the only free parameter of the model is the mass of the
Higgs boson itself;  there are, however, both experimental and theoretical
constraints on this fundamental parameter, as will be summarized below.

\begin{figure}[!h]
\vspace*{-2.2cm}
\begin{center}
\begin{minipage}{6.5cm}
\hspace*{-6mm}
\includegraphics[width=1.1\linewidth]
                {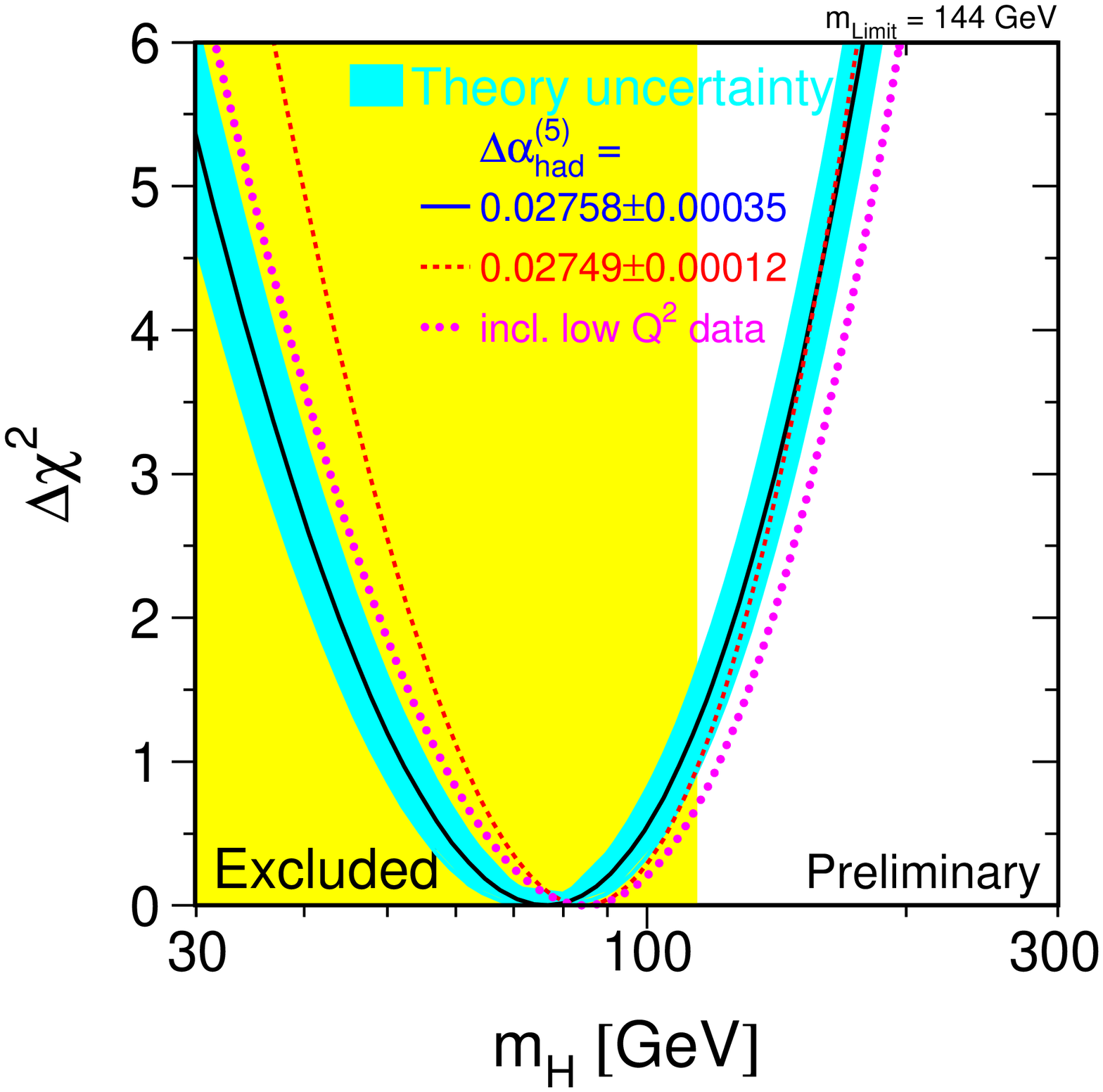}\hspace*{6mm}
\end{minipage}
\hspace*{.5cm}
\begin{minipage}{7cm}
\vspace*{-2.cm}		
\includegraphics[width=1.4\linewidth]
                {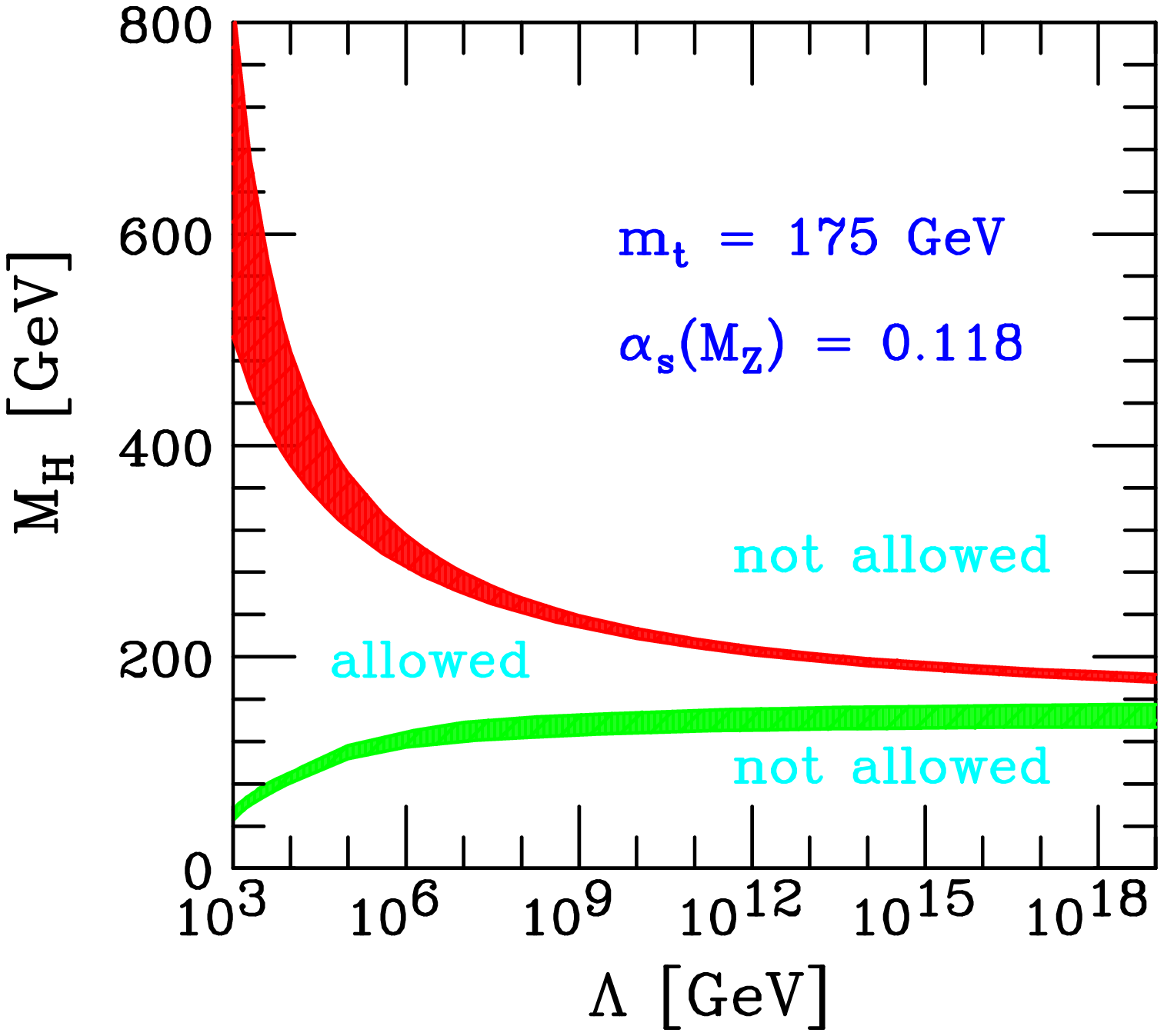}\hspace*{-10mm}
\end{minipage}
\end{center}
\vspace*{-3.4cm}
\caption[Fit to the precision data and theoretical bounds on the Higgs 
mass in the SM]
{Left:  Global fit to the electroweak precision data within the SM; the
excluded region form direct Higgs searches is also shown \cite{LEPEWWG}.
Right: theoretical upper and lower bounds on $M_H$ from the
assumption that the SM is valid up to the cut--off scale $\Lambda$
\cite{Hambye:1996wb}. }
\label{Hfig:constraint}
\vspace*{-0.3cm}
\end{figure}

The only available direct information on the Higgs mass is the lower limit $M_H
\gsim 114.4$ GeV at 95\% confidence level established at LEP2
\cite{Barate:2003sz}. The collaborations have also reported a small, $\lsim 2
\sigma$, excess of events beyond the expected SM backgrounds consistent with a
SM--like Higgs boson with a mass $M_H \sim 115$ GeV \cite{Barate:2003sz}. This
mass range can be tested soon at the Tevatron if high enough luminosity is
collected. Furthermore, the high accuracy of the electroweak data measured at
LEP, SLC and Tevatron \cite{PDG}  provides an indirect sensitivity to $M_H$: the
Higgs boson contributes logarithmically, $\propto \log (M_H/M_W)$, to the
radiative corrections to the $W/Z$ boson propagators. A recent analysis, which
uses the updated value of the top quark mass yields the value
$M_H=76^{+33}_{-24}$ GeV, corresponding to a 95\% confidence level upper limit
of $M_H \lsim 144$ GeV \cite{LEPEWWG}. The left--hand side of
Fig.~\ref{Hfig:constraint} shows  the  global fit to the electroweak data; the
Higgs fit has a probability of 15.1\%.  If the Higgs boson turns out to be
significantly heavier than 150 GeV, there should be an additional new ingredient
that is relevant at the EWSB scale which should be observed at the next round of
experiments.

From the theoretical side, interesting constraints can be derived from
assumptions on the energy range within which the SM is valid before perturbation
theory breaks down and new phenomena would emerge. For instance, if the Higgs
mass were larger than $\sim$ 1 TeV, the $W$ and $Z$ bosons would interact very
strongly with each other to ensure unitarity in their scattering at high
energies. Imposing the unitarity requirement in the high--energy scattering of
gauge bosons leads to the  bound $M_H \lsim 700$ GeV \cite{H-LQT}.  If the Higgs
boson were too heavy, unitarity would be violated in these processes at energies
above $\sqrt s \gsim  1.2$ TeV and new phenomena should appear to restore it.

Another important theoretical constraint comes from the fact that the quartic
Higgs self--coupling, which at the scale $M_H$ is fixed by $M_H$ itself, grows
logarithmically with the energy scale. If $M_H$ is small, the energy cut--off
$\Lambda$ at which the coupling grows beyond any bound and new phenomena should
occur, is large; if $M_H$ is large, the cut--off $\Lambda$ is small.  The
condition $M_H \lsim \Lambda$ sets an upper limit on the Higgs mass in the SM,
the triviality bound. A naive one--loop analysis assuming the validity of
perturbation theory as well as lattice simulations lead to an estimate of $M_H
\lsim 630$ GeV for this limit \cite{H-Lattice}.  Furthermore, loops involving
top quarks tend to drive the coupling to negative values for which the vacuum is
no longer stable. 

Requiring the SM to be extended to, for instance,
the GUT scale $\Lambda_{\rm GUT} \sim 10^{16}$ GeV and including the effect of
top quark loops on the running coupling, the Higgs boson mass should lie in the
range 130 GeV $\lsim M_H \lsim 180$ GeV \cite{Hambye:1996wb}; see the
right--hand side of  Fig.~\ref{Hfig:constraint}.

In fact in any model beyond the SM in which the theory is required to be weakly
interacting up to the GUT or Planck scales the Higgs boson should be lighter
than $M_H \lsim 200$ GeV. Such a Higgs particle can be produced at the ILC
already for center of mass energies of $\sqrt s\sim 300$ GeV. However, to cover
the entire Higgs mass range in the SM, $M_H \lsim 700$ GeV, c.m. energies close
to $\sqrt s=1$ TeV would be required.

\subsection{The Higgs particles in the MSSM}

It is well known that there are at least two severe problems in the SM, in
particular when trying to extend its validity to the GUT scale $\Lambda_{\rm
GUT}$. The first one is the so--called  naturalness problem: the Higgs boson
tends to acquire a mass of the order of these large scales [the radiative
corrections to $M_H$ are quadratically divergent]; the second problem is that
the running of the three gauge couplings of the SM is such that they do not meet
at a single point and thus do not unify at the GUT scale. Low energy
supersymmetry solves these two problems at once: supersymmetric particle loops
cancel exactly the quadratic divergences and contribute to the running of the
gauge couplings to allow their unification at $\Lambda_{\rm GUT}$.

The minimal supersymmetric extension of the SM (MSSM), which will be discussed
in chapter \ref{sec:susy}, requires the existence of two isodoublet  Higgs
fields to cancel anomalies and to give mass separately to up and down--type
fermions.  Two CP--even neutral Higgs bosons $h,H$, a pseudoscalar $A$ boson and
a pair of charged scalar particles, $H^\pm$, are introduced by this extension of
the Higgs sector \cite{HHG,Djouadi:2005gj}. Besides the four masses, two
additional parameters define the properties of these particles: a mixing angle
$\alpha$ in the neutral CP--even sector and the ratio of the two vacuum
expectation values $\tb$, which lies in the range $1 \lsim \tb \lsim m_t/m_b$.

\begin{figure}[h]
\begin{center}
\vspace*{-3.mm}
\includegraphics[width=14cm,bb=73 448 669 752]{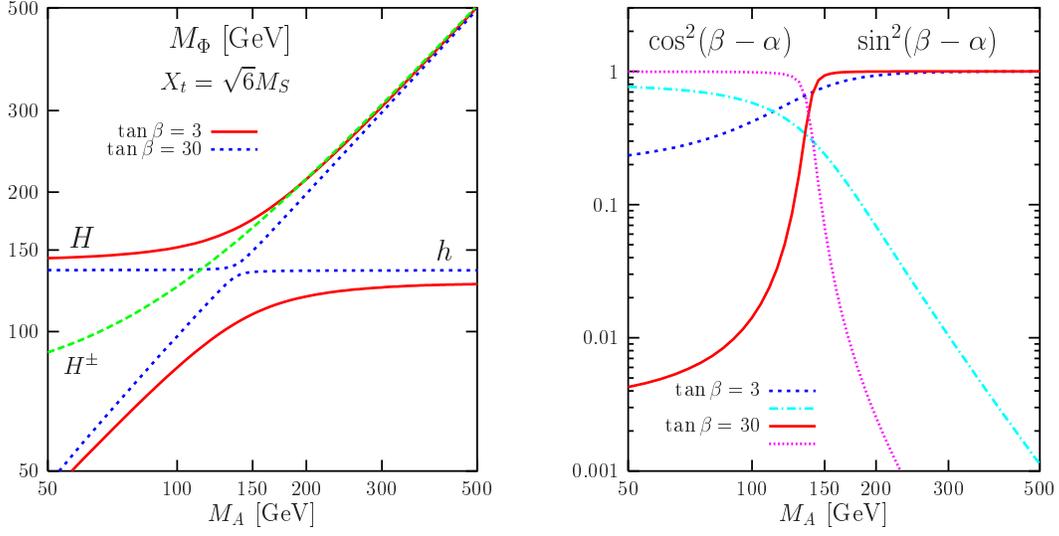}
\end{center}
\vspace*{-.8cm}
\caption[The masses and couplings to W/Z bosons of the Higgs bosons in the MSSM]
{The masses (left) and the couplings to gauge bosons (right) of the
MSSM Higgs bosons as a function of $M_A$ for $\tb=3,30$ with $M_S=2$ TeV and
$X_t=\sqrt 6 M_S$.}
\label{Hfig:SUSYmass}
\vspace*{-3mm}
\end{figure}

Supersymmetry leads to several relations among these parameters and only two of
them, taken in general to be $M_A$ and $\tb$, are in fact independent. These
relations impose a strong hierarchical structure on the mass spectrum, $M_h<M_Z,
M_A<M_H$ and $M_W<M_{H^\pm}$, which however is broken by radiative corrections
as the top quark mass is large; see Ref.~\cite{gigazsven} for a recent review.
The leading part of this correction grows as the fourth power of $m_t$ and
logarithmically with the SUSY scale  or common squark mass $M_S$; the mixing (or
trilinear coupling) in the stop sector $A_t$ plays an important role. For
instance, the upper bound on the mass of the lightest Higgs boson $h$ is shifted
from the tree level value $M_Z$ to $M_h \sim 130$--140 GeV in the maximal mixing
scenario where $X_t= A_t -\mu/\tb \sim 2M_S$ with $M_S={\cal O}(1$ TeV)
\cite{gigazsven}; see the left--handed side of  Fig.~\ref{Hfig:SUSYmass}. The masses
of the heavy neutral and charged Higgs particles are expected to range from
$M_Z$ to the SUSY breaking scale $M_S$.

The pseudoscalar Higgs boson $A$  has no tree level couplings to gauge bosons,
and its couplings to down (up) type fermions are (inversely) proportional to
$\tb$. This is also the case for the couplings of the charged Higgs boson to
fermions, which are admixtures of scalar and pseudoscalar currents and depend
only on $\tb$. For the CP--even Higgs bosons $h$ and $H$, the couplings to
down (up) type fermions are enhanced (suppressed) compared to the SM Higgs
couplings for $\tb >1$. They share the SM Higgs couplings to vector bosons as
they are suppressed by $\sin$ and $\cos(\beta-\alpha)$ factors, respectively for
$h$ and $H$; see the right--hand side of Fig.~\ref{Hfig:SUSYmass} where the
couplings to the $W^\pm, Z$ bosons are displayed.

If the pseudoscalar mass is large, the $h$ boson mass reaches its upper limit
[which, depending on the value of $\tb$ and stop mixing, is in the range
100--140 GeV] and its couplings to fermions and gauge bosons are SM--like; the
heavier CP--even $H$ and charged $H^\pm$ bosons become degenerate with the
pseudoscalar $A$ boson and have couplings to fermions and gauge bosons of the
same intensity. In this decoupling limit, which can be already reached for 
pseudoscalar masses $M_A \gsim 300$ GeV, it is very difficult to distinguish the
Higgs sectors of the SM and MSSM if only the lighter $h$ particle has been
observed.  

Finally, we  note that there are experimental constraints on the MSSM Higgs
masses, which mainly come from the negative LEP2 searches \cite{LEP2-HMSSM}. In
the decoupling limit where the $h$ boson is SM--like, the limit $M_h \gsim
114$ GeV from the Higgs--strahlung process holds; this constraint rules out
$\tb$ values smaller than $\tb \sim 3$. Combining all processes, one obtains
the absolute mass limits $M_h \sim M_A \gsim M_Z$ and $M_{H^\pm} \gsim M_W$
\cite{LEP2-HMSSM}.

\subsection{Higgs bosons in non--minimal SUSY models}

The Higgs sector in SUSY models can be more complicated than previously
discussed if some basic assumptions of the MSSM, such as the absence of new
sources of CP violation, the presence of only two Higgs doublet fields, or
R--parity conservation, are relaxed; see chapter \ref{sec:susy} for a
discussion. A few examples are listed below. 

In the presence of \underline{CP--violation in the SUSY sector}, which is
required if baryogenesis  is to be explained at the electroweak scale, the new
phases will enter the MSSM Higgs sector [which is CP--conserving at tree--level]
through the large radiative corrections.  The masses and the couplings of the
neutral and charged Higgs particles will be altered and, in particular, the
three neutral Higgs bosons will not have definite CP quantum numbers and will
mix with each other to produce the physical states $H_1,H_2,H_3$. The properties
of the various Higgs particles can be significantly affected; for reviews, see
e.g.~Refs.~\cite{SUSY-CPV,CPHmasses}.  Note, however, that there is a sum rule which forces
the three $H_i$ bosons to share the coupling of the SM Higgs to gauge bosons,
$\sum_i g_{H_iVV}^2 =g^2_{H_{\rm SM}}$, but only the CP--even component is
projected out. 

\begin{figure}[!h]
\begin{center}
\mbox{
\includegraphics[width=0.3\linewidth,height=5.0cm]{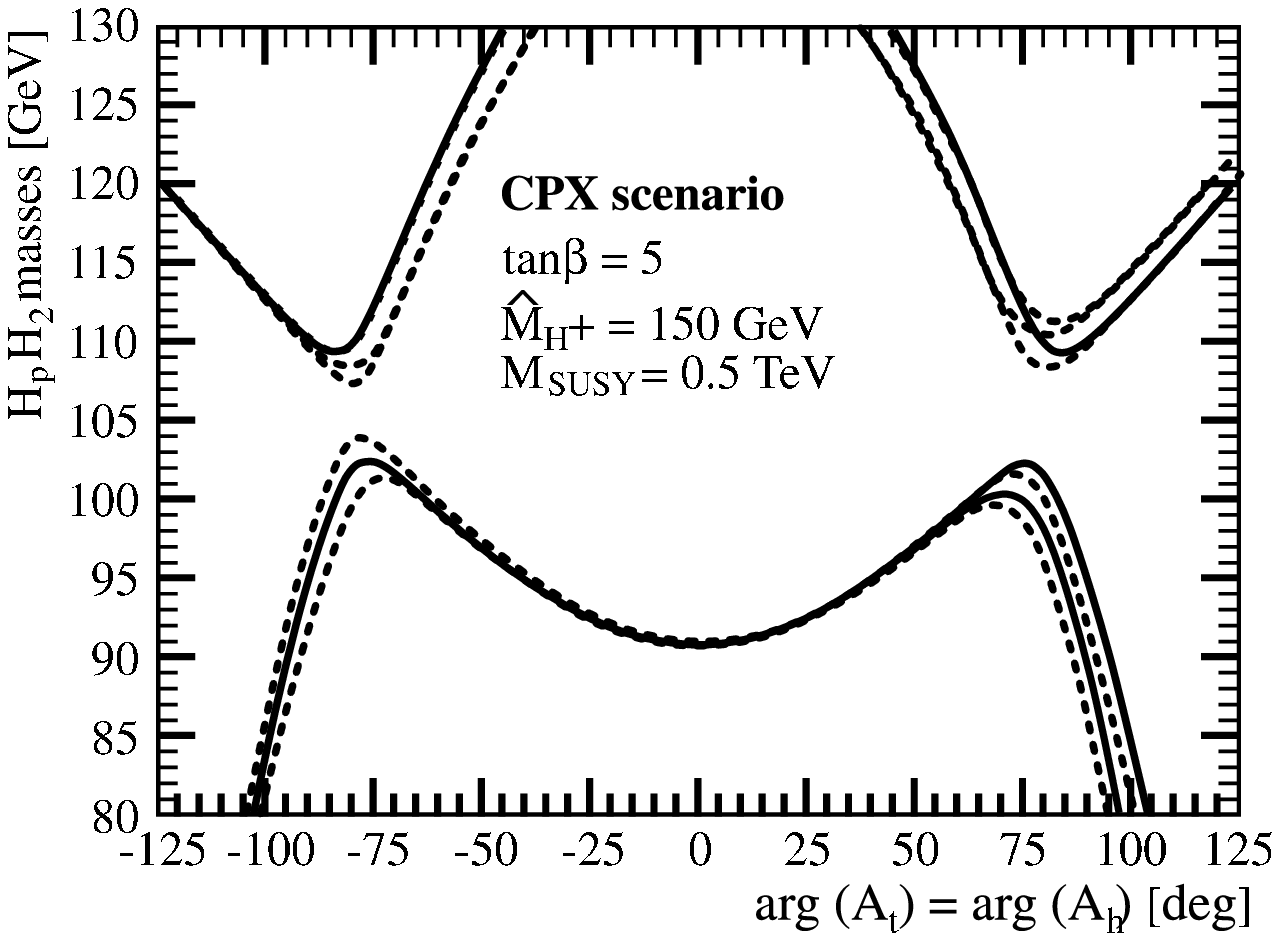}\hspace*{3mm}
\includegraphics[width=0.3\linewidth,height=5.0cm]{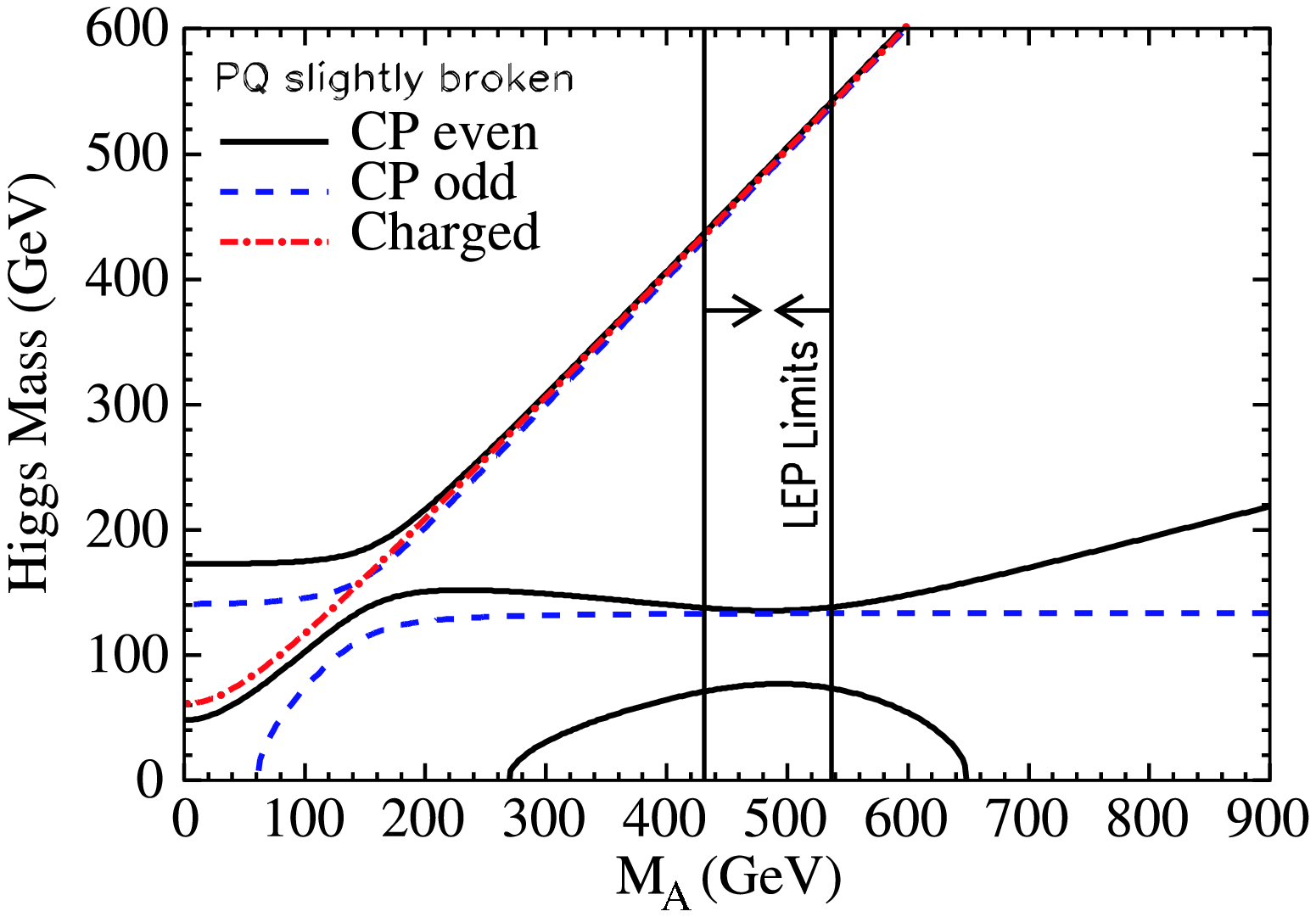}\hspace*{3mm}
\includegraphics[width=0.3\linewidth,height=5.0cm]{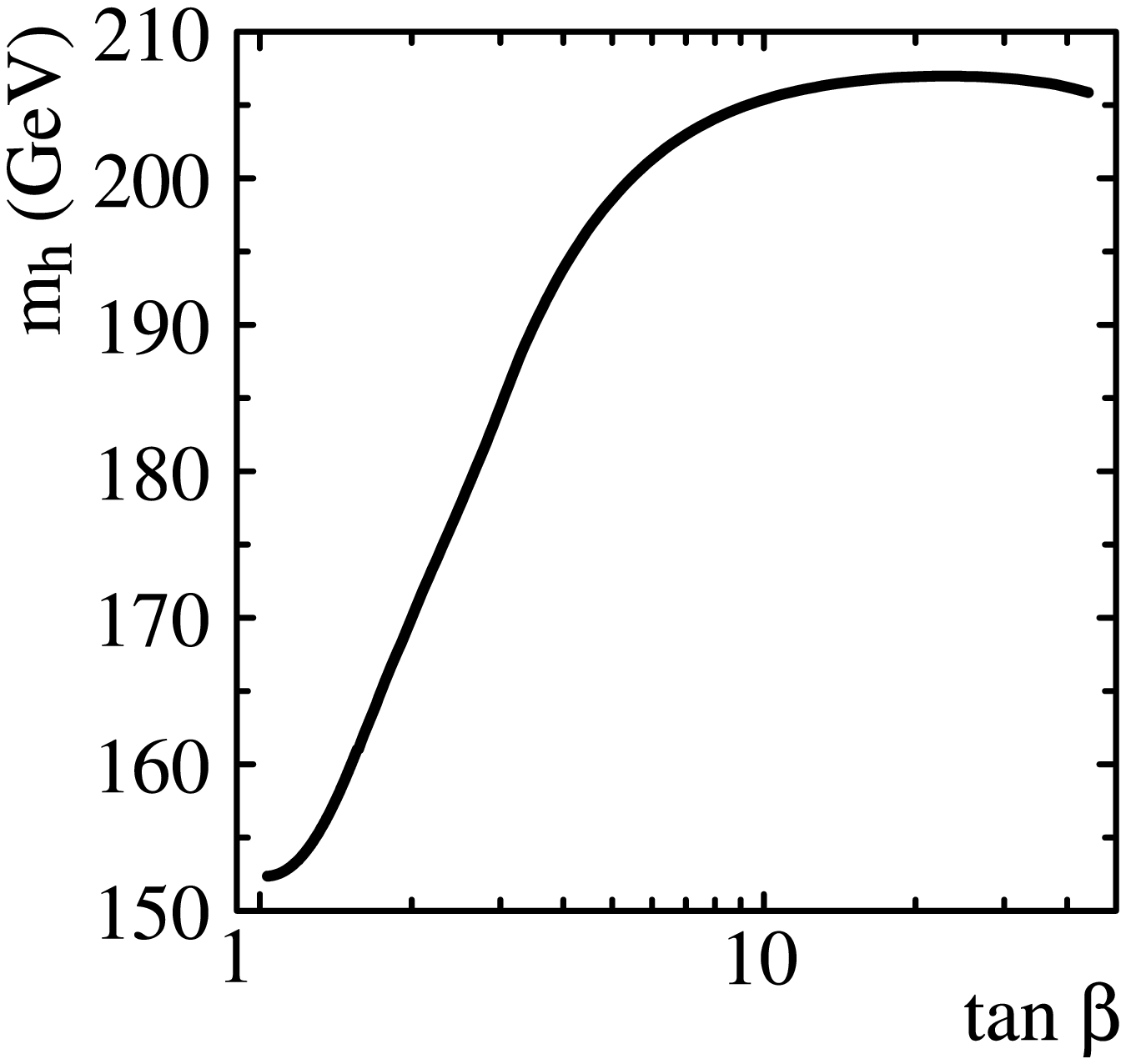}
}
\end{center}
\vspace*{-1cm}
\caption[The spectrum of neutral Higgs particles in extensions of the MSSM.]
{The spectrum of neutral Higgs particles in a CP--violating MSSM
scenario (for $\tb\!=\!5, M_{H^\pm}\!=\!150$ GeV and $M_S\!=\!0.5$ TeV)
\cite{CPHmasses} 
(left) typical Higgs mass spectrum in the NMSSM  as a function of $M_A$
\cite{HNMSSMp} (center) and the upper bound on the lighter Higgs mass
in a general SUSY model \cite{HSUSY-thbound}.}
\label{Hbeyond}
\vspace*{-3mm}
\end{figure}

As examples of new features compared to the usual MSSM, we simply mention the
possibility of a relatively light $H_1$ state with very weak couplings to the
gauge bosons which could have escaped detection at LEP2 \cite{HMSSMCPX} and the
possibility  of resonant $H/A$ mixing when the two Higgs particles are
degenerate in mass \cite{HCPR}; an example of the Higgs mass spectrum is shown
in  Fig.~\ref{Hbeyond} (left) as a function of the phase of the coupling $A_t$. 
These features have to be proven to be a result of CP--violation by, for
instance,  studying CP--odd observables.

\underline{The next--to--minimal SUSY extension, the NMSSM}, consists of simply
introducing a complex iso-scalar field which naturally generates a weak scale
Higgs--higgsino parameter $\mu$ (thus solving the $\mu$ problem); the model is
more natural than the MSSM and has less fine--tuning
\cite{HNMSSMp,SUSY-NMSSM,SUSY-NMSSMh}. The NMSSM Higgs sector is thus extended
to include an additional CP--even and a CP--odd Higgs particle and an example of
a Higgs mass spectrum is shown in Fig.~\ref{Hbeyond} (center).  The upper bound
on the mass of the lighter CP--even  particle slightly exceeds that of the MSSM
$h$ boson and the negative searches at LEP2 lead to looser constraints. 

In a large area of the parameter space, the Higgs sector of the NMSSM reduces to
the one of the MSSM but there is a possibility, which is not completely 
excluded, that one of the neutral Higgs particles, in general the lightest 
pseudoscalar $A_1$, is very light with a mass  of a few ten's of GeV. The  light
CP--even Higgs boson, which is SM--like in general, could then decay into pairs
of $A_1$ bosons, $H_1 \to A_1 A_1 \to 4b, 4\tau$, with a large branching
fraction.  

\underline{Higgs bosons in GUT theories.} A large variety of theories,  string
theories, grand unified theories,  left--right symmetric models, etc., suggest
an additional gauge symmetry which may be broken only at the TeV scale;  see
chapter \ref{sec:alternatives}. This leads to an extended particle spectrum and,
in particular, to additional Higgs fields beyond the minimal set of the MSSM.
Especially common are new U(1)' symmetries broken by the vev of a singlet field
(as in the NMSSM)  which leads to the presence of a $Z'$ boson and one
additional CP--even Higgs particle compared to the MSSM; this is the  case, for
instance, in the exceptional MSSM \cite{H-GUT-ESSM} based on the string inspired
$E_6$ symmetry. The secluded ${\rm SU(2)\times U(1) \times U(1)'}$ model
\cite{H-GUT-secluded}, in turn, includes four  additional singlets that are
charged under U(1)', leading to  6 CP--even and 4 CP--odd neutral Higgs states.
Other exotic Higgs sectors in SUSY models \cite{H:higheR} are, for instance,
Higgs representations that transform as SU(2) triplets or  bi--doublets under
the ${\rm SU(2)_L}$ and  ${\rm SU(2)_R}$ groups in left--right symmetric models,
that are motivated by the seesaw approach to  explain the small neutrino masses
and which lead e.g. to a doubly charged Higgs  boson $H^{--}$. These extensions,
which also predict extra matter fields, would lead to a very interesting
phenomenology and new collider signatures in the Higgs sector.   

In a general SUSY model, with an arbitrary number of singlet and doublet
scalar fields [as well as a matter content which  allows for the unification of
the gauge couplings], a linear combination of Higgs fields has to generate  the
$W/Z$ masses and thus, from the triviality argument discussed earlier, a Higgs
particle should have a mass below 200 GeV  and significant couplings to gauge
bosons \cite{HSUSY-thbound}. The upper bound on the mass of the lightest Higgs
boson in this most general SUSY model is displayed in Fig.~\ref{Hbeyond} (right)
as a function of $\tb$.

\underline{R--parity violating models.}    Models in which  R--parity is
spontaneously broken [and where one needs to either enlarge the SM symmetry or
the spectrum to include additional gauge singlets],  allow for an explanation of
the light neutrino data \cite{H-RparityV}.  Since $\not \hspace*{-1.5mm}R_p$
entails the breaking of the total lepton number $L$, one of the CP--odd scalars,
the Majoron $J$, remains massless being  the Goldstone boson associated to $\not
\hspace*{-1.5mm}L$. In these models, the neutral Higgs particles have also reduced
couplings to the gauge bosons. More importantly,  the CP--even Higgs particles
can decay into pairs of invisible Majorons, $H_i \to JJ$, while the CP--odd
particle can decay into a CP--even Higgs and a Majoron, $A_i \to H_i J$, and
three Majorons,  $A \to JJJ$ \cite{H-RparityV}.

\subsection{Higgs bosons in alternative models}

There are also many non supersymmetric extensions of the SM which might lead to
a different Higgs phenomenology. In some cases, the Higgs sector would consist
of one scalar doublet leading to a Higgs boson which would mimic the SM Higgs,
but the new particles that are present in the models might alter some of its
properties. In other cases, the Higgs sector is extended to contain  additional
scalar fields leading to the presence of new Higgs particles. Another
possibility is a scenario with a composite and strongly interacting Higgs,  or
where no Higgs particle is present at all, leading to strong interactions of  
the $W/Z$ bosons. Many of these models, such as e.g. extra--dimensional, little
Higgs and Higgsless models, will be discussed in chapter \ref{sec:alternatives}.
Here will simply  give a non exhaustive list of various possible scenarios. 

\underline{Scenarios with Higgs mixing.} In warped extra--dimensional models
\cite{WED} the fluctuations of the size of the extra  dimension about
its stabilized value manifest themselves as a single scalar  field, the radion.
In the Randall Sundrum model with a bulk scalar field,  it is expected that  the
radion is the lightest state beyond the SM fields with a mass probably  in  the
range between ${\cal O}$(10 GeV)  and $\Lambda={\cal O}$(TeV)
\cite{Hewett:2002nk,Dominici:2002jv}. The couplings of the radion are order of 
$1/\Lambda$ and are very similar to the couplings of the SM Higgs boson, 
except  for one important difference: due to the trace anomaly, the radion
directly  couples to massless gauge bosons at one loop.  Moreover, in the low
energy four--dimensional effective theory, the radion can mix with the  Higgs
boson.  This mixing  can lead to important  shifts in the Higgs couplings which
become apparent in the Higgs decay widths and production cross sections.  In
large extra dimension models \cite{LED}, mixing of the Higgs
boson with graviscalars  also occurs \cite{H-graviscalars}, leading to an
invisible decay width.  Mixing effects also occur if the SM is minimally
extended in a renormalizable way to contain a singlet scalar field $S$ that does
not couple to the other SM particles; its main effect would be to alter the
scalar potential and to  mix with the SM Higgs field \cite{NMSM} and, in such a
case, the Higgs could mainly decay into two invisible $S$ particles.

\underline{Scenarios with an extended Higgs/gauge/matter sector.}
Non--supersymmetric  extensions of the Higgs sector with additional singlet,
doublet and higher representation fields have also been advocated
\cite{H:higheR}. Examples are the minimal SM extension with a singlet discussed
above, two--Higgs doublet models which potentially include CP--violation,
triplet Higgs fields in models for light neutrino mass generation, etc...  These
extensions lead to a rich spectrum of Higgs particles which could be produced at
the ILC. In other extensions of the SM, new gauge bosons and new matter
particles are predicted and they can affect the properties of the SM--like Higgs
boson. For instance the new fermions present in little Higgs and
extra--dimensional  models might contribute to the loop induced Higgs couplings,
while new heavy gauge bosons could alter the Higgs couplings to $W$ and $Z$
bosons for instance.

\underline{Scenarios with a composite Higgs boson.} In little Higgs models
\cite{LHM},  the dynamical scale is around $\Lambda=10$ TeV, unlike the
traditional Technicolor model \cite{H-technicolor,Technicolor}.  A light Higgs
boson can be generated as a pseudo Goldstone boson  and its mass  of order 100
GeV is protected against  large radiative corrections individually in the boson
and the fermion sectors. The models predict a rich spectrum of new particles not
only at the scale $\Lambda$ but also at lower scales. Axion--type pseudoscalar
bosons may be associated with the spontaneous breaking of U(1) factors in the
extra global symmetries \cite{Kilian:2006eh}. These particles have properties
analogous to Higgs bosons and  can be produced in $\eei$ collisions; deviations
in the production and decay rates of the  SM--like Higgs boson can also be
induced by these particles. Note that, recently, a model--independent
description of a strongly  interacting light Higgs  has been given
\cite{H-SILH}.  

\underline{Higgless models and strong $W/Z$ interactions.} The problem of 
unitarity violation at high energies in the SM can also be solved, apart from
introducing a relatively light Higgs boson, by assuming the $W/Z$ bosons to
become  strongly interacting at TeV energies, thus damping the rise of the
elastic $W/Z$ scattering  amplitudes. Naturally, the strong forces between the
massive gauge bosons may be traced back to new fundamental interactions
characterized by a scale of order 1 TeV \cite{H-technicolor}. Also in theories
with extra space dimensions, the electroweak symmetries can  be broken without
introducing additional fundamental scalar fields, leading  also to Higgsless
theories \cite{Hless}. Such scenarios can be studied in massive gauge boson
scattering experiments, where the $W/Z$ bosons are radiated, as quasi--real
particles, off electrons and positrons in TeV linear colliders
\cite{Aguilar-Saavedra:2001rg}. This aspect will be discussed in chapter 
\ref{sec:alternatives}.

\subsection{The expectations at the LHC}

\begin{figure}[!h]
\vspace*{-.3cm}
\begin{center}
\epsfig{file=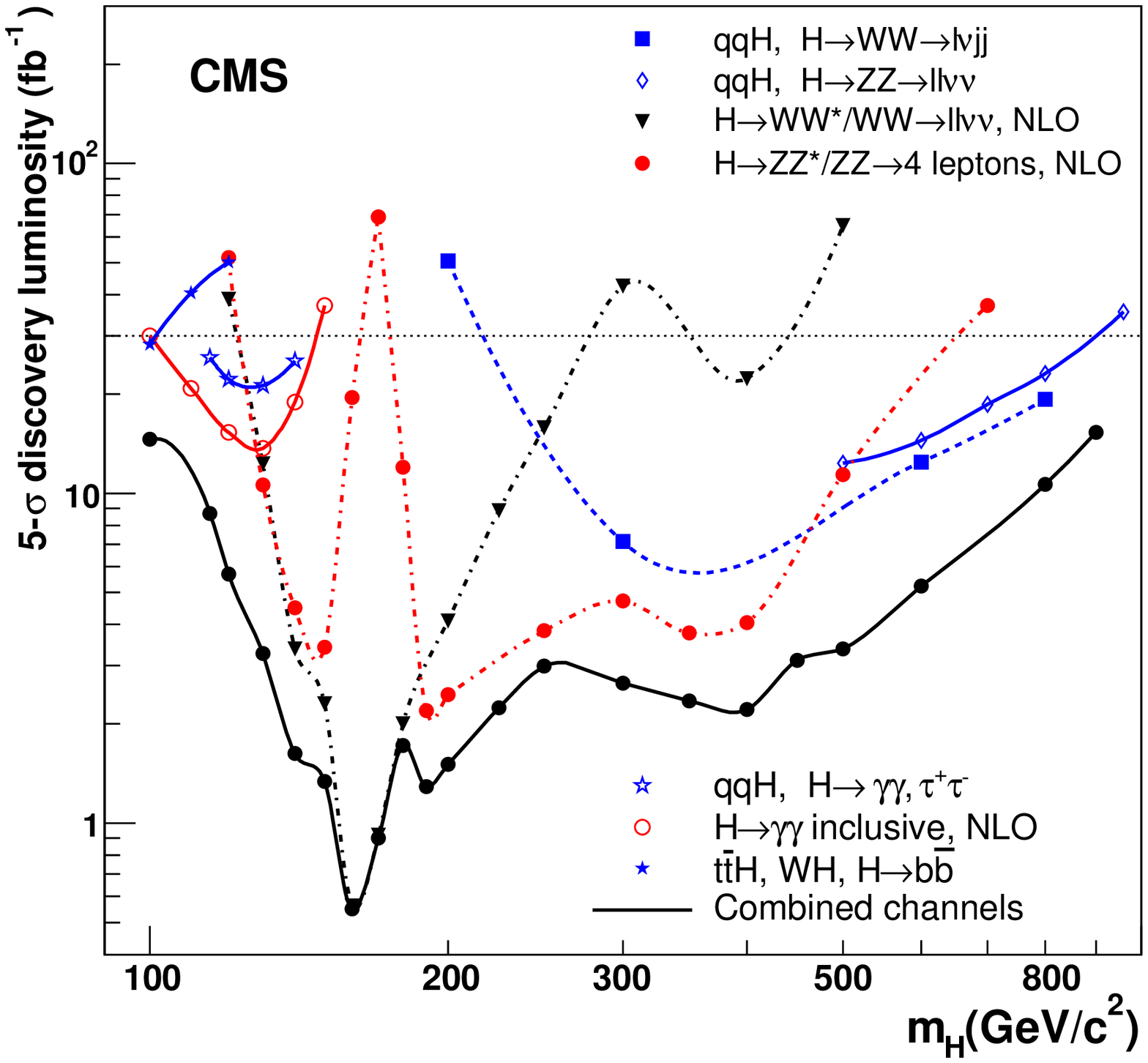,width=7.cm,height=6.3cm}
\epsfig{file=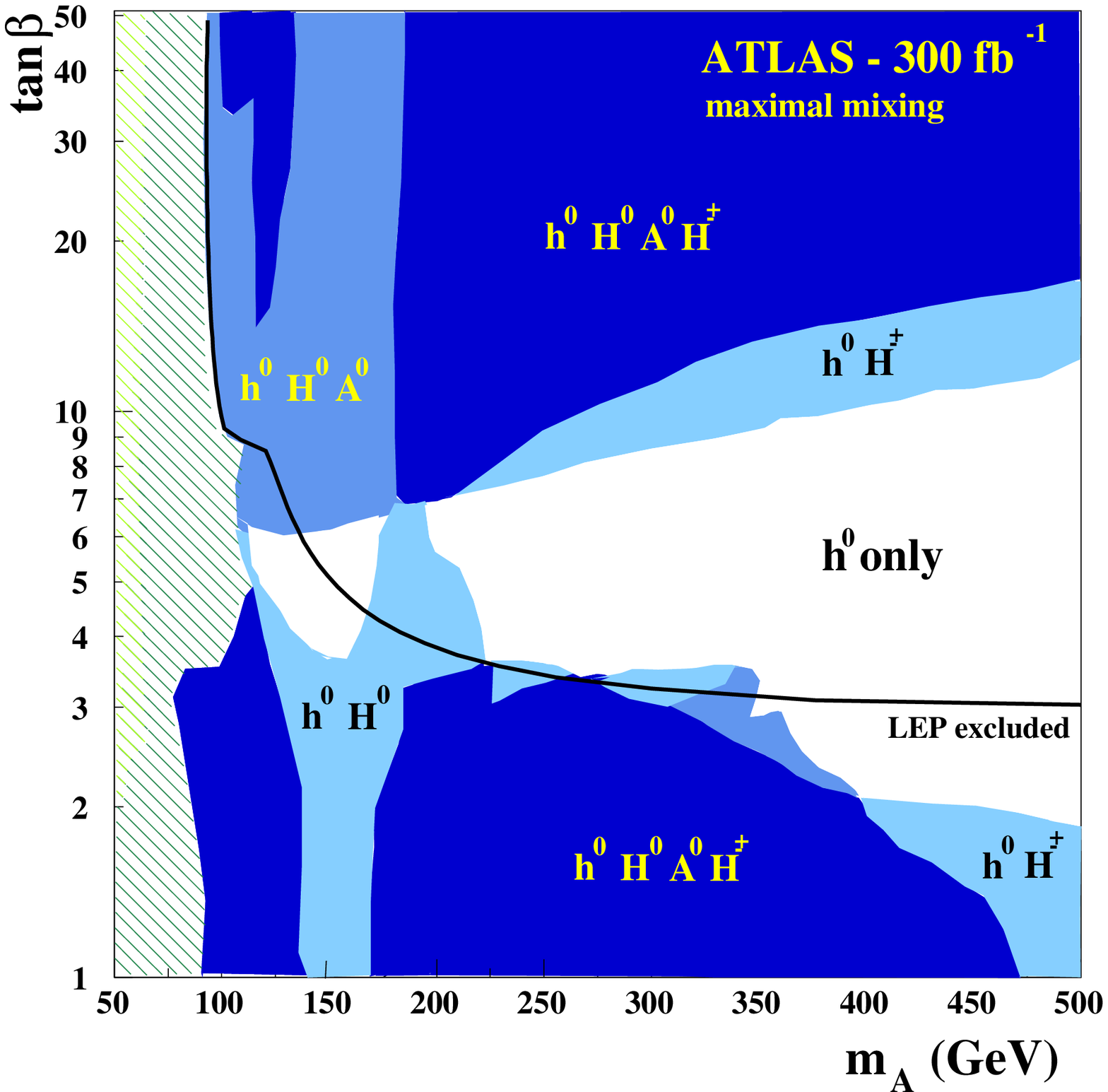,width=7.cm,height=5.8cm}
\end{center}
\vspace*{-.5cm}
\caption[The LHC discovery potential for Higgs bosons in the SM and the MSSM.]
{The required luminosity that is needed to achieve a $5\sigma$
discovery signal at LHC using various detection channels as a
function of $M_H$ \cite{CMSTDR} (left) and the number of  Higgs
particles that can be detected in the MSSM  $[\tb,M_A]$ parameter space
\cite{atlastdr} (right).}
\vspace*{-.3cm}
\label{Hfig:LHC}
\end{figure}

The search for the Higgs boson(s) is the one of the primary  tasks of the CMS
and ATLAS experiments at the LHC. For the SM Higgs boson, detailed studies have
been performed \cite{atlastdr,CMSTDR} with the conclusion that a
5$\sigma$ discovery is possible with an integrated luminosity  of 30 fb$^{-1}$
for the entire Higgs mass range. Several production and decay channels can be
used for this purpose; see Fig.~\ref{Hfig:LHC} (left). The spin--zero nature of
the Higgs boson can be determined and a preliminary probe of its  CP nature can
be performed. Furthermore,  information on the Higgs couplings to gauge bosons
and fermions can be obtained with a higher luminosity;  the estimated precision
for coupling ratios are typically ${\cal O }(10)$\%  with ${\cal L}\!=\!100$
fb$^{-1}$ \cite{Duhrssen:2004cv}. Because of  the small production rates and
large backgrounds, the determination  of the Higgs self--coupling is too
difficult and will require a  significantly higher luminosity.

In the MSSM, all the Higgs bosons can be produced for masses below 1 TeV and
large enough $\tb$ values if a large integrated luminosity, $\sim 300$
fb$^{-1}$,  is collected; Fig.~\ref{Hfig:LHC} (right). There is, however, a
significant region of the parameter space where only the light SM--like $h$
boson will be found. In such a case the mass of  the $h$ boson may be the only
characteristic information of the MSSM Higgs sector at the LHC. Nevertheless,
there are  some situations in which MSSM Higgs searches at the LHC could be
slightly more complicated. This is for instance the case  when Higgs decays into
SUSY particles such as charginos and (invisible) neutralinos are kinematically
accessible and significant. Furthermore,  in the so--called
intense coupling regime where the three neutral Higgs particles are very close
in mass and have strong couplings to $b$--quarks, not all  three states can be
resolved experimentally \cite{H:intense}.

The search of the Higgs particles can be more complicated in some extensions of
the MSSM. For instance, if CP--violation occurs, the lighter neutral $H_1$ boson
can escape observation in a small region of the parameter space with low $M_A$
and $\tb$ values, while the heavier $H,A$ and $H^\pm$ bosons  can be accessed in
smaller areas than in the usual MSSM \cite{SUSY-CPV}. In the NMSSM with a
relatively light pseudoscalar $A_1$ particle, the dominant decay of the lighter
CP--even $H_1$ boson could be $H_1 \to A_1 A_1 \to 4b$, a signature which is
extremely difficult to detect at the LHC \cite{SUSY-NMSSM}.  A possibility that
should not be overlooked is that in  several extensions of the Higgs sector,
such as non--minimal SUSY, extra--dimensional models and the extension with a
singlet scalar field, the Higgs boson might decay invisibly making its detection
at the LHC very challenging if possible at all. In addition, in some other SM
extensions, the rates for the dominant $gg\to H$ production can be  strongly
suppressed.

\section{The Higgs boson in the Standard Model}

\subsection{Higgs decays and production}

In the SM, the profile of the Higgs particle is uniquely determined once its
mass $M_H$ is fixed \cite{HHG,Djouadi:2005gi}. The decay width, the branching
ratios and the production cross sections are given by the strength of the
Yukawa couplings to fermions and gauge bosons, the scale of which is set by the
masses of these particles.  The trilinear and quartic Higgs self couplings are
also uniquely fixed in terms of the Higgs boson mass.

In the  ``low Higgs mass" range, $M_H \lsim 140$ GeV, the Higgs boson decays
into a large variety of channels. The main decay mode is by far the decay into
$b\bar{b}$ pairs with a branching ratio of ${\cal O}(80\%)$ followed by the
decays into $c\bar{c}$ and $\tau^+\tau^-$ pairs with fractions of ${\cal O}
(5\%)$.  Also of significance, the top--loop mediated Higgs decay into gluons
which for $M_H$ around 120 GeV occurs at the level of $\sim 5\%$. The top and
$W$--loop mediated $\gamma\gamma$ and $Z \gamma$ decay modes are very rare the
branching fractions being of ${\cal O }(10^{-3})$.  However, these decays  are,
together with $H \to gg$,  theoretically interesting being sensitive to new
heavy states such as SUSY particles.

\begin{figure}[h!]
\begin{center}
\includegraphics[width=0.92\linewidth,bb=73 465 600 720]{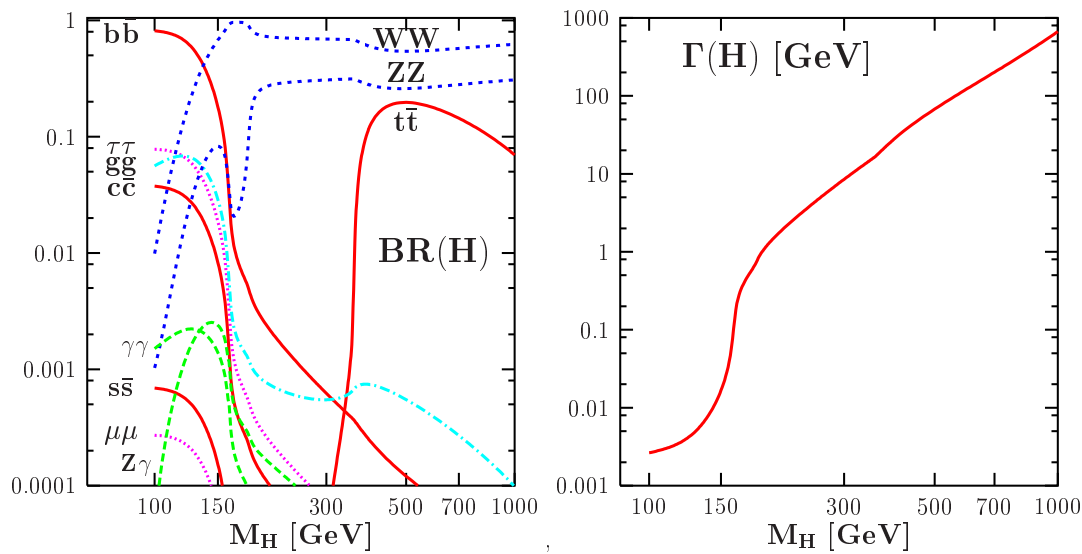} 
\end{center}
\vspace{-8.mm}
\caption[The decay branching ratios and the total width of the SM Higgs 
boson]{The decay branching ratios (left) and the total decay width (right)
of the SM Higgs boson as a function of its mass $M_H$; from
Refs.~\cite{Djouadi:1997yw,H-decay}.}
\label{Hfig:brth}
\vspace{-1mm}
\end{figure}

In the ``high Higgs mass" range, $M_H \gsim 140$ GeV, the Higgs
bosons decay mostly into $WW^{(*)}$ and $ZZ^{(*)}$ pairs, with one
of the gauge bosons being virtual if below the $WW$ threshold. Above
the $ZZ$ threshold, the Higgs boson decays almost exclusively into
these channels with a branching ratio of $\frac23$ for $H\to WW$ and
$\frac13$ for $H\to ZZ$ decays. The opening of the $t\bar{t}$
channel for $M_H \gsim 350$ GeV does not alter this pattern
significantly as BR$(H\to t\bar t)$ does not exceed the level of
10--15\% when kinematically accessible.

In the low mass range, the Higgs boson is very narrow $\Gamma_H<10$ MeV, but
the width becomes rapidly wider for masses larger than 140 GeV, reaching
$\Gamma_H \sim 1$ GeV at the $ZZ$ threshold. For large masses, $M_H \gsim 500$
GeV, the Higgs becomes obese since its total width is comparable to its mass,
and it is hard to consider it as a resonance.\smallskip

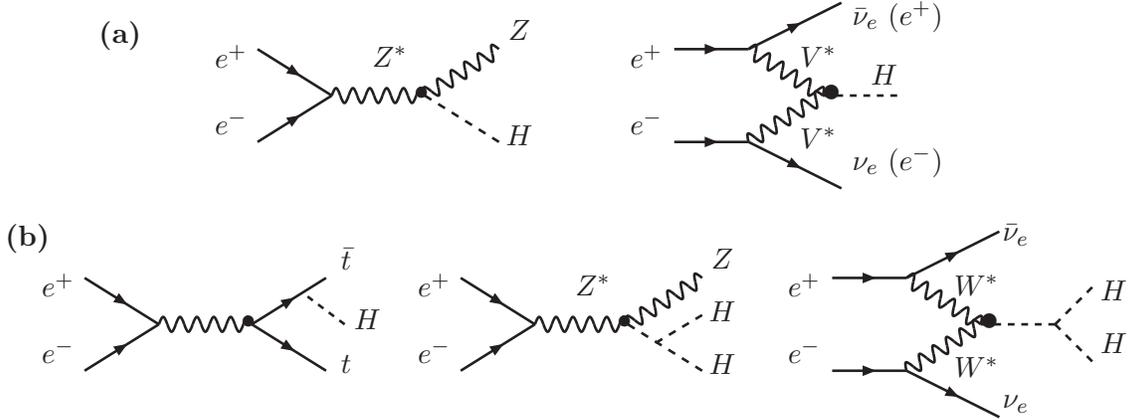
\begin{figure}[h]
\vspace*{-1.4cm}
\begin{center}
\begin{picture}(300,100)(0,0)
\SetScale{0.7}
\SetWidth{1.4}
\hspace*{1cm}
\ArrowLine(0,25)(40,50)
\ArrowLine(0,75)(40,50)
\Photon(40,50)(90,50){4}{5.5}
\DashLine(90,50)(130,25){4}
\Text(62,36)[]{$\bullet$}
\Photon(90,50)(130,75){4}{5.5}
\Text(-10,25)[]{$e^-$}
\Text(-10,50)[]{$e^+$}
\Text(50,50)[]{$Z^*$}
\Text(99,20)[]{$H$}
\Text(99,60)[]{$Z$}
\put(-60, 55){\bf (a)}
\hspace*{0.6cm}
\ArrowLine(200,25)(240,25)
\ArrowLine(200,75)(240,75)
\ArrowLine(240,25)(290,0)
\ArrowLine(240,75)(290,100)
\Photon(240,25)(280,50){4}{5.5}
\Photon(240,75)(280,50){4}{5.5}
\DashLine(280,50)(320,50){4}
\Text(200,36)[]{\Large $\bullet$}
\Text(130,25)[]{$e^-$}
\Text(130,50)[]{$e^+$}
\Text(195,19)[]{$V^*$}
\Text(195,50)[]{$V^*$}
\Text(220,43)[]{$H$}
\Text(225,10)[]{$\nu_e \ (e^-)$}
\Text(225,65)[]{$\bar{\nu}_e \ (e^+)$}
\end{picture}
\end{center}
\vspace*{-1.cm}
\begin{center}
\begin{picture}(300,100)(0,0)
\SetScale{0.7}
\SetWidth{1.4}
\hspace*{-1.3cm}
\ArrowLine(0,25)(40,50)
\ArrowLine(0,75)(40,50)
\Photon(40,50)(90,50){4}{5.5}
\ArrowLine(90,50)(130,25)
\ArrowLine(90,50)(130,75)
\DashLine(120,65)(140,50){4}
\Text(62,36)[]{$\bullet$}
\Text(-10,25)[]{$e^-$}
\Text(-10,50)[]{$e^+$}
\Text(107,37)[]{$H$}
\Text(99,20)[]{$t$}
\Text(99,60)[]{$\bar t$}
\put(-30, 65){\bf (b)}
\hspace*{5.cm}
\ArrowLine(0,25)(40,50)
\ArrowLine(0,75)(40,50)
\Photon(40,50)(90,50){4}{5.5}
\DashLine(90,50)(130,25){4}
\DashLine(105,40)(130,55){4}
\Text(62,36)[]{$\bullet$}
\Photon(90,50)(130,75){4}{5.5}
\Text(-10,25)[]{$e^-$}
\Text(-10,50)[]{$e^+$}
\Text(50,50)[]{$Z^*$}
\Text(99,20)[]{$H$}
\Text(99,40)[]{$H$}
\Text(99,60)[]{$Z$}
\ArrowLine(200,25)(240,25)
\ArrowLine(200,75)(240,75)
\ArrowLine(240,25)(290,0)
\ArrowLine(240,75)(290,100)
\Photon(240,25)(280,50){4}{5.5}
\Photon(240,75)(280,50){4}{5.5}
\DashLine(280,50)(320,50){4}
\DashLine(320,50)(340,70){4}
\DashLine(320,50)(340,30){4}
\Text(200,36)[]{\Large $\bullet$}
\Text(130,25)[]{$e^-$}
\Text(130,50)[]{$e^+$}
\Text(195,19)[]{$W^*$}
\Text(195,50)[]{$W^*$}
\Text(247,48)[]{$H$}
\Text(247,28)[]{$H$}
\Text(210,5)[]{$\nu_e$}
\Text(210,69)[]{$\bar{\nu}_e$}
\end{picture}
\end{center}
\vspace*{-7mm}
\caption[Feynman diagrams for the various Higgs production  mechanisms at ILC.]
{Diagrams for the dominant (a) and subleading (b) Higgs production 
mechanisms at ILC.}
\label{Hfig:proc}
\vspace*{-.2cm}
\end{figure}
\begin{figure}[!h]
\begin{center}
\includegraphics[width=0.99\linewidth,bb=73 490 600 745]{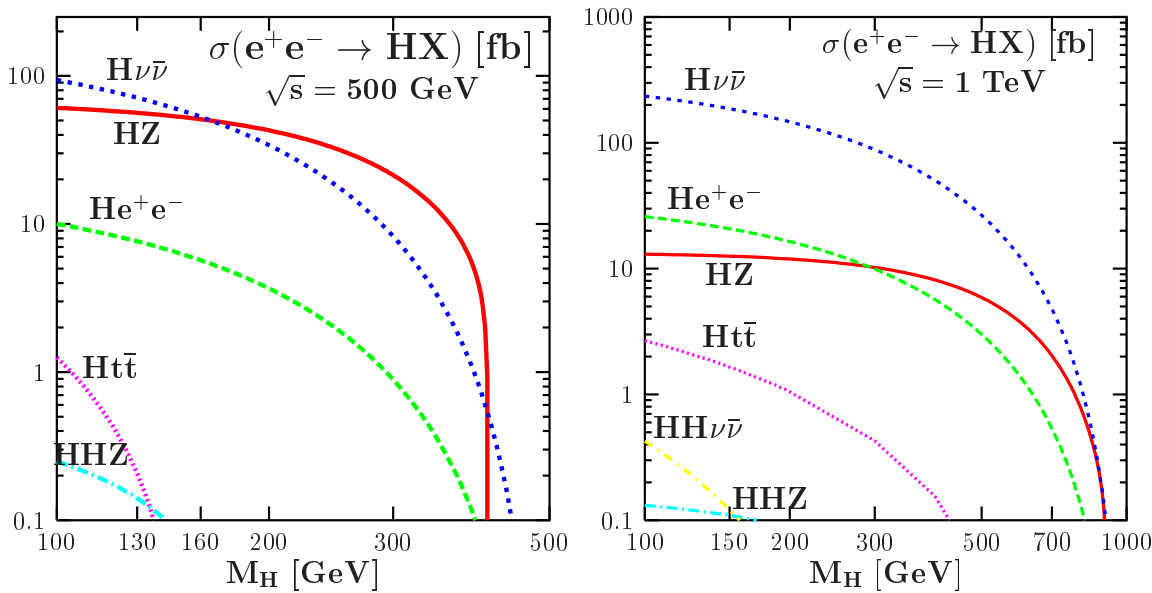} 
\end{center}
\caption[Production cross sections of the SM Higgs boson at 500 GeV and 1 TeV 
ILC] {Production cross sections of the SM Higgs boson at the ILC as a
function of $M_H$ for $\sqrt s=500$ GeV (left) and $\sqrt s=1$ TeV (right);
from Ref.~\cite{Djouadi:2005gi}.}
\vspace{-.3cm}
\label{Hfig:xs}
\end{figure}

In $\eei$ collisions, the main production mechanisms for the SM Higgs particles
are, Fig.~\ref{Hfig:proc}a, the Higgs--strahlung \cite{H-LQT,Higgs:R1} and the
$WW$ fusion \cite{Higgs:R2} processes
\begin{eqnarray}
\eei \to ZH \to f\bar f H  \ \  {\rm and} \ \ \eei \to \bar{\nu}_e \nu_e H
\end{eqnarray}
The final state $H\nu \bar \nu$ is generated in both the fusion and
Higgs--strahlung processes.  Besides the $ZZ$ fusion mechanism \cite{Higgs:R2}
$\eei \to \eei H$ which is similar to $WW$ fusion but with an order of magnitude
smaller cross section, sub--leading Higgs production channels,
Fig.~\ref{Hfig:proc}b, are associated production with top quarks $\eei \to
t\bar{t}H$  \cite{Higgs:R3} and double Higgs production  \cite{Higgs:R4,
Higgs:R4b} in the Higgs--strahlung $\eei \to ZHH$ and fusion  $\eei \to
\bar{\nu} \nu HH$ processes. Despite the smaller production rates,  the latter
mechanisms are very useful when it comes to the study of the Higgs fundamental
properties. The production  rates for all these processes are shown in
Fig.~\ref{Hfig:xs} at energies $\sqrt s\!=\!500$ GeV and $\sqrt s\!=\!1$ TeV as
a function of $M_H$. Other sub--leading processes such as associated production
with a photon $\eei \to H\gamma$ and loop induced pair production $\eei \to HH$
have even smaller rates and will not be discussed here.

The cross section for Higgs--strahlung scales as $1/s$ and therefore dominates
at low energies, while the one of the $WW$ fusion mechanism rises like
$\log(s/M_H^2)$ and becomes more important at high energies. At $\sqrt{s} \sim
500$ GeV, the two processes have approximately the same cross sections, ${\cal
O} (50~{\rm fb})$ for the interesting Higgs mass range 115 GeV $\lsim M_H \lsim$
200 GeV favored by high--precision data.  For the expected ILC integrated
luminosity ${\cal L} \sim 500$ fb$^{-1}$, approximately 30000 and 40000 events
can be collected in, respectively, the $\eei \to HZ$ and $\eei \to \nu \bar \nu
H$ channels for $M_H \sim 120$ GeV.  This sample is more than enough to observe
the Higgs particle at the ILC and to study its properties in great detail.

Turning to the sub--leading processes, the $ZZ$ fusion mechanism $\eei \to
H\eei$ is similar to $WW$ fusion but has a cross section that is one order of
magnitude smaller as a result of the smaller neutral couplings compared to the
charged current couplings. However, the full final state can be reconstructed in
this case. Note that at $\sqrt s\gsim 1$ TeV, the cross section for this 
process is larger than that of Higgs--strahlung for $M_H\lsim 300$ GeV.

The associated production with top quarks has a very small cross section at
$\sqrt s =500$ GeV due to phase space suppression but, at $\sqrt s=800$ GeV, it
can reach the level of a few femtobarns.  The $t\bar{t}H$ final state is
generated almost exclusively through Higgs--strahlung off top quarks and the
process allows thus the determination of the important $g_{Htt}$ Yukawa coupling
in an almost unambiguous way. The electroweak and QCD corrections are known and
are moderate \cite{HO-Htt}, except near the production threshold where large
coulombic corrections occur and double the production rate
\cite{Farrell:2006xe}. For $M_H \lsim 140$ GeV, the main signal $t \bar t H \to
W^+W^- b\bar{b} b \bar{b}$ is spectacular and $b$--quark tagging as well as the
reconstruction of the Higgs mass peak are essential to suppress the large
backgrounds. For larger Higgs masses, $M_H \gsim 140$ GeV, the process leads
mainly to $H t \bar t \to 4W b\bar b$ final states which give rise to ten jets
if all $W$ bosons are allowed to decay hadronically to increase the
statistics.    

The cross section for double Higgs production in the strahlung process, $\eei
\to HHZ$, is at the level of $\sim \frac12$ fb at $\sqrt{s}=500$ GeV for  a
light Higgs boson, $M_H \sim 120$ GeV, and is smaller at higher energies
\cite{Higgs:R4b}. It is rather sensitive to the trilinear Higgs--self coupling
$\lambda_{HHH}$: for $\sqrt{s}\!=\!500$ GeV and $M_H\!=\!120$ GeV for instance,
it varies by about 20\% for a 50\% variation of $\lambda_{HHH}$.  The
electroweak corrections to the process have been shown to be moderate
\cite{HO-HHH}.  The characteristic signal  for $M_H \lsim 140$ GeV consists of
four $b$--quarks to be tagged and a $Z$ boson which needs to be reconstructed in
both leptonic and hadronic final states to increase the statistics.  For higher
Higgs masses, the dominant signature is $Z+4W$ leading to multi--jet (up to 10)
and/or multi--lepton final states.  The rate for double Higgs production in 
$WW$ fusion, $\eei \to \nu_e \bar{\nu}_e HH$, is extremely small at $\sqrt{s}
=500$ GeV but reaches the level of $\frac12$ fb at 1 TeV; in fact, at high
energies, only the latter process can be used.


Finally, future linear colliders can be turned to $\gamma \gamma$
colliders, in which the photon beams are generated by Compton
back--scattering of laser light with c.m.~energies and integrated
luminosities only slightly lower than that of the original $\eei$
collider. Tuning the maximum of the $\gamma \gamma$ spectrum to the
value of $M_H$, the Higgs boson can be formed as $s$--channel
resonances, $\gamma \gamma \to H$, decaying mostly into $b\bar{b}$
pairs and/or $WW^*,ZZ^*$ final states. This allows precise
measurement of the Higgs couplings to photons, which are mediated by
loops possibly involving new particles \cite{ggtdr} as well as the 
CP nature of the Higgs particle \cite{HCPR,Niezurawski:2003ik}.

\subsection{Higgs detection at the ILC}

\begin{figure}[!h]
\vspace*{-1.cm}
\begin{center}
\hspace*{-.2mm}
\begin{minipage}{7cm}
\epsfig{file=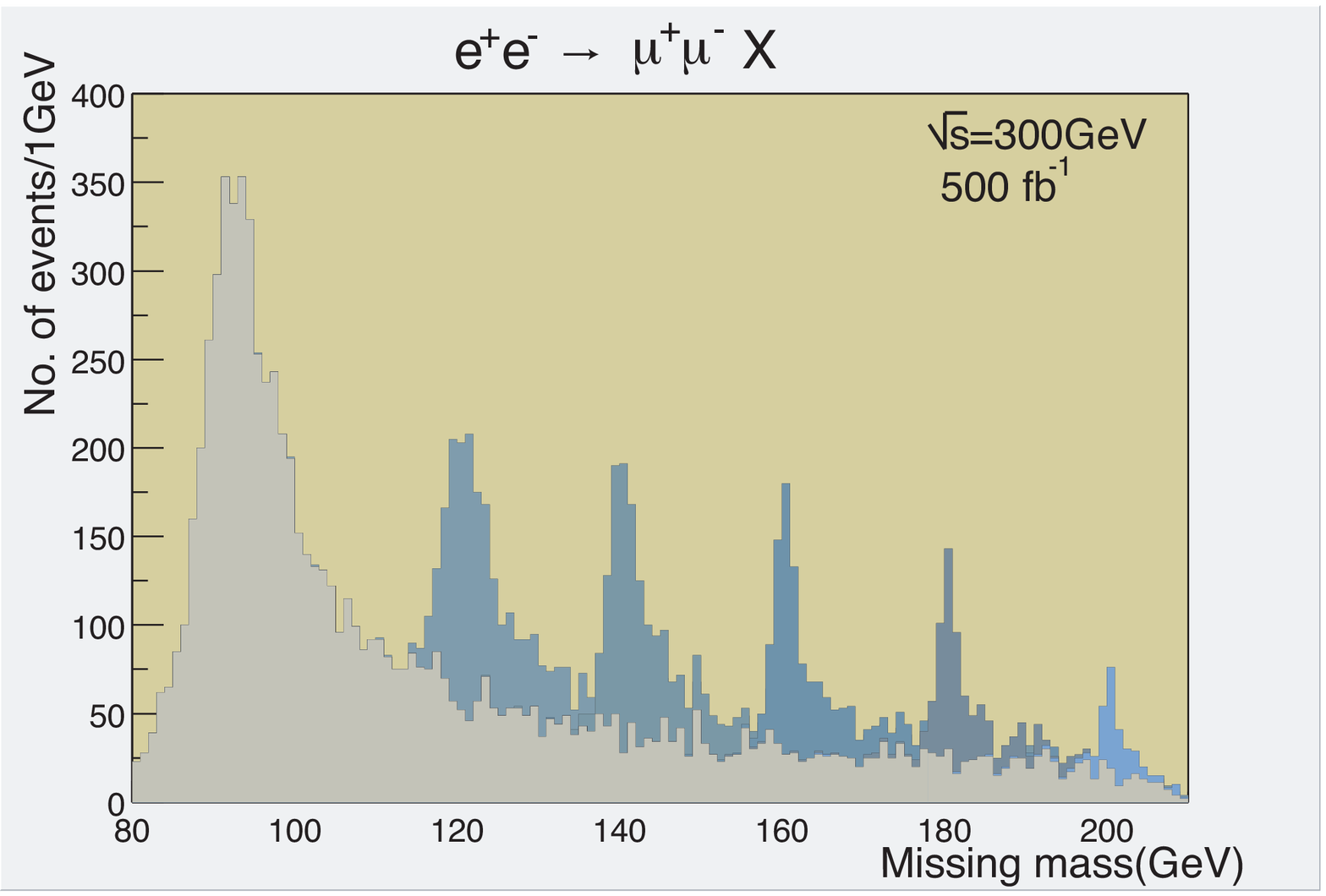,width=9cm,height=6.8cm,angle=0}
\end{minipage}
\hspace*{2.1cm}
\begin{minipage}{6cm}
\vspace*{.9cm}
\epsfig{file=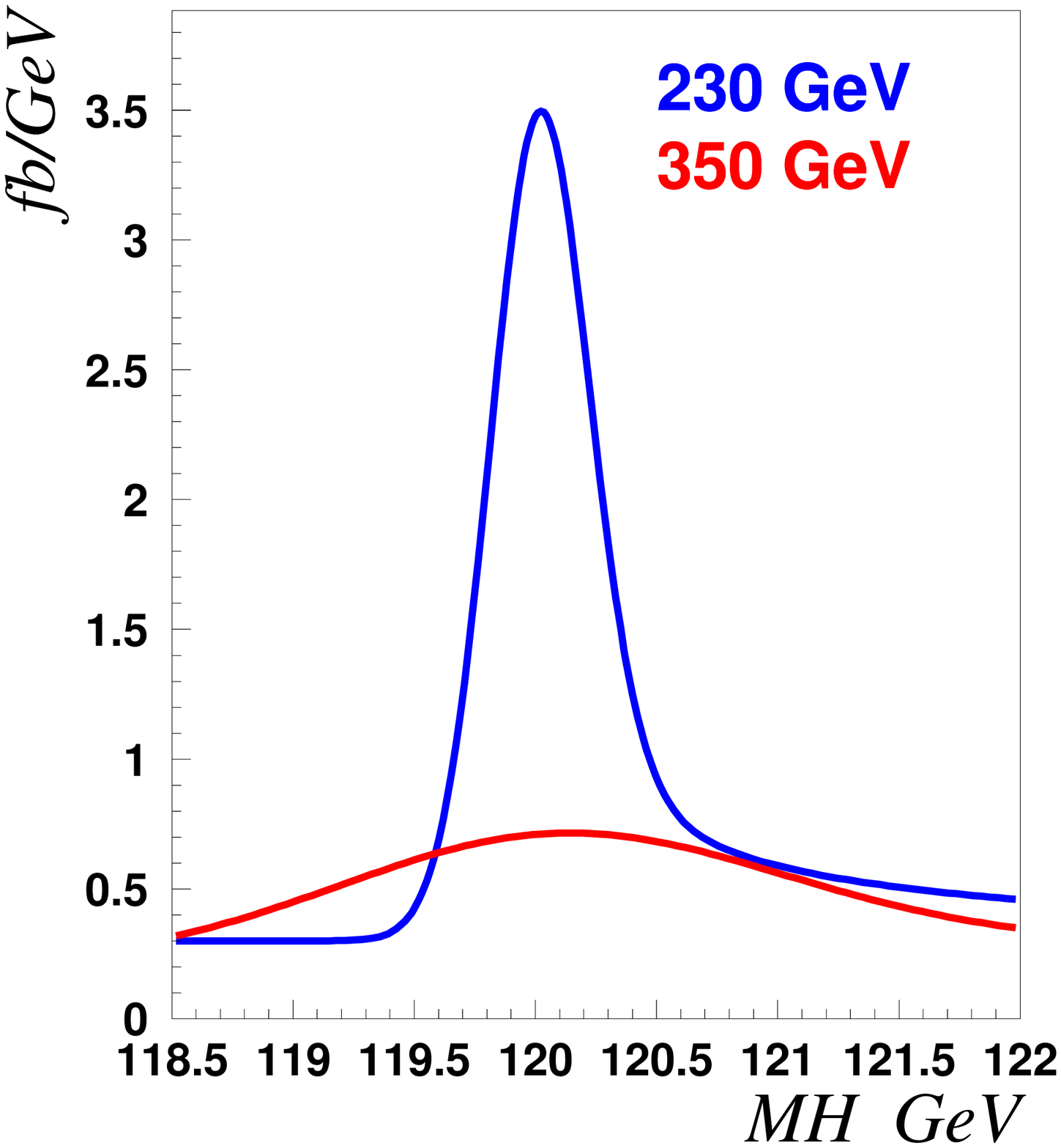,width=6.8cm}
\hspace*{-2.5cm}
\end{minipage}
\end{center}
\vspace*{-15mm}
\caption[Distributions of the dimuons recoiling against a SM Higgs boson at the
ILC.] 
{Left: distribution of the $\mu^+\mu^-$ recoil mass in $e^+e^- \to
\mu^+ \mu^- X$; the background from $Z$ pair production and the SM Higgs signals
with various masses are shown \cite{Abe:2001gc}. Right: differential cross
section for $\ee \to HZ \to H\mu^+ \mu^-$ for two different c.m. energies
with $M_H=120$ GeV \cite{Richard}.}
\label{Hfig-GLC-missingmass}
\vspace*{-.3cm}
\end{figure}

In Higgs--strahlung, the recoiling $Z$ boson is mono--energetic and the Higgs
mass can be derived from the $Z$ energy since the initial $e^\pm$ beam energies
are sharp when beamstrahlung is ignored (the effects of beamstrahlung must be
thus suppressed as strongly as possible).  The $Z$ boson can be tagged through
its clean $\ell^+ \ell^-$ decays ($\ell\!=\!e,\mu$) but also through decays
into quarks which have a much larger statistics. Therefore, it will be easy to
separate the signal from the backgrounds, Fig.~\ref{Hfig-GLC-missingmass}
(left). In the low mass range, $M_H\!\lsim\!140$ GeV, the process leads to
$b\bar{b}q\bar{q}$ and $b\bar{b}\ell \ell$ final states, with the $b$ quarks
being efficiently tagged by  micro--vertex detectors. For $M_H\!\gsim\!140$ GeV
where the decay $H \to WW^*$ dominates, the Higgs boson can be reconstructed by
looking at the $\ell \ell + \,$4--jet or 6--jet final states, and using the
kinematical constraints on the fermion invariant masses which peak at $M_W$ and
$M_H$, the backgrounds are efficiently suppressed. Also the $\ell \ell q\bar q 
\ell \nu$ and $q\bar q q\bar q  \ell \nu$ channels are easily accessible.

It has been shown in detailed simulations \cite{Aguilar-Saavedra:2001rg,H-Desch}
that only a few fb$^{-1}$ data are needed to obtain a 5$\sigma$ signal for a
Higgs boson with a mass $M_H \lsim 150$ GeV at a 500 GeV collider, even if it
decays invisibly (as it could happen e.g.~in the MSSM).  In fact, for such small
masses, it is better to move to lower energies where the Higgs--strahlung cross
section is larger and the reconstruction of the $Z$ boson is better
\cite{Richard}; for $M_H \sim 120$ GeV, the optimum energy is $\sqrt s=230$
GeV as shown in Fig.~\ref{Hfig-GLC-missingmass}  (right). Moving to higher
energies, Higgs bosons with masses up to $M_H\sim 400$ GeV can be discovered in
the Higgs--strahlung  process at an energy of 500 GeV and with a luminosity of
500 fb$^{-1}$. For even larger masses, one needs to increase the c.m. energy of
the collider and, as a rule of thumb, Higgs masses up to $\sim 80$\% $\sqrt{s}$
can be probed. This means that a 1 TeV collider can probe the entire Higgs mass
range that is theoretically allowed in the SM, $M_H \lsim 700$ GeV.

\begin{figure}[!h]
\vspace*{-1mm}
\begin{center}
\mbox{
\epsfig{file=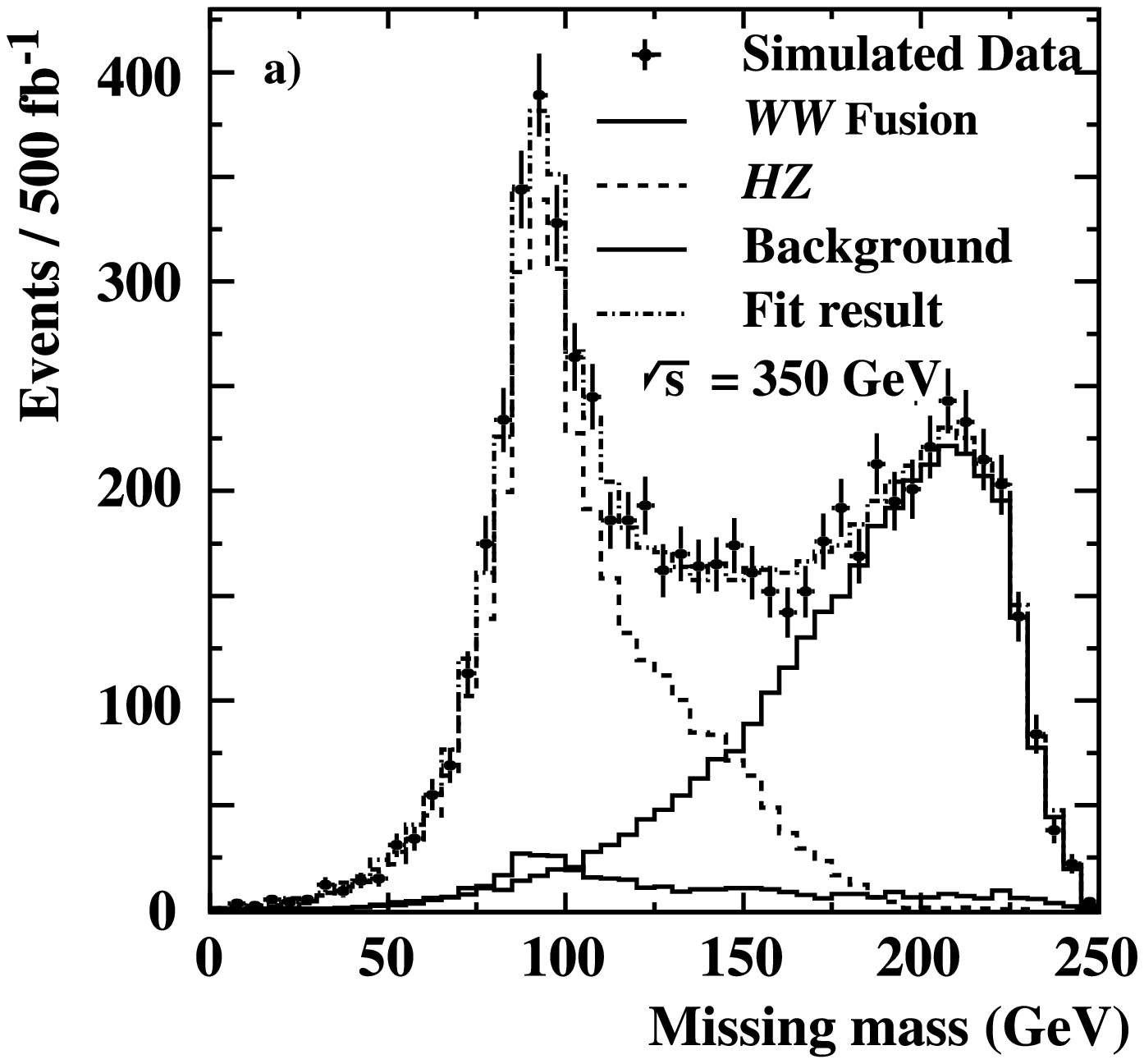,width=0.42\linewidth}\hspace*{1cm}
\epsfig{file=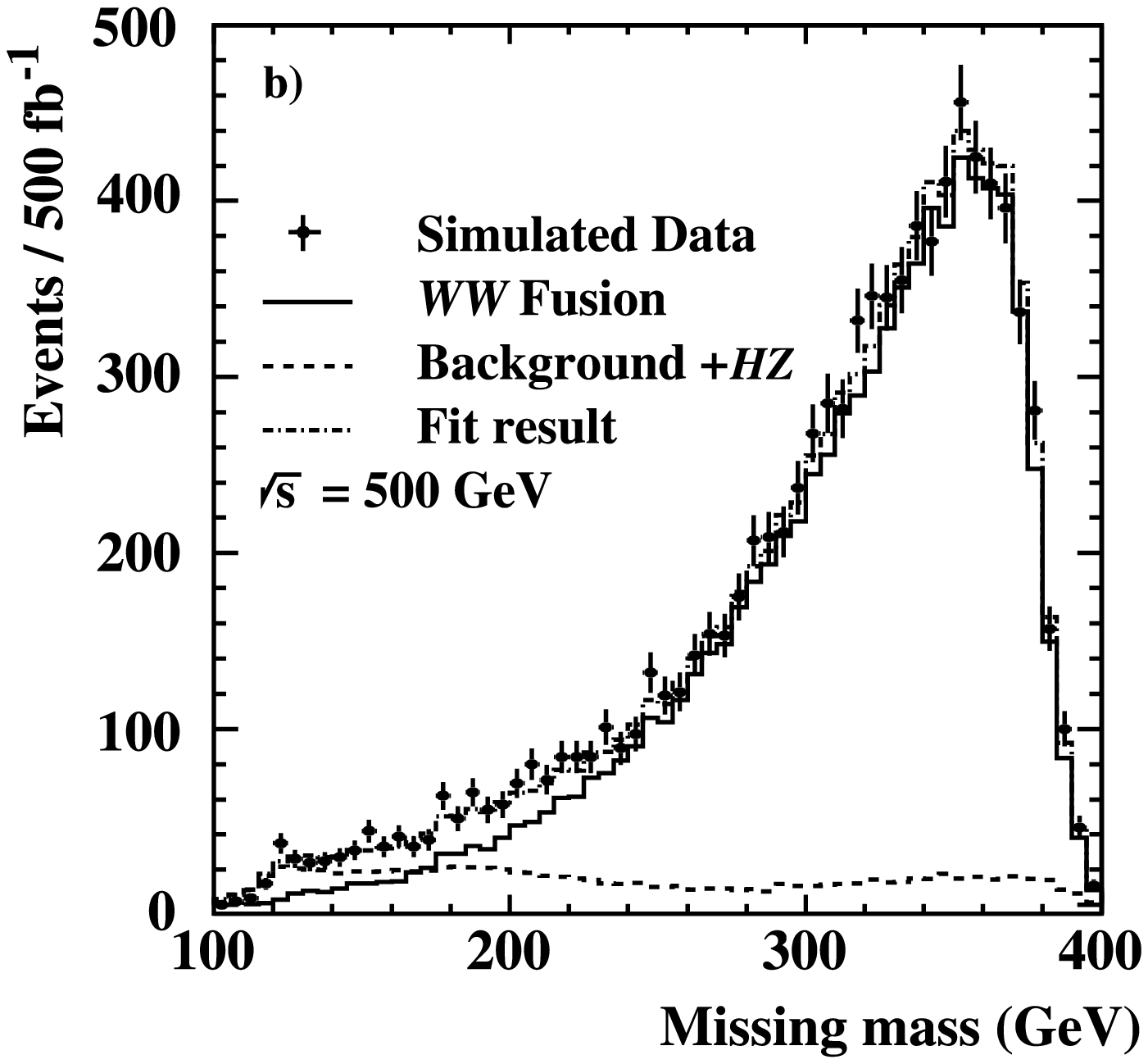,width=0.42\linewidth}
}
\end{center}
\vspace*{-6mm}
\caption[The missing mass in ${\nu \bar{\nu} b \bar b}$ final states from WW fusion and Higgs-strahlung]
{The missing mass distribution in the $\nu \bar{\nu} b \bar b$ final state at
$\sqrt{s}\!=\!350\,{GeV}$ (a) and 500 GeV (b) for $M_H\!=\!120$ GeV in $WW$
fusion, Higgs-strahlung, the interference, as well as for the background
\cite{Aguilar-Saavedra:2001rg}.}
\label{Hfig-nunua}
\vspace*{-.5cm}
\end{figure}

The $WW$ fusion mechanism offers a complementary production channel. For low
$M_H$  where the decay $H\to b\bar{b}$ is dominant, flavor tagging plays an
important role to suppress the 2--jet plus missing energy background.  The $\eei
\to H\bar{\nu}\nu \to b\bar{b}\bar{\nu}\nu$ final state can be separated
\cite{Aguilar-Saavedra:2001rg} from the corresponding one in the
Higgs--strahlung process, $\eei \to HZ \to b\bar{b}\bar{\nu}\nu$, by exploiting
their different characteristics in the $\nu \bar{\nu}$ invariant mass which are
measurable through the missing mass distribution; Fig.~\ref{Hfig-nunua}.  The
polarization of the electron and positron beams, which allows tuning of the $WW$
fusion contribution, can be very useful to control the systematic
uncertainties.  For larger Higgs masses, when the decays $H \to
WW^{(*)},ZZ^{(*)}$ and even $t\bar t$ are dominant, the backgrounds can be
suppressed using kinematical constraints from the reconstruction of the Higgs
mass peak and exploiting the signal characteristics.

\subsection{Determination of the SM Higgs properties}

Once the Higgs boson is found it will be of great importance to explore all its
fundamental properties. This can be done in great detail in the clean
environment of $\eei$ linear colliders: the Higgs boson mass, its spin  and
parity quantum numbers and its couplings to fermions, massive and massless gauge
bosons as well as its trilinear self--couplings can be measured with very high
accuracies. The measurements would allow to probe in all its facets the
electroweak symmetry breaking mechanism in the SM and probe small manifestations
of new physics. 

\subsubsection*{\underline{The Higgs mass}}

\begin{figure}[ht!]
\vspace*{-0mm}
\begin{center}
\begin{tabular}{c c}
{{\epsfig{file=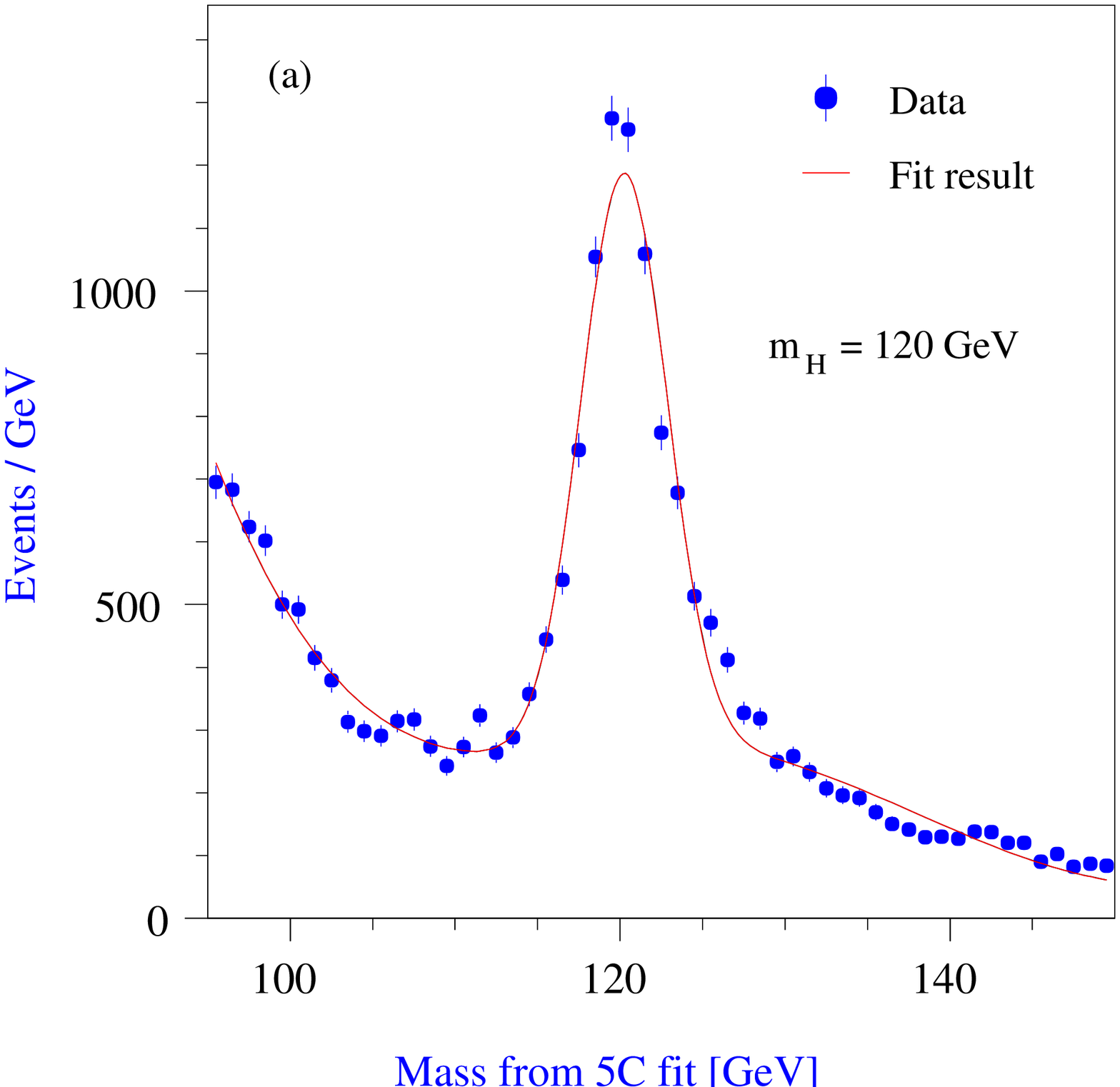,width=0.45\linewidth}}} \ & \
{{\epsfig{file=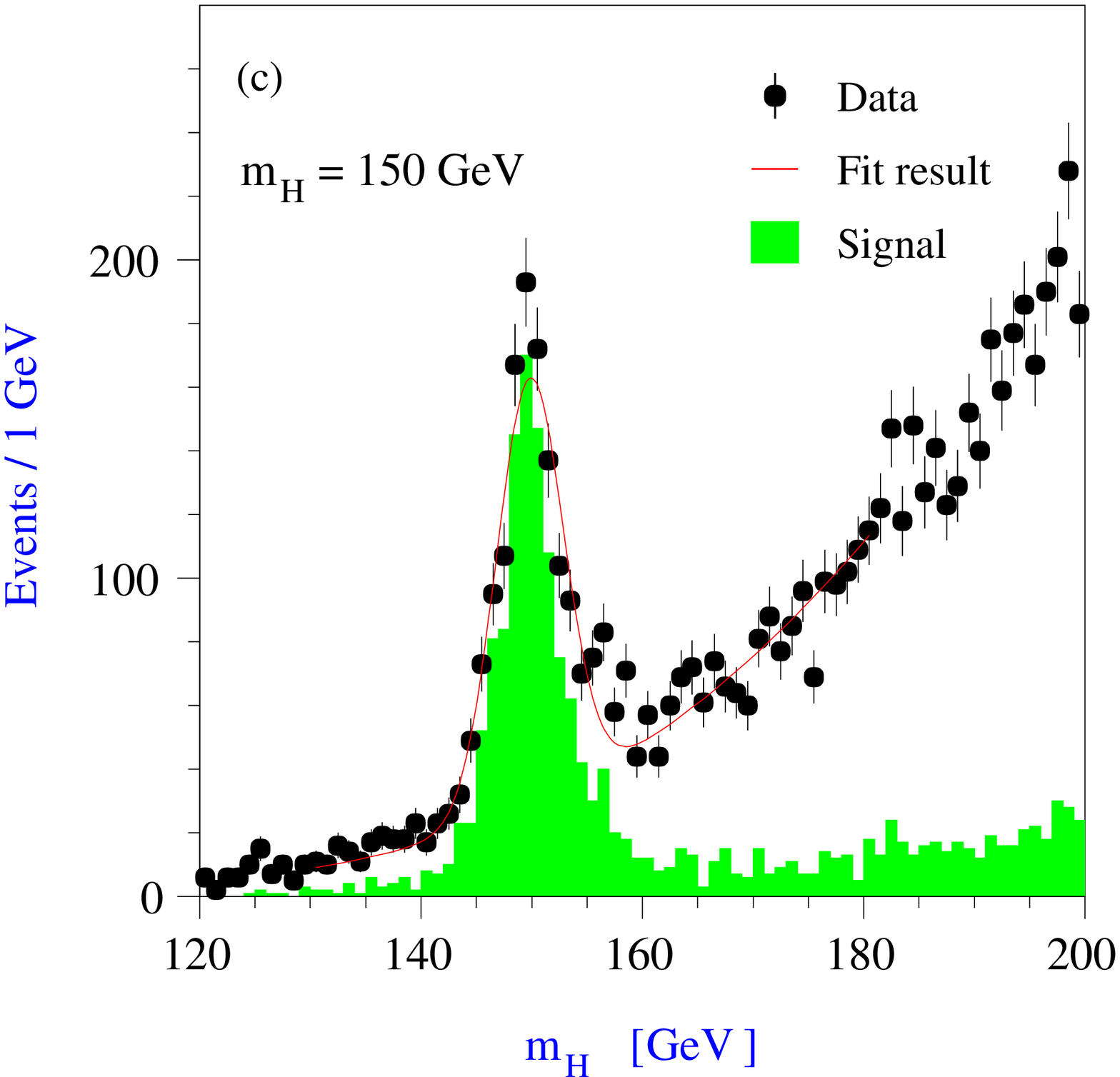,width=0.45\linewidth}}} \\
\label{Hfig:mass}
\end{tabular}
\end{center}
\vspace*{-1.3cm}
\caption[Higgs mass peaks reconstructed in different channels with constrained fits]
{The Higgs mass peaks reconstructed in different channels with
constrained fits for two values of $M_H$, a luminosity of 500\,fb$^{-1}$ and
$\sqrt{s} =350~{GeV}$: $HZ \rightarrow b \bar b q \bar q$ at $M_H = 120~{GeV}$
(left) and $HZ \rightarrow WW^* q \bar q$ at $M_H = 150~{GeV}$ (right); from
Ref.~\cite{Aguilar-Saavedra:2001rg}.}
\vspace*{-6mm}
\end{figure}

Many of the properties of the SM Higgs boson can be determined in a model
independent way by exploiting the recoil mass technique in the Higgs--strahlung
process, $\eei \to HZ$.  The measurement of the recoil $\ell^+ \ell^-$ mass in
$\eei \to ZH\to H\ell \ell$ allows a very good determination of the Higgs  mass
\cite{Garcia-Abia:1999kv}. At $\sqrt{s}=350$ GeV and with ${\cal L}= 500$
fb$^{-1}$, a precision of $ \Delta M_H \sim 70$ MeV can be reached for $M_H \sim
120$ GeV. The precision can be increased to $\Delta M_H \sim 40$ MeV by using
the hadronic decays of the $Z$ boson in addition \cite{Garcia-Abia:2005mt}. Note
that here, running at energies $\sqrt s \sim M_H+100$ GeV is more adequate as
the production cross section is largest and the resolution on the $Z\to \ell
\ell$  decays is better \cite{Richard}. For $M_H=150$--180 GeV when the Higgs
boson decays mostly into gauge bosons, accuracies of the same order can also be
reached.   The reconstructed Higgs mass peaks are shown in Fig.~\ref{Hfig:mass}
at a c.m. energy of $\sqrt s=350$ GeV in the channels $HZ \rightarrow b \bar b q
\bar q$ and $HZ \rightarrow WW^* q \bar q$.

\subsubsection*{\underline{The Higgs spin and parity}}

\begin{figure}[h!]
\begin{center}
\begin{minipage}{6.5cm}
\psfig{figure=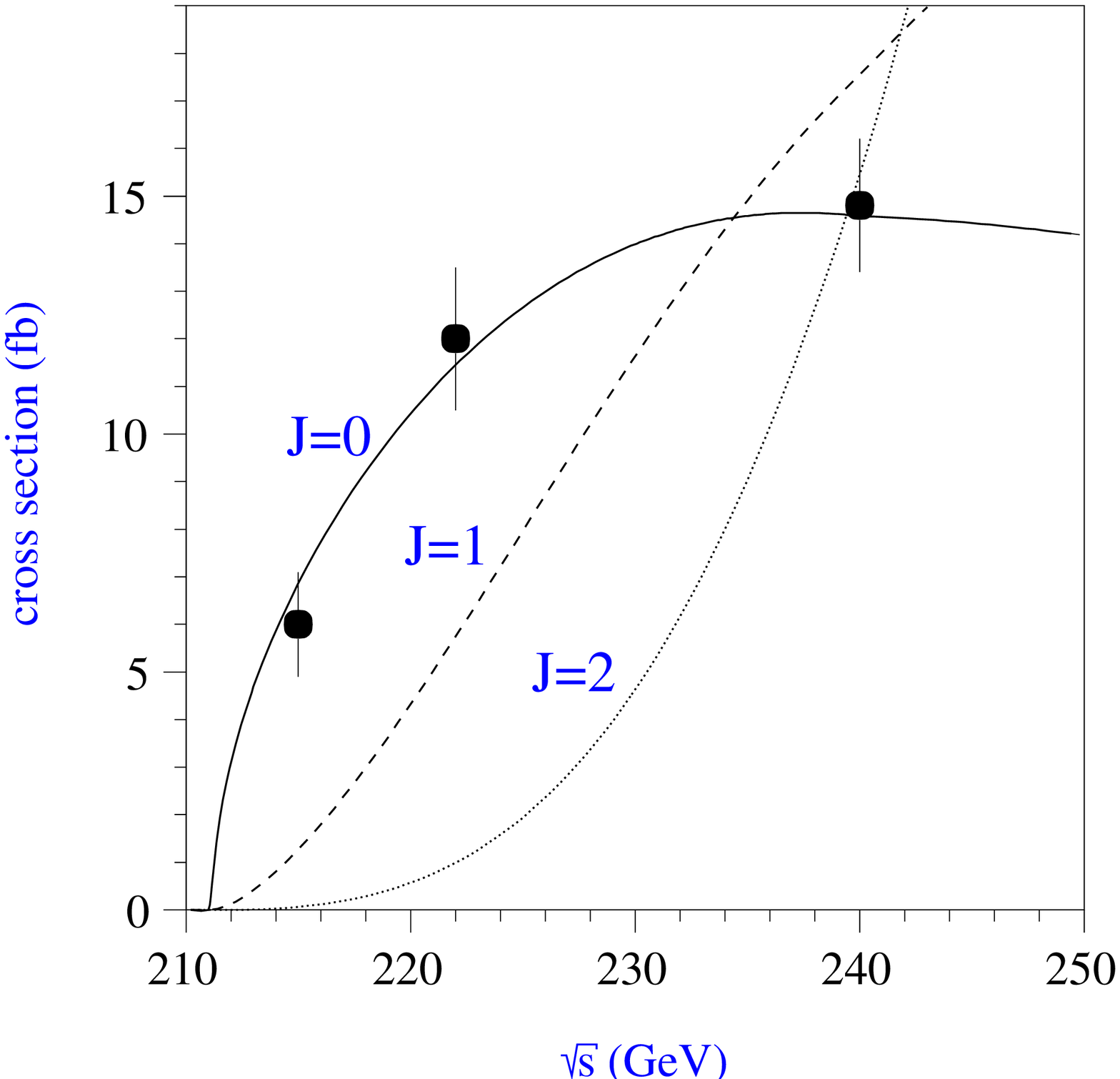,width=6.99cm}
\end{minipage}
\hspace*{5mm}
\begin{minipage}{7cm}
\vspace*{-10.mm}
\psfig{figure=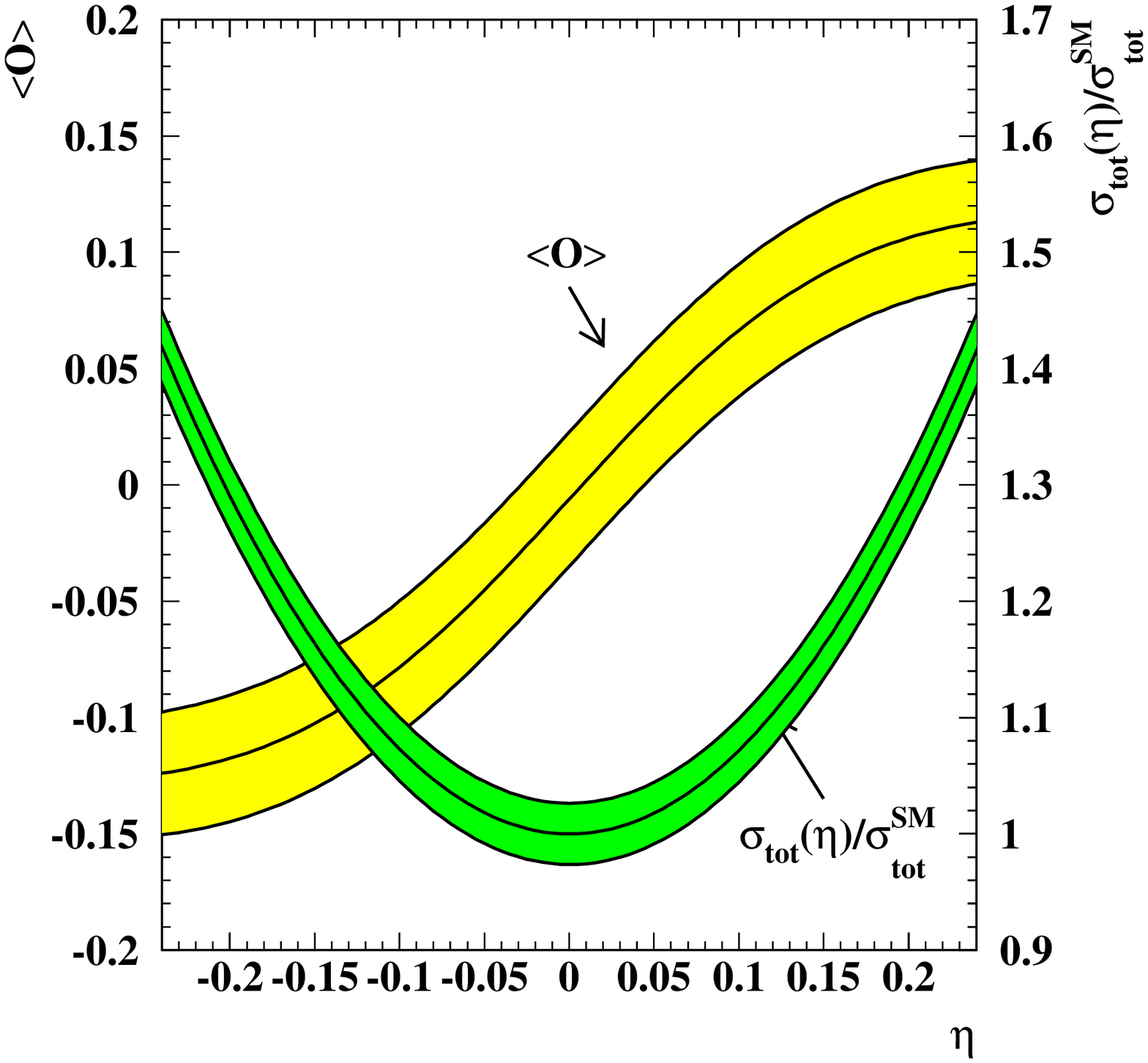,width=6.99cm}
\end{minipage}
\end{center}
\vspace*{-8.mm}
\caption[The determination of the spin and CP--quantum numbers of a SM Higgs]
{The $\eei \to ZH$ cross section energy dependence near
threshold for $M_H=120$ GeV for spin $0^+, 1^-$ and $2^+$ bosons
\cite{Dova:2003py} (left); the determination of the CP mixture $\eta$ with the
bands showing the $1\sigma$ errors at $\sqrt{s}=$ 350 GeV and 500 fb$^{-1}$
\cite{Schumacher:2001ax} (right).}
\label{Hfig:spin}
\vspace*{-3.mm}
\end{figure}

The determination of the $J^{\rm P}=0^{+}$ quantum number of the SM Higgs boson
can also be performed in the Higgs--strahlung process. The measurement of the
rise of the cross section near threshold, $\sigma (\eei \to HZ) \propto
\lambda^{1/2}$, rules out $J^{\rm P}=0^{-}, 1^{-}, 2^{-}$ and higher spin
$3^\pm,  \cdots$, which rise with higher powers of the velocity $\lambda^{1/2}$;
the possibilities $1^{+}, 2^{+}$ can be ruled out by studying angular
correlations  \cite{H-spin}. A threshold scan with a luminosity of 20 fb$^{-1}$
at three c.m. energies is sufficient to distinguish the various
behaviors; Fig.~\ref{Hfig:spin} (left). The angular distribution of the $Z/H$
bosons in Higgs--strahlung is also sensitive to the spin--zero of the Higgs
particle: at high--energies, the $Z$ is longitudinally polarized and the
distribution follows the $\sim \sin^2 \theta$ law which unambiguously
characterizes the production of a $J^P=0^+$ particle. Assuming that the Higgs
particle is a mixed CP--even and CP--odd state with $\eta$ parameterizing the
mixture, the angular distribution can be checked experimentally;
Fig.~\ref{Hfig:spin} (right).  The Higgs $J^{\rm PC}$ quantum numbers can also
be checked by looking at correlations in the production $\eei \to HZ \to 4f$ or
in the decay $H \to WW^*, ZZ^* \to 4f$ processes \cite{Barger:1993wt}.

The CP nature of the Higgs boson would be best tested in the couplings to
fermions, where the scalar and pseudoscalar components might have comparable
size. Such tests can be performed in the decay channel $H \to \tau^+ \tau^-$ for
$M_H \lsim 140$ GeV by studying the spin correlations between the final decay
products of the two $\tau$ leptons \cite{Kramer:1993jn}.  The acoplanarity angle
between the decay planes of the two $\rho$ mesons produced from $\tau^+$ and
$\tau^-$, which can be reconstructed in the Higgs rest frame using the $\tau$
lifetime information, is a very sensitive probe, allowing a discrimination
between a CP--even and CP--odd state at the 95\% CL; additional information from
the $\tau$ impact parameter is also useful. The CP quantum numbers of the Higgs
boson can be determined unambiguously in associated production with top quark 
pairs either by looking at regions of phase space which single out the different
mass effects generated by scalar and pseudoscalar Higgs production  or simply
from the very different threshold behavior of the cross section as well as the
polarization of the final top quarks \cite{HCP-Htt}. 

\subsubsection*{\underline{The Higgs couplings to gauge bosons}}

The fundamental prediction that the Higgs couplings to $W/Z$ bosons are
proportional to the masses of these particles can be easily verified
experimentally since  these couplings can be directly determined by measuring
the production cross sections in the Higgs--strahlung and fusion processes.
$\sigma(\eei \to HZ \to H \ell^+ \ell^-)$ can be measured by analyzing the
recoil mass against the $Z$ boson and provides a determination of the couplings
$g_{HZZ}$ independently of the Higgs decay modes. Adding the two lepton
channels, one obtains an accuracy of less than 3\% at $\sqrt{s}\!\sim\! 350$ GeV
with ${\cal L}\!=\!500$ fb$^{-1}$ \cite{Garcia-Abia:1999kv}. The coupling
$g_{HWW}$ for $M_H\!\lsim\! 2M_W$ can be determined, once the branching ratio of
a  visible channel is available,  from the measurement of $\sigma(\eei \to H\nu
\bar{\nu})$ which, as mentioned previously, can be efficiently separated from
the $\eei \to HZ \to H \nu \bar \nu$ channel and from the backgrounds; a
precision of less than 3\% can also be achieved for $M_H=120$ GeV, but at a
slightly higher energy $\sqrt{s}\!\sim\! 500$ GeV, where the production rate is
larger \cite{Meyer:2004ha}. The precision on the Higgs couplings is half of
these errors, since the cross sections scale as $g_{HVV}^2$ and, thus, a
measurement of the $HVV$  couplings can be performed at the statistical level of
1 to 2\% and would allow probing the quantum corrections.

\subsubsection*{\underline{The Higgs decay branching ratios}}

The measurement of the branching ratios of the Higgs boson
\cite{Abe:2001gc,Hildreth:1993qt,Schumacher:2003ss,Battaglia:1999re,
Brient:2002sk,Borisov:1999mu,Boos:2000bz} is of utmost importance.  For Higgs
masses below $M_H \lsim 140$ GeV, a large variety of branching ratios can be
measured at the ILC, since the $b\bar b, c\bar c$ and $gg$ final states have
significant rates and can be very efficiently disentangled by means of
micro--vertex detectors.  The $b\bar{b}, c\bar{c}$  and $\tau^+ \tau^-$
fractions allow to measure the relative couplings of the Higgs boson to these
fermions and to check the prediction of the Higgs mechanism that they are indeed
proportional to fermion masses. In particular, BR$(H \to \tau^+ \tau^-) \sim
m_{\tau}^2/3\bar{m}_b^2$ allows such a test in a rather clean way.  The gluonic
branching ratio is indirectly sensitive to the $t\bar{t}H$ Yukawa coupling and
would probe the existence of new strongly interacting particles that couple to
the Higgs boson and which are too heavy to be produced directly. The branching
ratio of the loop induced $\gamma \gamma$ and $Z\gamma$ Higgs decays are 
sensitive to new heavy particles and their measurement is thus very important.
The branching ratio of the Higgs decays into $W$ bosons starts to be significant
for $M_H \gsim 120$ GeV and allows measurement of the $HWW$ coupling in a model
independent way. In the mass range 120 GeV $\lsim M_H \lsim 180$ GeV, the $H \to
ZZ^*$ fraction is too small to be precisely measured, but for higher masses it
is accessible and allows an additional determination of the $HZZ$ coupling.

\begin{table}[htbp]
\renewcommand{\arraystretch}{1.3}
\caption[Expected accuracy on the Higgs branching ratio measurements at the ILC.]
{Expected precision of the Higgs branching ratio measurements
at ILC for $M_H=120$ GeV and a luminosity of 500 $fb^{-1}$. Ranges of results
from various studies are shown with c.m. energies of 300 GeV \cite{Abe:2001gc},
350 GeV \cite{Battaglia:1999re,Brient:2002sk,Borisov:1999mu} and 350/500 GeV
\cite{Boos:2000bz}.
}
 \begin{center}
  \begin{tabular}{|c|c|c|}
    \hline
     Decay mode  & Relative precision (\%)   &   References \\
    \hline
      $b\bar{b}$ &1.0--2.4  & \cite{Abe:2001gc}\cite{Battaglia:1999re}
       \cite{Brient:2002sk}\cite{H:KlausBR}
   \\
    \hline
      $c\bar{c}$ &8.1--12.3 &  \cite{Abe:2001gc}\cite{Battaglia:1999re}
       \cite{Brient:2002sk}\cite{H:KlausBR}
  \\
    \hline
      $\tau^+\tau^-$ &4.6--7.1 & \cite{Abe:2001gc} \cite{Battaglia:1999re}
      \cite{Brient:2002sk}
  \\
    \hline
      $gg$ & 4.8--10  &\cite{Abe:2001gc} \cite{Battaglia:1999re}
       \cite{Brient:2002sk}\cite{H:KlausBR}
   \\
    \hline
      $WW$ & 3.6--5.3 & \cite{Abe:2001gc}\cite{Battaglia:1999re}
      \cite{Brient:2002sk} \cite{Borisov:1999mu}\\
    \hline
      $\gamma \gamma$ &23--35  & \cite{Brient:2002sk} \cite{Boos:2000bz}
   \\
    \hline
  \end{tabular}
 \end{center}
 \label{tab-higgs-br}
\vspace*{-2mm}
\end{table}

\begin{figure}[h] \centering
    \includegraphics[width=10cm,height=7cm,angle=0]{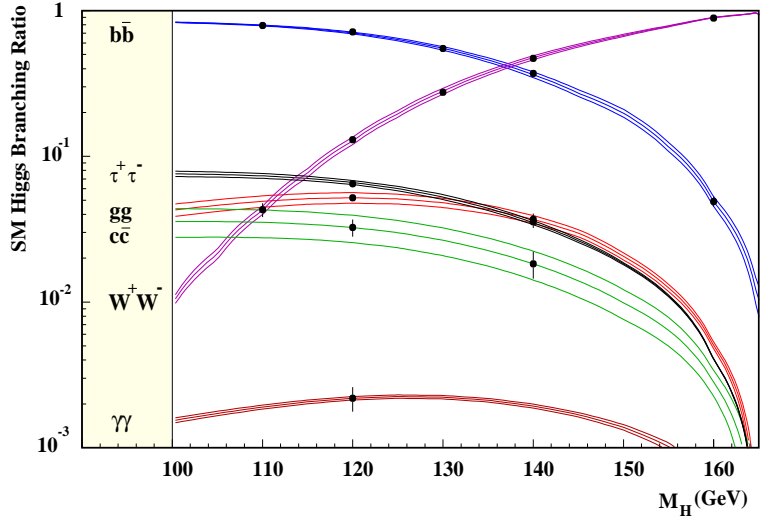}
\vspace*{-5mm}
    \caption[The expected sensitivity on the SM Higgs branching ratios at the ILC] 
    {The branching ratio for the SM Higgs boson
    with the expected sensitivity at ILC. A luminosity of
    500 fb$^{-1}$ at a c.m. energy of 350 GeV
    are assumed; from Ref.~\cite{Battaglia:1999re}.}
\label{fig-TESLA-fig2204}
\vspace*{-5mm}
\end{figure}

There are several studies on the sensitivity of the Higgs branching
ratios for a light SM Higgs boson at ILC. Although each analysis is
based on slightly different assumptions on detector performance,
center-of-mass energy, and analysis method, overall consistent
results are obtained. The accuracies of the branching ratio
measurements for a SM Higgs boson with a mass of 120 GeV are listed
in Tab.~\ref{tab-higgs-br}, while for $M_H=$120, 140 and 160 GeV
from the simulation study of Ref.~\cite{Battaglia:1999re}, they are
shown in Fig.~\ref{fig-TESLA-fig2204}. For $M_H \gsim 180$ GeV, the
available decay modes are limited as the Higgs boson predominantly
decays into two gauge bosons. In such cases, the measurement of at
least one Higgs--fermion coupling is important for establishing the
fermion mass generation mechanism.  The $H \to b\bar{b}$ branching
ratio can be determined with a 12\%, 17\% and 28\% accuracy for,
respectively,  $M_H=180, 200$ and 220 GeV, assuming an integrated
luminosity of 1 ab$^{-1}$ at $\sqrt s =800$ GeV
\cite{Battaglia:2002av}.

Note that invisible Higgs decays can also be probed with a very good accuracy,
thanks to the missing mass technique. One can also look directly for the
characteristic signature of missing energy and momentum.  Recent studies show
that in the range 120 GeV $\lsim M_H \lsim$ 160 GeV, an accuracy of $\sim 10\%$
can be obtained on a $5\%$ invisible decay and a  $5\sigma$ signal can be seen
for a branching fraction as low as 2\%  \cite{Schumacher:2003ss}.

\subsubsection*{\underline{The Higgs total decay width}}

The total decay width of the Higgs boson is large enough,  for $M_H \gsim 2M_W$
GeV, to be accessible directly from the reconstruction of the Higgs  boson
lineshape. For this purpose, it is better to run the ILC at relatively low
energies. It has been shown in  Ref.~\cite{Richard} that, for $M_H=175$ GeV, a
measurement of the width $\Gamma_H\sim 0.5$ GeV to a precision of 10\%  requires
100 fb$^{-1}$ data at $\sqrt s=290$ GeV, while at $\sqrt s=500$ GeV, one needs 5
times more luminosity. 

For smaller Higgs masses, $\Gamma_H$  can be determined indirectly by exploiting
the relation between the total and partial decay widths for some given final
states. For instance, in the decay $H\to WW^*$, the  width is given by $\Gamma_H
= \Gamma(H \to WW^*)/{\rm BR}(H \to WW^*)$ and one can combine the direct
measurement of BR($H \to WW^*)$ and use the information on the  $HWW$ coupling
from $\sigma(\eei \to H\nu \nu)$ to determine the partial width $\Gamma (H\to
WW^*)$.  Alternatively, on can exploit the measurement of the $HZZ$ coupling
from $\sigma(\eei \to HZ)$ for which the mass reach is higher than in $WW$
fusion,  and assume SU(2) invariance to relate the two couplings,
$g_{HWW}/g_{HZZ} = 1/\cos\theta_W$. The accuracy on the total decay width
measurement follows then from that of BR($H \to WW^{(*)})$ and $g_{HWW}$. In the
range 120 GeV\,$\lsim M_H \lsim$\,160 GeV, an accuracy  ranging from 4\% to 13\%
can be achieved on $\Gamma_H$ if $g_{HWW}$ is measured in the fusion process; 
Tab.~\ref{Htab:width}.  This accuracy greatly improves for higher $M_H$ values 
by assuming SU(2) universality and if in addition one measures BR($H\to WW)$  at
higher energies.

\begin{table}[hbt]
\vspace{-.4cm}
\renewcommand{\arraystretch}{1.0}
\caption[Relative precision in the determination of the SM Higgs total decay width.]
{Relative precision in the determination of the SM Higgs decay
width with $\int {\cal L}=500$ fb$^{-1}$ at $\sqrt{s} = 350$ GeV
\cite{Aguilar-Saavedra:2001rg};
the last line shows the improvement which can be obtained when using in
addition measurements at $\sqrt{s}\sim 1$ TeV with $\int {\cal L}=1$ ab$^{-1}$
\cite{H:Barklow}.}
\begin{center}
\begin{tabular}{|c|c|c|c|}
\hline
Channel & $M_H=120$ GeV & $M_H=140$ GeV & $M_H=160$ GeV \\ \hline
$g_{HWW}$ from $\sigma(\eei \to H \nu \nu)$& 6.1\% & 4.5\% & 13.4 \%  \\
$g_{HWW}$ from $\sigma(\eei \to H Z)      $ & 5.6\% & 3.7\% & 3.6 \%  \\
\hline \hline
BR$(WW)$ at $\sqrt{s}=1$ TeV & 3.4\% & 3.6\% & 2.0 \%  \\ \hline
\end{tabular}
\end{center}
\vspace{-.5cm}
\label{Htab:width}
\end{table}

Note that the same technique would allow extraction of the total
Higgs decay width using the $\gamma \gamma$ decays of the Higgs
boson together with the cross section from $\gamma \gamma \to H \to
b\bar b$ as measured at a photon collider. This is particularly true
since the measurement of BR($ H\to \gamma \gamma)$ at $\sqrt s \sim
1$ TeV is rather precise, allowing the total width to be determined
with an accuracy of $\sim 5\%$ with this method for $M_H=120$--140
GeV.

\subsubsection*{\underline{The Higgs Yukawa coupling to top quarks}}

The Higgs Yukawa coupling to top quarks, which is the largest coupling in the
electroweak SM, is directly accessible in the process where the Higgs is
radiated off the top quarks, $\eei \to t\bar{t}H$. Because of the limited phase
space, this measurement can only be performed at high energies $\sqrt{s} \gsim
500$ GeV.  For $M_H \lsim 140$ GeV, the Yukawa coupling  can be measured in the
channel $W Wb\bar{b}b\bar{b}$ with the $W$ bosons decaying both leptonically and
hadronically; $b$--tagging is essential in this mass range
\cite{HttYukawa,agay,juste}.  For higher Higgs masses,  $M_H
\gsim 140$ GeV, the complicated channels with $b\bar b+4W$ have to be
considered, with again, at least two $W$ bosons decaying hadronically, leading
to 2 leptons plus 6 jets and one lepton plus 8 jets, respectively
\cite{agay}.  The next--to--leading QCD corrections to $\sigma(e^+e^- \to t
\bar{t} H)$ have been recently calculated and, at $\sqrt{s}= 500$ GeV, it has
been shown that the total cross section is enhanced by a factor of two by
threshold dynamics \cite{Farrell:2006xe}.

\begin{figure}[!h]
\vspace*{-.2cm}
\centerline{\psfig{figure=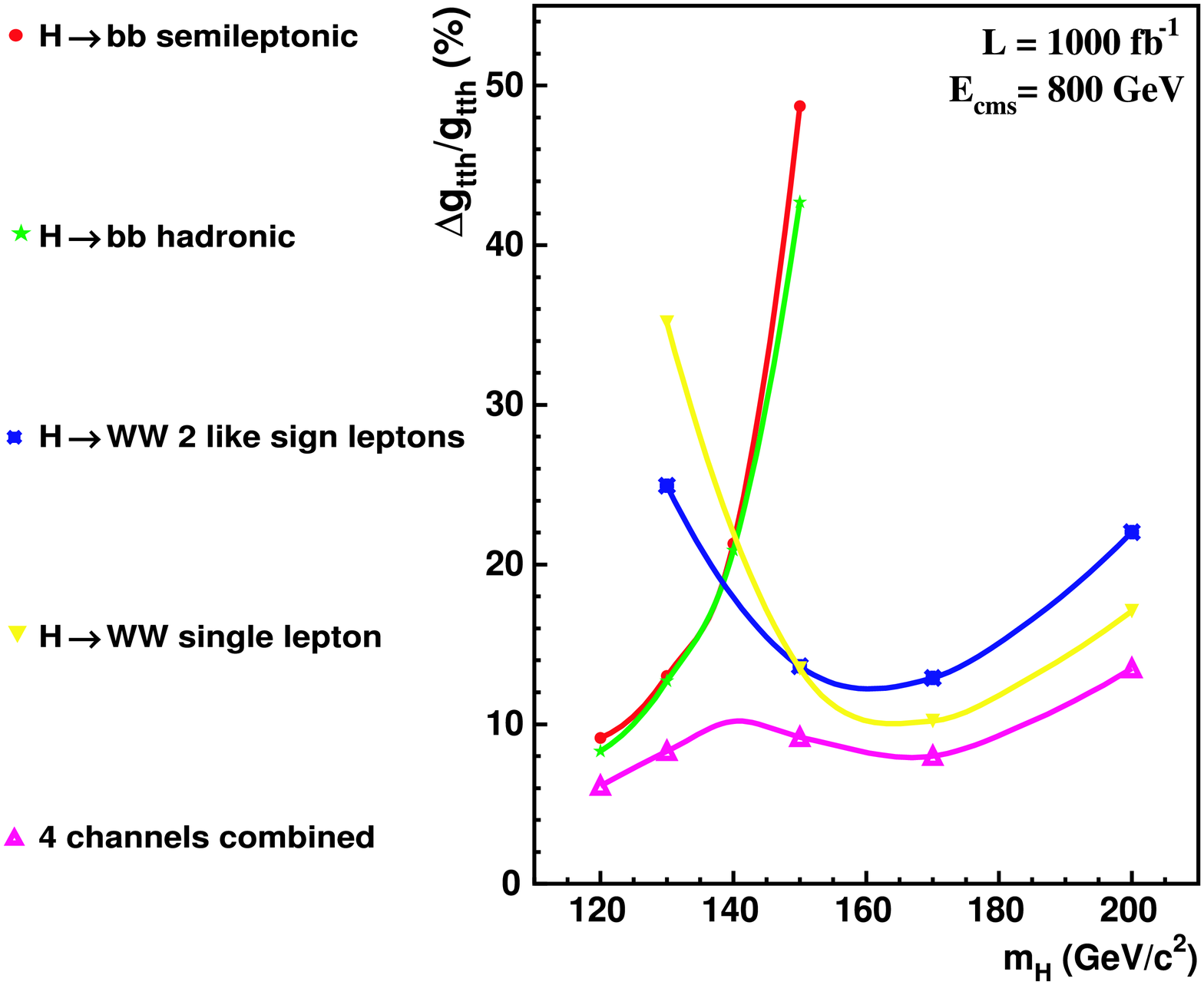,width=12cm,,height=7.5cm}}
\vspace*{-.7cm}
\caption[Expected accuracies for the measurement of the top--Higgs Yukawa coupling]
{Expected accuracies for the measurement of the $Ht\bar t$
coupling as a function of $M_H$ in $\eei \to t \bar t H$ for
$\sqrt{s}= 800$ GeV and 1 ab$^{-1}$ in various decay channels
\cite{agay}.}
\label{Hfig:top}
\vspace*{-.4cm}
\end{figure}

The expected accuracies on the $H t \bar t$ Yukawa coupling are shown in
Fig.~\ref{Hfig:top} as a function of the Higgs mass, for $\sqrt s = 800$ GeV and
a luminosity of 1 ab$^{-1}$. Assuming a 5\% systematical uncertainty on the
normalization of the background, accuracies on the $Ht\bar t$ Yukawa coupling of
the order of 5\% can be achieved for Higgs masses in the low mass range, $M_H
\lsim 140$ GeV, when the $H\!\to\! b\bar b$ decays are dominant; in this case  a
500 GeV ILC can reach an accuracy at the 10\% level \cite{juste}.  A 10\%
measurement of the Yukawa coupling is possible at $\sqrt s=800$ GeV up to Higgs
masses of the order of 200 GeV, when the  $H\!\to\! WW$ channel takes over. Note
that the measurement of this coupling is rather difficult at the LHC; see
chapter \ref{sec:top}. 

For large masses, $M_H \gsim 350$ GeV, the $Ht \bar{t}$ coupling can be derived
by measuring the ratio BR($H \to t\bar{t}$) with the Higgs boson produced in the
Higgs--strahlung and $WW$ fusion processes \cite{Htt-Ztt}. A detailed simulation
\cite{Aguilar-Saavedra:2001rg} shows that once the $t\bar t$ and $\eei t\bar t$
backgrounds are removed, an accuracy of 5\% (12\%) for $M_H=400$ (500) GeV can
be achieved on $g_{Htt}$, again at a c.m. energy of $\sqrt{s}= 800$ GeV and with
${\cal L} \sim 1$  ab$^{-1}$ data \cite{Alcaraz:2000xr}.

\subsubsection*{\underline{The trilinear Higgs coupling}}

\begin{figure}[!h]
\vspace*{-.1cm}
\centerline{
\includegraphics[width=7cm,height=5.8cm]{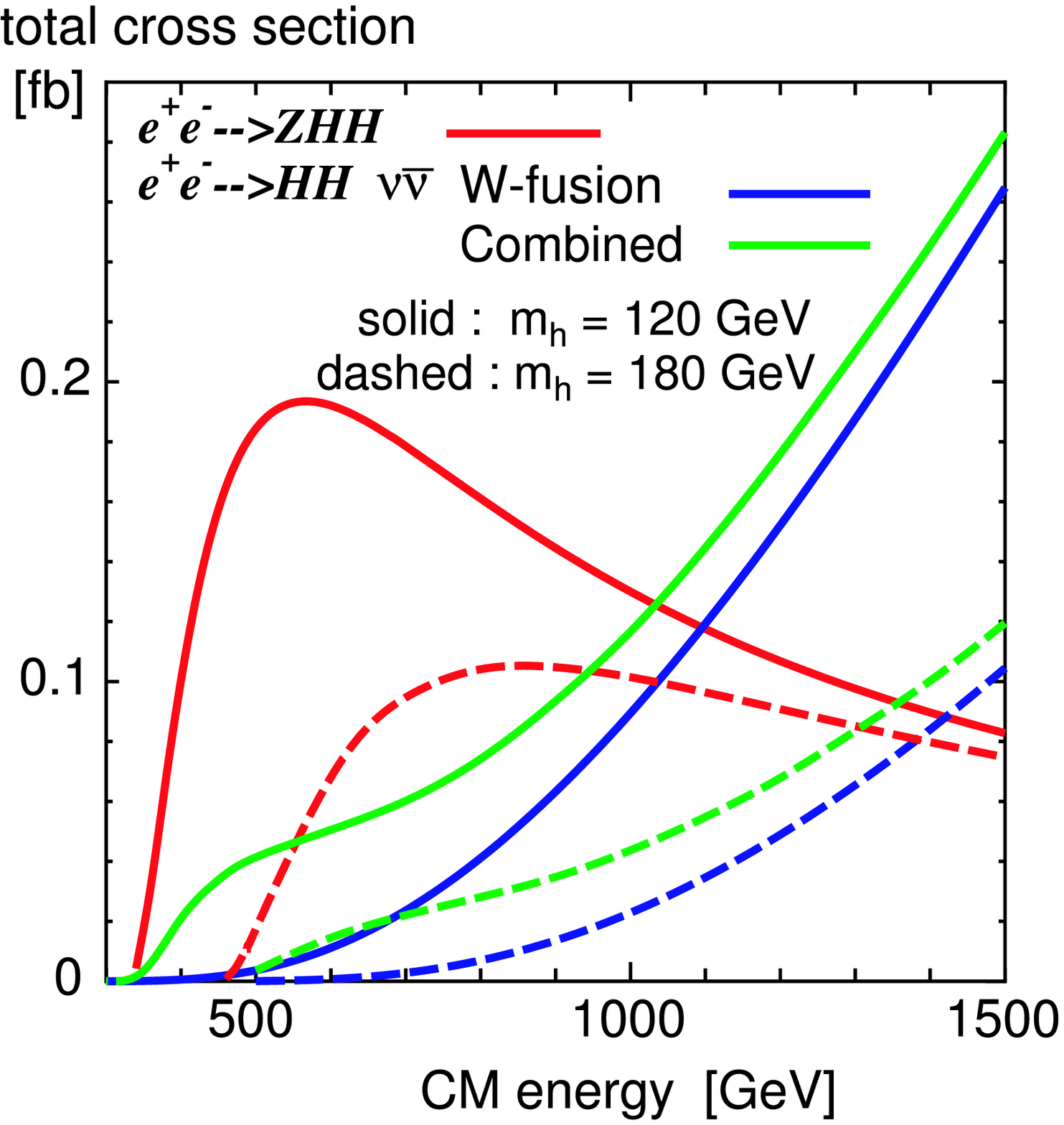}\hspace*{5mm}
\psfig{figure=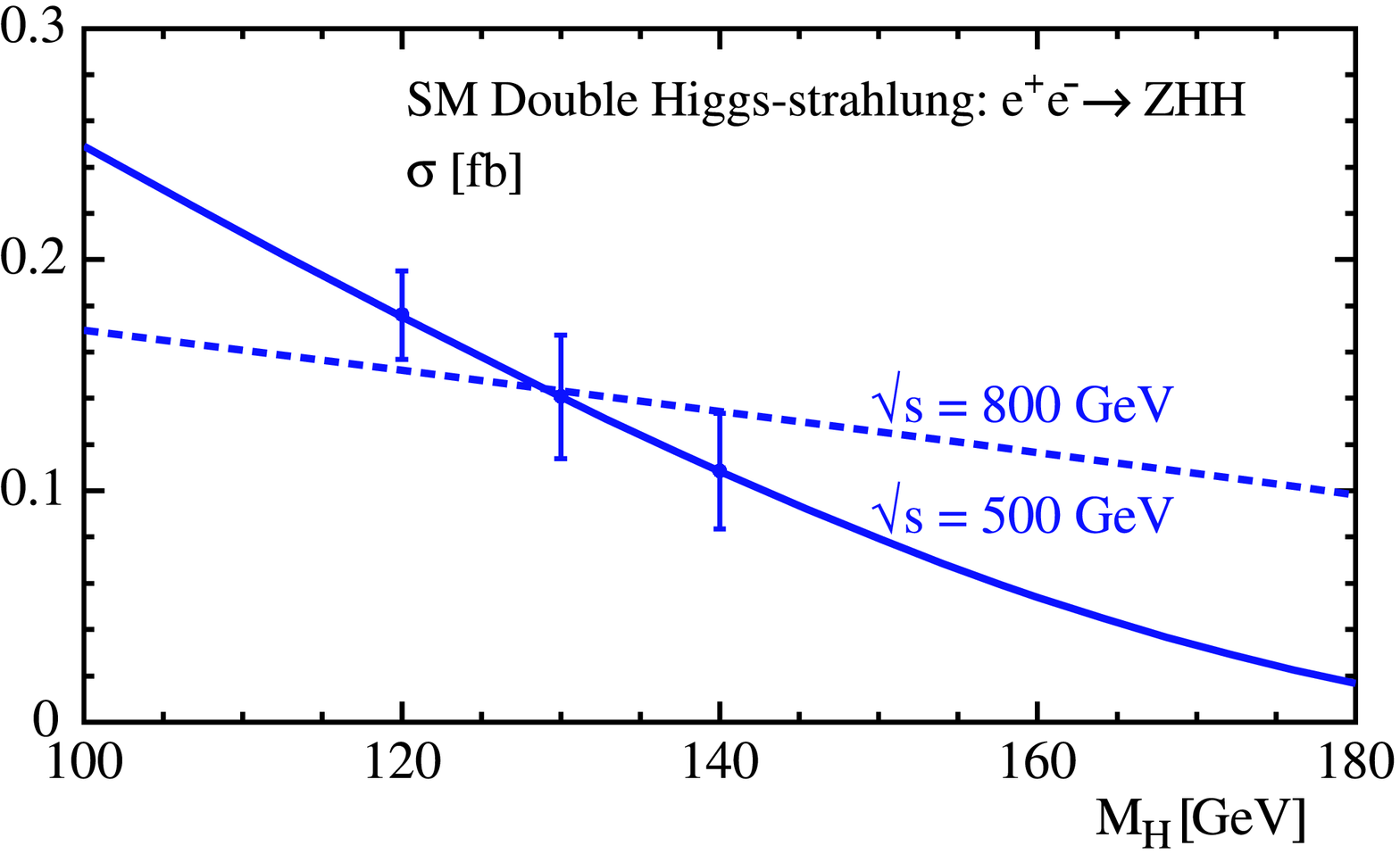,width=7.cm,height=5.8cm}}
\vspace*{-7.mm}
\caption[Cross sections for double Higgs production and their determination
at ILC]
{The separate and combined production cross sections for the $ZHH$ 
and $\nu \bar{\nu}HH$ processes as a function of $\sqrt s$ \cite{Yasui:2002se} 
(left) and the accuracy in the determination of $\sigma(\eei \to
HHZ)$ for several Higgs masses at $\sqrt{s} = 500$ GeV with ${\cal L} =1$
ab$^{-1}$ \cite{Castanier:2001sf} (right).}
\label{Hfig:trilinear}
\vspace*{-4mm}
\end{figure}

The measurement of the trilinear Higgs self--coupling, which  is the first
non--trivial probe of the Higgs potential and, probably, the most decisive test
of the EWSB mechanism, is possible in the double Higgs--strahlung process. For
Higgs masses in the range $M_H \lsim 140$ GeV, one has to rely on the $b\bar b$
decays and the cross section in the $\eei \to HHZ \to \bar{b}b \bar{b}b+\ell^+
\ell^-$ or $q\bar{q}$ channels is rather small, while the four and six fermion
background are comparatively very large. The excellent $b$--tagging efficiencies
and the energy flow which can be achieved at ILC makes it possible to overcome
the formidable challenge of suppressing the backgrounds, while retaining a
significant portion of the signal.  Accuracies of about 20\% can be obtained on
the measurement of $\sigma(\eei \to HHZ)$ in the mass range below 140 GeV; see
Fig.~\ref{Hfig:trilinear}. Neural network analyses allow to improve the accuracy
from 17\% to 13\% at $M_H=120$ GeV and to obtain a 6$\sigma$ significance for
the signal \cite{Castanier:2001sf}; see also
Ref.~\cite{Baur:LHC-LC,Battaglia:2001nn}.

Since the sensitivity of the process $\eei \to HHZ$ to the trilinear Higgs
coupling is diluted by the additional contributions originating from diagrams
where the Higgs boson is emitted from the $Z$ boson lines, only an accuracy of
$\Delta \lambda_{HHH} \sim 22\%$  can be obtained for $M_H=120$ GeV at
$\sqrt{s}\sim 500$ GeV with a luminosity of ${\cal L} \sim 1$  ab$^{-1}$. The
accuracy becomes worse for higher Higgs masses, when the decays $H\to WW^*$
must be used.  In this case, one can proceed to higher energy and take
advantage of the fusion process $\eei\to HH \nu \bar \nu$ \cite{Yasui:2002se}
which has a larger cross section, in particular with longitudinally polarized
$e^\pm$ beams.  The sensitivity of the triple coupling constant is dominated by
Higgs--strahlung at low energy and $WW$ fusion for $\sqrt{s} \gsim 700$ GeV. A
recent simulation at $\sqrt{s}=1$ TeV which combines both the $\eei \to HHZ$ and
$\eei \to HH\nu \bar \nu$ processes with  $HH\to 4b$ final states, assuming a
80\% $e^-_L$ polarization and a luminosity of 1 ab$^{-1}$, shows that an
accuracy of $\Delta \lambda_{HHH}/\lambda_{HHH} \sim 12\%$ for $M_H=120$ GeV
could be be achieved if $\lambda_{HHH}$ is SM--like \cite{Yamashita:lcws2004}.
The relative phase of the coupling and its sign, may be also measured from the
interference terms \cite{Yasui:2002se,Yamashita:lcws2004}.

Note that this coupling is not accessible at the LHC unless the integrated 
luminosity is significantly increased. The quartic Higgs self--coupling is not
accessible at both the LHC and ILC as a result of the very small cross sections
for tripe Higgs  production. 

\subsubsection*{\underline{The two--photon Higgs coupling}}

\begin{figure}[!h]
\vspace*{-.3cm}
\hspace*{.5cm}
\epsfig{figure=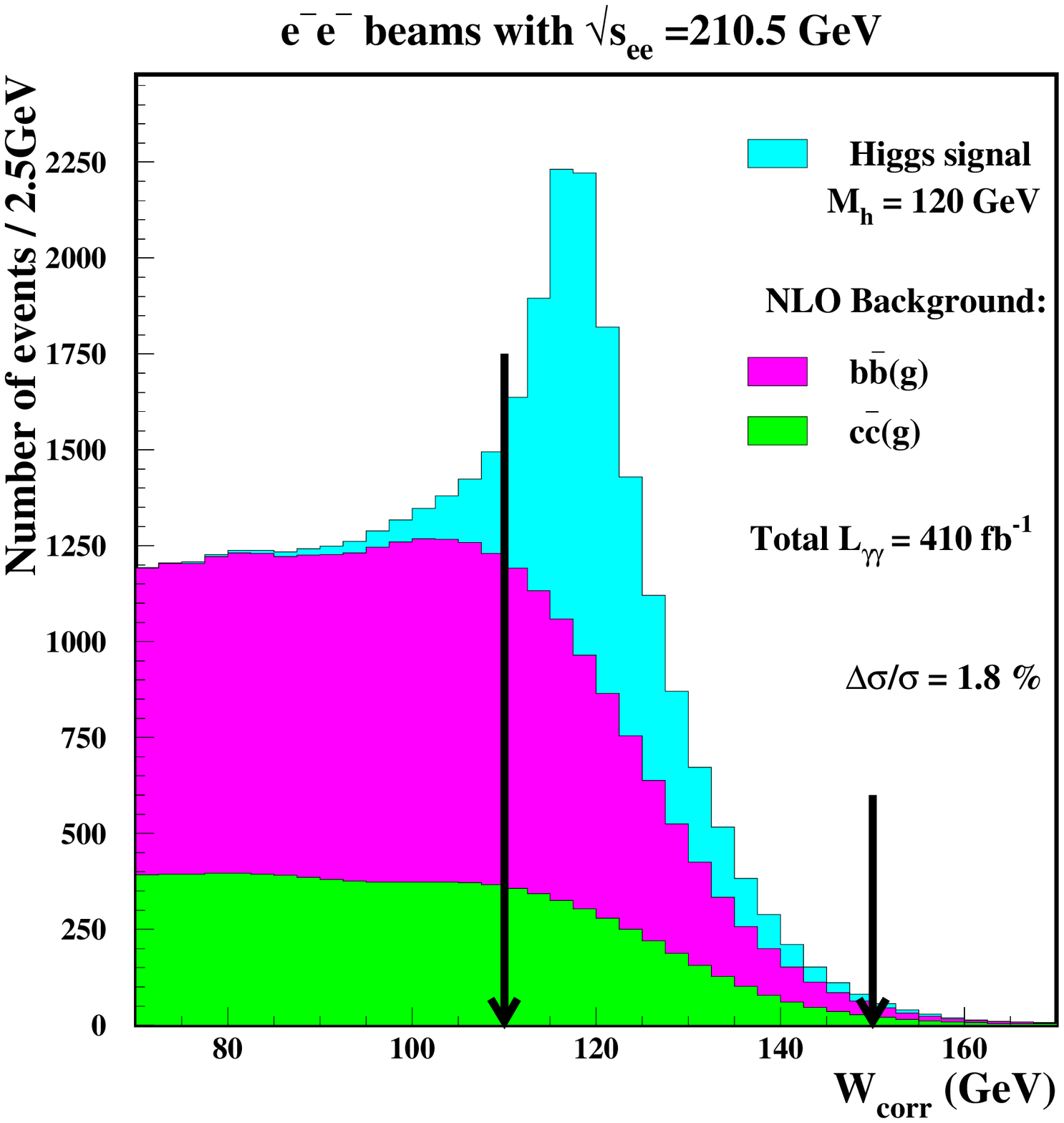,width=7cm,height=6.5cm}
\epsfig{figure=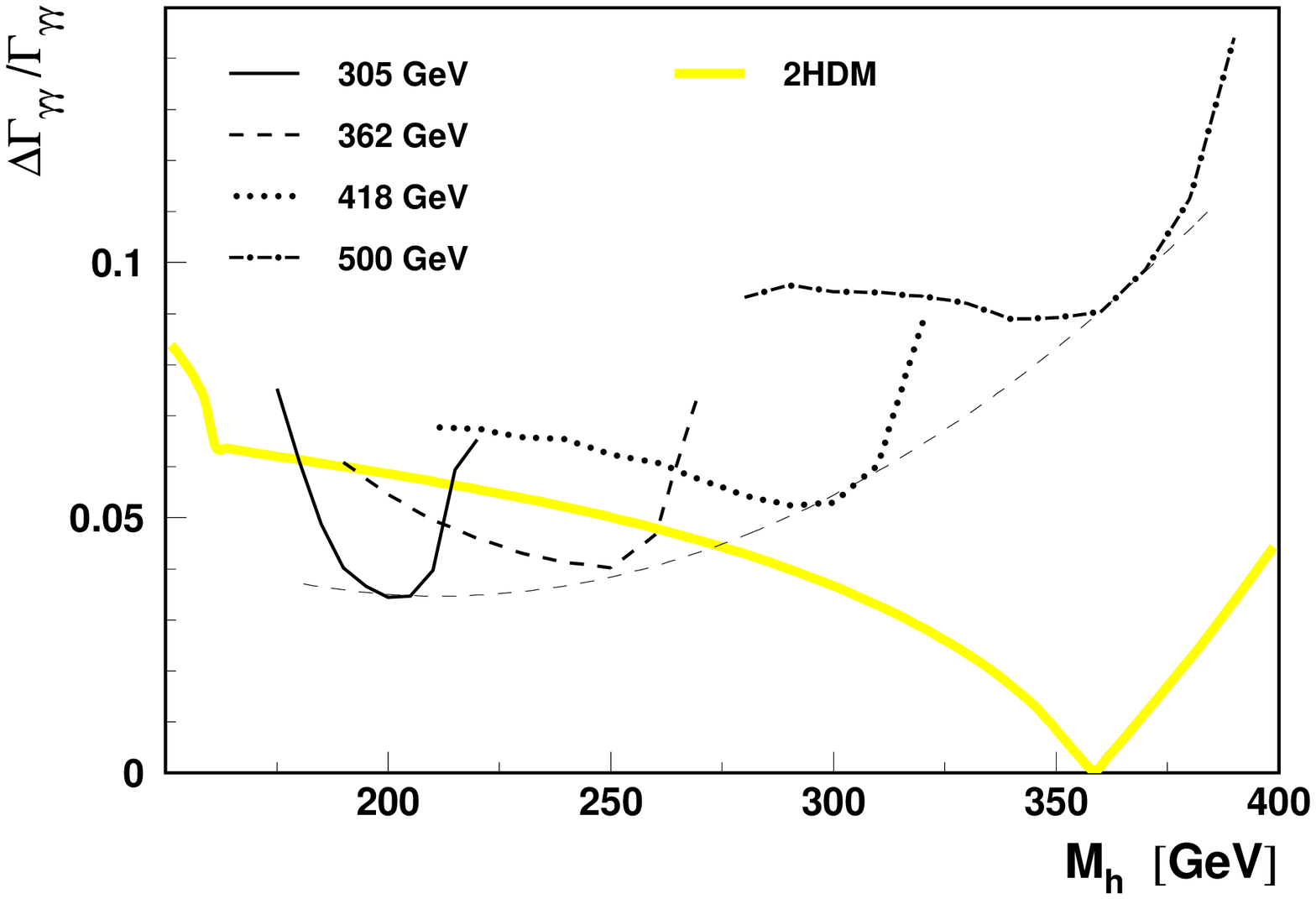,width=7cm,height=6.cm,clip=}
\vspace*{-.7cm}
\caption[Determination of the Higgs two--photon coupling in 
${\rm \gamma \gamma \to H \to b\bar b, WW/ZZ}$.]
{The reconstructed invariant mass distribution of the $\gamma \gamma
\to H\to b\bar b$ signal and the $b\bar b (g)$ and $c \bar c (g)$ backgrounds
\cite{Niezurawski:2003ik} (left) and the expected statistical errors in the
determination of the $H\gamma \gamma$ coupling in $\gamma \gamma \to H \to
WW/ZZ$  (right) with the yellow (thick light) band showing the prediction in
a general two--Higgs doublet model \cite{Niezurawski:2003ik}.}
\label{Hgamma}
\vspace*{-.7cm}
\end{figure}

At the $\gamma \gamma$ option of the ILC, when the energy is tuned to $M_H$, the
Higgs boson can be formed as an $s$--channel resonance, $\gamma \gamma \to $
Higgs. This allows a very precise measurement of the loop induced two--photon
Higgs coupling. For a low mass Higgs boson, when the decays $H \to b\bar{b}$ are
dominant, the main background $\gamma \gamma \to b\bar{b}$ can be suppressed by
choosing proper helicities for the initial $e^\pm$ and laser photons which
maximizes the signal cross section, and eliminating the gluon radiation by
taking into account only two--jet events.  Clear signals can be obtained
\cite{Niezurawski:2003iu} which allow the measurement of $\Gamma (H \to \gamma
\gamma) \times {\rm BR} (H\to b \bar{b})$ with a statistical accuracy of 2\% for
$M_H=120$ GeV at an energy $\sqrt{s_{ee}}=210$ GeV and a luminosity ${\cal
L}_{\gamma \gamma}=410$ fb$^{-1}$; Fig.~\ref{Hgamma} (left). Because of the
smaller $H \to b\bar b$ branching ratio, the accuracy drops to  7\% for $M_H=$
160 GeV.  For heavier Higgs particles decaying into $WW/ZZ$ final states, the
two--photon width can be measured with a precision $\Delta \Gamma_{\gamma
\gamma} \simeq 3\%$--10\% for $M_H=200$--350 GeV \cite{Niezurawski:2003ik};
Fig.~\ref{Hgamma} (right). The relative phase of the coupling can also be
measured and, for $M_H=200$ GeV, one obtains an accuracy of $\Delta \phi_{\gamma
\gamma} \sim 35$ mrad \cite{Niezurawski:2003ik}.

\subsubsection*{\underline{Impact of Higgs coupling measurements}}

\begin{figure}[!h] 
\centering
\vspace*{-.3cm}
\includegraphics[width=8cm,angle=0]{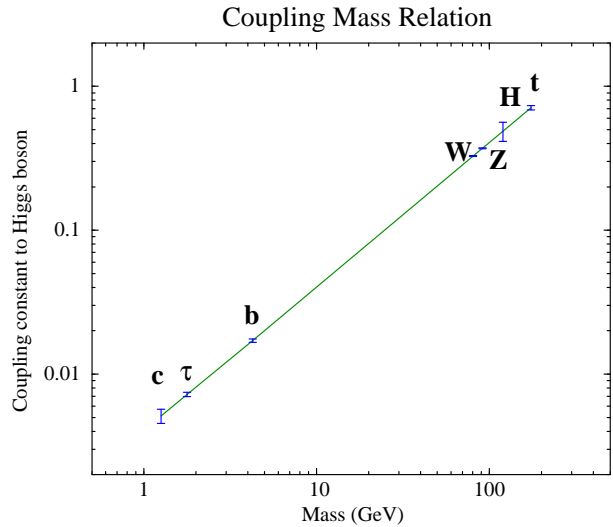}
\vspace*{-.6cm}
\caption[ILC determination of the relation between particle couplings and masses] {The relation between the Higgs couplings and the particle masses as
determined from the high--precision ILC measurements \cite{:2003mg}; on the $y$ 
axis, the coupling $\kappa_i$  of the  particle $i$ with mass $m_i$  is 
defined in a such a way that the relation $m_i= v \kappa_i$ with $v \simeq 
246$ GeV holds in the SM.}
\label{fig-cmrel3}
\vspace*{-.5cm}
\end{figure}

If we combine the Higgs--strahlung and $WW$ fusion processes for single Higgs
production, the decay branching ratio measurements, associated Higgs  production
with top quark pairs and double Higgs production in the strahlung and $WW$
fusion processes, the various couplings  associated with the Higgs particle can
be determined rather accurately. We can then compare the magnitudes of these
couplings with the the SM and check the fundamental  prediction that they are
indeed proportional to the particle masses.  Relations between various Higgs
couplings and particle masses are shown in Fig.~\ref{fig-cmrel3} for the case of
a 120 GeV SM Higgs boson with accuracies corresponding to ${\cal L}=500$
fb$^{-1}$ at $\sqrt{s}$=300 GeV for the $c,\tau, b, W$ and $Z$ couplings,
$\sqrt{s}= 500$ GeV for the $\lambda_{HHH}$ self--coupling and $\sqrt{s}=700$
GeV for the $t\bar{t} H$ Yukawa coupling. A summary of the various precision
measurements at ILC is given in Table \ref{tab:Hcplg-sum}

An important feature of ILC experiments is that absolute values of these
coupling constants can be determined in a model--independent way. This is
crucial in establishing the mass generation mechanism for elementary  particles
and very useful to explore physics beyond the SM. For instance,  radion-Higgs
mixing in warped extra dimensional models could reduce the  magnitude  of the
Higgs couplings to fermions and gauge bosons  in a  universal way
\cite{Hewett:2002nk, Dominici:2002jv} and such effects can be  probed only if
absolute coupling measurements are possible. Another example is related to the
electroweak baryogenesis scenario to explain the baryon number of the universe:
to be successful, the SM Higgs sector has to  be extended to realize a strong
first-order phase transition and the change of the Higgs potential can lead to
observable effects in the triple Higgs coupling
\cite{Grojean:2004xa,Kanemura:2004ch}. Finally, the loop induced gluonic and
photonic decay channels are sensitive to scales far beyond the Higgs mass and
can  probe new particles that are too heavy to be produced directly 
\cite{H-RSHgg}.

\begin{table}[!h]
\caption[Summary of the predictions of the SM Higgs couplings at the ILC] 
{Precision of the Higgs couplings determination for various particles at the
    ILC from a global fir for $M_H\!=\!120$ GeV with 500 fb$^{-1}$ data. For $c,
    \tau,b,W,Z$ couplings, $\sqrt s\!=\!500$ GeV is assumed while
    $\sqrt{s}\!=\!500$ (800) GeV is taken for the  $HHH$ $(t\bar{t} H)$
    couplings and 1 ab$^{-1}$ data is assumed (the measurement of $\lambda_{HHH}$ 
    can be improved by a factor of two at $\sqrt s=1$ TeV). 
    The accuracy for the determination of the Higgs boson mass, total decay
    width and CP--mixture at $\sqrt s=350$ GeV with 500 fb$^{-1}$ data, 
    are also shown. From Ref.~\cite{Aguilar-Saavedra:2001rg}.} 
\label{tab:Hcplg-sum}
\renewcommand{\arraystretch}{1.3}
\begin{tabular}{|l|c|c|c|c|c|c|c|}
\hline
coupling &
$\lambda_{HHH}$& 
$g_{HWW}$& 
$g_{HZZ}$& 
$g_{Htt}$& 
$g_{Hbb}$&
$g_{Hcc}$&
$g_{H\tau\tau}$ \\ \hline
accuracy  
& $\pm 0.22  $
& $\pm 0.012 $              
& $\pm 0.012 $             
& $\pm 0.030 $          
& $\pm 0.022 $           
& $\pm 0.037 $       
& $\pm 0.033 $   
 \\  \hline 
\end{tabular}
\begin{center}
\begin{tabular}{|c|c|c|c|}\hline
observable &  $M_H$ & $\Gamma_H$  & CP--mixture\\ \hline
accuracy   &  $\pm 0.00033$  & $\pm 0.061$  & $\pm 0.038$\\ \hline
\end{tabular}
\end{center}
\end{table}

\section{The Higgs bosons in SUSY theories}

\subsection{Decays and production of the MSSM Higgs bosons}

\begin{figure}[!h]
\vspace*{-1mm}
\begin{center}
\includegraphics[width=0.48\linewidth,bb=73 463 579 754]{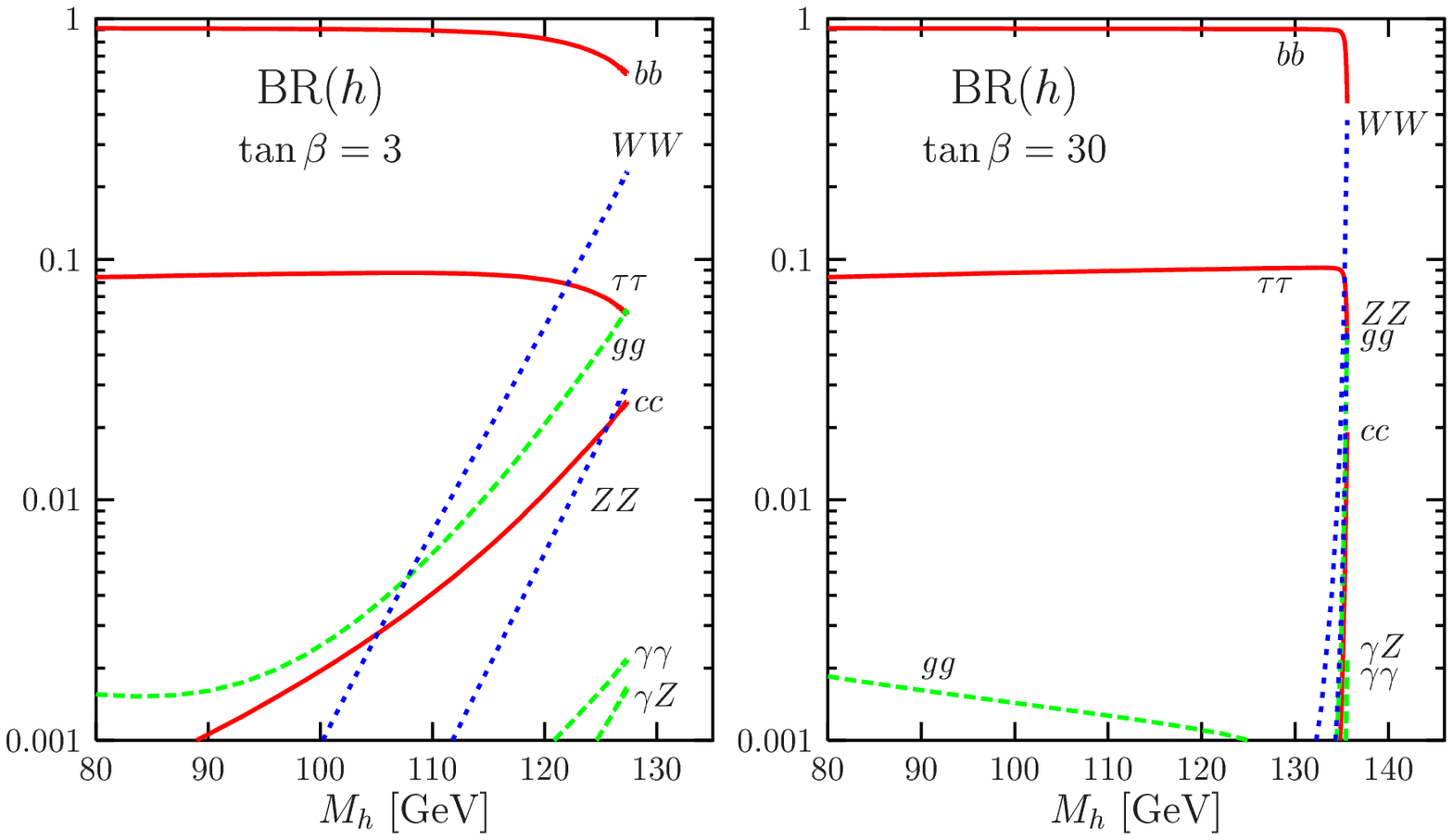}
\includegraphics[width=0.48\linewidth,bb=73 463 579 754]{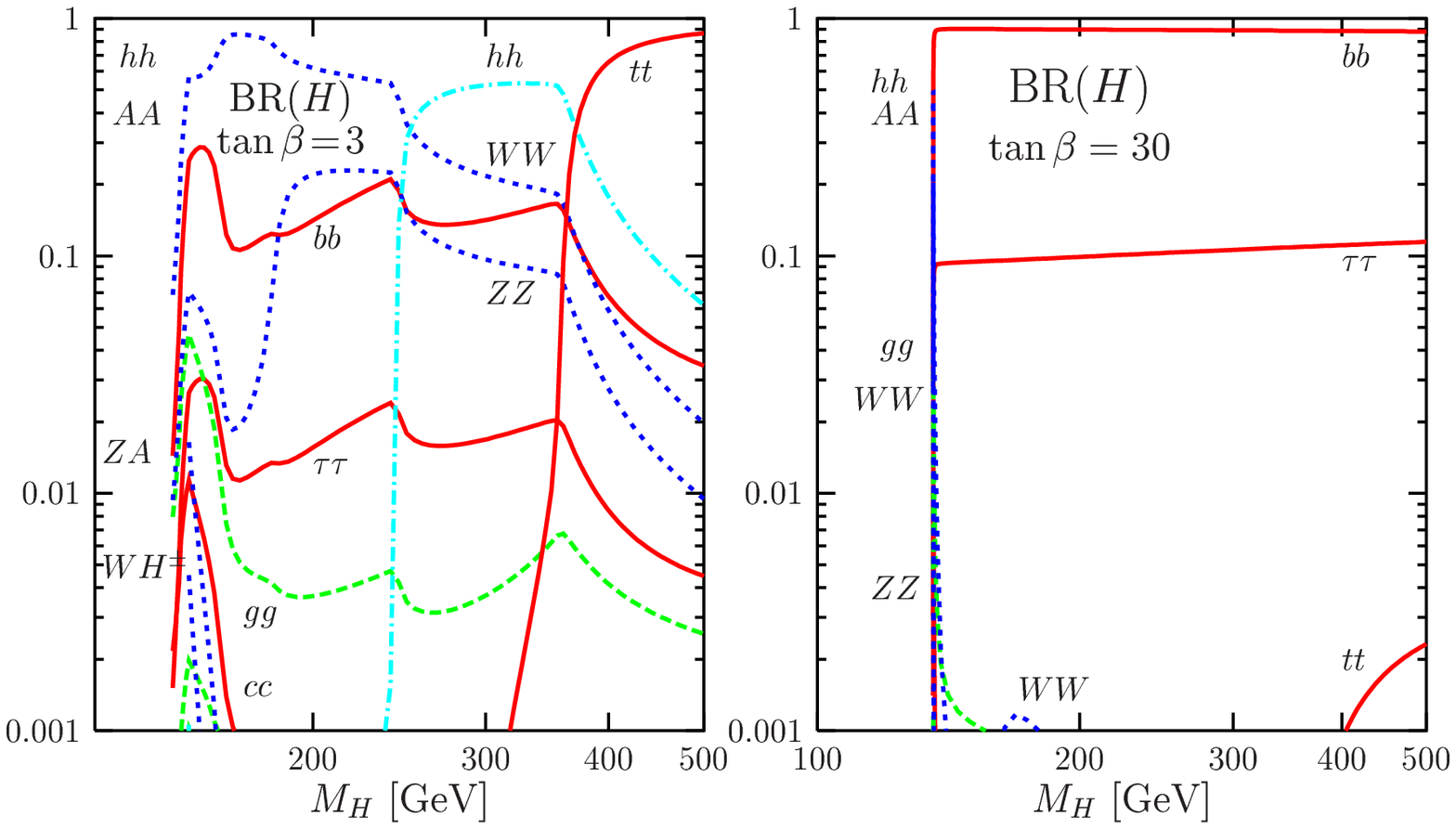}
\includegraphics[width=0.48\linewidth,bb=73 463 579 754]{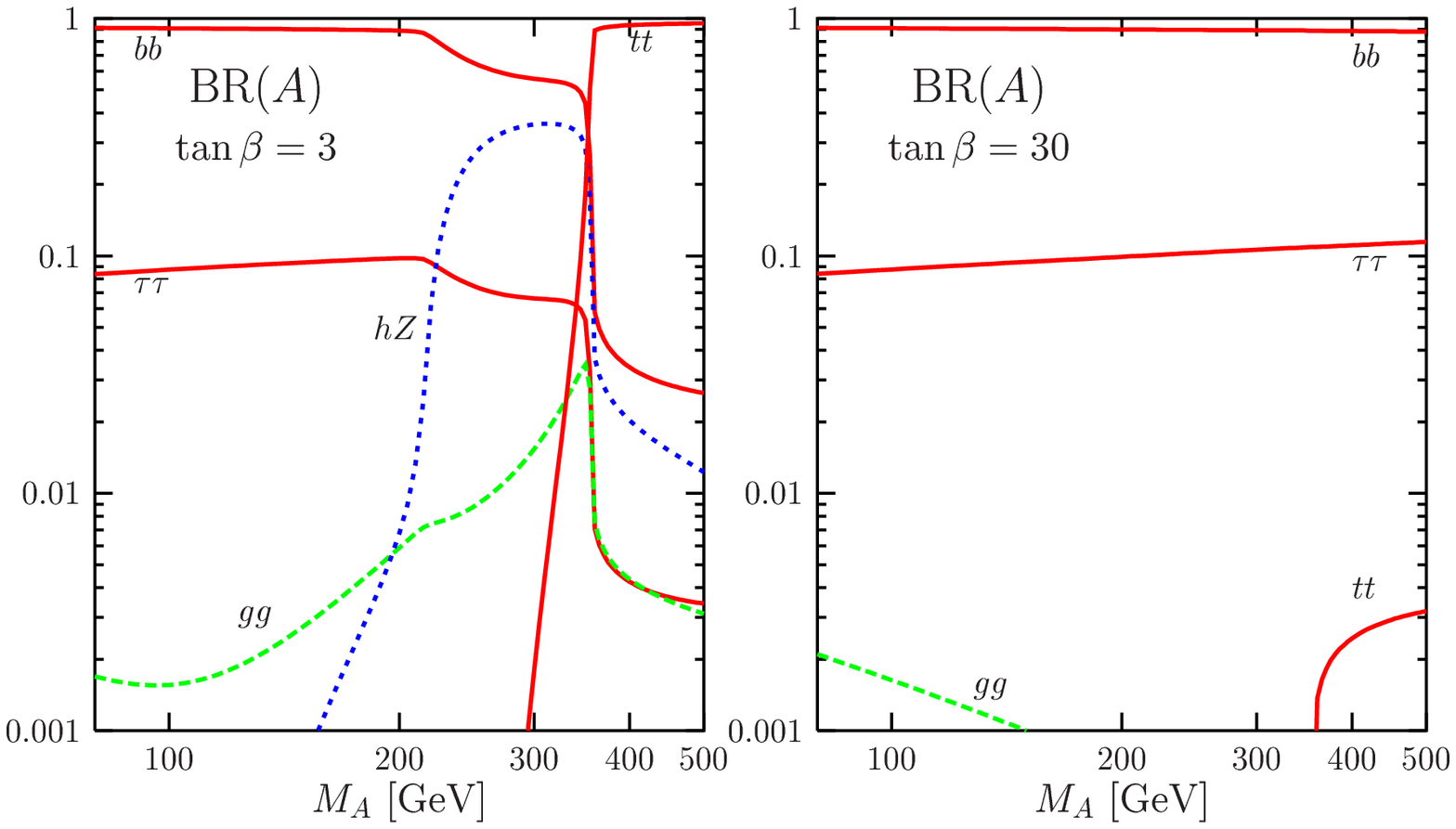}
\includegraphics[width=0.48\linewidth,bb=73 463 579 754]{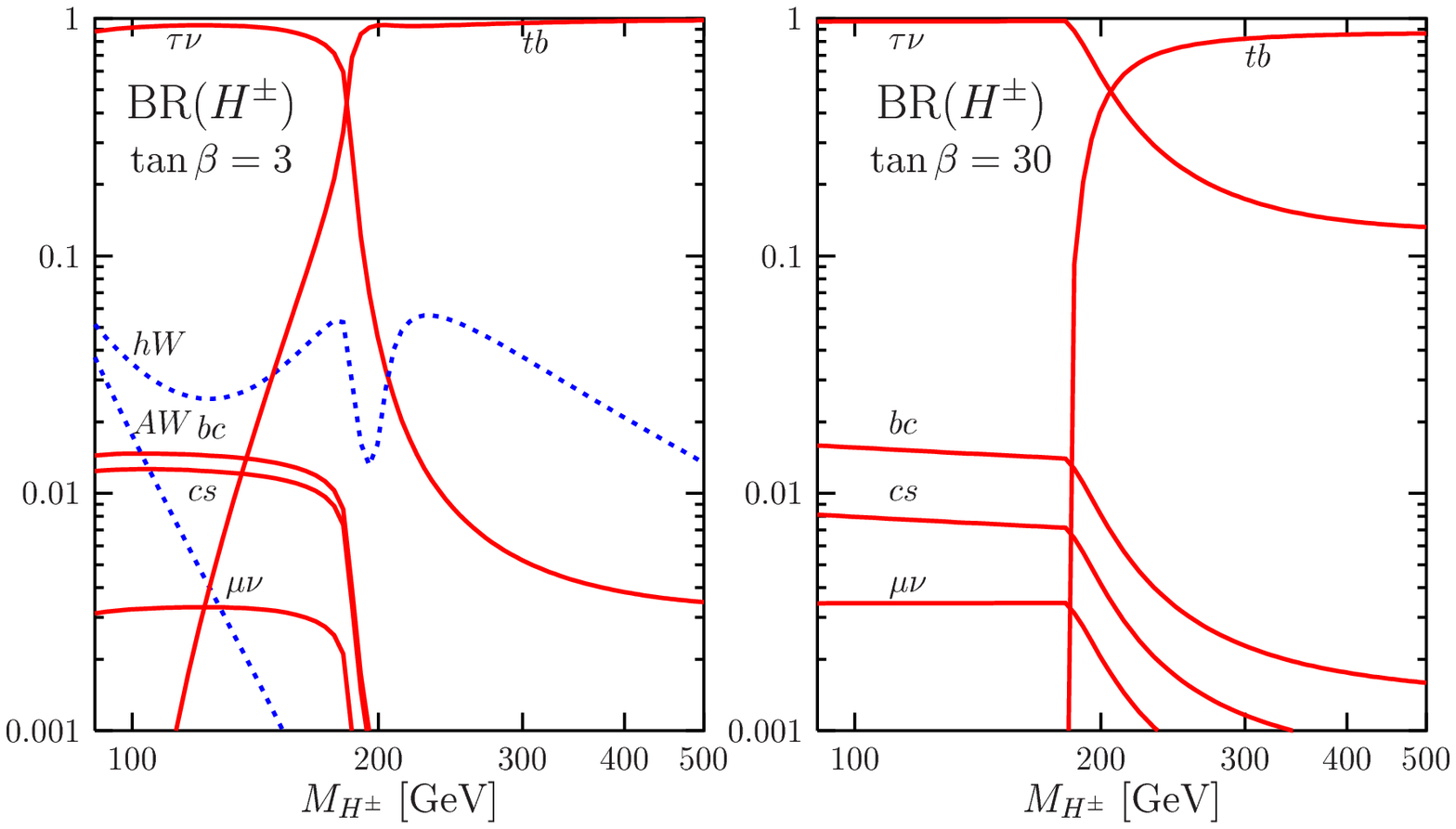}
\end{center}
\vspace*{-7mm}
\caption[The decay branching ratios of the MSSM Higgs bosons]
{The decay branching ratios of the MSSM Higgs bosons as functions of
their masses for $\tb=3$ and 30 in the maximal mixing
scenario with $M_S=2$ TeV.}
\label{Hfig:BR-MSSM}
\vspace*{-1mm}
\end{figure}

The decay pattern of the Higgs bosons of the MSSM \cite{Djouadi:2005gj} is more
complicated than in the SM and depends strongly on the value of $\tb$ and the
Higgs masses; see Fig.~\ref{Hfig:BR-MSSM} where the branching ratios are shown
for $\tb=3$ and 30.  The lightest $h$ boson will decay mainly into fermion pairs
since its mass is smaller than $\sim$~140~GeV, except in the decoupling limit in
which it decays like the SM--Higgs boson and thus the $WW$ decays can be
dominant. The fermionic channels are in general also the dominant decay modes of
the heavier scalar $H$ and pseudoscalar $A$ bosons, except for the $H$ boson  when
it is SM--like. For values of $\tb$ much larger than unity, the main decay modes
of the three neutral Higgs bosons are decays into $b \bar{b}$ and $\tau^+
\tau^-$ pairs with the branching ratios being of order $ \sim 90\%$ and $10\%$,
respectively. For large masses, the top decay channels $H, A \rightarrow
t\bar{t}$ open up, yet for large $\tb$ these modes remain suppressed.  If the
masses are high enough, the heavy $H$ boson can decay into gauge bosons or light
$h$ boson pairs and the pseudoscalar $A$ particle into $hZ$ final states.
However, these decays are strongly suppressed for $\tb \gsim 3$--$5$ as is is
suggested by the LEP2 constraints.  The charged Higgs particles decay into
fermions pairs: mainly $t\bar{b}$ and $\tau \nu_{\tau}$ final states for $H^\pm$
masses, respectively, above and below the $tb$ threshold.  If allowed
kinematically and for small values of $\tb$, the $H^\pm$ bosons decay also into
$hW$ final states for $\tb \lsim 5$.

Adding up the various decay modes, the widths of all five Higgs bosons remain
very narrow. The total width of one of the CP--even Higgs particles will be
close to the SM Higgs boson width, while the total widths of the other Higgs
particles will be proportional to $\tb$ and will be of the order of 10~GeV even
for large masses and large $\tb$ values. 

Other possible decay channels for the MSSM bosons, in particular the heavy $H,
A$ and $H^\pm$ states, are decays into supersymmetric particles. In addition to
light sfermions, decays into charginos and neutralinos could eventually be
important if not dominant. Decays of the lightest $h$ boson into the lightest
neutralinos (LSP) can be also important in some parts of the SUSY parameter
space; see Ref.~\cite{Djouadi:2005gj} for a recent review. These decays can
render the search for Higgs particle rather difficult, in particular at hadron
colliders.

\begin{figure}[hbtp]
\vspace*{-2mm}
\begin{center}
\includegraphics[width=0.88\linewidth,bb=73 430 600 745]{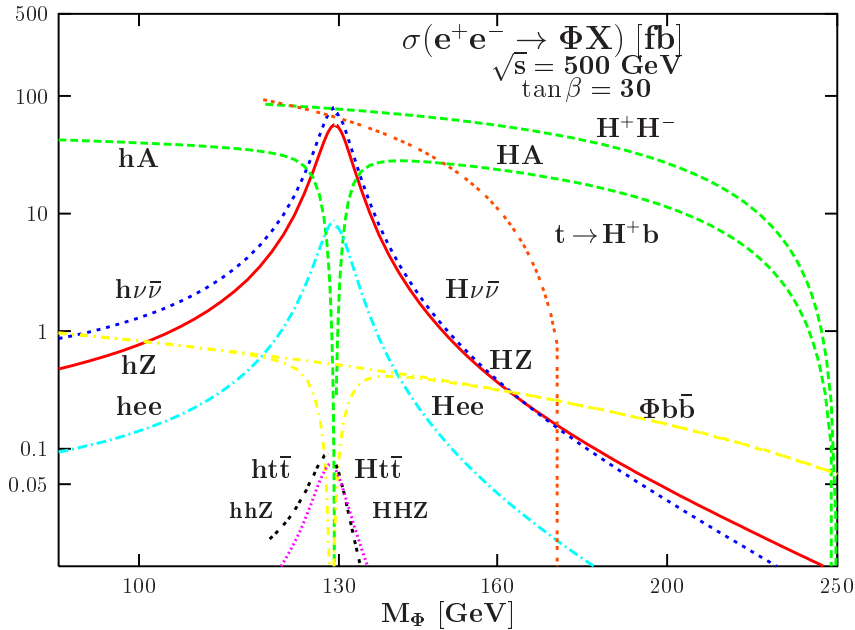} 
\end{center}
\vspace*{-7mm}
\caption[Production cross sections of the MSSM Higgs bosons at a 500 GeV ILC]
{Production cross sections of the MSSM Higgs bosons in $e^+ e^-$ collisions
as functions of the masses for $\tb=30$ and $\sqrt{s}=500$ GeV; from
Ref.~\cite{Djouadi:2005gj}.}
\label{Hfig:ee-MSSM}
\vspace*{-2mm}
\end{figure}

At the ILC, besides the usual Higgs--strahlung and fusion processes for
$h$ and $H$ production, the neutral Higgs particles can also be
produced pairwise: $\eei \to A + h/H$ \cite{H:eeSUSY}. The cross
sections for the Higgs--strahlung and the pair production as well as the
cross sections for the production of $h$ and $H$ are mutually
complementary, coming either with a coefficient $\sin^2(\beta-
\alpha)$ or $\cos^2(\beta -\alpha)$; Fig.~\ref{Hfig:ee-MSSM}.  The
cross section for $hZ$ production is large for large values of $M_h$,
being of ${\cal O}(100$ fb) at $\sqrt{s}=500$ GeV; by contrast, the
cross section for $HZ$ is large for light $h$ [implying small $M_H$].
In major parts of the parameter space, the signals consist of a $Z$
boson and $b\bar{b}$ or $\tau^+ \tau^-$ pairs, which is easy to
separate from the backgrounds with $b$--tagging. For the associated
production, the situation is opposite: the cross section for $Ah$ is
large for light $h$ whereas $AH$ production is preferred in the
complementary region.  The signals consists mostly of four $b$ quarks
in the final state, requiring efficient $b$--quark tagging; mass
constraints help to eliminate the QCD jets and $ZZ$ backgrounds. The
CP--even Higgs particles can also be searched for in the $WW$ and $ZZ$
fusion mechanisms.

In $\eei$ collisions, charged Higgs bosons can be produced pairwise, $\eei \to
H^+H^-$, through $\gamma,Z$ exchange. The cross section depends only on the
charged Higgs mass; it is large almost up to $M_{H^\pm} \sim \frac12 \sqrt{s}$.
$H^\pm$ bosons can also be produced in top decays; in the range $ 1 < \tb <
m_t/m_b$, the $t \to H^+b$ branching ratio and the $t\bar{t}$ production cross
sections are large enough to allow for their detection in this mode as will
be discussed in chapter \ref{sec:top}.

The discussion of SUSY Higgs production at the ILC can be briefly summarized in
the following three points.

\begin{figure}[!h]
\vspace*{-2.mm}
\centerline{\epsfig{file=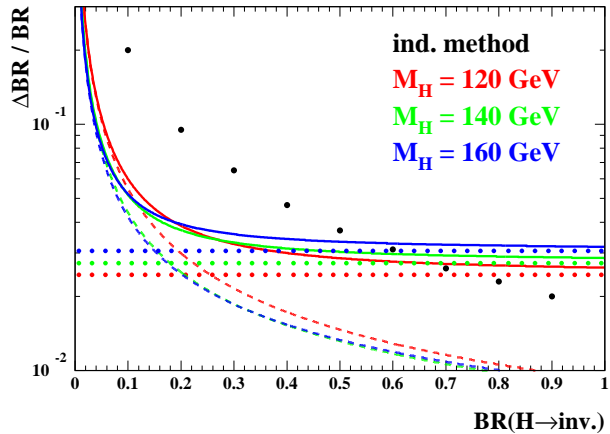,width=8cm}}
\vspace*{-7mm}
\caption[Accuracy in the determination of invisible decays of the MSSM 
Higgs bosons]  
{The expected accuracy on the invisible decay rate as a function of the 
branching ratio at $\sqrt {s}=350$ GeV with 500 fb$^{-1}$ data (full
lines). The other lines indicate the individual contributions from the 
measurement of the invisible rate (dashed lines) and the total Higgs--strahlung 
cross section (dotted lines); the large dots are the result of the indirect 
method \cite{Aguilar-Saavedra:2001rg}; from Ref.~\cite{Schumacher:2003ss}.}
\label{fig:H-invisible}
\vspace*{-5mm}
\end{figure}

-- The Higgs boson $h$ can be detected in the entire range of the MSSM parameter
space, either through the Higgs--strahlung [and $WW$ fusion] process or
associated production with the pseudoscalar $A$ boson.  In fact, this conclusion
holds true even at a c.m. energy of 250 GeV and with a luminosity of a few
fb$^{-1}$.  Even if the decay modes of the $h$ boson are very complicated,
missing mass techniques allow for their detection. For instance, the branching 
ratios for the invisible $h$ boson decays into the LSP neutralinos can be
measured at the percent level as exemplified in Fig.~\ref{fig:H-invisible} for a
350 GeV ILC. The accuracy can be substantially improved by running at lower c.m.
energies \cite{Richard}. The same very detailed tests and precision measurements
discussed previously for the SM Higgs boson can be performed for the MSSM $h$
boson, in particular in the decoupling limit.

\begin{figure}[h] \centering
\vspace*{-2mm}
\begin{tabular}{cc}
    \includegraphics[width=6.1cm,angle=0]{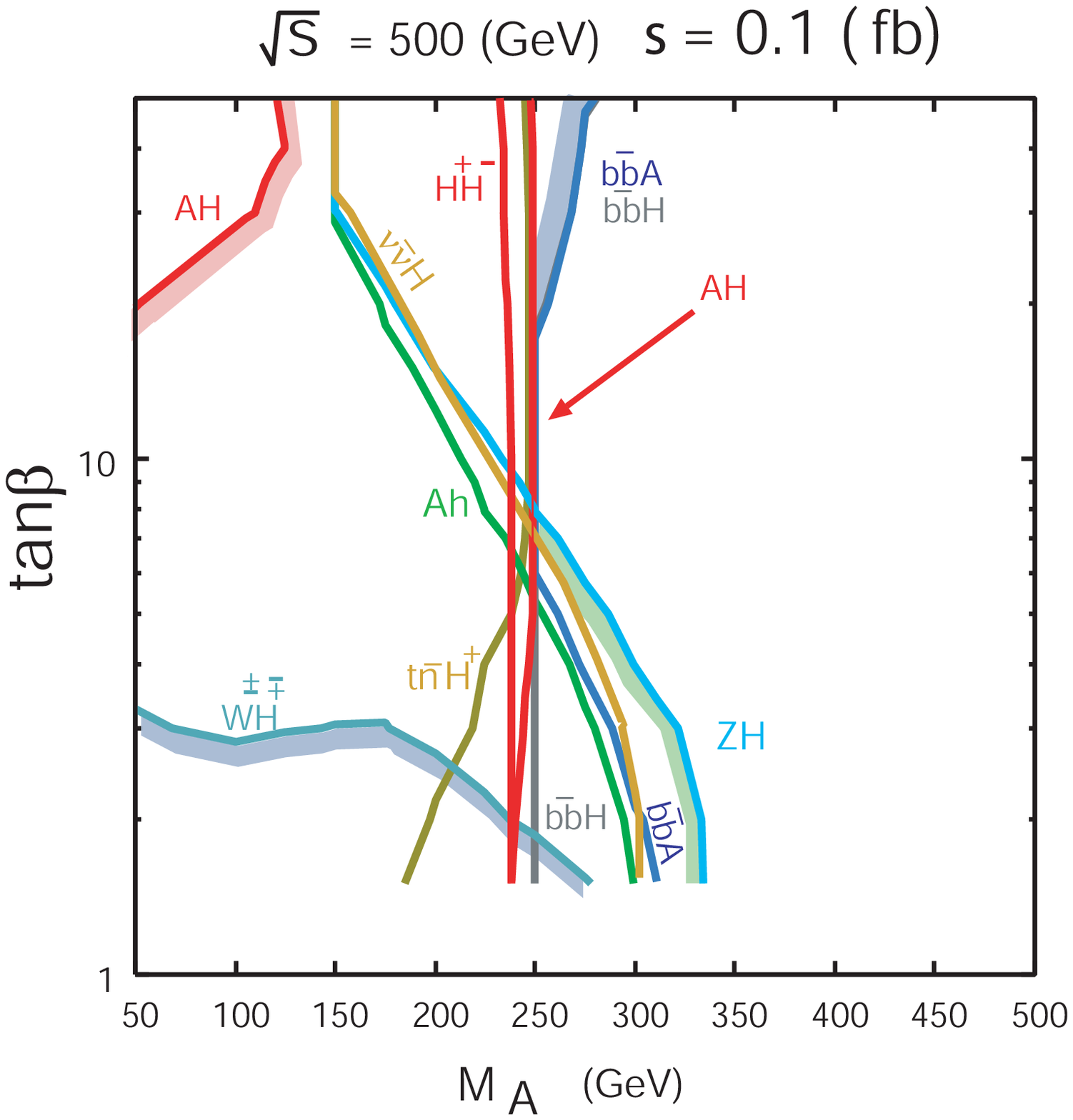}&
    \includegraphics[width=6.1cm,angle=0]{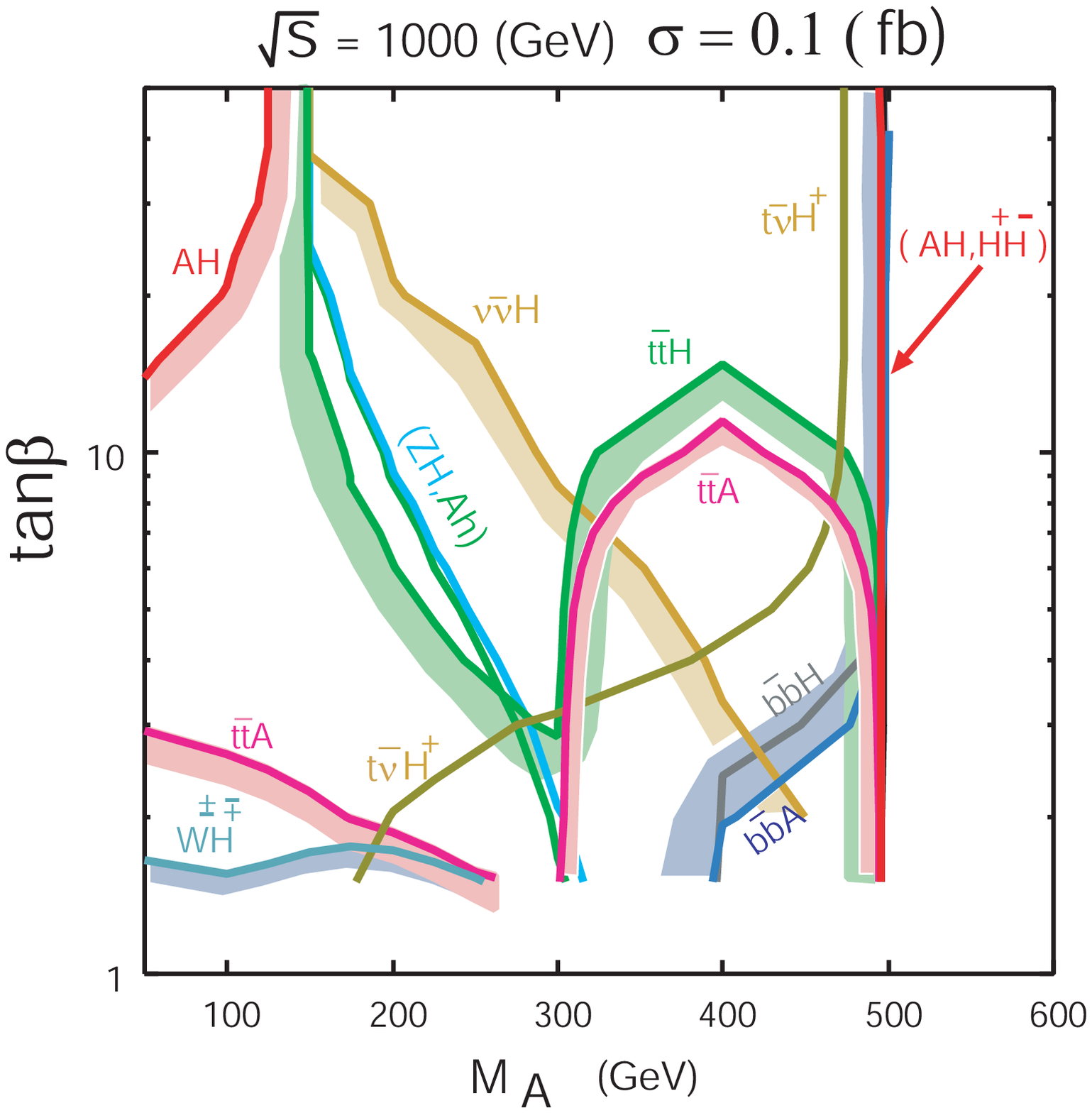}
\end{tabular}
\vspace*{-8mm}
    \caption[Cross section contours from various MSSM Higgs production processes]
    {Cross section contours  of various heavy MSSM Higgs
    production processes in the plane $[M_A, \tan{\beta}]$
    for $\sqrt{s}=500$ GeV  and 1 TeV \cite{Kiyoura:2003tg}.}
\label{fig-roots1000.1fx.ACFA}
\vspace*{-2mm}
\end{figure}

-- All SUSY Higgs bosons can be discovered at an $\eei$ collider if the $H,A$
and $H^{\pm}$ masses are less than the beam energy; for higher masses, one
simply has to increase the c.m. energy, $\sqrt{s} \gsim 2M_A$. The various cross
section contours for heavy MSSM Higgs production processes are shown in
Fig.~\ref{fig-roots1000.1fx.ACFA} in the $[M_A, \tan\beta]$  plane for
$\sqrt{s}=500$ GeV and 1 TeV \cite{Kiyoura:2003tg}. As can be seen, several 
channels might be observable depending on the value of $\tan{\beta}$. Note that
the additional associated neutral Higgs production processes with  $t\bar{t}$
and $b\bar{b}$ allow for the measurement of the Yukawa couplings. In particular,
$\eei \to b\bar{b}+h/H/A$ for high $\tb$ values allow for the determination of 
this important parameter for low $M_A$ values.

\begin{figure}[!h]
\vspace*{-4mm}
\centerline{\resizebox{110mm}{!}{\psfig{file=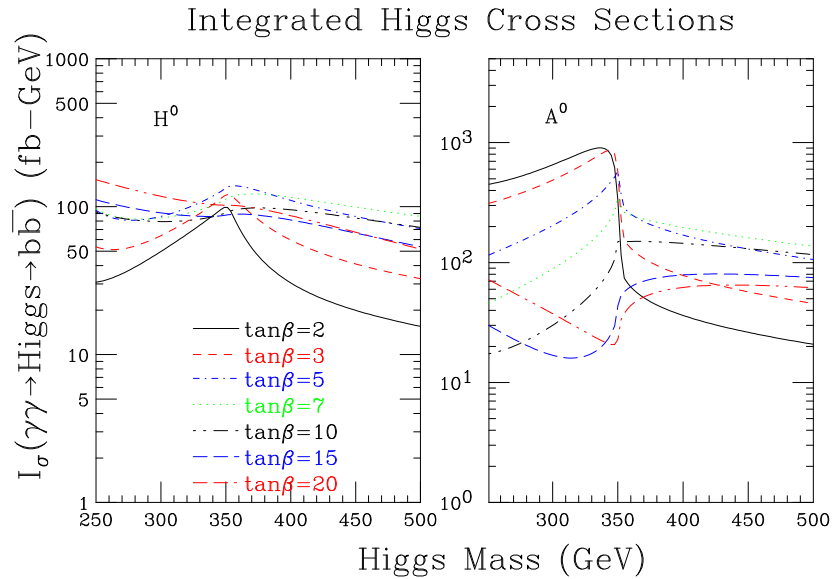,height=6.5cm}}}
\vspace*{-7mm}
\caption[Production cross sections  neutral MSSM Higgs 
bosons in $\gamma \gamma$ collisions]
{Effective cross sections for the production of the heavier
CP--even (left) and the CP--odd (right) Higgs bosons in $\gamma \gamma$
collisions, $\sigma( \gamma \gamma \to H/A \to b\bar b)$
for several $\tb$ values; from Ref.~\cite{Asner:2001ia}.}
\label{fig:Hasner}
\vspace*{-5mm}
\end{figure}

-- If the energy is not high enough to open the $HA$ pair production threshold,
the photon collider option may become the discovery machine for the heavy Higgs
bosons \cite{Hgamgam,Asner:2001ia}. Since the $A,H$ bosons are produced as
$s$--channel resonances, the mass reach at a photon collider is extended
compared to the $e^+e^-$ mode and masses up to 80\% of the original c.m. energy
can be probed.  It has been shown in Ref. \cite{Asner:2001ia} that the whole
medium $\tan{\beta}$ region up to about 500 GeV, where only one light Higgs
boson can be found at the LHC, can be covered by the photon collider option with
three years of operation with an $e^-e^-$ c.m.~energy of 630 GeV; see
Fig.~\ref{fig:Hasner}. The photon collider mode is also important to determine
the CP properties of the heavy Higgs bosons, either by studying angular
correlation of Higgs decay products or by using initial beam polarization. The
discrimination between the scalar and pseudoscalar particles can be performed
and CP violation  can be unambiguously probed. 

\subsection{Measurements in the MSSM Higgs sector}

\begin{figure}[ht!]
\vspace*{-1mm}
\begin{center}
\begin{tabular}{c c}
\epsfig{file=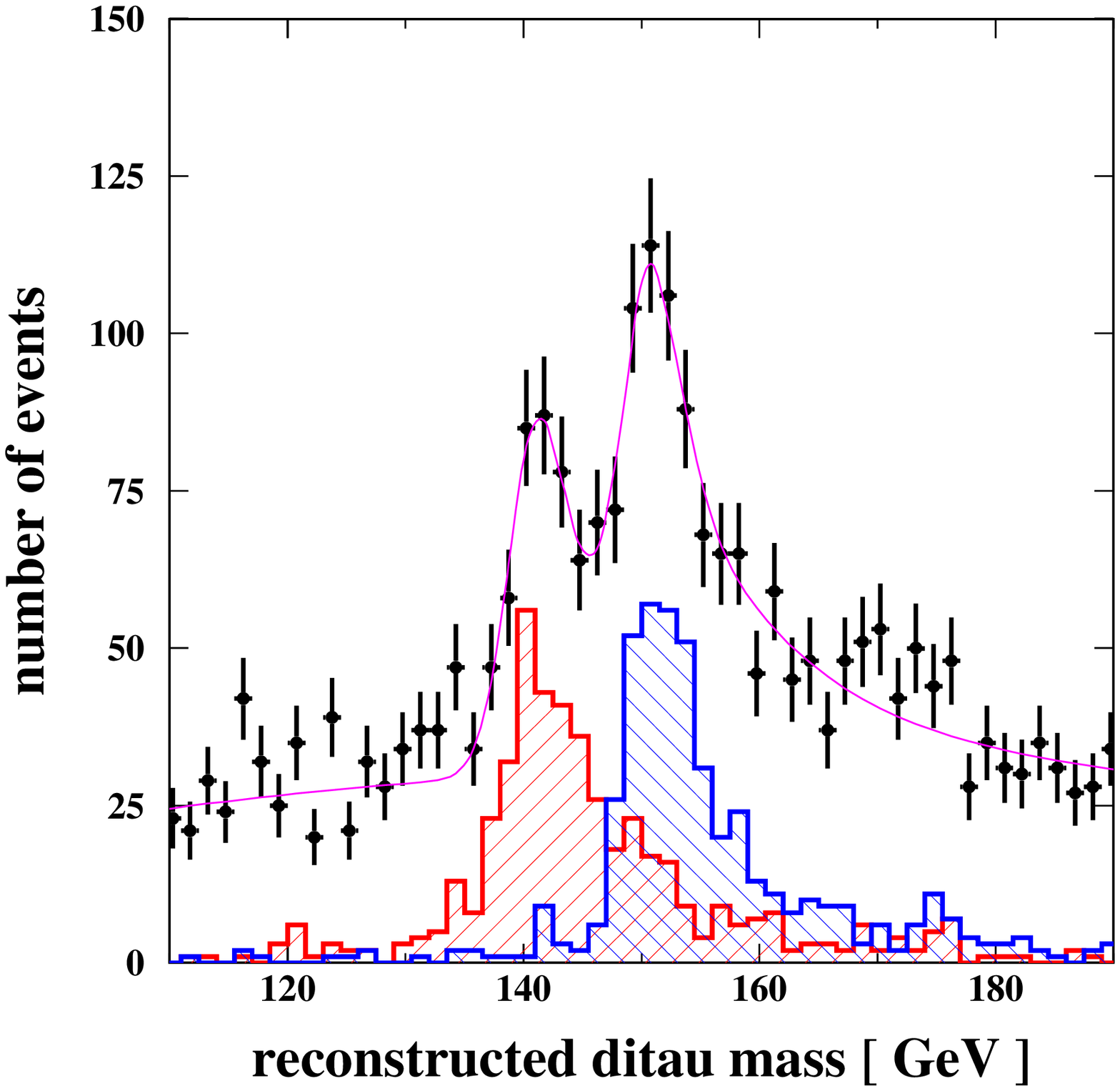,height=5.9cm,width=7cm,clip}
&
\epsfig{file=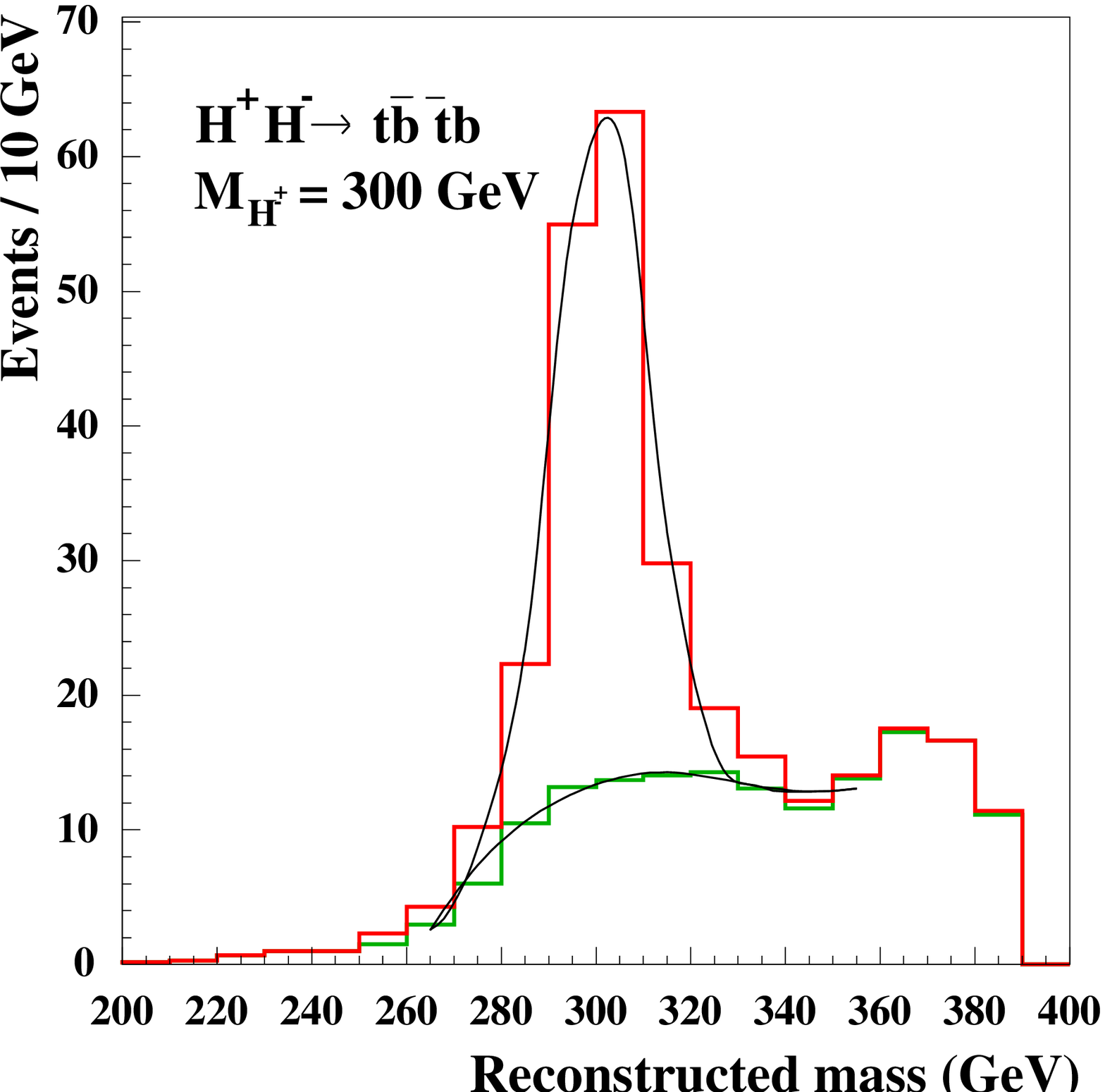,height=5.9cm,width=7cm,clip}
\end{tabular}
\end{center}
\vspace*{-7mm}
\caption[Detection of the heavy neutral and charged MSSM Higgs bosons at the ILC]
{The reconstructed $\tau\tau$ invariant mass from a kinematic fit in
 $\eei\! \to\! HA\! \to\! b \bar b\tau^+ \tau^-$ for $M_A\!=\!140$ 
GeV and $M_H\!=\! 150$ GeV at $\sqrt{s}\!=\!500$ GeV with 500 fb$^{-1}$ data 
\cite{HAee-desch} (left) and the di--jet invariant mass distribution for 
$e^+ e^-\! \to\! H^+H^-\! \to\! t \bar b \bar t b$  for $M_{H^\pm}\!=\!300$ GeV
after applying the intermediate $W,t$ and the equal mass final state
constraints for 500 fb$^{-1}$ data at $\sqrt{s}\!=\! 800$ GeV
\cite{Aguilar-Saavedra:2001rg}
(right).}
\vspace*{-2mm}
\label{Hfig:MSSM-heavy}
\end{figure}

A number of very important measurements can be performed at the ILC in the MSSM
Higgs sector. If the heavier $H,A$ and $H^\pm$ states are kinematically
accessible, one can measure their masses and cross sections times decay
branching ratios with a relatively good accuracy. In the pair production process
$\eei \to HA$, a precision of the order of $0.2\%$ can be achieved on the $H$
and $A$ masses, while a measurement of the cross sections can be made at the
level of a few percent in the $b\bar b b\bar b$ and ten percent in the $b\bar b
\tau^+ \tau^-$ channels; see Fig.~\ref{Hfig:MSSM-heavy} (left). 

For the charged Higgs boson, statistical uncertainties of less than 1 GeV on its
mass and less than 15\% on its production cross section times branching ratio
can be achieved in the channel $e^+e^- \to H^+H^- \to t \bar b \bar t b$ for
$M_{H^\pm} \sim 300$ GeV with high enough energy and luminosity;
Fig.~\ref{Hfig:MSSM-heavy} (right).  These measurements allow the determination
of the most important branching ratios, $b\bar b$ and $\tau^+\tau^-$ for the
$H/A$ and $tb$ and $\tau \nu$ for the $H^\pm$ particles, as well as the total
decay widths which can be turned into a determination of the value of $\tb$,
with an accuracy of 10\% or less.  The spin--zero nature of the particles can be
easily checked by looking at the angular distributions which should go as
$\sin^2\theta$. Several other measurements, such as the spin--parity of the
Higgs particles in $H/A \to \tau^+ \tau^-$ decays and, in favorable regions of
the parameter space, some trilinear Higgs couplings, can be made.

The profile of the lighter Higgs boson can be entirely  determined. This is
particularly the case close to the decoupling regime where the $h$ boson behaves
like the SM Higgs particle but with a mass below $M_h \sim 140$ GeV.  This is,
in fact, the most favorable mass range for precision measurements as the Higgs
boson has many decay channels that are accessible in this case.  This has been
shown in the previous section when we reviewed the precision studies for a SM
Higgs boson at the ILC.

\begin{figure}[h!]
\vspace*{-.6cm}
\begin{center}
\mbox{
\epsfig{file=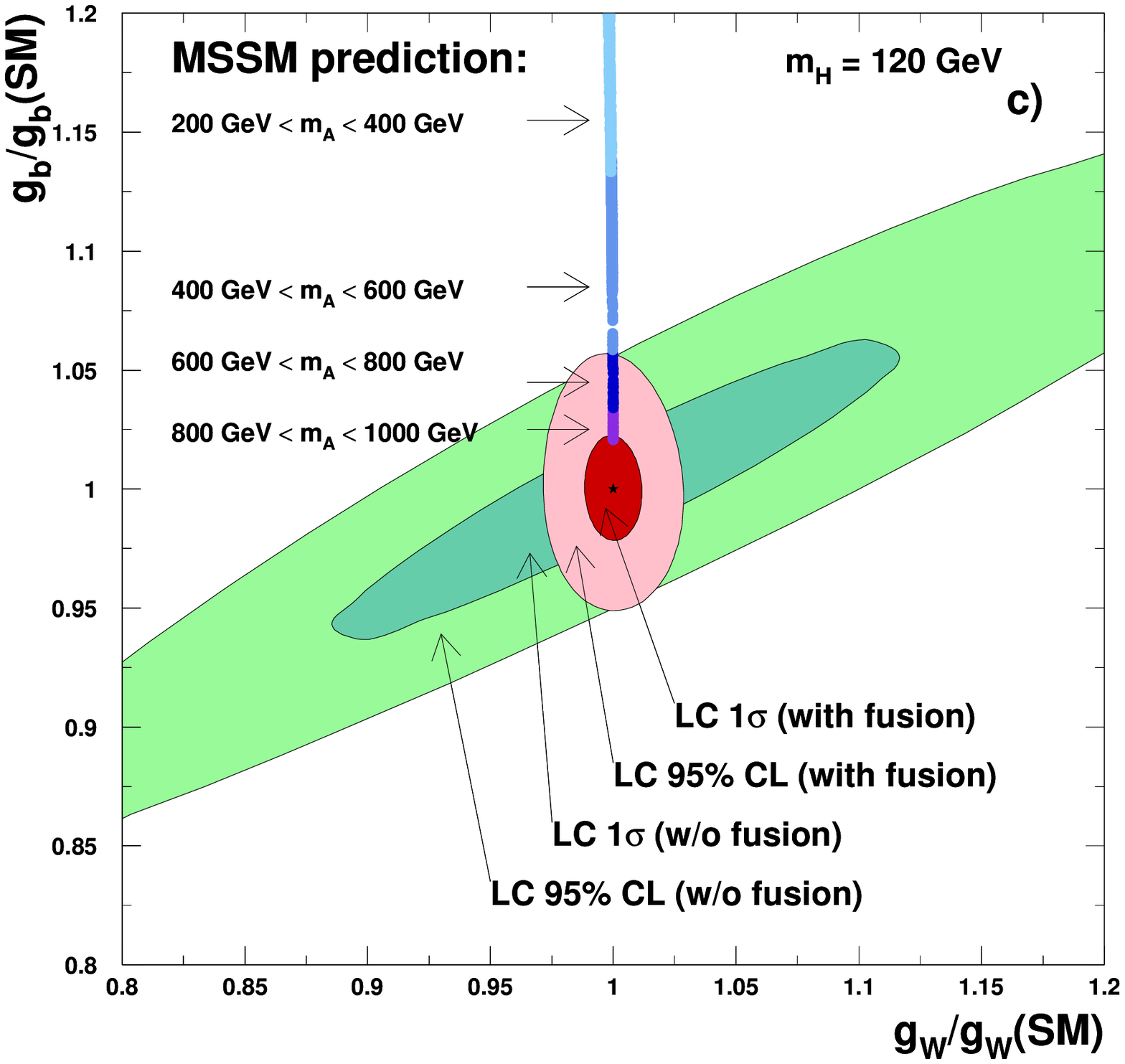,width=0.35\linewidth}\hspace*{-4mm}
\epsfig{file=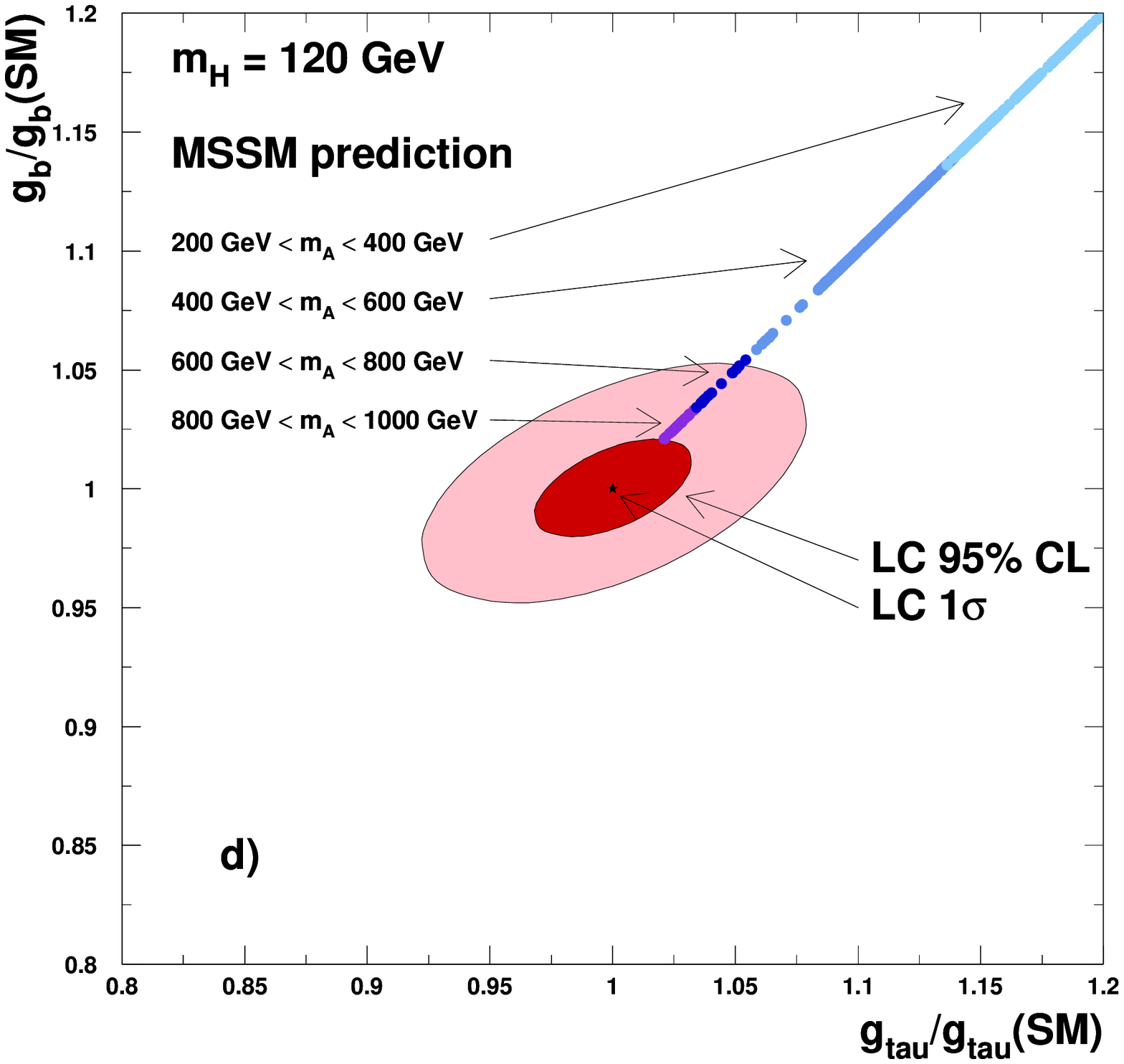,width=0.35\linewidth}\hspace*{-4mm}
\epsfig{file=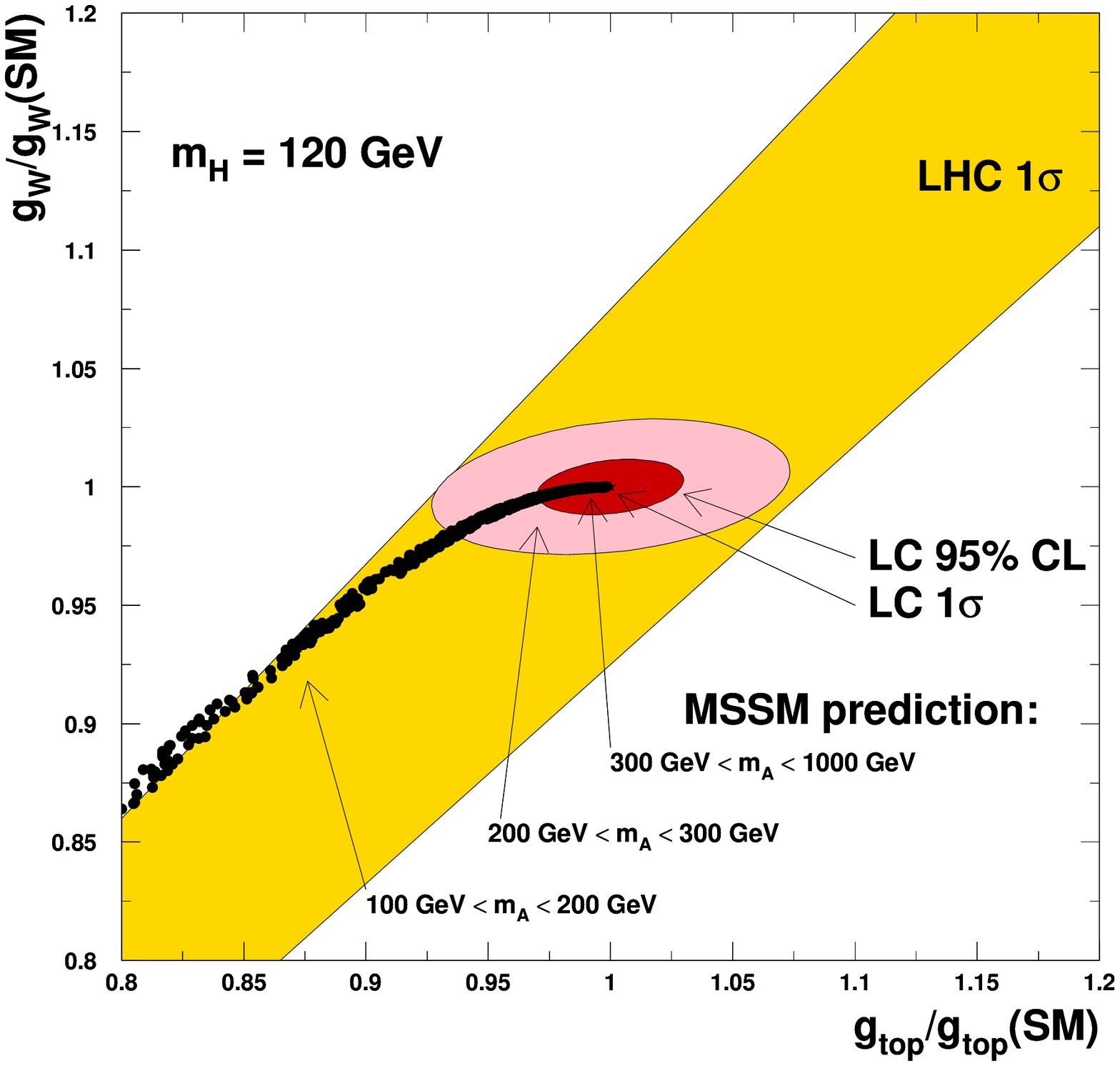,width=0.32\linewidth}
}
\end{center}
\vspace*{-.6cm}
\caption[Determination of the couplings of a SM--like Higgs and MSSM interpretation]
{Determination of the couplings of a SM--like Higgs boson at the ILC
and the interpretation within the MSSM. The contours are the couplings of
a 120 GeV Higgs boson as measured with 500\,fb$^{-1}$ data at $\sqrt{s}=350$
GeV except for $g_{Htt}$ which uses 800 GeV (here the expectation at the
LHC is also shown); from Ref.~\cite{Aguilar-Saavedra:2001rg}.}
\vspace*{-.4cm}
\label{Hfig:cpl-TESLA}
\end{figure}

A detailed analysis of the deviations of the couplings of the $h$ boson with a
mass $M_h=120$ GeV, from the predictions in the SM  has been performed in
Ref.~\cite{Aguilar-Saavedra:2001rg} using a complete scan of the MSSM $[M_A,
\tb]$ parameter space, including radiative corrections. In
Fig.~\ref{Hfig:cpl-TESLA}, shown are the  1$\sigma$ and 95\% confidence level
contours for the fitted values of various pairs of ratios of couplings, assuming
the experimental accuracies at the ILC discussed in the previous section. From a
$\chi^2$ test which compares the deviations, 95\% of all MSSM solutions can be
distinguished from the SM case for $M_A \lsim 600$ GeV and this number reduces
to only 68\% for $M_A \lsim 750$ GeV. In some cases, one is sensitive to MSSM
effects even for masses $M_A\!\sim\!1\;$TeV, i.e. beyond the LHC mass reach. If
the deviations compared to the SM are large, these precision measurements would
also allow for an indirect determination of $M_A$; for instance, in the mass
range $M_A=300$--600 GeV an accuracy of 70--100 GeV is possible on the $A$ mass.

This type of indirect determination cannot be made in a convincing way at the
LHC as the experimental errors in the various measurements are worse than at the
ILC; see Fig.~\ref{Hfig:cpl-TESLA} where the $g_{hWW}$ and $g_{htt}$ contours
are displayed. While at the ILC, MSSM effects can be probed for  masses close to
$M_A=1$ TeV, there is practically no sensitivity at the LHC. However, the
precision measurements at the ILC can gain enormously from other measurements
that can be performed only at the LHC. Indeed, the various Higgs couplings are
not only sensitive to the tree--level inputs $M_A$ and $\tb$ but also, on
parameters that enter through radiative corrections such as the stop and sbottom
masses which could be accessible only at the LHC. If, in addition, the $A$ boson
is seen at the LHC [which means that $\tb$ is large, $\tb \gsim 15$] and its
mass is measured at the level of 10\%, the only other important parameter
entering the Higgs sector at one--loop is the trilinear coupling $A_t$ [and to a
lesser extent, $A_b$ and $\mu$] which will be only loosely constrained at the
LHC. Nevertheless, using this knowledge and the fact that the top mass can be
measured with a precision of 100 MeV at the ILC, one can vastly improve the
tests of the MSSM Higgs sector that can be performed at the LHC or at the ILC
alone; see Ref.~\cite{Weiglein:2004hn} for a discussion on the LHC--ILC
complementarity.

\subsection{The Higgs sector beyond the MSSM}

In the MSSM with CP--violation, the three neutral Higgs bosons $H_1, H_2, H_3$
are mixtures of CP--even and CP--odd states. Because of the sum rule for the
Higgs couplings to gauge bosons, $\sum_i g_{H_iVV}^2 =g^2_{H_{\rm SM}}$, the
production cross sections in the Higgs--strahlung and $WW$ fusion processes
should be large for at least one of the particles and there is a complementarity
between $H_i$ single and $H_j H_k$ pair production.  In fact, similar to the
usual MSSM, the  normalized couplings are such that $|g_{H_1VV}| = |g_{H_2H_3V}|
\sim 1$ in the decoupling limit $M_{H^\pm} \gsim 200$ GeV and at least $H_1$ is
accessible for $\sqrt s \gsim 300$ GeV,  since $M_{H_1}\lsim 130$ GeV.  If two
or the three Higgs particles  are very close in mass, the excellent energy and
momentum resolution on the recoiling $Z$ boson in the Higgs--strahlung process
would allow to resolve  the coupled Higgs systems, e.g. from an analysis of the
lineshape. The presence of CP--violation can be   unambiguously checked by
studying the spin--spin correlations in Higgs decays into tau lepton pairs or
controlling the  beam polarization of the colliding photon beams at the $\gamma
\gamma$ option of the ILC; see Ref.~\cite{SUSY-CPV} for instance.

\begin{figure}[h!]
\vspace*{-.2cm}
\begin{center}
\mbox{
\epsfig{file=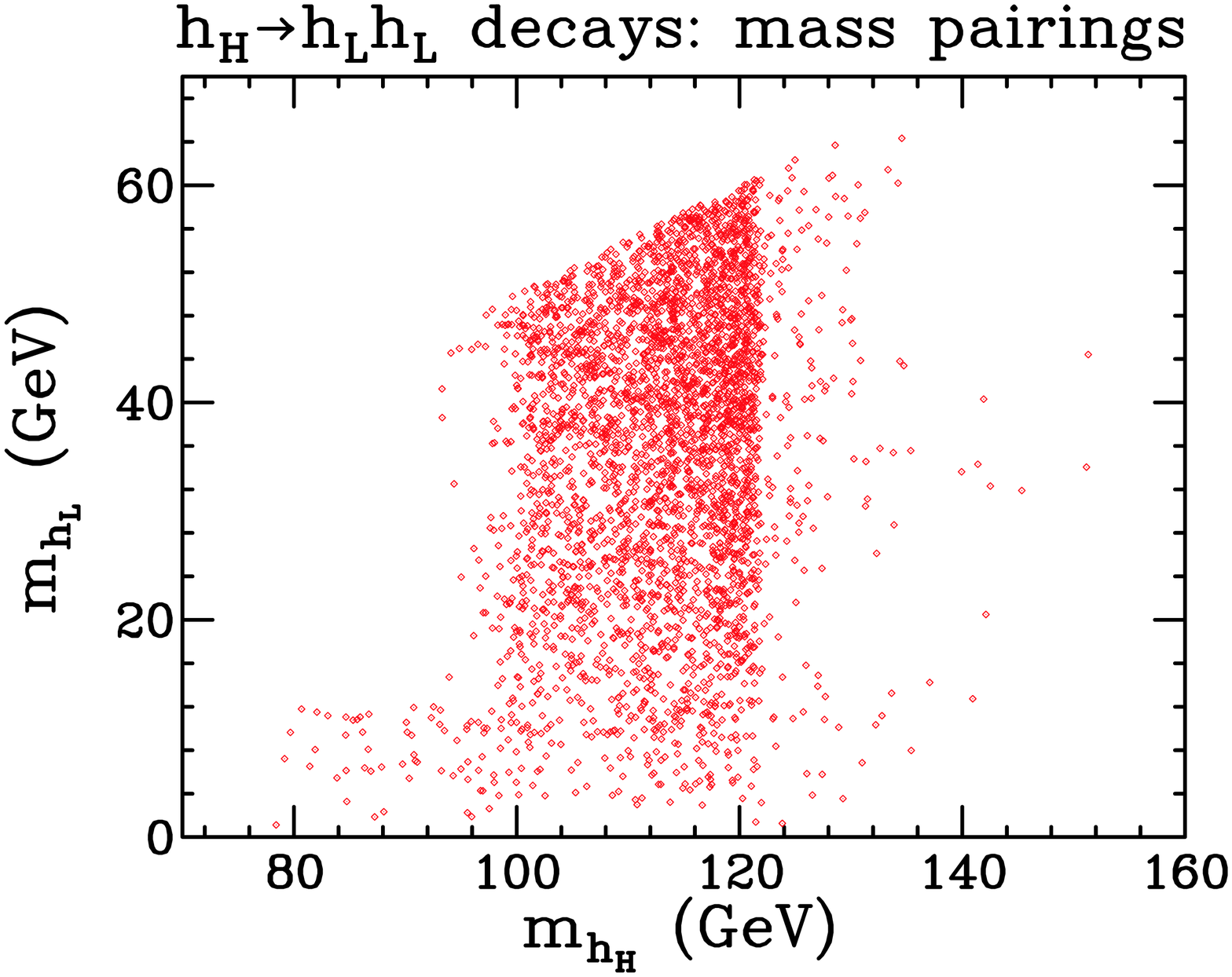,width=5.cm,height=5cm}
\epsfig{file=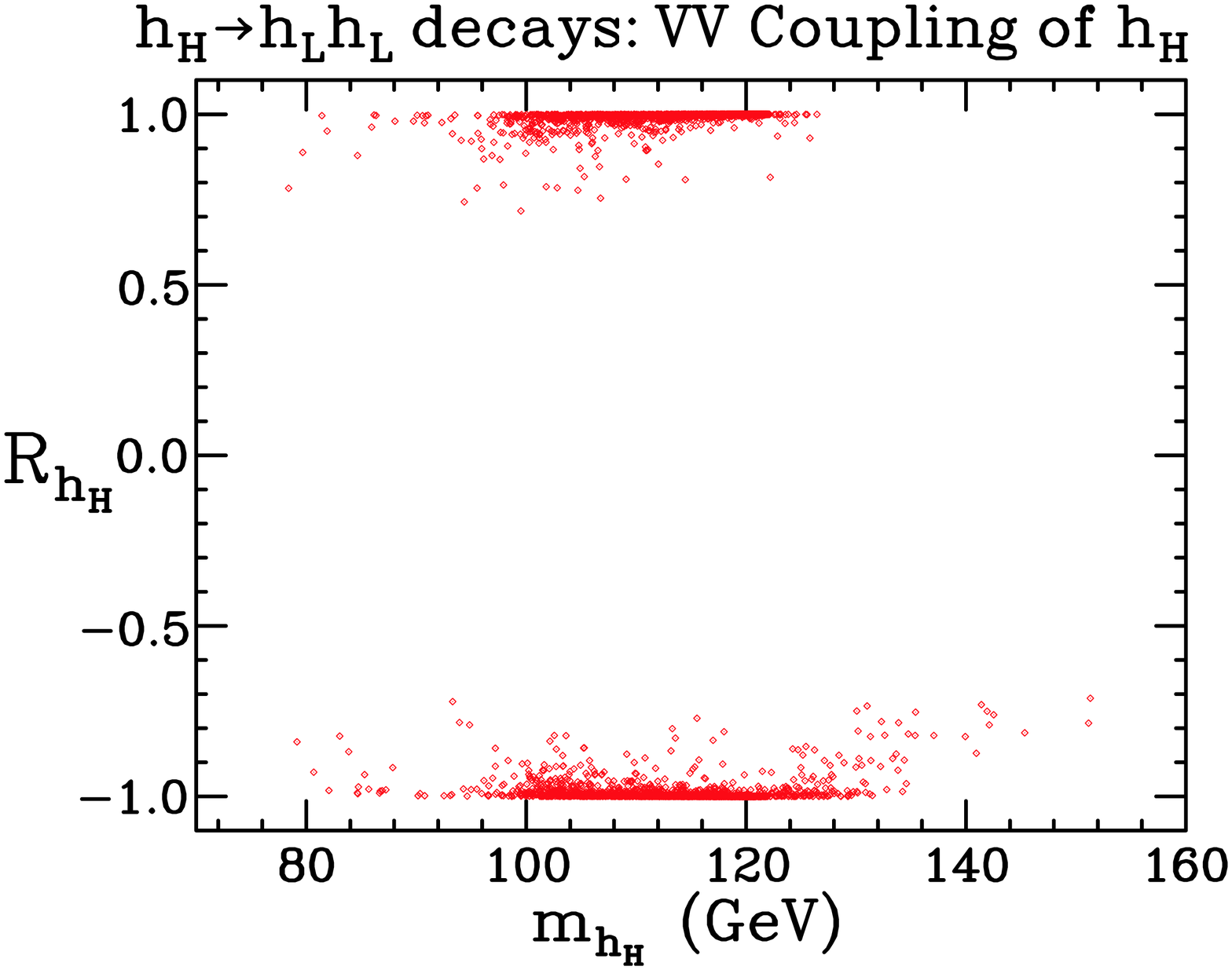,width=5.cm,height=5cm}
\epsfig{file=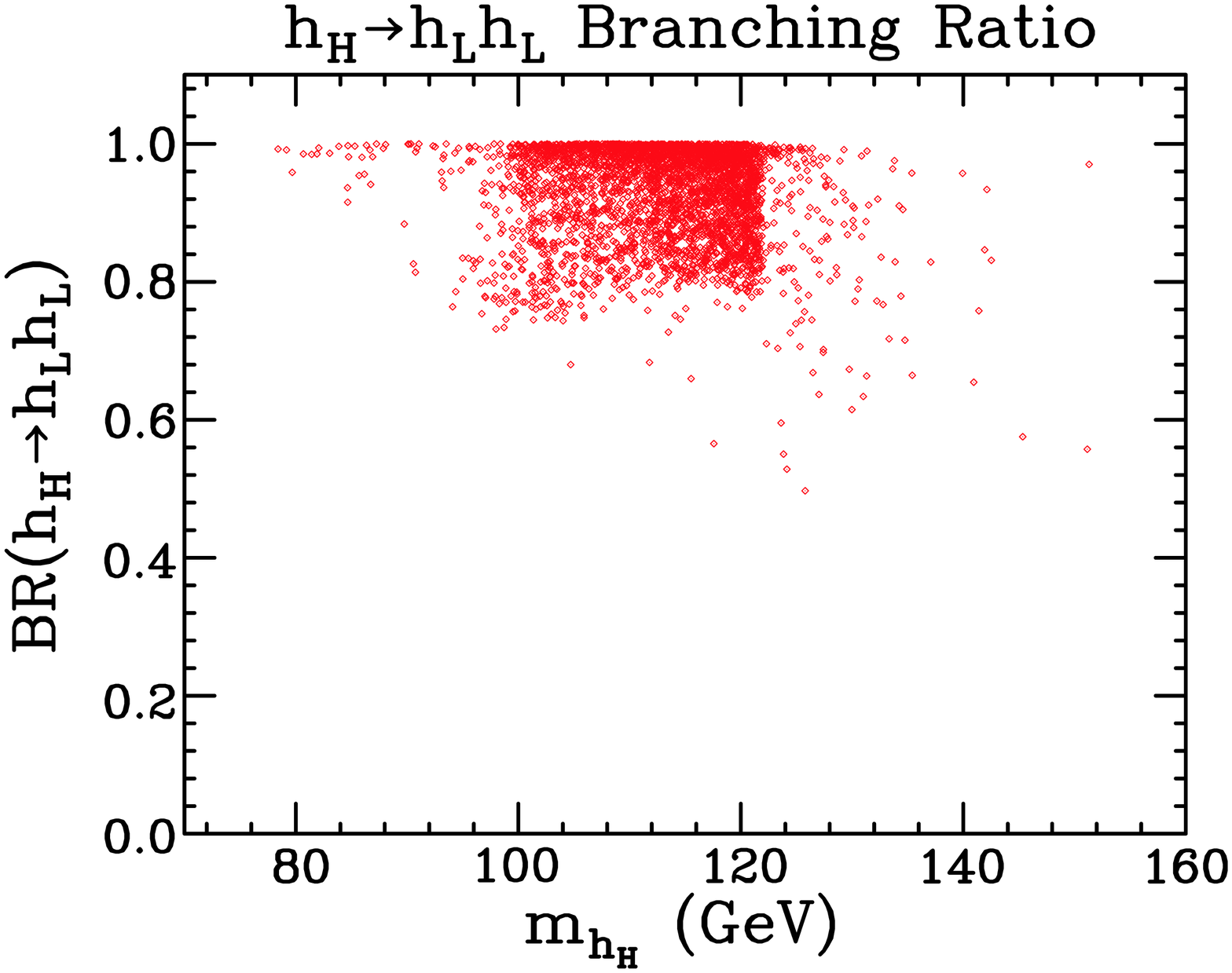,width=5.cm,height=5cm}
}
\end{center}
\vspace*{-.9cm}
\caption[Masses, couplings and branching ratios for some NMSSM Higgs bosons] 
{Scatter plots for the mass of the $h_H$ and $h_L$ boson (left), the
normalized couplings to the $h_H$ boson (middle) and the branching ratio 
of its decays to lighter $h_L$  bosons (right) as function of the Higgs mass;
they have been obtained in an NMSSM scan for regions with  $h_H \to h_L h_L$ 
decays; from~\cite{SUSY-NMSSM}.}
\vspace*{-.3cm}
\label{H-NMSSMscan}
\end{figure}

In the NMSSM, where a complex iso-scalar field is introduced, leading to an
additional pair of scalar and pseudoscalar Higgs particles,  the axion--type or
singlino  character of the pseudoscalar $A_1$ boson  makes it preferentially
light and decaying into $b$ quarks or $\tau$ leptons \cite{SUSY-NMSSMh}.
Therefore, in some areas of the NMSSM parameter space, the lightest scalar Higgs
bosons may dominantly decay into a pair of light pseudoscalar $A_1$ bosons
generating four $b$ quarks or $\tau$ leptons  in the final state. In fact, it is
also possible that  $H_1$ is very light with small $VV$ couplings, while $H_2$
is not too heavy and plays the role of the SM--like  Higgs particle; the decays
$H_2\to H_1 H_1$ can also be substantial and will give the same signature as
above.   This is exemplified in Fig.~\ref{H-NMSSMscan} where shown are scatter
plots for the mass of the SM--like Higgs boson $(h_H$) and the
pseudoscalar--like ($h_L)$ boson, the ratio of $h_H$ coupling to $Z$ bosons
($R_H$) compared to the SM Higgs coupling, and the branching ratio of the heavy
to light Higgs decay $(h_H \to h_L h_L)$ \cite{SUSY-NMSSM}. As seen previously, 
Higgs--strahlung allows for the detection of the CP--even Higgs particles
independently of their decay modes, provided that their couplings to the $Z$
boson are substantial, as it occurs for one CP--even Higgs boson  as exemplified
in the middle plot of  Fig.~\ref{H-NMSSMscan}. In fact,  thanks to the usual 
sum rule which relates the CP--even Higgs couplings to the those of the SM
Higgs, a ``no--lose theorem" for discovering at least one Higgs state has been
established for ILC while the situation is presently less clear for the
LHC and all Higgs particles could escape detection~\cite{SUSY-NMSSM,SUSY-NMSSMh}.

In a general SUSY model, with an arbitrary number of isosinglet and isodoublet
scalar fields (as well as a matter content which allows for the unification of
the gauge coupling constants), a linear combination of Higgs fields has to
generate the $W/Z$ boson masses and thus, from the triviality argument discussed
earlier, a Higgs particle should have a mass below 200 GeV and significant
couplings to gauge bosons \cite{HSUSY-thbound}. This particle should be
therefore kinematically accessible at the ILC with a c.m. energy $\sqrt s \gsim
350$ GeV. It can be detected in the Higgs--strahlung process independently of
its (visible or invisible) decay modes. If its mass happens to be in the high
range, $M_h \sim 200$ GeV, at least its couplings to $W,Z$ bosons and
$b$--quarks (eventually $t$--quarks at high energies and luminosities), as well
as the total decay widths and the spin--parity quantum numbers can be 
determined. 

We should stress again that even in scenarios with invisible Higgs decays, as
would be the case for instance of spontaneously broken R--parity scenarios in
which the Higgs particles could decay dominantly into escaping Majorons, $H_i
\to JJ$, at least one CP--even Higgs boson is light and has sizable couplings to
the gauge bosons and should be observed by studying the recoil mass spectrum
against the $Z$ boson in the Higgs--strahlung process. 

From the previous discussions, one can thus conclude that the ILC is the ideal 
machine for the SUSY Higgs sector, whatever scenario nature has chosen.

\section{The Higgs sector in alternative scenarios}

As discussed in the introductory section, several non--supersymmetric scenarios
beyond the SM predict new features which might significantly affect the Higgs
sector. To illustrate the large impact that such models can have, we will take
as an example  the effects of a radion in warped extra dimensional models. 
Other possibilities will be discussed in chapter \ref{sec:alternatives}. 

In Randall--Sundrum models \cite{WED}, a
scalar radion field is introduced to stabilize the distance between the SM and
the gravity brane. Carrying the same quantum numbers, the Higgs and radion
fields can mix and the properties of the Higgs boson will be altered
\cite{Hewett:2002nk,Dominici:2002jv}. In particular,  Higgs--radion mixing
can lead to important  shifts in the Higgs couplings which become apparent in
the various decay widths. These shifts depend on the radion and Higgs masses,
the mixing parameter $\xi$ which is expected to be of order unity  and the ratio
of the Higgs vacuum expectation value   $v$ to the effective new scale $\Lambda
\sim 1\;$TeV. 

\begin{figure}[h!]
\vspace*{5mm}
\begin{minipage}{7cm}
\includegraphics[width=7cm,height=6cm]{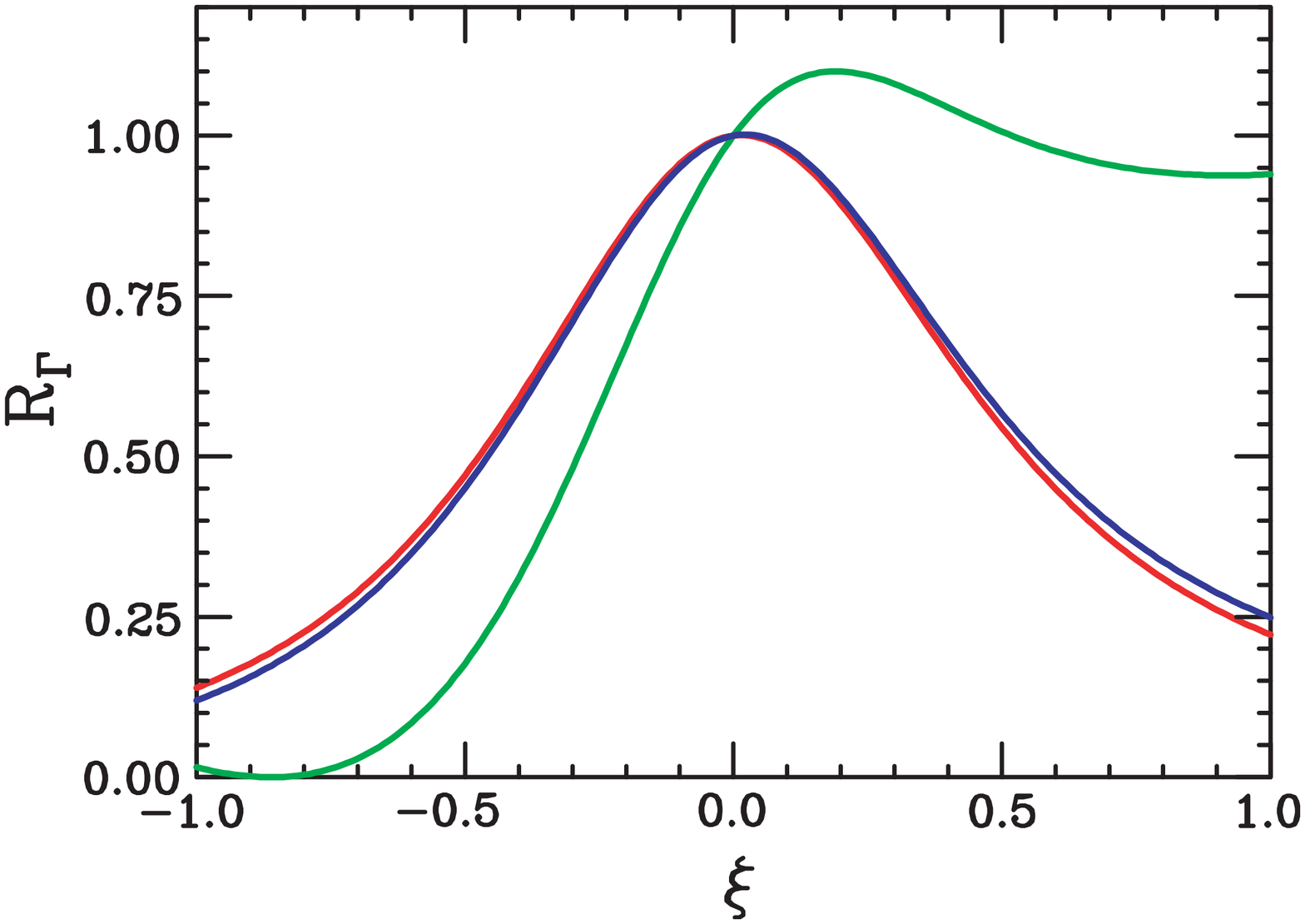}
\end{minipage}
\hspace*{-2cm}
\begin{minipage}{1.cm}
\hspace*{.7cm} $gg$ \vspace*{.9cm}

\hspace*{.8cm} $\gamma \gamma$\\
\hspace*{-.5cm}$ff/VV$\\
\end{minipage}
\hspace*{1cm}
\hspace*{1cm}
\begin{minipage}{5cm}
\vspace*{-5mm}
\includegraphics[width=5.2cm,height=6cm]{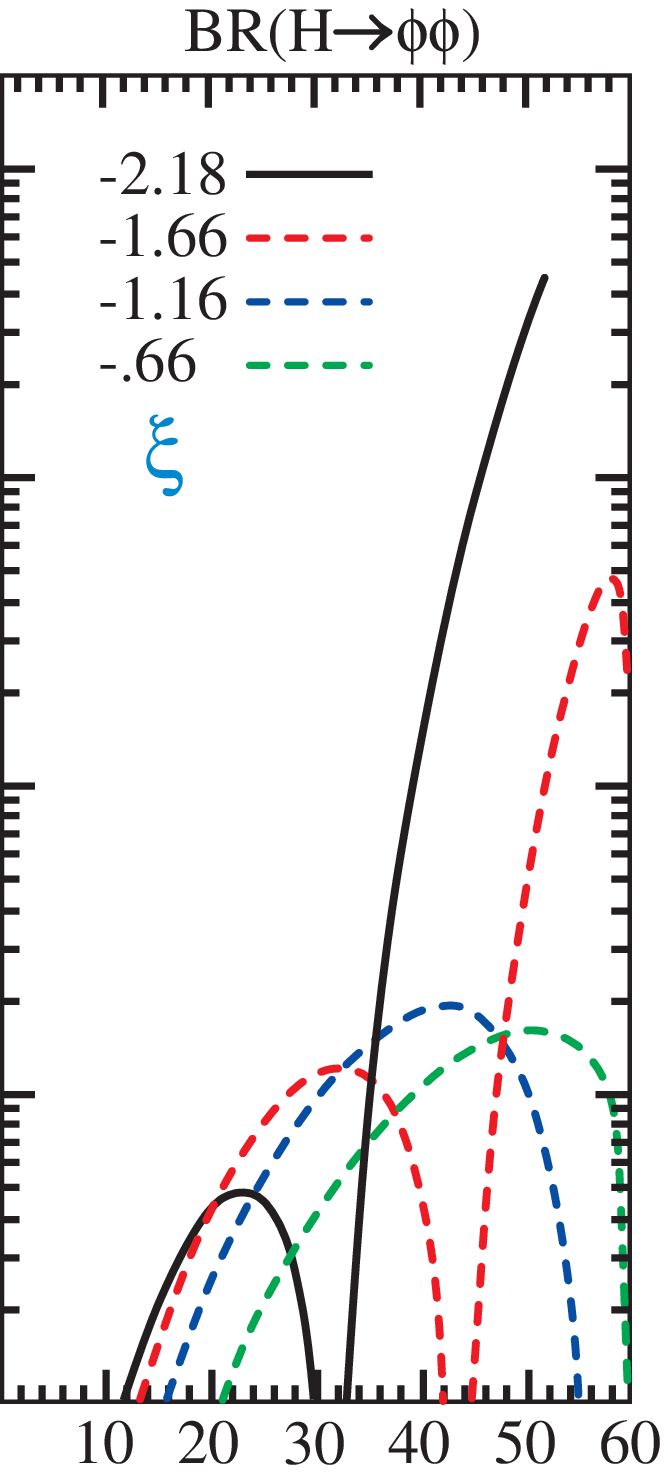}
\end{minipage}
\hspace*{-.1cm}
\begin{minipage}{1cm}
\hspace*{1.5cm} \\  \vspace*{.7cm} $1$\\ \vspace*{.7cm} $10^{-1}$\\  \vspace*{.7cm} 
 $10^{-2}$\\  \vspace*{.7cm} $10^{-3}$\\  \vspace*{.7cm} $10^{-4}$
 \hspace*{-1cm}
\end{minipage}
\vspace*{-6mm}
\caption[Some decay widths and branching ratios for a Higgs mixing with a 
radion]  
{Left: the ratio $R_\Gamma$ of Higgs partial widths to their SM values, 
as a function of the mixing parameter $\xi$ with $M_H=125$ GeV,
$M_\phi=300$ GeV and $v/\Lambda=0.2$ \cite{Hewett:2002nk}. Right: the  
branching fractions for the decays $H\to \phi \phi$ as a function of $M_\phi$ 
for different $\xi$ values and $M_H=120$ GeV, $\Lambda=5$ TeV 
\cite{Dominici:2002jv}.}
\label{fig:Hradion}
\vspace*{-4mm}
\end{figure}

The ratio of Higgs partial decay widths in these models to their SM values is
illustrated in the left--hand side of Fig.~\ref{fig:Hradion} for  $M_H=125$ GeV
and various values of the radion mass  $M_\phi$ and  the ratio $v/\Lambda$
\cite{Hewett:2002nk}. As can be seen, while the shifts in the  $f \bar f/VV$ and
$\gamma \gamma$ widths are rather similar, the shift in the $H\to gg$ partial
decay  width is different; the width  can become close to zero for some values
of the mixing.  The impact of mixing in $f\bar f$ and $VV$ final states is in
general smaller and the branching ratios will not be significantly affected as
these decays are dominant. This implies that it will be imperative to perform a
precise measurement of the Higgs total decay width in order to probe the mixing 
with radions. At the ILC, the shift in the photon couplings can be probed in
$\gamma \gamma \to H$ production while in the $\eei$ option, the $H\to gg$  
width can be precisely measured. Since the total decay width can be also
measured, the absolute values of the Higgs couplings can be unambiguously
determined.  

The suppression of the $Hgg$ loop induced coupling  can occur in other
extensions of the SM as well. Besides the MSSM with light top squarks and large
trilinear $A_t$ couplings, the ${\rm SU(2)_R}$ partner of the right--handed top
quark in warped extra dimensional models with an extended  left--right symmetric
structure will also contribute to the  $Hgg$ vertex and could interfere
destructively with the top quark contribution, leading to a much smaller
coupling \cite{H-RSHgg}. In the strongly interacting light Higgs scenario
proposed recently \cite{H-SILH}, the  Higgs couplings to gluons, as well as  the
couplings to fermions and gauge bosons, are also suppressed.  Note that the
suppression of the $Hgg$ coupling  would lead to a decrease  of the cross
section for the dominant Higgs production mechanism in proton collisions, $gg
\to H$, and would make the Higgs search more complicated at the LHC.  

Another important consequence of radion mixing is the decays of the Higgs boson
into a pair of radions. Indeed, if the radion is relatively light, the decays
$H\to \phi \phi$ might be kinematically accessible and, for some mixing values,
the branching fractions might be substantial. This is exemplified in the
right--hand side of  Fig.~\ref{fig:Hradion} where BR($H\to \phi\phi$)  is
displayed as a function of the mixing parameter $\xi$ for $M_H=120$ GeV and
$\Lambda=5$ TeV \cite{Dominici:2002jv}. As can be seen, the rate can be very
large, in particular for the largest  $|\xi|$  values when $M_\phi$ is close to
$\frac12 M_H$. The detection of the $H\to \phi\phi$ decay mode could provide the
most striking evidence for the presence of non--zero $\xi$ mixing. In the
considered mass range, $M_\phi \lsim 60$ GeV, the radion will mainly decay into
$b\bar b$ and $gg$ final states, while the $\gamma \gamma$ branching ratio is
very small. Observing these final states will be rather difficult at the LHC
while in Higgs--strahlung at the ILC, the final state $ZH \to Z \phi \phi \to
Z+4\;$jets should be easily detectable.  Finally, the reverse decay process
$\phi \to HH$ is also possible for radion masses  larger than $M_\phi \gsim 230$
GeV. The branching fractions, when this decay occurs, can be rather large. For
$M_H \sim 120$ GeV, the process $\eei \to Z\phi \to Z HH \to Z+4b$ would
dramatically increase the $ZHH$ production rate at the ILC and would lead to
spectacular events; see chapter \ref{sec:alternatives}. 

Note that in models with large extra dimensions \cite{LED}, the
interaction  of the Higgs field and the Ricci scalar curvature of the induced
four--dimensional metric also generates a mixing term with the closest
Kaluza--Klein graviscalar fields \cite{H-graviscalars}. This mixing results in
an effective Higgs decay width, $\Gamma(H \to  {\rm graviscalar})$, which is
invisible as the graviscalars are weakly interacting and mainly reside in the
extra dimension while the Higgs is on the TeV brane. These invisible Higgs
decays can be largely dominating. In addition, there is the possibility of Higgs
decays into a pair of graviscalars, but the rates are smaller than the ones from
mixing. These decays will complicate the Higgs search at the LHC, while they can
be easily detected in Higgs--strahlung at the ILC and the branching fractions
precisely measured.  

Other models also predict large rates for invisible decays of the Higgs boson. 
An example, besides decays into the lightest neutralinos and Majorons 
\cite{H-RparityV} in  non minimal SUSY models,  is again given by extra
dimensional models in which the Higgs bosons decay into the lightest
Kaluza--Klein particles which are supposed to form the dark matter in the
universe \cite{DM-LZP}. Finally, in the minimal extension of the Higgs sector
with a singlet field $S$, invisible $H\to SS$ decays occur and could be the
dominant channels \cite{NMSM}. 

Thus, one can conclude that also in alternative scenarios to supersymmetry, 
the ILC will be a valuable tool to unravel the electroweak symmetry breaking
mechanism.

%
\chapter{Couplings of gauge bosons}
\label{sec:couplings}

The Standard Model has been thoroughly tested in the last two decades
with the high-precision measurements of LEP, SLC and the Tevatron
which have firmly established that it describes correctly the
electroweak and strong interactions of quarks and leptons. However,
many important aspects of the model, besides the electroweak symmetry
breaking mechanism for particle mass generation, need more
experimental investigation. This can be done at the ILC in the
production of fermion antifermion pairs as well as electroweak gauge
bosons, in particular single and pair production of $W$ bosons, which
provide the largest cross sections leading to event samples of a few
million each with the ILC expected luminosity.

An important task is to measure the interactions amongst gauge bosons
much more precisely than it was possible at LEP and the Tevatron and
will be possible at the LHC, for instance, determine the trilinear
self-couplings of the $W$ and $Z$ bosons at the per-mille level. 
Anomalous values of these couplings are most precisely measured in the
clean environment of an $\ee$ collider and at the highest possible
c.m.~energy $\sqrt s$. The ILC thus allows to constrain new physics at
scales far above the direct reach of the collider through quantum
corrections and, alternatively, to probe small effects from operators
in an effective Lagrangian that are suppressed by powers of
$s/\Lambda^2$ where $\Lambda$ is the scale at which the new physics
sets in. The measurement of the quartic gauge boson self-couplings is
of utmost importance, especially if no Higgs particles have been
observed at the LHC and ILC. In this scenario, the interactions
between massive gauge bosons become strong at energies close to $1
\TeV$ and the effective scale for the new interactions needed to restore
quantum-mechanical unitarity can be extracted from a precise
measurement of anomalous values of these self-couplings.

Another important task, once the top quark and the Higgs
boson masses are accurately known, is to measure the value of
the effective weak mixing angle $\stl$ and the $W$  boson mass $\MW$
and to test more precisely their quantum corrections and the
consistency of the model in an unambiguous way. These parameters
can be determined with an accuracy that is far better than the one
presently available by running the high-luminosity ILC near the $Z$
boson resonance and near the $WW$ threshold and this
test can be performed at an unprecedented level of precision. Then,
and only then, virtual effects of new physics beyond the SM can be
probed in an unambiguous way.  Furthermore, observables in fermion
pairs produced in $\ee$ collisions at high energy are sensitive to new
physics far beyond the center of mass energy.  As one example, an ILC
running at $500 \GeV$ is sensitive to effects of a heavy $Z^\prime$ boson, that
is predicted in many SM extensions, beyond the reach of the LHC and it can, 
if such a particle has been observed at the LHC, measure its
couplings and thus distinguish between the various models where this
new $Z^\prime$ boson occurs.

Finally, the ILC offers the possibility of testing QCD at high energy
scales in the experimentally clean and theoretically tractable $\ee$
environment. In particular, it allow a more precise
determination of the strong coupling $\alpha_s$, which is presently
known with an error of several percent \cite{PDG}, and the measurement
of its evolution with the energy scale.  Since the weak and
electromagnetic couplings are known with a much higher accuracy, this
measurement is very important as the present error on $\alpha_s$
represents the dominant uncertainty on the prediction of the scale for
grand unification of the strong, weak and electromagnetic~forces.

\section{Couplings of gauge bosons to fermions}

In the SM, fermion pair production, $\ee \to \ff$ for ${\rm f} \ne
{\rm e}$, proceeds at tree-level via the exchange of photons and $Z$
bosons in the $s$-channel.  These processes can thus be used to
measure the couplings of fermions to gauge bosons.  All cross sections
are given by the product of the initial state $\ee V$ and the final
state $\ff V$ couplings. Assuming universality, lepton pair production
thus measures the leptonic couplings while quark production measures
the product of the leptonic and the quark couplings.

Since weak interactions violate parity, the vector- ($\gvf$) and the
axial-vector- ($\gaf$) couplings can vary independently in general.
However they can be disentangled experimentally without major
problems. The total cross section is proportional to the squared sum
of the couplings ($\gvf^2 + \gaf^2$) while several asymmetries like
the left-right asymmetry $A_{LR}^f$ with polarized beams or the
forward-backward asymmetry $A_{FB}^f$ measure their ratio
$\gvf/\gaf$.

The fermion couplings to the $Z$ boson have already been measured with
great success at LEP and SLD on the $Z$-boson resonance
\cite{Z-Pole}. The comparison of their precise measurements with
accurate calculations led to the prediction of the top quark mass
before it was actually discovered \cite{lepew93} and to the current
prediction that the Higgs boson should be light \cite{Z-Pole}.

At $\sqrt{s} \sim 500 \GeV$, $\ee \rightarrow \ff$ samples of a few million
events are expected so that the couplings can be measured at the per-mille
level accuracy. The main interest in fermion pair production lies in limits on 
physics beyond the SM. Apart from photons and $Z$ bosons, all other particles 
that couple to electrons and the final state fermions can be exchanged and
thus contribute to the cross section.  In a more model independent approach,
the virtual effects of new physics can be parameterized in terms of contact
interactions using the effective helicity-conserving Lagrangian, with the  
interaction strength set to $g_*^2 / 4\pi = 1$,
\begin{equation}
{\cal L}_{\rm eff} = \sum_{i,j=L,R} \eta_{ij} \, \frac{4\pi}{\Lambda_{ij}^2}  
\, \bar{e}_i \gamma^\mu e_i \cdot \bar{f}_j \gamma_\mu f_j\; . 
\end{equation}
Here, one assumes that the masses of the exchanged particles are so heavy, that
details of the propagator are not felt and only the Lorenz structure of the
couplings remains visible. 

\begin{figure}[htbp]
\vspace*{-4mm}
  \mbox{
\hspace*{-3mm} 
  \includegraphics[height=9cm]{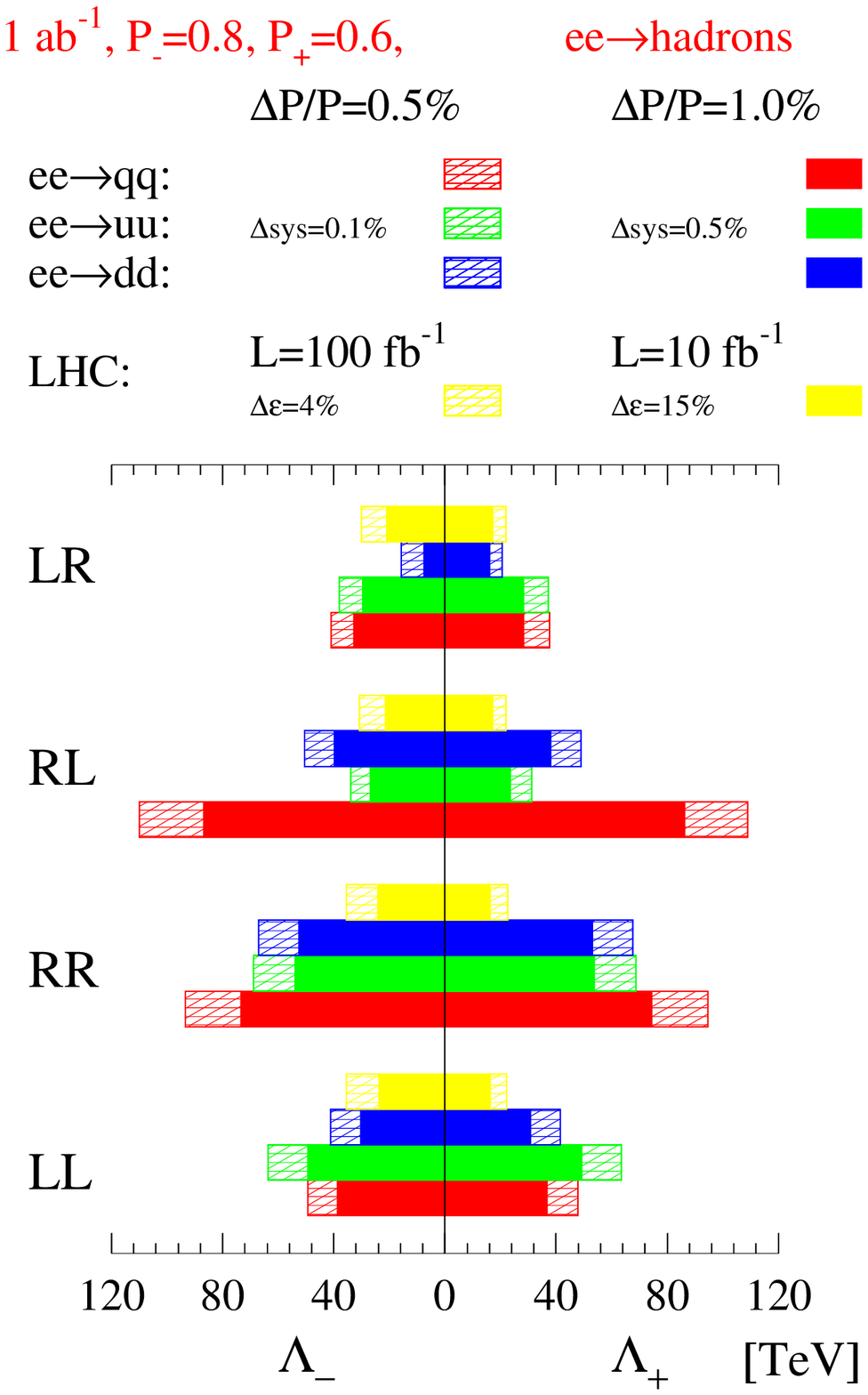}\hspace*{-1cm}
  \includegraphics[height=9cm]{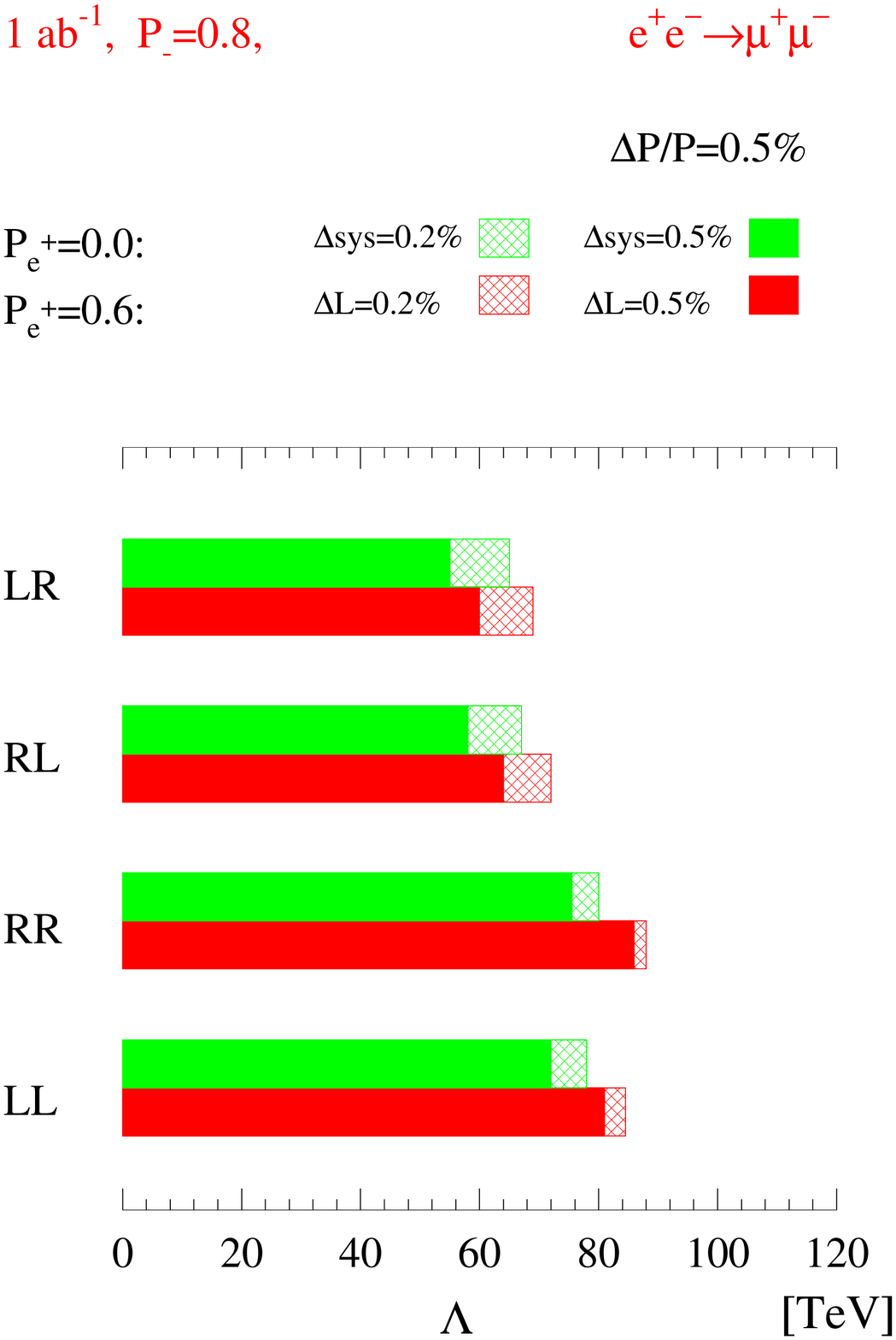}
}
\vspace*{-8mm}
  \caption[Limits on contact interactions from fermion couplings
  at the ILC]
    {Sensitivities at the 95\% CL of a $500 \GeV$ ILC to contact interaction 
    scales $\Lambda$ for different helicities in $e^+e^-\to$ hadrons (left) and
$e^+e^-\to \mu^+\mu^-$ (right) including beam polarization 
\cite{contact_sabine}. }
  \label{fig:contact}
\vspace*{-4mm}
\end{figure}

In a detailed experimental analysis it has been shown that fermion
pair production at the ILC provides a large sensitivity to the contact
interaction scales $\Lambda_{ij}$ \cite{contact_sabine}. The limits on
the scales that one can extract from the precision measurements are
shown in Fig.~\ref{fig:contact} for quark (left) and muon (right) pair
production at $\rts=500 \GeV$ using 1\,ab$^{-1}$ of data, ${\rm e}^-$
polarization and various assumptions for the systematical errors; for
muon final states, the significant improvement using ${\rm e}^+$
polarization is also displayed. As can be seen, scales of the order of
$\Lambda\!=\!20$ to $100 \TeV$ can be reached at this energy,
significantly higher than those obtainable at the LHC; this is shown
in the $\ee \to q\bar q$ case as the LHC cannot probe
$\ee \mu^+\mu^-$ couplings. At $\sqrt{s}\!=\!1 \TeV$, the limits are
expected to be approximately 50\% larger.



A model dependent application of the precision measurements of fermion
pair production, besides probing for instance fermion compositeness
and/or anomalous couplings, leptoquarks, etc., is the search for
heavy neutral $Z^\prime$ vector bosons. The fermion cross sections and
asymmetries are altered by the virtual exchange of the $Z^\prime$ boson and
are thus sensitive to its mass and couplings. In general, the ILC
precision measurements at $\rts = 500 \GeV$ are more or equally
sensitive to the $Z^\prime$ mass as the LHC direct mass reach and more
sensitivity is gained at a $1 \TeV$. If a $Z^\prime$ boson with a mass
$M_{{\rm Z}^\prime} \lsim 3-4 \TeV$ has been observed at the LHC, the ILC
allows to determine the model origin.  A more detailed discussion of
$Z^\prime$ effects and other applications of ILC precision measurements is
given in chapter~\ref{sec:alternatives}.

Another possibility to measure the fermion couplings to the $Z$ boson
is to return to the $Z$-resonance in the GigaZ option of the ILC
\cite{gigaz}. With a luminosity around ${\cal L} = 5 \cdot 10^{33}
{\rm cm}^{-2} {\rm s}^{-1}$, a billion $Z$ decays can be collected
within a few months of running. The most sensitive observable to
measure the $Z$-fermion couplings is the left-right polarization
asymmetry $\ALR = \frac{1}{\pol} \frac{\sigma_L - \sigma_R}{\sigma_L +
  \sigma_R}$, where $\sigma_{L,R}$ denotes the cross section for
left/right handed polarized electron beams and $\pol$ the beam
polarization.  This asymmetry is sensitive to the ratio of the vector
to axial-vector coupling of the electron to the $Z$ boson, $\ALR = 2
\gve \gae/(\gve^2 + \gae^2)$, which in turn measures the effective
weak mixing angle in $Z$ decays, $\gve/\gae = 1 - 4 Q_e \stl$.

If ${\rm e}^\pm$ polarization is available, the cross section for a given
beam polarization is given by \begin{equation} \sigma = \sigma_u
  \left[ 1 - \ppl \pmi + \ALR (\ppl - \pmi ) \right] .
  \label{eq:blondel} \end{equation} If the sign of the electron and
positron polarization can be flipped independently, four measurements
with four unknowns are possible, so that $\ALR$ can be measured
without the need for absolute polarimetry. Polarimeters are, however,
still needed to measure a possible polarization difference between the
left- and the right-handed state and to track any time dependences
of the polarization which enters in the polarization product of
equation (\ref{eq:blondel}).  $\ALR$ can be measured with a
statistical accuracy of about $\Delta \ALR = 3 \cdot 10^{-5}$. The
largest systematic uncertainty by far comes from the knowledge of the
beam energy. The slope close to the $Z$-peak is d$\ALR/$d$\sqrt{s} =
2 \cdot 10^{-2}/\GeV$ and is due to the $\gamma-$$Z$ interference. 
Not to be dominated by this effect the center of mass
energy needs to be known to $1\MeV$ relative to the $Z$-mass which has to
be calibrated by frequent scans.
If the beamstrahlung is the same in the peak running
and in the scans for energy calibration, its effect cancels out and
beamstrahlung does not contribute to the systematic uncertainty.

Conservatively, a final error of $\Delta \ALR = 10^{-4}$ will be assumed
corresponding to $\Delta \stl =1.3 \cdot 10^{-6}$. This is an improvement of
more than one order of magnitude compared to the value obtained at LEP/SLD. To
achieve this precision, one also needs to know the fine structure constant at
the scale $\MZ$, $\alpha(\MZ^2)$, with a much better precision than presently. 
Measuring the cross section $\sigma(\ee \rightarrow {\rm hadrons)}$ to 1\%
roughly up to the $J/\Psi$ resonance would reduce the uncertainty of the $\stl$
prediction to the level of the experimental error \cite{fred}. With modest
upgrades this is possible using present machines.

If absolute values of the couplings are to be measured, one needs to
obtain the $Z$ boson leptonic width $\Gll$. The peak cross section
$\sigma(\ee \rightarrow \lept)$ for $\sqrt{s} = \MZ$ is proportional
to $\Gll^2/\Gtot^2$. Thus, to measure $\Gll$, apart from the cross
section, the total width of the $Z$ boson needs to be determined from
a scan. Many systematic uncertainties enter the determination of
$\Gll$ and the relative knowledge of the beam energy affects the
determination of $\Gtot$ while the knowledge of the total luminosity
and the selection efficiency directly enter the cross section
measurement. The most severe systematics are expected to come from the
beam energy spread and from beamstrahlung. Because the second
derivative of a Breit-Wigner distribution at the peak is very large,
the effective peak cross section is strongly reduced by these effects,
which may well limit the $\Gll$ measurement. A probably optimistic
estimate \cite{gigaz} shows a possible improvement of a factor two
relative to the LEP measurement.

The b-quark, the isospin partner of the top quark, plays a special
role in many models. Its forward-backward asymmetry as measured at
LEP is one of the few observables that deviates from the SM prediction
by more than two standard deviations \cite{Z-Pole}, a deviation that
can be explained, e.g.~in extra-dimensional models
\cite{Djouadi:2006rk}. At GigaZ, the asymmetry parameter ${\cal A}_b =
\frac{2 \gvb \gab}{\gvb^2 + \gab^2}$ can be measured one order of
magnitude better than at LEP/SLD and without a dependence on the $Zee$
couplings, revealing if the current deviation is real or simply a
statistical fluctuation. Also the measurement of the fraction of
$\bb$ events in hadronic $Z$ decays, $R_{\rm b}$, which is proportional to
$\gvb^2 + \gab^2$ can be improved by a factor five.

In addition to the fermion-$Z$ couplings, the $W$ boson mass can be
measured at the ILC with a threshold scan to a precision around $6
\MeV$ \cite{wscan}.  Because of a similar structure of the radiative
corrections, this observable is usually interpreted together with the
coupling measurements. Within a wide range of models, the measurement
of $\MW$ can replace the one of $\Gll$ which is not accurately
determined as mentioned above. However, this measurement takes one
year of running at $\sqrt{s}\!\sim\!160 \GeV$, where not many physics
issues can be addressed.

\begin{figure}[htbp]
  \centering
  \includegraphics[width=0.45\linewidth]{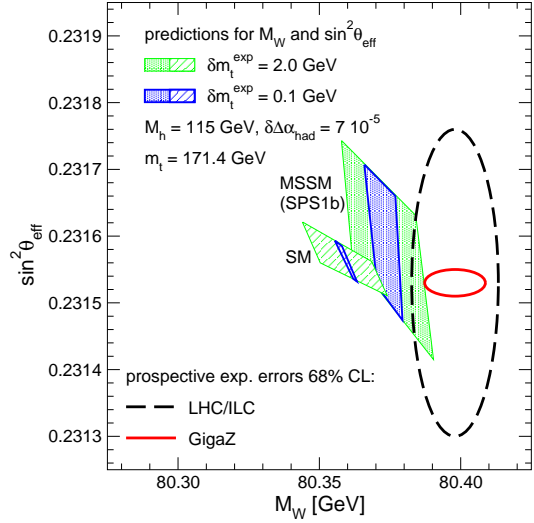}
\vspace*{-5mm}
  \caption[$\stl$ vs $\MW$ measurements and compared SM and MSSM predictions]
    {$\stl$ versus $\MW$ for different experimental assumptions
    compared to the predictions from the SM and the MSSM \cite{gigazsven}.}
  \label{fig:gigazsven}
\end{figure}

As a possible application of the precision measurements discussed above, 
Fig.~\ref{fig:gigazsven} displays the projected $\stl$ and $\MW$ measurements
under different assumptions compared to the prediction of the SM and its
supersymmetric extension, the MSSM \cite{gigazsven}. Within the SM, a stringent
test of the model is possible while for the MSSM the sensitivity is good enough
to constrain some of its parameters. It can also be seen that the precise top
quark mass measurement at the ILC is needed for an optimal sensitivity of the
comparison.

\section{Couplings among gauge bosons}

\subsection{Measurements of the triple couplings}

The couplings among the electroweak gauge bosons are directly given by
the structure of the gauge group. This structure can thus directly be
determined by a measurement of the gauge boson interactions.  $W$-boson
pair production is an especially interesting process in this respect.
Without gauge interactions, ${W^+ W^-}$ pairs are produced in
$\ee$ collisions via neutrino $t$-channel exchange.  This mechanism
violates unitarity and is regulated by the photon and $Z$ boson
$s$-channel exchange processes which involve the triple gauge boson
couplings.  Since the exact values of the self-couplings, as predicted
by the ${\rm SU(2)_L \times U(1)_Y}$ gauge structure, are needed for
unitarity restoration, small changes lead to large variations of the
cross section. For this reason, the $\ee \to {W^+ W^-}$ process is much
more sensitive to the triple gauge boson couplings than one would
naively expect from cross section estimates.

The triple gauge boson couplings are conventionally parameterized as
\cite{wwlag}:
\begin{eqnarray}
  \nonumber
  {\cal L}_{{WWV}} & = & g_{{WWV}} \bigg[ 
  i g_1^{V}  V_\mu \left( W^{-}_{\nu}  W^{+}_{\mu\nu}
  - W_{\mu\nu}^- W^{+}_{\nu} \right)
  + i \kappa_{V} W_\mu^- W_\nu^+ V_{\mu \nu }
  + i \frac{\lambda_{V}}{\MW^2}
  W_{\lambda\mu}^- W^{+}_{\mu\nu} V_{\nu\lambda} \\
  \nonumber
  & & + \, g_4^{V} W_\mu^- W_\nu^+
  \left( \partial_\mu V_\nu + \partial_\nu V_\mu \right)
  + g_5^{V} \epsilon_{\mu\nu\lambda\rho}
  \left( W_\mu^- \partial_\lambda W_\nu^+ -
    \partial_\lambda W_\mu^- W_\nu^+ \right) V_\rho \\
  \label{eq:hagiwara}
  & & + \,  i \tilde \kappa_{V} W_\mu^- W_\nu^+ 
       \tilde V_{\mu\nu}
  + i \frac{\tilde\lambda_V}{\MW^2} 
  W_{\lambda\mu}^-
  W^{+}_{\mu\nu} \tilde V_{\nu\lambda} \bigg]\,,
\end{eqnarray}
using the antisymmetric combinations $V_{\mu\nu}\!=\!\partial_\mu V_\nu\!-\!
\partial_\nu V_\mu$ and their duals ${\tilde V}_{\mu\nu}\!=\!\frac12 \epsilon_{
\mu\nu\rho\sigma}V_{\rho\sigma}$.  The overall coefficients are $g_{WW\gamma}\!=
\!e$ and $g_{WWZ}\!=\!e\cot\theta_W$.  Electromagnetic gauge invariance requires
that $g_1^{\gamma}=1$ and $g_5^{\gamma}=0$ at zero momentum transfer. In the SM,
one has {$g_1^V = \kappa_V = 1$}, all other couplings are equal to zero.  Among
the different couplings $g_1,\,\kappa$ and $\lambda$ are C- and P-conserving,
$g_5$ is C and P-violating but  CP-conserving while $g_4,\,\tilde{\kappa},\,
\tilde{\lambda}$ violate CP symmetry.

Experimentally, the different types of couplings can be disentangled
by ana\-ly\-sing the production angle distribution of the $W$ boson
and the $W$ polarization structure which can be obtained from the
decay angle distributions.  Anomalous $WW\gamma$ and $WWZ$ couplings
give similar signals in the final state distributions. However they
can be disentangled easily at the ILC using beam polarization. Because
of the strong dominance of the left-handed electron state, high
polarization values are needed for this analysis. This can also be
achieved by increasing the effective polarization using polarized
positron beams.

An analysis using a fast simulation has been performed at the two
energies $\sqrt{s} = 500\GeV$ and $800 \GeV$ \cite{tgc_wolfgang} and
the results for single parameter fits are shown in Table
\ref{tab:tgc}. For the multi-parameter fits, the correlations are
modest at $\sqrt s=800\GeV$ so that the errors increase by at most
20\%, while at $\sqrt s=500\GeV$ they are much larger and the errors
increase by about a factor two in the multi-parameter fit of the C,P
conserving parameters. For the C or P violating parameters, the
correlations are small at both energies \cite{tgc_wolfgang}. In
scenarios in which there is no Higgs boson and new strong interactions
at high energies occur, the anomalous triple gauge couplings translate
into a mass scale for the new physics around $10\TeV$, i.e. far beyond
the energy where unitarity breaks down in this case
\cite{Aguilar-Saavedra:2001rg}.

\newcommand{\Cdgz}{\ensuremath{\Delta g^\mathrm{Z}_1}}
\newcommand{\Cdgg}{\ensuremath{\Delta g^\mathrm{\gamma}_1}}
\newcommand{\Cdkz}{\ensuremath{\Delta \kappa_\mathrm{Z}}}
\newcommand{\Cdkg}{\ensuremath{\Delta \kappa_{\gamma}}}
\newcommand{\Ckg}{\ensuremath{\kappa_{\gamma}}}
\newcommand{\Ckz}{\ensuremath{\kappa_{\mathrm{Z}}}}
\newcommand{\Clg}{\ensuremath{\lambda_{\gamma}}}
\newcommand{\Clz}{\ensuremath{\lambda_{\mathrm{Z}}}}
\newcommand{\Cgv}[1]{\ensuremath{g^V_{#1}}}
\newcommand{\Cgz}[1]{\ensuremath{g^Z_{#1}}}
\newcommand{\Cgg}[1]{\ensuremath{g^{\gamma}_{#1}}}
\newcommand{\Ckzt}{\ensuremath{\tilde{\kappa}_\mathrm{Z}}}
\newcommand{\Clzt}{\ensuremath{\tilde{\lambda}_\mathrm{Z}}}
\newcommand{\Ckgt}{\ensuremath{\tilde{\kappa}_{\gamma}}}
\newcommand{\Clgt}{\ensuremath{\tilde{\lambda}_{\gamma}}}
\begin{table}
\caption[Results of parameter fits to the triple gauge boson couplings at the ILC]
  {Results of the single parameter fits ($1 \sigma$) to the different 
    triple gauge couplings at the ILC for $\sqrt{s}=500 \GeV$ with ${\cal L}=
    500\fbi$ and $\sqrt{s}=800 \GeV$ with ${\cal L}=1000\fbi$;
    $\pmi = 80\%$ and $\ppl = 60\%$ has been used.}
\label{tab:tgc} 
\centering
\renewcommand{\arraystretch}{1.2}
\begin{tabular}[c]{|c|c|c|}
\hline
coupling & \multicolumn{2}{|c|}{error $\times 10^{-4}$} \\
\cline{2-3}
         & $\sqrt{s}=500\GeV$ & $\sqrt{s}=800\GeV$ \\
\hline
  \Cdgz  &$ 15.5 \phantom{0} $&$ 12.6 \phantom{0} $\\
  \Cdkg  &$  3.3 $&$  1.9 $\\
  \Clg   &$  5.9 $&$  3.3 $\\
  \Cdkz  &$  3.2 $&$  1.9 $\\
  \Clz   &$  6.7 $&$  3.0 $\\
\hline
  \Cgz{5}&$ 16.5 \phantom{0} $&$ 14.4 \phantom{0} $\\
  \Cgz{4}&$ 45.9 \phantom{0} $&$ 18.3 \phantom{0} $\\
  \Ckzt  &$ 39.0 \phantom{0} $&$ 14.3 \phantom{0} $\\
  \Clzt  &$  7.5 $&$  3.0 $\\
  \hline
\end{tabular}
\end{table}

\begin{figure}[htbp]
  \centering
  \includegraphics[width=0.4\linewidth,bb=33 17 492 468]{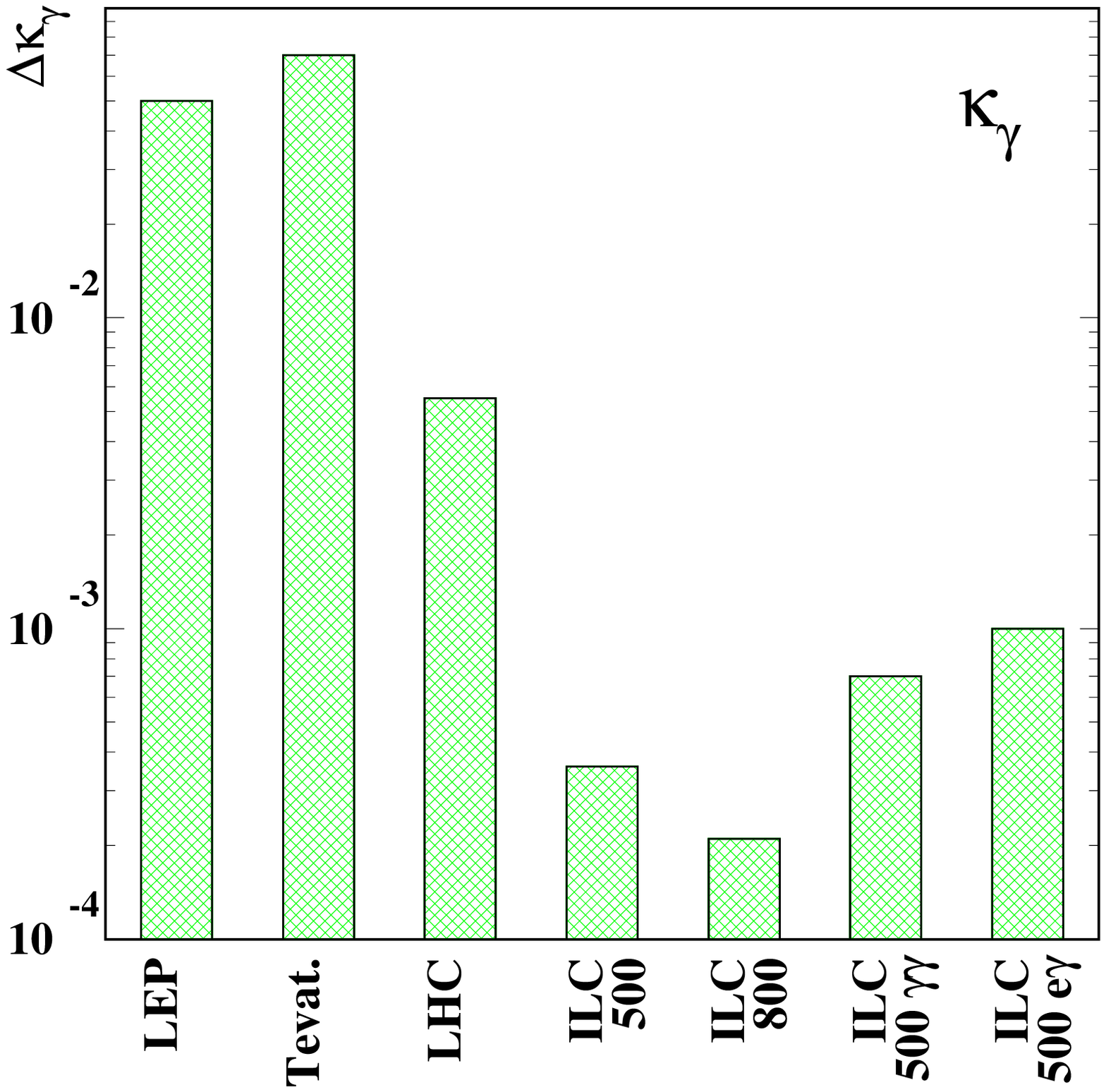}
  \hspace*{5mm}
  \includegraphics[width=0.4\linewidth,bb=33 17 492 468]{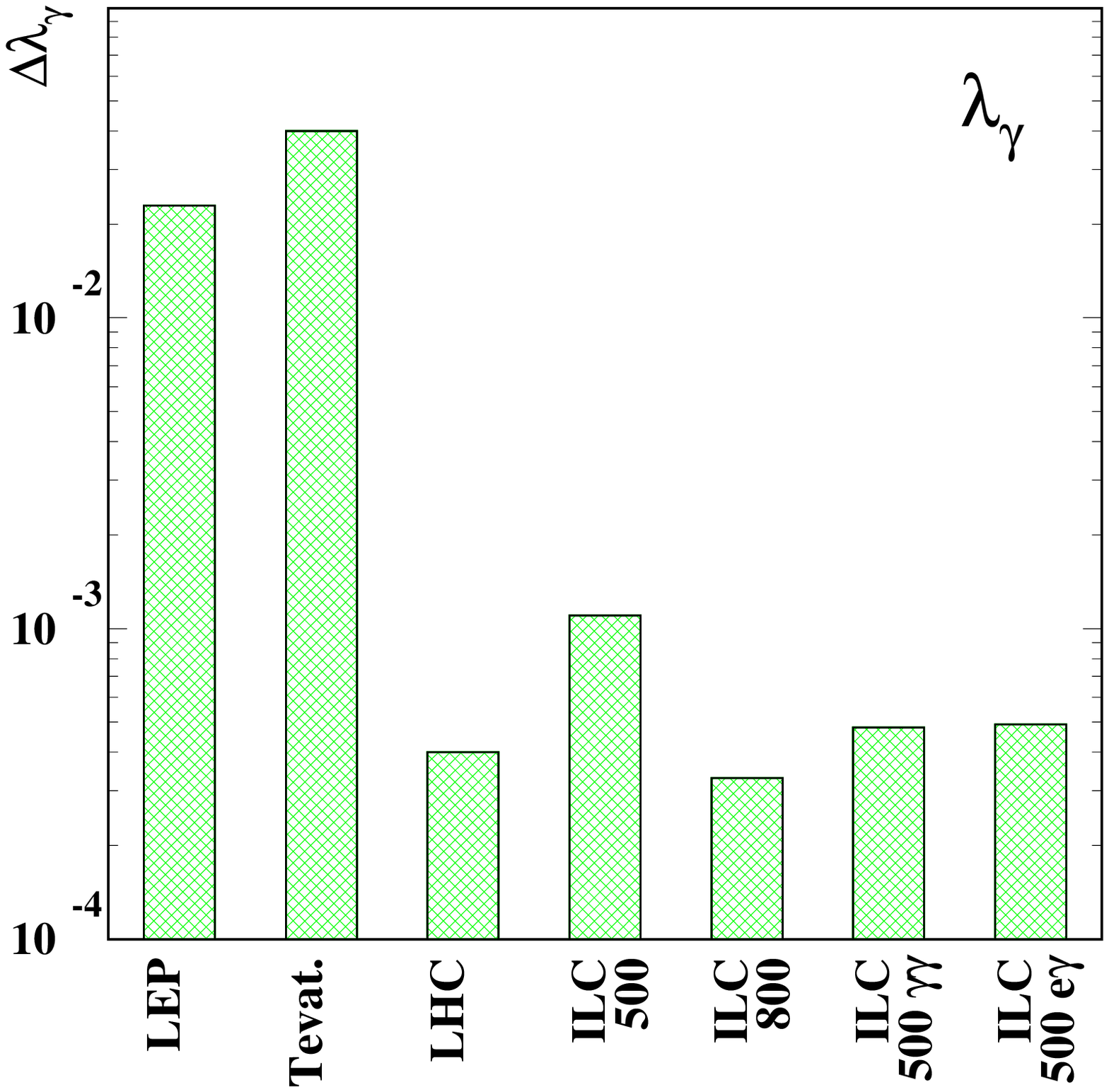}
\vspace*{-5mm}
  \caption[Measurement of anomalous gauge boson couplings at different machines.]
    {Comparison of $\Delta \kappa_\gamma$ and $\Delta
      \lambda_\gamma$ at different machines. For LHC and ILC three
      years of running are assumed (LHC: $300\fbi$, ILC
      $\sqrt{s}=500\GeV$: $500 \fbi$, ILC $\sqrt{s}=800\GeV$: $1000
      \fbi$). If available the results from multi-parameter fits have
      been used. }
 \label{fig:tgccomp}
\end{figure}
 
Additional information on the triple gauge couplings can be obtained
from the $e\gamma$ and $\gamma \gamma$ options of the ILC. In this
case, only the $WW\gamma$ couplings can be measured without
ambiguities from the $WWZ$ couplings. It is often claimed that these
options are particularly sensitive because of the large cross sections
and because the leading contributions depend on the triple gauge
couplings. However, in $e \gamma\! \rightarrow\!  {W}^- \nu$ and $\gamma
\gamma\! \rightarrow\! { W^+ W^-}$, no gauge cancellations occur so that
the sensitivity is reduced. Detailed studies have shown that for the
coupling $\kappa_\gamma$, the $\ee$ mode is by far superior, while for
the coupling $\lambda_\gamma$ competitive results can be obtained
\cite{jaga1,jaga2}. Figure \ref{fig:tgccomp} compares the $\kappa_\gamma$ and
$\lambda_\gamma$ measurements at different machines. Particularly for
the coupling $\kappa$ which, because of its lower mass dimension is
interesting to study, the measurement at the ILC is an order of
magnitude better than the one at the LHC.

\subsection{Measurements of the quartic couplings}

In addition to the triple electroweak gauge boson couplings, the ILC
is also sensitive to the quartic couplings. Two processes are
important in this context: triple gauge boson production, $\ee
\rightarrow VVV$, and vector boson scattering, $\ee \rightarrow \ell_1
\ell_2 VV^\prime$ with $\ell_{1,2} = e,\nu$ and $V,V^\prime = {W,Z}$. In
vector boson scattering, the underlying process is the quasi-elastic
scattering $V_1 V_2 \rightarrow V_3 V_4$. The subprocesses with initial
$Z$ bosons are, however, suppressed as a result of the small $Zee$
couplings. Nevertheless ${ WZ \rightarrow WZ}$ and ${ ZZ
  \rightarrow ZZ}$ are of some use in the case where no custodial
$SU(2)$ invariance is assumed.

In the SM in which a light Higgs boson is absent, unitarity requires that the
interaction among gauge bosons becomes strong at high energies. In this case,
the physics of EWSB below the symmetry breaking scale is described by the most
general effective Lagrangian for the Goldstone bosons required by the
spontaneous ${\rm SU (2)_L\times U (1)_Y \to U(1)_Q}$ breaking. This Lagrangian
describes the physics of longitudinal gauge bosons and its parameters can be
probed in their interactions. The most general C and P conserving effective
Lagrangian contains 10 dimension-four interactions $L_{1,..,10}$
\cite{Appelquist}.  As the SM accounts for the small deviation of the $\rho 
=\MW^2/(\cos^2\theta_W \MZ^2)$ parameter from unity, a custodial  ${\rm
SU(2)_c}$ symmetry appears to be conserved  and, in a first step,  one can
restrict the analyses to the five $\mathrm{SU}(2)_c$ invariant and  linearly
breaking operators. Three of them  contribute to the triple gauge
boson couplings, while the remaining two contribute only to the quartic
couplings, 
\begin{equation}
L_{4}  = {\alpha_{4}} \mathop{\mathrm{tr}}
\Bigl( V_\mu V_\nu \Bigr) \mathop{\mathrm{tr}} \Bigl( V^\mu V^\nu \Bigr) \ ,\ \
L_{5} = {\alpha_{5}} \mathop{\mathrm{tr}} \Bigl( V_\mu V^\mu
\Bigr) \mathop{\mathrm{tr}} \Bigl( V_\nu V^\nu \Bigr)\,. 
\end{equation} 
where  $V_\mu$ simplifies to $-i g\frac{\sigma^i}{2}W^i_\mu +ig^\prime\frac{\sigma^3}
{2}B_\mu$ ($B$ is the hypercharge gauge boson) in the unitarity gauge. The
coefficients $\alpha_i$  are related to scales of new physics~$\Lambda^*_i$ by
naive dimensional analysis,  $\alpha_i=  (v/\Lambda^*_i)^2$. In the
absence of resonances that are lighter than  $4\pi v$, one expects a strongly
interacting symmetry breaking sector at a  scale $\Lambda^*_i \approx  4\pi v
\approx 3\TeV$ which means the  coefficients $\alpha_i$ are of order
$1/16 \pi^2$ unless they are suppressed by some symmetry. 

Thus, the quartic electroweak gauge couplings can be parameterized in
an almost model-independent way (only the custodial SU(2) symmetry can
be assumed for simplicity) by the operators $L_4$ and $L_5$ and their
coefficients $\alpha_4$ and $\alpha_5$ can be determined or
constrained by studying, for instance, quasi-elastic gauge boson
scattering at high energies. In fact, the sensitivity of the quartic
couplings to the two parameters rises strongly with energy and useful
results can be obtained only with the upgrade of the ILC to the energy
of $1 \TeV$.

\begin{figure}[t]
\begin{minipage}{0.45\textwidth}
    \epsfig{figure=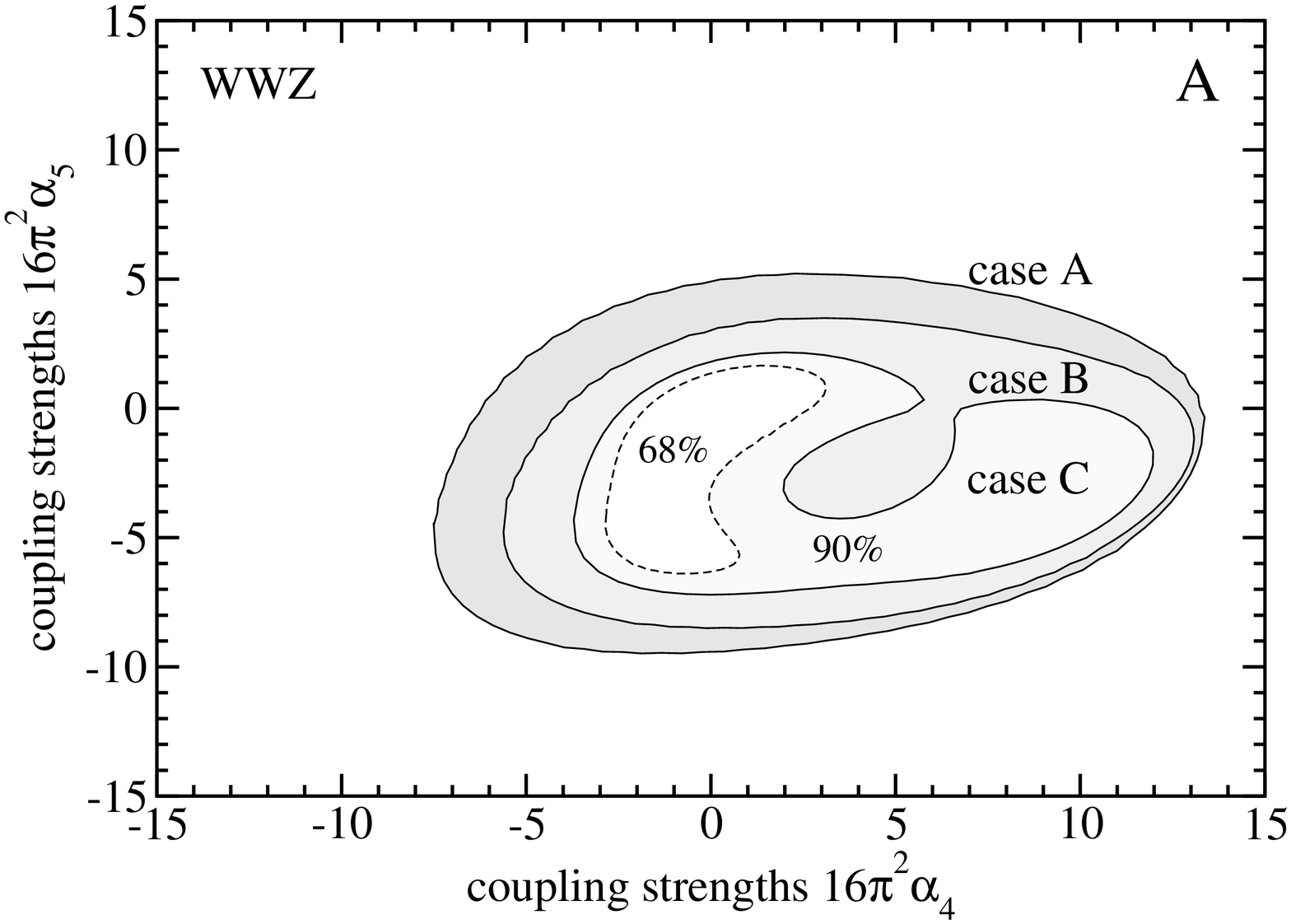,width=\textwidth}
\end{minipage}\hfill
\begin{minipage}{0.45\textwidth}
    \epsfig{figure=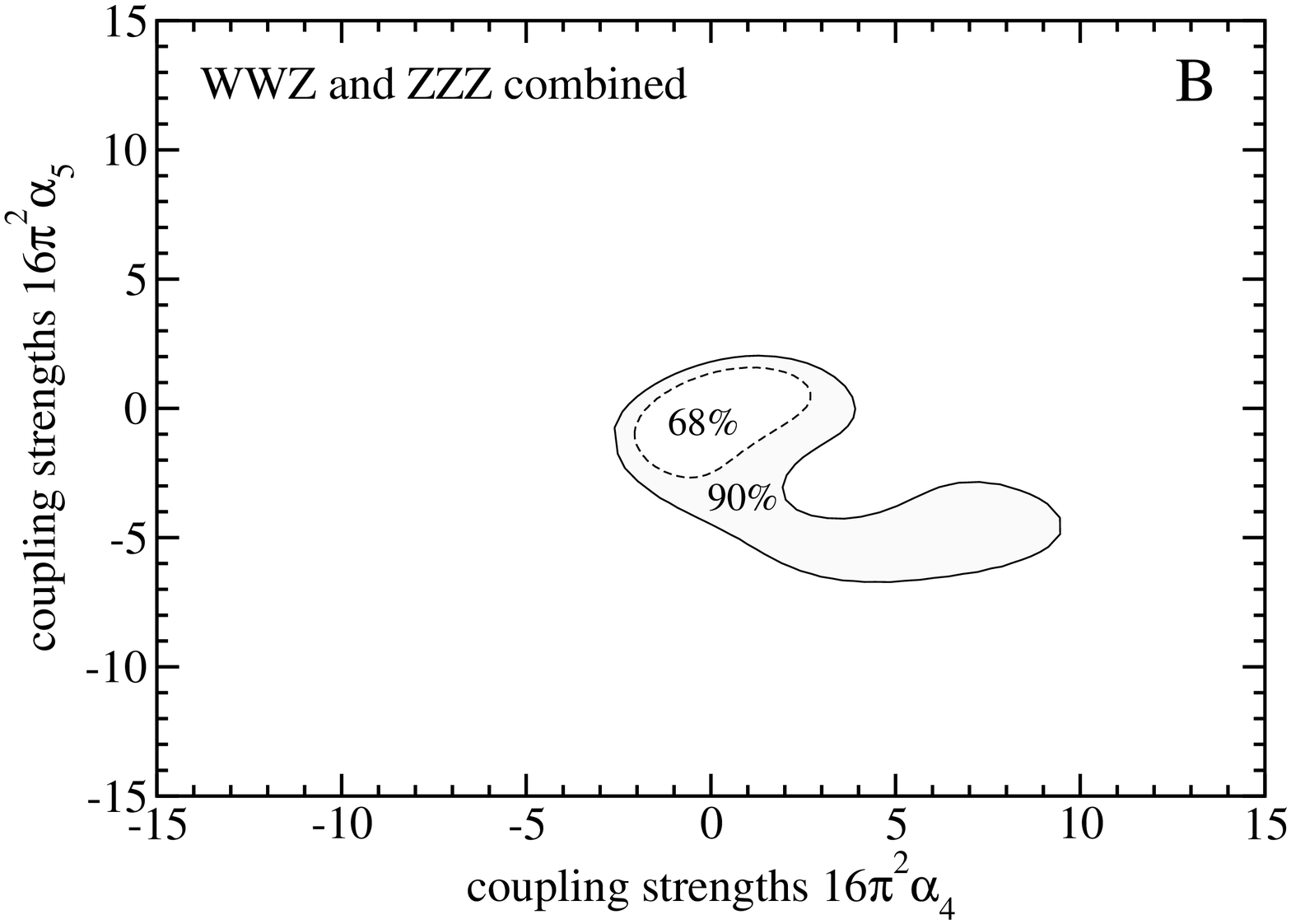,width=\textwidth}
\end{minipage}
\vspace*{-5mm}
\caption[Sensitivity on effective weak gauge boson parameters in no Higgs
scenarios]
{Expected sensitivity for $\alpha_4,\alpha_5$  at the ILC with 
$\sqrt{s}=1 \TeV$ and  1 ab$^{-1}$ from the $\ee\to VVV$ channels 
\cite{predrag}. Left: the $WWZ$ channel for unpolarized (A), only ${\rm e}^+$
polarized (B) and both $e^\pm$ polarized (C) beams. Right: combined 
fit using $WWZ$ and $ZZZ$ for ${\rm e}^\pm$ polarized beams. Lines represent $90\%$ 
The outer (inner) line represents 90\% (68\%) confidence level.}
\label{fig:VVVfit}
\vspace*{-5mm}
\end{figure}

\begin{figure}[htbp]
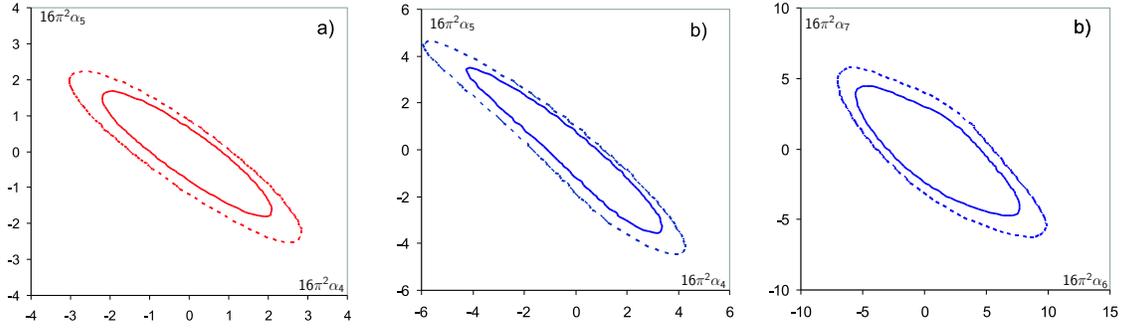

  \mbox{
  \includegraphics[width=0.31\linewidth]{qgc_cust.eps}\hspace*{3mm}
  \includegraphics[width=0.63\linewidth]{qgc_nocust.eps}
}
\vspace*{-6mm}
  \caption[Limits on the quartic coupling parameters from the 
   VVV and VV$\ell \ell$ processes]
  {Limits of $\alpha_4$, $\alpha_5$ assuming $SU(2)_c$
    conservation (a) and $\alpha_4$ - $\alpha_7$ without this
    assumption (b) from a combined analysis of three-vector-boson
    production and of vector-boson scattering assuming $1000\,\fbi$
    at $\sqrt{s}=1\TeV$.  The dashed line represent 90\% c.l. and the
    solid line 68\%.}
  \label{fig:predrag}
\vspace*{-4mm}
\end{figure}

Within the generic effective-field theory context discussed above,
all processes that contain quasi-elastic weak boson scattering, $\ee
\to \ell \ell V^* V^* \to \ell \ell VV$, and triple weak boson
production, $\ee \to VVV$, have been recently reanalyzed
\cite{predrag}. The study uses complete six-fermion
matrix elements in unweighted event samples, fast simulation of the
ILC detector and a multidimensional parameter fit of the set of
anomalous couplings.  It also includes a study of triple weak boson
production which is sensitive to the same set of anomalous couplings.
In the case where the simplifying assumption of custodial symmetry is
used, the results are illustrated in Figs.~\ref{fig:VVVfit} for the
$\ee \to {WWZ,ZZZ}$ channels and Fig.~\ref{fig:predrag}a for the 
combination of both channels
assuming a $1 \TeV$ ILC with 1 ab$^{-1}$ of data.
As can be seen, an accuracy of the order of $1/(16\pi^2)$ can be
obtained on the coefficients $\alpha_4$ and $\alpha_5$.

With the assumption of conserved ${\rm SU(2)_c}$ symmetry, the LHC obtains
similar limits as those shown above. However, since the ILC can, contrary to the
LHC,  tag the initial and final state gauge bosons, the separation of couplings
is possible without the need of this assumption. An example of constraints in
this case,  including the four-dimension operators $L_6$ and $L_7$ which break
the custodial symmetry, is shown in Fig.~\ref{fig:predrag}b where the same
energy and luminosity as above is  assumed. Despite of the increase of the
parameter  space, the constraints are only a factor of two to three worse than
in the conserved SU(2) case.   

Note that  the limits on the parameters $\alpha_i$ can be interpreted in terms
of heavy resonances; the constraints on the masses of these resonances depend
strongly on the assumptions and vary between 1 and 4\,TeV \cite{predrag}. This
aspect will be discussed in chapter \ref{sec:alternatives}.

\section{The strong interaction coupling}

Precision measurements in strong interaction processes will be part of
the physics program of the ILC. Among the many aspects of perturbative
QCD which can be studied at the collider, the measurement of the
strong coupling $\alpha_s$ will represent one of the most important
outcome.

The strong coupling $\alpha_s$ can be determined from event shape
observables in $\ee \to q\bar q g$ that are sensitive to the
three-jet nature of the particle flow; examples of such observables
are the thrust, jet masses and jet rates.  In this method, one usually
forms a differential distribution, applies corrections for detector
and hadronization effects and fits a perturbative QCD prediction to
the data, allowing $\alpha_s$ to vary. 
Measurements from LEP and SLC have shown that statistical errors below
0.001 can be obtained with samples of a few tens of thousands hadronic
events. With the current ILC design luminosities, hundreds of
thousands of $\ee \to q\bar q$ events can be produced each year and
a statistical error on $\alpha_s(\MZ)$ below 0.0005 can be achieved
\cite{FQCD-otmar,Aguilar-Saavedra:2001rg}. The systematic error,
however, is at present a factor ten larger than this value and it is
not clear, how much it can be improved by higher order calculations. 

The GigaZ option also provides the possibility for a very accurate 
determination of the value of $\alpha_s(\MZ)$ via the measurement of
the inclusive ratio of the $Z$ boson decay widths
$\Rhad = \Ghad/\Gll$.
The current LEP data sample of $1.6 \cdot 10^6$ $Z$ bosons provides an
accuracy $\Delta \alpha_s(\MZ) =0.0025$ from the ratio $\Rhad$
\cite{PDG}. At GigaZ, the statistical error can be lowered to the
level of $0.0004$ but systematic errors arising from the hadronic and
leptonic event selection will probably limit the precision to $\Delta
\alpha_s (\MZ)=0.0008$ \cite{Winter:2001av}. This would be a very
precise and reliable measurement from a single and clean observable
which is subject to very small theoretical uncertainties. Especially
$\Rhad$ is unaffected by any non-perturbative corrections.

\begin{figure}[h]
  \begin{center}
  \includegraphics[width=8.cm]{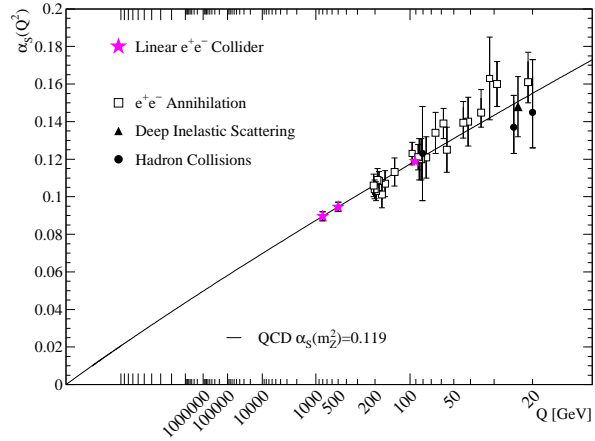}
  \end{center}
  \vspace*{-9mm}
\caption[Determination of the evolution of $\alpha_s$ at various ${\rm e^+ e^-}$ energies.] 
{The evolution of $\alpha_s$  with $1/\ln Q$ from various measurements;
the data points are from present ones and the stars denote simulated 
ILC measurements for $\sqrt s=91, 500$ and $800 \GeV$.}
\label{fig_alpha_s}
\vspace*{-3mm}
\end{figure}

The translation of the measurements of $\alpha_s(\MZ)$ discussed above
to other energies, $\alpha_s(Q)$ with $Q \neq \MZ$, requires the
assumption that the running of the coupling is determined by the QCD
$\beta$ function. Since the logarithmic decrease of $\alpha_s$ with
energy is an essential component of QCD, reflecting the underlying
non-Abelian dynamics of the theory, it is important also to test this
energy dependence explicitly. Such a test would be particularly
interesting if new colored particles were discovered, since deviations
from QCD running would be expected at energies above the threshold for
pair production of the new particles. Furthermore, extrapolation of
$\alpha_s$ to very high energies of the order of $M_U=10^{16} \GeV$ can
be combined with corresponding extrapolations of the weak and
electromagnetic couplings in order to constrain the coupling
unification or the GUT scale.  Hence, it would be desirable to measure
$\alpha_s$ in the same detector, with the same technique and by
applying the same treatment to the data, at a series of different
energies $Q$, so as to maximize the lever-arm for constraining the
running. This is shown in Fig.~\ref{fig_alpha_s} where simulated
measurements of $\alpha_s(Q)$ at $Q = 91, 500$ and $800 \GeV$ are
displayed, together with existing measurements in the range $20\leq
Q\leq 200 \GeV$ \cite{FQCD-otmar,Aguilar-Saavedra:2001rg}.

\begin{figure}[!h]
\begin{center}
\includegraphics[width=0.42\linewidth,bb=100 374 365 654]{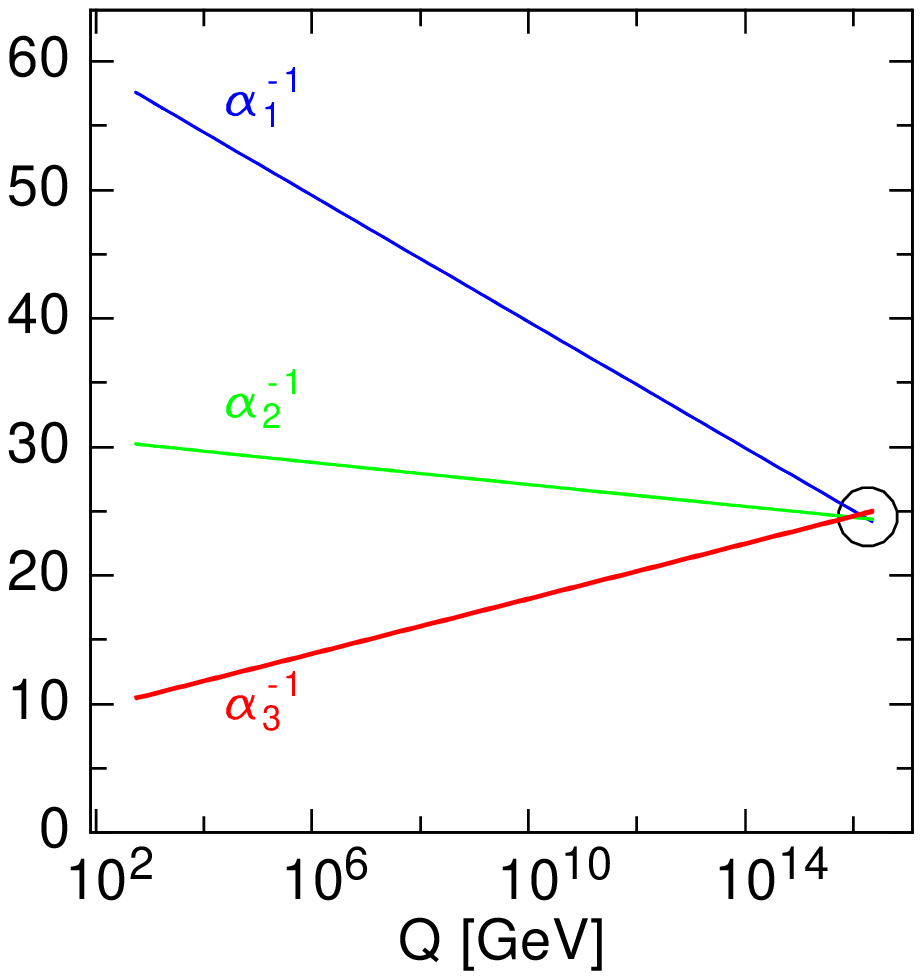}
\hspace{1cm}
\includegraphics[width=0.42\linewidth,bb=100 374 365 654]{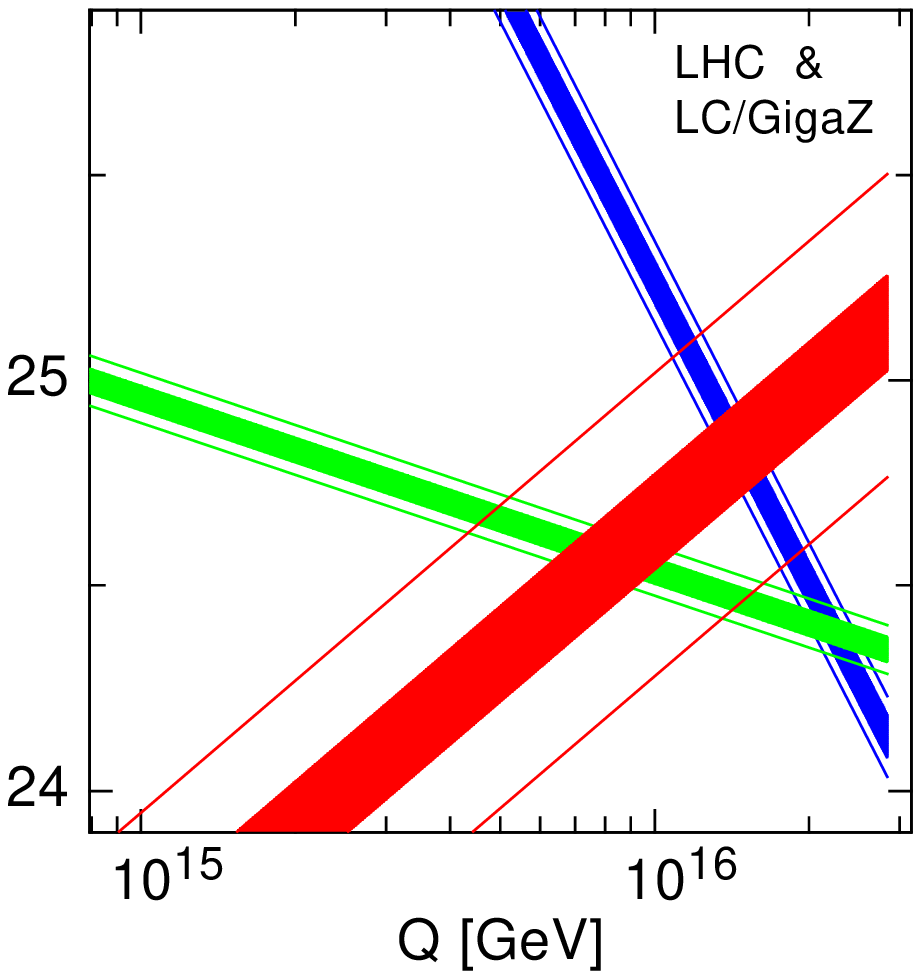}
\end{center}
\vspace*{-9mm}
\caption[Extrapolations of the gauge couplings from GigaZ to the unification scale]
{Extrapolations of the gauge couplings as measured at ILC to the 
unification scale \cite{F-gauge-unif}.}
\label{fig_gauge-uni}
\vspace*{-3mm}
\end{figure}

It is therefore clear that ILC data adds significantly to the
lever-arm in the energy evolution of $\alpha_s$ and allows a
substantially improved extrapolation to the GUT scale. This is
exemplified in Fig.~\ref{fig_gauge-uni} where the evolution of the
three gauge couplings is displayed. The measurements at GigaZ will
support unification at a scale $M_U \simeq 2\times 10^{16}$~GeV, with
a precision at the percent level. However, the couplings are not
expected to meet exactly because of the high threshold effects at the
scale $M_U$.  The quantitative evaluation of the discrepancy will
provide important constraints on the particle content at the GUT
scale.

Many other aspects of QCD can be addressed at the ILC. In particular,
the $\gamma \gamma$ and $e\gamma$ options offer a broad new area of
QCD studies in two-photon interactions at high energy and luminosity.
Examples are (see also chapter \ref{sec:top} for QCD studies in the
process $\ee \to t\bar t$) \cite{Aguilar-Saavedra:2001rg} the total
cross section, the photon structure function and the annihilation of
virtual photons as a test of BFKL dynamics.

\chapter{Top quark physics}
\label{sec:top}

The top quark is the heaviest particle in the Standard Model and,
thus,  the most strongly coupled to the electroweak symmetry
breaking sector; it is therefore expected to play a  fundamental
role in the dynamics behind the symmetry  breaking mechanism. It
might also hold clues in  solving the longstanding flavor problem
and provide clear indications on new physics beyond the Standard
Model. For instance,  if the Higgs mechanism should be verified, the
measurement of the top quark  Yukawa coupling would help to
discriminate between SM and non--standard Higgs scenarios. If the
new physics beyond the SM is sufficiently decoupled, shifts in the
production and decay properties of a SM--like top quark may be the
only evidence for it. With the precision ILC measurements, one could
have  sensitivity to new physics at mass scales far above the
electroweak symmetry breaking  scale. For example, it has been
shown~\cite{Djouadi:2006rk,depree} that in warped extra--dimensional
models,  as the top quark has a wavefunction that is near the TeV
brane, its  production cross section at the ILC can reveal
Kaluza--Klein excitations of gauge bosons with masses up to
10--100~TeV.

Precise and model--independent measurements at the ILC of the top couplings to
weak gauge bosons will be sensitive to interesting sources of non--SM physics as
many models predict anomalous top quark couplings. In Technicolor  and other
models with a strongly--coupled Higgs sector, non--standard CP--conserving
couplings may be induced at the 5--10\% level~\cite{top-TC}.  In supersymmetric
and multi--Higgs models, CP--violating couplings  may be induced at the one-loop
level, with predictions in the range $10^{-3}$--$10^{-2}$~\cite{sumino}. Little
Higgs or top--seesaw models predict definite shifts in  the top quark couplings
to the $W$ and $Z$ bosons.

High--precision measurements of the properties and interactions of the top
quark  are  therefore mandatory. The ILC will have broad capabilities to
outline the top quark profile with high precision and in a model--independent
way. In particular, the $t\bar{t}$ threshold holds the promise of very precise
measurements of the top quark mass  and total decay width. Both at threshold and
in the continuum, the neutral and charged  current interactions of the top quark
can be very precisely determined.  Its vector and axial--vector couplings to the
$Z$ boson in the production  vertex and to the $W$ boson in the decay vertex, as
well as its magnetic and electric dipole  moments, could be measured at the one
percent level. The high luminosity expected at the ILC will allow to determine
the important top quark  Yukawa coupling to the Higgs boson with a precision
greatly exceeding that foreseen at the LHC.

Finally, if the threshold of new physics is nearby, new decay channels of the
top quark,  such as decays into a charged Higgs boson in  supersymmetric or
multi--Higgs doublet models, may be observed and studied in detail in the clean
environment of the ILC.

\section{The top quark mass and width}

The top quark mass is a fundamental parameter of the SM and also a crucial
ingredient of the electroweak precision measurement program, hence the
importance to measure it as accurately as possible~\cite{Heinemeyer:2003ud}.  In
many extensions to the SM in which the Higgs boson mass can be calculated,  the
theoretical prediction for $M_H^2$  depends sensitively on $m_t$. For instance,
in the minimal supersymmetric extension  of the SM, the radiative corrections
grow as $m_{t}^4$~\cite{gigazsven}. In this case, the expected LHC precision of
1~GeV on $m_t$ translates into a similar uncertainty for the predicted value  of
the lighter Higgs boson mass $M_h$~\cite{gigazsven}.  The anticipated accuracy
at the ILC is more than an order of magnitude better, obtaining a parametric
error small enough to allow for a very incisive comparison of theory and
measurement. A smaller uncertainty on $m_t$ also improves the sensitivity to new
physics causing anomalous $W$ and $Z$ couplings~\cite{peskin,sumino}.

\begin{figure}[htbp]
\vspace*{-2mm}
\centerline{ \epsfig{file=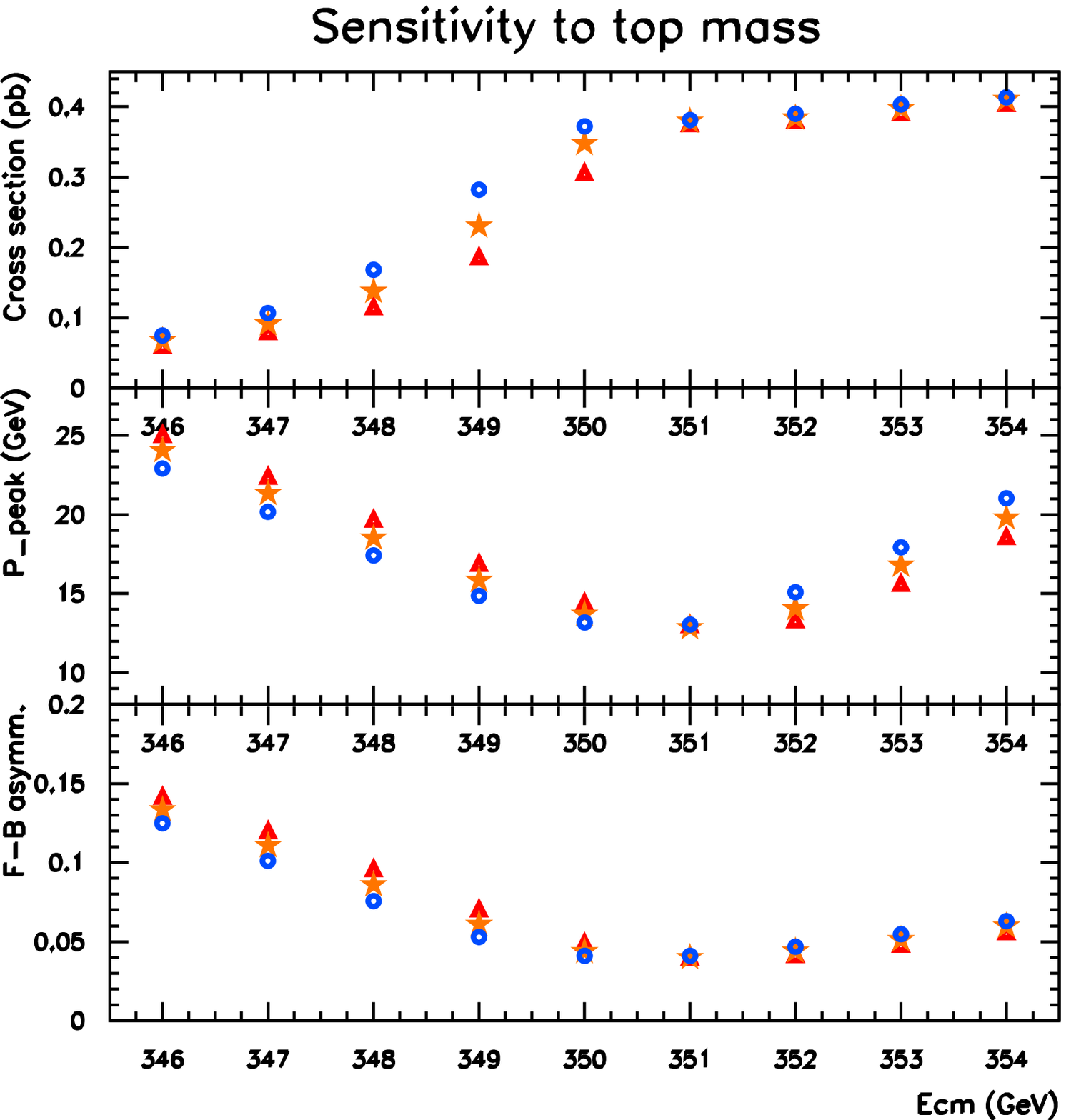,width=3.in}\hspace*{3mm}
\epsfig{file=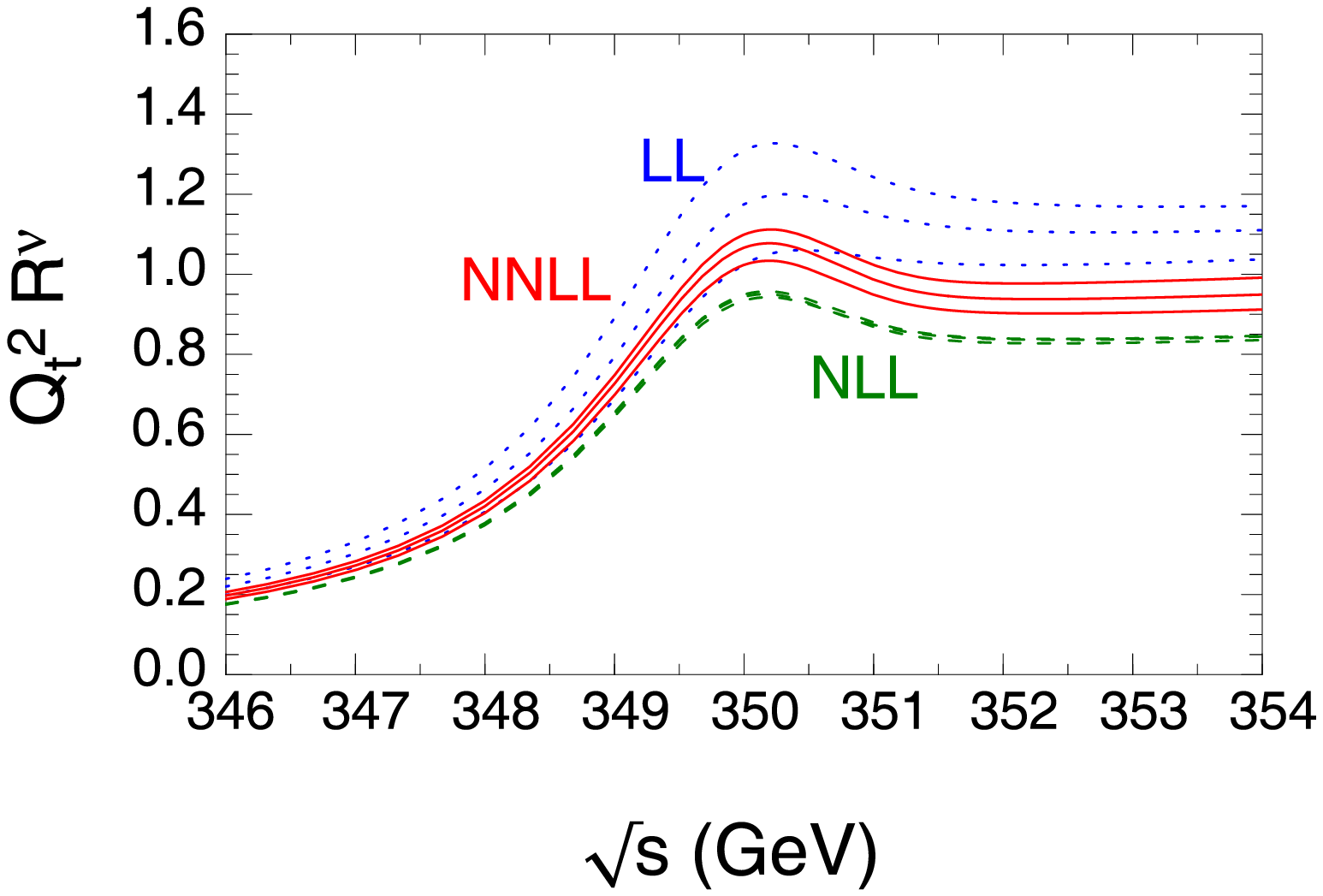,width=2.8in,height=3in} }
\vspace*{-5mm}
\caption[Sensitivity of observables to the top mass in a scan around the ${\rm t\bar t}$ threshold]
{Left: sensitivity of the observables to the top mass in a c.m.energy scan
around the $t\bar{t}$ threshold with the different symbols denoting 200~MeV
steps in top mass \cite{mm}. Right: dependence of the $\eei \to t\bar t$ cross
section on the c.m.energy in various approximations for QCD corrections \cite{hoang}.}
\label{fig:sensitivitymt}
\vspace*{-2mm}
\end{figure}

Because of its large width, $\Gamma_t\sim 1.5$~GeV, the top quark will decay
before it hadronizes, thus non-perturbative effects are expected to be highly
suppressed. As a result, the energy dependence of the cross section
$\sigma_{t\bar{t}}$ for $e^+e^- \to t\bar{t}$ can be computed reliably, with an
expected increase in rate by a factor of ten as the center-of-mass (CM) energy
is varied by 5~GeV around the threshold energy. The location of the rise of the
cross section can be used to extract the value of $m_t$, while the shape and
normalization yield information about the total width $\Gamma_t$,  the strong
coupling  $\alpha_s$ and eventually, the $t\bar t H$ Yukawa coupling $g_{ttH}$
~\cite{fms}. In Ref.~\cite{mm}, three threshold observables:
$\sigma_{t\bar{t}}$, the peak of the top momentum distribution, and the
forward--backward charge asymmetry, were simultaneously fitted to obtain
measurement uncertainties on $m_t$, $\Gamma_t$, $\alpha_s$ of 19~MeV, 32~MeV,
and 0.0012, respectively. However this study did not include a complete
evaluation of important systematic uncertainties, such as e.g. the determination
of the luminosity spectrum or theoretical uncertainties on differential
observables. Figure~\ref{fig:sensitivitymt} (left) demonstrates the sensitivity
of the top mass measurement to these observables. It is expected that the top
mass can be measured with a statistical uncertainty of 40~MeV in a modest scan
of 10~fb$^{-1}$, a small fraction of a year at typical design luminosities.  A
longer scan of about 100~fb$^{-1}$ can determine the top width to 2\%.

The threshold cross section has been calculated including some of the
next-to-next-to-leading logarithmic (NNLL) QCD corrections, as shown in
Fig.~\ref{fig:sensitivitymt} (right) \cite{tmass,hoang}. The full  NNLL
contribution is not yet available, but the large size of the corrections
relative to the NLL terms~\cite{hoang2} suggests that the theoretical
uncertainty on the cross section will ultimately be approximately
$\delta\sigma_{t\bar{t}}/\sigma_{t\bar{t}}\sim \pm 3\%$, but the effect on the
mass determination is small.

The high-precision measurements of the ILC at the $t\bar{t}$ threshold will
determine a ``threshold'' (or resonance) mass parameter with an
accuracy significantly below 100~MeV. This threshold mass can then be
translated into another short-distance mass that is useful as a theory
input, such as the $\overline{\rm MS}$ mass. This translation will give rise to an
additional theoretical uncertainty. The current estimate for the combined
experimental and theoretical uncertainty in the determination of the
top-quark mass is about 100 MeV~\cite{mt-determ}.

A threshold scan will require precise knowledge of the average c.m.  energy and
the shape of the luminosity spectrum d${\cal L}$/d$E$
\cite{tt-spectrum}. Schemes for precision measurement of
$\langle E_{cm} \rangle$ include the use of beam spectrometers or using physics
processes such as $Z$ boson pair production or radiative returns to the $Z$. The
luminosity spectrum is determined by  the beam spread,  beamstrahlung and
initial state radiation (ISR).  All three effects will lead to a smearing of the
$t\bar t$ threshold cross section, resulting in a significant reduction of the
effective luminosity and hence the observed cross section,
$\sigma^{\rm obs}(\sqrt{s}) = {\cal L}_0^{-1} \int_0^1 {\cal L}(x)\,
\sigma(x\sqrt{s})\,{\rm d}x $.

\begin{figure*}[htbp]
\begin{center}
{\includegraphics[bb=0 10 567 350,width=7cm]{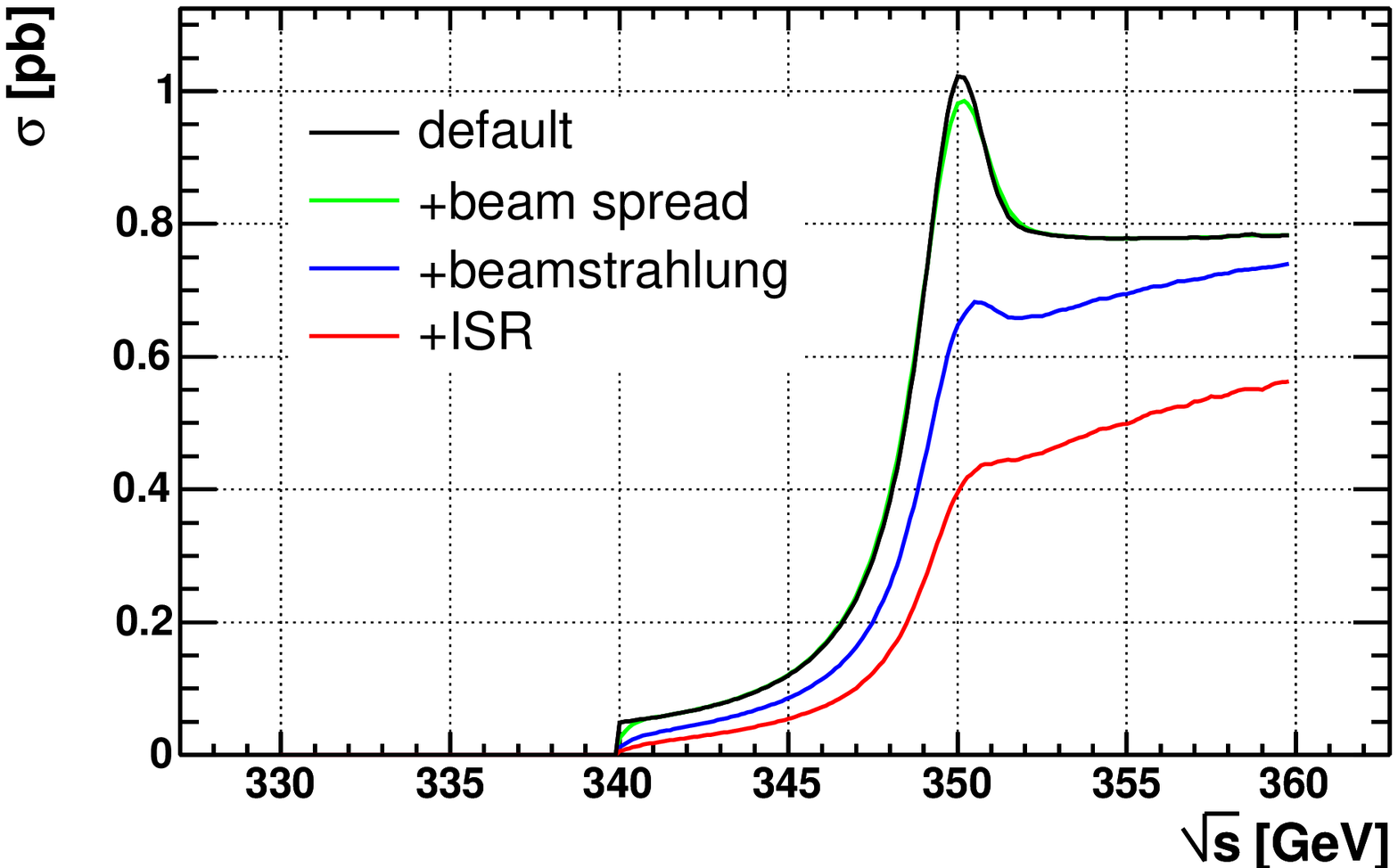}\hspace*{5mm}
\includegraphics[bb=0 10 567 350,width=7cm]{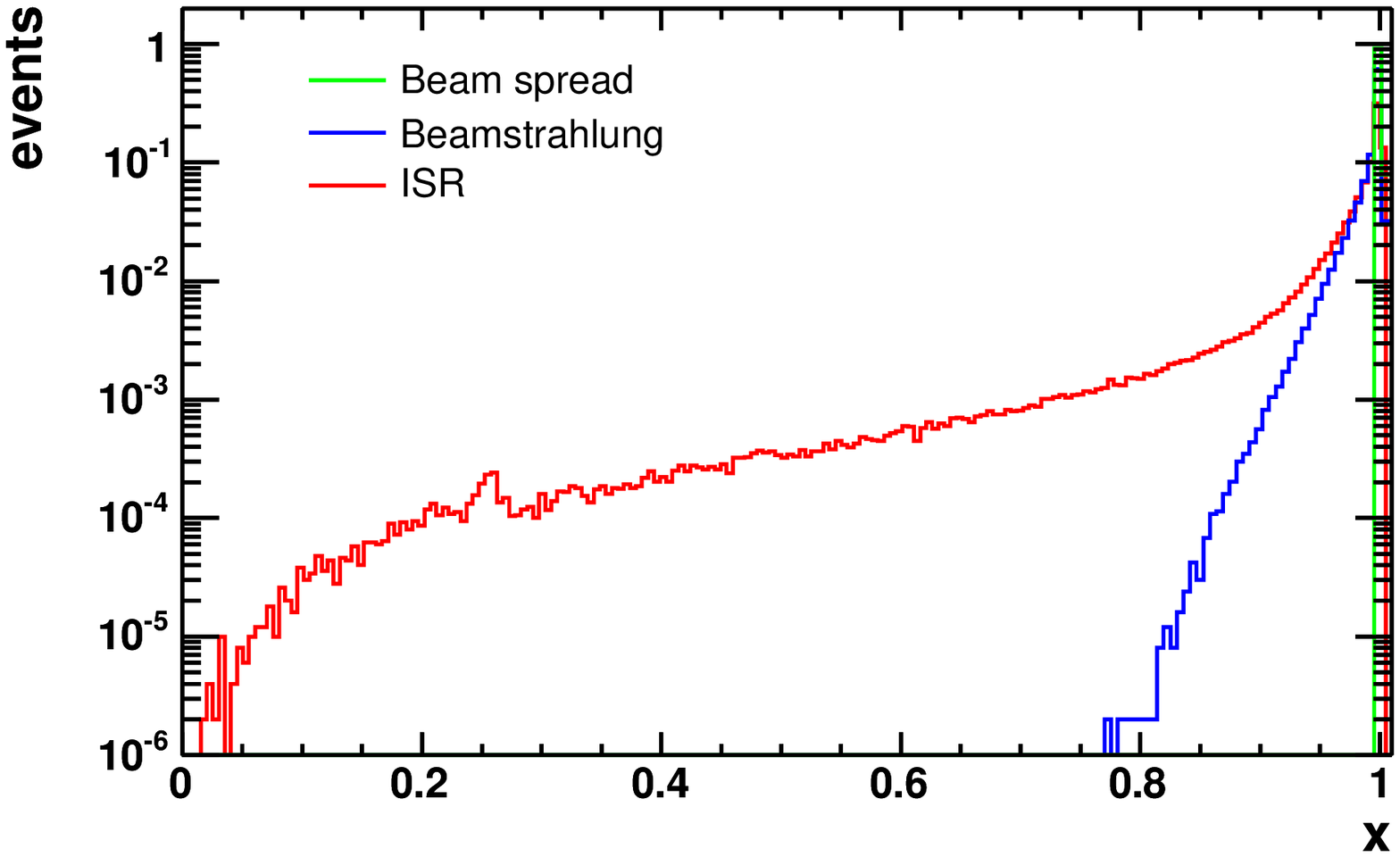}}
\vspace*{-5mm}
\caption[Beam spread, beamstrahlung and ISR effects on the top quark cross section.]
{Left: Smearing of the theoretical $t\bar t$ cross section (`default') by beam
  effects and initial state radiation.  Right panel: Simulation of beam spread,
  beamstrahlung and ISR as distributions of $x=\sqrt{s}/\sqrt{s_0}$ (where
  $\sqrt{s_0}$ is the nominal c.m. energy of the machine). From
  Ref.~\cite{Boogert:Snowmass05}.}
\label{fig:sigmaexp}
\end{center}
\vspace*{-5mm}
\end{figure*}

The influence of the three effects is demonstrated in Fig.~\ref{fig:sigmaexp}.
The beam spread will typically be $\sim 0.1\%$  and will cause comparably little
smearing (though additional beam diagnostics may be required to measure and
monitor the beam spread), but beamstrahlung and ISR are very important.  The
luminosity spectrum will lead to a systematic shift in the extracted top mass
which must be well understood; otherwise it could become the dominant systematic
error. The proposed method is to analyze the acollinearity of (large angle)
Bhabha scattering events, which is sensitive to a momentum mismatch between the
beams but insensitive to the absolute energy scale~\cite{Monig:2000bm}. For
this, the envisioned high resolution of the forward tracker will be very
important to achieve an uncertainty on the order of 50~MeV.

Including all these contributions, a linear collider operating at
the $t \bar t$ threshold will be able to
measure $m_t$ with an accuracy of $\sim 100-200$~MeV. This can be
compared with the current accuracy of $\sim 2$~GeV at the
TeVatron and possibly $\sim 1$~GeV at LHC~\cite{atlastdr}.

\section{Top quark interactions}

\subsection{The coupling to the Higgs boson}

Near threshold, the $t\bar{t}$ pair interacts, in addition to the QCD potential,
via a Yukawa potential associated with Higgs boson exchange. For a low Higgs
boson mass, the $t\bar{t}$ cross section is {\it a priori} sensitive  to the
top-Higgs Yukawa coupling, $g_{ttH}$~\cite{mm}. Even more sensitive is the
measurement of the $e^+e^- \to t\bar{t}H$ cross section in the continuum, which
is essentially proportional to $g_{ttH}^2$ as discussed in chapter
\ref{sec:higgs}.

At the ILC with energies larger than 500 GeV, the process $e^+e^-\to t\bar{t}H$
with the Higgs decaying to $W^{+}W^{-}$ or $b\bar{b}$ has the relatively clean
signature of $\geq 6$ jets in final state, with 4 $b$--jets and multi--jet
invariant mass constraints, but with backgrounds about three orders of magnitude
larger. The dominant backgrounds are radiative top production and/or decay
($t\bar{t}$+jets) and irreducible $t\bar{t}Z(Z\to b\bar{b})$ \cite{moretti}. For
Higgs bosons with 120--200~GeV masses,  studies with events processed through a
realistic detector simulation and involving rather sophisticated event selection
procedures, have been performed \cite{juste,agay}. They demonstrate that one can
measure $g_{ttH}$ to 6--10\% precision at $\sqrt{s}=800$~GeV with 1~ab$^{-1}$
data \cite{agay}. However, even a 500~GeV ILC can significantly improve our knowledge of the
the $ttH$ Yukawa coupling and accuracies up to 10\% can be  achieved
in the low Higgs mass range \cite{juste}.

A recent reexamination of the LHC measurement of the coupling suggests it will
be challenging to reach this level of precision. However, when combined with ILC
results at $\sqrt s=500$ GeV, LHC does better. ILC precision measurement of
BR($H \rightarrow W^{+}W^{-}$) and BR($h \rightarrow b\bar{b}$) replaces theory
assumptions in the LHC measurements and leads to a better combined uncertainty
of 10-15\% or better for a large range of $M_H$
values~\cite{Weiglein:2004hn,Duhrssen:2004cv,juste}. Therefore, for a number of
years, the combination of results at the LHC and ILC (500 GeV) would yield the
most precise determination of the top quark Yukawa coupling.

\subsection{Couplings to electroweak gauge bosons}

Since the charged electroweak current is involved in the top decay,
$t \bar t$ production in $e^+e^-$ collisions is sensitive to both
the neutral and charged gauge boson couplings of the top quark, and
in the neutral case, directly sensitive to both the $t\bar{t}\gamma$
and $t\bar{t}Z$ vertices. Because the top quark width, $\Gamma_t$,
is much larger than $\Lambda_{\rm QCD}$, the decay process is not
influenced by fragmentation effects and decay products will provide
useful information.

The most general $t\bar{t}(\gamma,Z)$ couplings can be written
as~\cite{Stefano,hioki}
\begin{equation}
  \Gamma^\mu_{t\bar{t}(\gamma,Z)} =
   i \, e \;
   \left\{ \gamma^\mu \;
    \left[ F^{\gamma,Z}_{1V} \; + F^{\gamma,Z}_{1A} \, \gamma^5 \right] +
    {(\;p_t^{}-p_{\overline t}^{})^\mu  \over 2 \; m_t^{} } \;
    \left[ F^{\gamma,Z}_{2V} \; + F^{\gamma,Z}_{2A} \, \gamma^5 \right] \,
  \right \} \ ,
\end{equation}
where the only form factors different from zero in the SM are
\begin{equation}
F^\gamma_{1V}={2\over 3} \ , \ F^Z_{1V}={1\over
4\sin\theta_W\cos\theta_W} \,
        \left(1-{8\over 3} \sin^2\theta_W^{} \right)  \ ,\
F^Z_{1A} = -{1 \over 4\sin\theta_W\cos\theta_W }.
\end{equation}
$ \left({e/m_t}\right) \cdot F^\gamma_{2A} $ is the electric dipole moment form
factor of the top quark and $\left({e/m_t}\right) \cdot F^Z_{2A} $ the weak
electric dipole moment; $ \left({e/m_t}\right) \cdot F^{\gamma,Z}_{2V} $ are the
electric and weak magnetic dipole moments. In the SM, the electric and dipole
moment terms violate CP and receive contributions only at the three--loop level
and beyond. The CP--conserving form factors are zero at tree--level but receive
non--zero ${\cal O}(\alpha_s)$ QCD corrections.

In Table~\ref{tab:topone} is shown the $1\sigma$ sensitivity limits
for the real parts of the $t\bar{t}(\gamma,Z)$ form factors obtained
from an analysis of the process $e^+e^-\to t\bar{t}\to\ell^\pm+$jets
at $\sqrt{s}=500$~GeV~\cite{Abe:2001nq}. Top quarks are selected and
reconstructed, and $b$ quarks are tagged using a detector model with
combined efficiency of 20\%, and purity of 88\%. To extract limits
on $F_{1V}^{\gamma,Z}$ and $F_{1A}^{\gamma,Z}$, the angular
distribution of the reconstructed top quark can be used. At the the
ILC limits on $F_{2A}^{\gamma,Z}$ may be obtained from CP--violating
angular asymmetries of the decay leptons, without assuming the $tbW$
couplings to be vanishing \cite{sdr1}. Longitudinal $e^-$ beam
polarization can be used to enhance the sensitivity, as well as to
obtain independent limits on $F_{2A}^{\gamma}$ and $F_{2A}^{Z}$,
when both are simultaneously kept nonzero. Combinations of decay
lepton energy and angular asymmetries can be made sensitive to
anomalous couplings either in the production or the decay by a
suitable choice of cuts on the lepton energy
\cite{Grzadkowski+Hioki}.

 $F_{1V}^{\gamma,Z}$ and $F_{2V}^{\gamma,Z}$ are derived from the left--right
polarization asymmetry $A_{LR}$ and $F_{2A}^{\gamma,Z}$ from the angular
distribution of the reconstructed top quark and the decay angles of the $t$ and
$\bar t$. The limits shown in Table~\ref{tab:topone} could be strengthened with
positron beam polarization, mostly from the increased $t \bar t$ cross section:
with ${\cal P}_{e^+}=0.5$, $\sigma(t \bar t)$  is about a factor 1.45 larger,
improving the precision in the measurement of  $A_{LR}$ by nearly a factor of
3~\cite{gudi}. Increasing the c.m. energy to $\sqrt{s}=800$~GeV improves the
limits by a factor 1.3--1.5~\cite{breuther}.

\begin{table}
\caption[Accuracies in the determination of the top quark couplings to gauge bosons.]
{\label{tab:topone} The $1\sigma$ statistical uncertainties
for the real parts of the $(\gamma,\,Z)t \bar t$ form factors
obtained from an analysis of the process $e^+e^-\to t \bar
t\to\ell^\pm +$~jets for $\sqrt{s}=500$~GeV. Only one coupling at a
time is varied. }
\renewcommand{\arraystretch}{1.25}
\begin{center}
\begin{tabular}{|c|c|c|c|c|} \hline\hline
Coupling & LO SM Value & ${\cal P}(e^-)$ & $\int\!{\cal L} dt$
(fb$^{-1}$) & $1\sigma$ sensitivity \\ \hline
$F_{1A}^\gamma$ & 0      & $\pm 0.8$ & 100 & 0.011 \\
$F_{1A}^Z$      & $-0.6$ & $-0.8$    & 100 & 0.013 \\
$F_{1V}^\gamma$ & $2/3$  & $\pm 0.8$ & 200 & 0.047 \\
$F_{1V}^Z$      & $0.2$  & $\pm 0.8$ & 200 & 0.012 \\
$F_{2A}^\gamma$ & 0      & $+0.8$    & 100 & 0.014 \\
$F_{2A}^Z$      & 0      & $+0.8$    & 100 & 0.052 \\
$F_{2V}^\gamma$ & 0      & $\pm 0.8$ & 200 & 0.038 \\
$F_{2V}^Z$      & 0      & $\pm 0.8$ & 200 & 0.009 \\
\hline\hline
\end{tabular}
\end{center}
\vspace*{-5mm}
\end{table}

The most general $tbW$ couplings can be parameterized in the
form~\cite{hioki}
\begin{equation}
  \Gamma^\mu_{tbW} =
  - {g\over \sqrt 2 } \, V_{tb}^{} \,
  \left\{ \gamma^\mu \;
    \left[  f^{L}_{1} \, P_L^{} +
            f^{R}_{1} \, P_R^{} \right] -
    { i \, \sigma^{\mu\nu }\over M_W^{} } \, (p_t^{}-p_b)_\nu^{} \,
    \left[  f^{L}_{2} \, P_L^{} +
            f^{R}_{2} \, P_R^{} \right] \,
   \right\} \ ,
\end{equation}

where $P_{R,L}\!=\!\frac12 (1\!\pm\!\gamma_5^{})$. In the limit $m_b\to 0$,
$f_1^R$ and $f_2^L$ vanish and, in the SM, $f_1^L=1$ and all other form factors
are zero  at tree--level. The $\bar{t}\bar{b}W$ vertex can be parameterized
similarly.

The $f_2^R$ coupling, corresponding to a V+A $tbW$ interaction,
can be measured in $t\bar{t}$ decays with a precision of about 0.01 for
$\sqrt{s}=500$~GeV and 500~fb$^{-1}$ if
electron and positron beam polarization are available~\cite{hioki}.
This quantity can also be measured at the LHC, but the expected
limit is a factor three to eight weaker~\cite{Beneke:2000hk}.

The ILC can measure the $tbW$ interaction to
significant precision by studying $t \bar{t}$ production {\it below}
threshold~\cite{BatraTait}. At c.m. energies below $2 m_t$ but still
above $m_t$, the total rate for $e^+e^- \to W^+ W^- b \bar{b}$ is dominated by
contributions from the virtual $t \bar t$ diagrams in a kinematic configuration
where one top is on-shell and the other is off-shell. Other contributions
include single top quark production and, to a smaller extent, non-resonant
interfering backgrounds. The rate becomes very sensitive to the $tbW$ interaction,
essentially because the narrow width approximation is no longer valid when
the top momentum is off-shell.

For simplicity, the analysis focuses on the case of all couplings
but $f^{L}_{1}$ equal to zero and defines the effective V--A
coupling as $g_{tbW} = g V_{tb} f^{L}_{1}$. Only the semi-leptonic
six-body final state where one W boson decays to a pair of jets and
the other into an readily tagged lepton ($e$, $\mu$ or $\tau$), is
considered. Combining the below-threshold cross section measurement
with the $\Gamma_t$ extracted from the threshold scan permits
extraction of $g_{tbW}$ and $\Gamma_t$ independently. Under the
assumption that the width is measured to an accuracy of 100 MeV,
$g_{tbW}$ can be measured to the $ 3\%$ level, which would represent
better than a factor of two improvement compared to the LHC.

\begin{figure}[htbp]
\centerline{\includegraphics[width=4in]{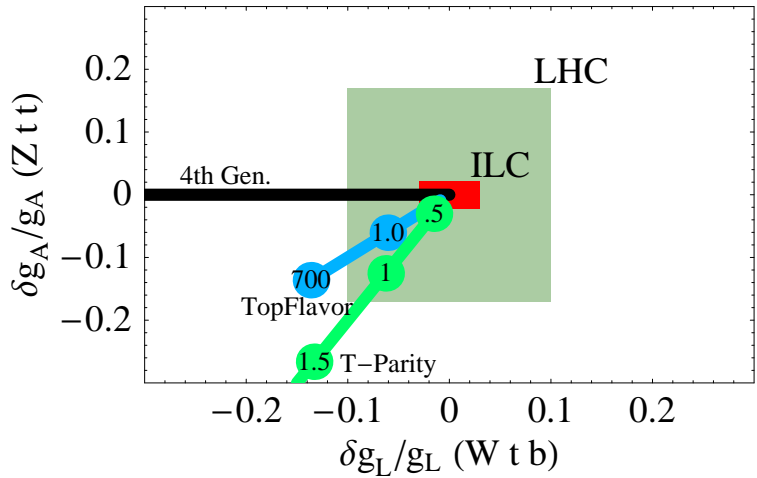}}
\vspace*{-5mm}
\caption[Expected bounds on top quark anomalous couplings from the ILC and LHC.]
{Expected bounds on  axial $t\bar{t}Z$ and left--handed $tbW$ couplings
from direct LHC (olive) and ILC (red) measurements;  superimposed are predicted
deviations from representative models \cite{BatraTait}.}
\label{fig:limits}
\vspace*{-5mm}
\end{figure}

Figure~\ref{fig:limits} shows the expected bounds on the SM--like top axial
$t\bar{t}Z$ and left--handed $tbW$ interactions and the discriminating power the
bounds can place on new physics models. Included in the plot are the 1$\sigma$
constraints on the independently varied axial $t\bar{t}Z$ coupling from the LHC
and ILC \cite{Abe:2001nq}, and the direct constraints on the left-handed $tbW$
coupling from the LHC \cite{Beneke:2000hk}.  Predicted deviations from a few
representative models are also superimposed: a Little Higgs model with T-parity,
a model of top-flavor, and a model with a sequential fourth generation whose
quarks mix substantially with the third family. The little Higgs model with
T--parity  has a heavy top quark partner $T$ with a  mass assumed to be
$m_T=500$~GeV (the numbers on the plot indicate the strength of the $hTt$
interaction); the top--flavor model has a mixing angle $\sin{\phi}=0.9$ (numbers
indicate the mass of the heavy $Z^\prime$). Top--seesaw models generate the same
mixing effect as the little Higgs models and, thus, trace out the same line in
the plane of deviations in the $t\bar{t}Z$ and $tbW$ as the seesaw model
parameters are varied.

Finally, the ILC has excellent reach for the measurement of the tensor coupling
$tZq$; see for instance Ref.~\cite{0102197}.  At the ILC, both the anomalous
production $e^{+}e^{-} \rightarrow tq$ and decay $e^{+}e^{-} \to t\bar{t}, t
\rightarrow Vq$ mechanisms can be explored, permitting sensitivity to flavor
changing neutral current  interactions. With 45\% positron and 80\% electron
polarization at $\sqrt s=500$~GeV, 100~fb$^{-1}$ of data would result e.g. in a
sensitivity to BR($t \rightarrow \gamma q$) of $2 \times 10^{-5}$.  The search
sensitivity might be significantly increased if the ILC runs in the
$\gamma\gamma$ mode~\cite{Cao:2002si}.

\subsection{Couplings to gluons}

The ILC can be competitive with and complementary to the LHC in the measurement
of the strong top quark coupling to gluons and would allow more refined tests of
perturbative QCD \cite{Aguilar-Saavedra:2001rg}. Hard gluon radiation in $t\bar
t$   events \cite{T-arnd} would allow several tests of the strong dynamics of
the top  quark: test of the flavour--independence of strong interactions,
limits on anomalous chromo-electric and/or chromo-magnetic dipole moments
\cite{T-rizzo} and the determination of the running top quark mass.  In turn,
soft gluon radiation in $t\overline{t}$ events  is expected to be strongly
regulated by the large top mass and width and  would provide  additional
constraints on the total decay width $\Gamma_t$ \cite{T-lynne}. Color
reconnection and Bose-Einstein correlations are also important to study
precisely \cite{T-bose} as they may affect the precision with which the  top
quark mass  can be reconstructed kinematically via their multijet decays.

Furthermore, polarized electron and positron beams can be exploited  to test
symmetries using multi--jet final states. For polarized $\ee$ annihilation to
three hadronic jets, one can define the triple product ${\bf S}_e\cdot({\bf
k}_1\times {\bf k}_2)$, which correlates the $e^-$ beam polarization vector
${\bf S}_e$ with the normal to the three--jet plane defined by ${\bf k}_1$ and
${\bf k}_2$, the momenta of the two quark jets. If the jets are ordered by
momentum (flavour), the triple--product is CP--even (odd) and T--odd
\cite{T-dixon}. In the SM, the contributions to the T--odd  form are expected to
be very small and limits  have been set for the $b\overline{b}g$ system. At the
ILC,  these observables will provide an additional possibility to search for
anomalous effects in the $t\overline{t}g$ system.


\section{New decay modes}

Besides the standard channel $t\to bW$, new decays of the top quark can occur in
some extensions of the SM. The prominent example is the top quark decay into a
charged Higgs boson, $t \to bH^+$, in supersymmetric extensions of the SM or in
multi--Higgs doublet extensions. This channel has been mentioned in chapter
\ref{sec:higgs} in the context of the MSSM and in this case, the coupling of the
$H^\pm$ bosons to top and bottom quarks is a mixture  of scalar and pseudoscalar
currents and depend only on the ratio of the {\it vev}'s of the two Higgs
doublet fields $\tb$,
\begin{equation}
g_{H^-tb} \sim m_b \tan\beta (1+\gamma_5) + m_t{\rm cot}\beta
(1-\gamma_5)
\end{equation}

\begin{figure}[!h]
\vspace{-.3cm}
\begin{center}
\includegraphics[width=0.9\linewidth,bb=73 505 600 745]{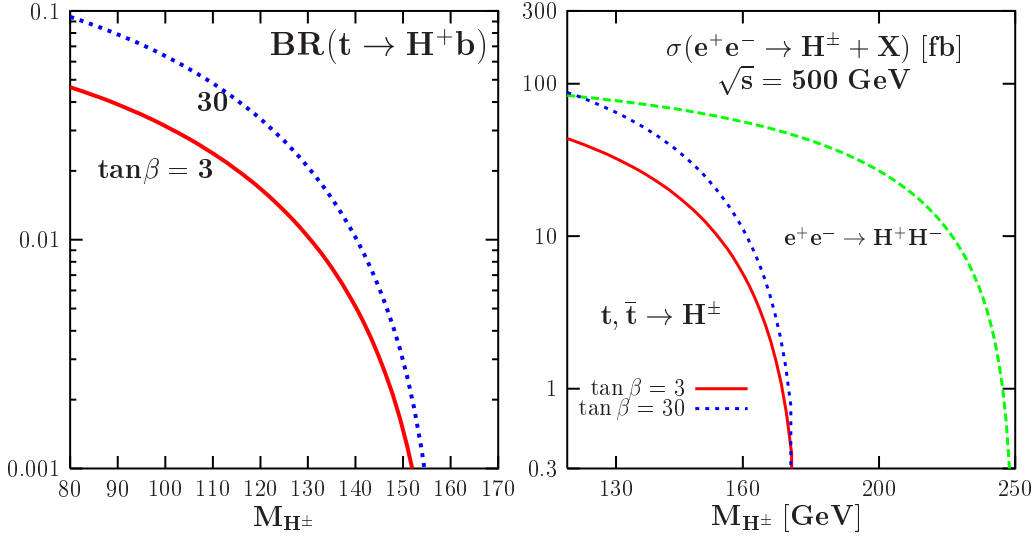}
\end{center}
\caption[Charged Higgs bosons from top decays: branching ratios and production rates]
{Left: the branching ratio for the decay $t\to H^+b$ as a function
of $M_{H^+}$ for $\tb=3$ and 30 in the MSSM. Right: the cross sections times
branching ratio for the production of one charged Higgs boson from top decays,
$\ee t\bar t$ and $t \to H^+b$, at the ILC with $\sqrt s=500$ GeV; the
direct $\ee \to H^+ H^-$ cross section is shown for comparison. From
Ref.~\cite{Djouadi:2005gj}.}
\vspace{-.3cm}
\label{fig:BR-tH+}
\end{figure}

The coupling is therefore very strong for small or large $\tb$ values for which
the $m_t$ component is not suppressed or the $m_b$ component is strongly
enhanced. The branching ratio ${\rm BR} (t\to bH^+) = \Gamma (t\to bH^+) /
[\Gamma (t\to bW) + \Gamma (t\to bH^+)]$ is displayed in the left--hand side of
Fig.~\ref{fig:BR-tH+} as a function of $M_{H^+}$ for two values $\tb=3$ and 30.
As can been seen, it is rather substantial being still at  the per--mille level
for $H^+$ masses as large as 150 GeV.

Since the cross section for top quark pair production is of the order of
$\sigma (\ee \to t\bar t)\sim 0.5$ pb at a $\sqrt{s}=500$ GeV ILC, the  cross
section times the branching ratio for the production of one charged  Higgs boson
is rather large if $M_{H^\pm}$ is not too close to $m_t$ for the
decay not to be suppressed by the small phase space.  This is shown in the
right--hand side of Fig.~\ref{fig:BR-tH+} where on can see that, for $M_{H^\pm}
\lsim 150$ GeV, the rates are of the same order of magnitude as the ones from
direct pair production, $\ee \to H^+ H^-$, which is displayed for comparison.

In the $M_{H^\pm}$ range under consideration,  the main two--body decays of the
charged Higgs boson will be into $\tau \nu_\tau$ and $c\bar{s} $ pairs with the
former being largely dominating for the chosen $\tb$ values; see
Fig.~\ref{Hfig:BR-MSSM}. This results in a surplus of $\tau$ final states over
$e, \mu$ final states, an apparent breaking of lepton universality. For low
values of $\tb$, the three--body decay modes $H^\pm \to hW^* , A W^*  \to b\bar
b W$ will lead to multi $b$ and $W$ final states.  These signals will be rather
easy to be disentangled from the backgrounds in the clean ILC environment.

This signal will be first observed at the LHC as it is one of the main
discovery channels for charged Higgs bosons. However, the ILC will provide a
very important information:   the precise measurement of the $t\to H^+b$
branching ratio would allow to determine the parameter $\tb$ which is known to
be rather difficult to access otherwise; see chapter \ref{sec:susy}.

In supersymmetric models, another possible and interesting decay mode of the top
quark would be into its scalar partner $\tilde t_1$ and the lightest neutralino
$\chi_1^0$ which is supposed to form the dark matter in the universe, $t \to
\tilde t_1 \chi_1^0$. In the minimal supersymmetric extension with universal
masses for the superpartners of the gauge bosons at the high GUT scale, the
phase space for this decay is squeezed by the constraints on the $\tilde t_1$
and $\chi_1^0$ masses from LEP and the Tevatron. In non minimal extensions, the
decay might be kinematically allowed and, in this case, branching ratios of the
order  of a few percent would be possible. Since the main decay modes of the top
squark in this mass range are the loop induced $\tilde t_1 \to c\chi_0^1$ and
the four--body $\tilde t_1 \to b f \bar f \chi_0^1$ channels,  the signal will
consist on the missing energy due to the escaping neutralinos. While it is
overwhelmed by huge QCD backgrounds at the LHC, this signature should be easy to
detect at the ILC.

Finally, flavor changing neutral current (FCNC) decays of the to quark may be
also observed. If new quark species exist and do not belong to the standard
doublet/singlet assignments of isospin multiplets, they will mix with the top
quark, breaking the GIM mechanism and allowing for FCNC top--charm couplings of
order $\sqrt{m_t m_c / M^2_X}$ to be induced. In this case,  besides breaking
the universality of the $V-A$ chiral $Wtb$ current,  FCNC top quark decays  such
as $t \rightarrow c \gamma$ or $t \to cZ$ may occur at the level of a few
permille and can be detected at the ILC \cite{T-blair}. However, the large
number of top quarks produced at the LHC  allows to search for these rare FCNC
decays down to branchings ratio less than $10^{-4}$.

%
\newcommand{\epem}{e^+e^-}
\chapter{Supersymmetry}
\label{sec:susy}

\section{Introduction}

\subsection{Motivations for supersymmetry}

Despite its enormous success in describing almost all known experimental data
available today, the Standard Model (SM) is widely believed to be an effective
theory valid only at the presently accessible energies. Besides the fact that it
does not say anything about the fourth fundamental force of nature, the
gravitational force, and does not explain the pattern of fermion masses, it has
at least three severe problems which call for new physics. Given the
high--precision data and the particle content of the SM, the energy evolution of
the gauge coupling constants is such that they fail to meet at a common point,
the grand unification (GUT) scale.  Moreover, the SM does not include any
candidate for a particle that is absolutely stable, fairly massive, electrically
neutral and having only weak interactions, which accounts for the cold dark
matter (DM) that makes up $\approx 25\%$ of the present energy of the universe.
Finally, in the SM, the radiative corrections to the Higgs boson mass squared
are quadratically divergent and $M_H$, which is expected to lie in the range of
the electroweak symmetry breaking scale, ${\cal O}(100)$ GeV, prefers to be
close to the cut--off scale beyond which the theory ceases to be valid, the very
high GUT or Planck scales.

Supersymmetry (SUSY) \cite{SUSY-SUSY}, which predicts the existence of a partner
to every known particle that differs in spin by $\frac12$, is widely considered
as the most attractive extension of the SM. Firstly, SUSY has many theoretical
virtues \cite{SUSY-Bagger}: it is the first non--trivial extension of the
Poincar\'e group in quantum field theory which, when made local, necessarily 
includes Einsteins's theory of gravity,  and it appears naturally in
superstring theories. These features may help to reach the ultimate goal of
particle physics: the unification of all forces including gravity.  However, the
most compelling arguments for SUSY are phenomenological ones: when it is
realized at low energies, it can  solve  at once all the above three  problems
of the SM. Indeed, the main reason for introducing low energy supersymmetric
theories in particle physics is their ability to solve naturally the
fine--tuning problem \cite{SUSY-Witten}: SUSY prevents $M_H$ from acquiring very
large radiative corrections as the quadratic divergent loop contributions of the
SM particles are exactly canceled by the corresponding loop contributions of
their supersymmetric partners. In fact, SUSY allows one to understand the origin
of the electroweak symmetry breaking itself  in terms of radiative corrections
triggered by SUSY breaking \cite{SUSY-REWSB}, which  must occur as
the newly predicted superparticles have not been observed up  to now and must be
thus heavy. In addition, the new SUSY particle spectrum contributes to the
evolution of the three gauge couplings and allows their unification at a scale 
$M_{\rm GUT} \simeq 2\!\cdot\! 10^{16}$ GeV \cite{SUSY-Unif}. Finally, a
discrete symmetry called $R$--parity \cite{SUSY-Fayet} can be naturally present
with the major consequence that the lightest supersymmetric particle (LSP) is
absolutely stable;  in many cases, this particle has the right properties and
the required cosmological relic density to account for the cold DM
\cite{DM-Ellis,DM-review}.

\subsection{Summary of SUSY models}

The most economical low--energy globally supersymmetric extension of the SM is
the minimal supersymmetric Standard Model (MSSM) \cite{SUSY-MSSM}. In this
model, one assumes the minimal (SM) gauge group, the minimal particle content
[i.e., three generations of fermions  and their spin--zero partners as well as
two Higgs doublet superfields to break the electroweak symmetry in a consistent
manner], and $R$--parity conservation, which makes the LSP absolutely stable. 
In order to explicitly break SUSY, a collection of soft terms is added to the
Lagrangian: mass terms for the gauginos, the SUSY  spin--$\frac12$ partners of
the gauge bosons, mass terms for the sfermions, the spin--0 partners of the SM
fermions, mass and bilinear terms for the two Higgs fields and trilinear
couplings between sfermion and  Higgs fields.

In the most general case, the soft SUSY--breaking terms will introduce a huge
number of unknown parameters, ${\cal O}(100)$. However, in the absence of
complex phases and intergenerational sfermion mixing and if the universality of
the two first generations of sfermions is assumed, to cope in a simple way with
the severe experimental constraints, this number reduces to ${\cal O} (20)$. 
Furthermore, if the soft SUSY--breaking parameters obey a set of boundary
conditions at a high energy scale, all potential phenomenological problems of
the general MSSM can be solved with the bonus that, only a handful of new free
parameters are present.  The underlying assumption is that SUSY--breaking occurs
in a hidden sector which communicates with the visible sector only
``flavor--blind'' interactions, leading to universal soft breaking terms.  This
is assumed to be the case in the celebrated minimal supergravity (mSUGRA) model
\cite{SUSY-mSUGRA} or constrained MSSM (cMSSM) which is often used as a
benchmark scenario  in phenomenological analyses.

Besides the GUT scale which is derived from the unification of the
three gauge coupling constants, the cMSSM  has only four free
parameters plus a sign:

\centerline{$m_0, \ m_{1/2}, \ A_0, \ \tb, \ {\rm sign}({\mu})$  ,}

\noindent where $m_0,m_{1/2}$ and $A_0$ are, respectively, the common soft terms
of all scalar (sfermion and Higgs) masses, gaugino (bino, wino and gluino)
masses and trilinear scalar interactions, all defined at the GUT scale. $\tb$ is
the ratio  of the vacuum expectation  values (vev's) of the two Higgs doublets
at the weak scale and $\mu$ is the supersymmetric Higgs(ino) mass parameter. As
in the MSSM in general, all soft SUSY--breaking parameters at the weak scale are
then obtained via known Renormalization Group Equations (RGEs). The masses of
the physical states, the spin--$\frac12$ charginos $\chi_{1,2}^\pm$ and
neutralinos $\chi_{1,2,3,4}^0$ which are mixtures of the SUSY partners of the
gauge and Higgs bosons, the two scalar partners $\tilde f_{1,2}$ of the SM
fermions and the five MSSM Higgs bosons $h,H,A$ and $H^\pm$ are then obtained by
diagonalyzing the relevant mass matrices.  In this scenario, the LSP is in
general the lightest neutralino $\chi_1^0$. 

There are also other constrained MSSM scenarios with only a few
basic input parameters, two of  them being the anomaly (AMSB) \cite{SUSY-AMSB}
and gauge (GMSB) \cite{SUSY-GMSB} mediated models in which SUSY--breaking also
occurs in a hidden sector but is transmitted  to the visible one by anomalies or
by the SM gauge interactions; in the later case, a very light gravitino is the
LSP\footnote{In fact, in mSUGRA--like models, one can also have the gravitino
being the LSP in large areas of the parameter space \cite{DM-grav0}; this issue
will be discussed in the cosmology chapter.}.  On the other hand, one can
slightly depart from the restrictive minimality of the MSSM and interesting
examples are the  CP violating MSSM \cite{SUSY-CPV} where some SUSY parameters
can be complex, the NMSSM \cite{SUSY-NMSSM} in which the spectrum is extended to
include a singlet superfield and $R$--parity violating models \cite{SUSY-RPV} in
which the LSP is not stable.  

The Terascale is a mystery that will be revealed by the LHC and the ILC and both
machines will have an important role to play in deciphering it. In  particular
the high precision of the ILC will be necessary to understand the new physics,
no matter which scenario nature has chosen.  In this chapter, we will mainly
focus on the unconstrained and constrained MSSMs  defined above as they are very
well defined and have been studied in great detail. These models provide us with
an  excellent testground for the opportunities offered by the high--energy
colliders, the ILC in particular, in reaching out to new physics domains. 

\subsection{Probing SUSY and the role of the ILC}

To prove and to probe supersymmetry, one not only needs to produce the new 
particles but also, and this is equally important,  to verify its most 
fundamental predictions in a model independent way.  A detailed investigation
of  the properties of the SUSY and Higgs particle spectrum is thus required
and,  in particular, one needs to: 

\begin{itemize}
\vspace*{-2.5mm}
\item[$\bullet$] measure the masses and mixings of the newly produced 
particles, their decay widths and branching ratios, their production cross 
sections, etc...; 
\vspace*{-2.5mm}
\item[$\bullet$] verify that there are indeed the superpartners of the SM 
particles and, thus, determine their spin and parity, gauge quantum numbers 
and their couplings;
\vspace*{-2.5mm}
\item[$\bullet$] reconstruct the low--energy soft--SUSY breaking parameters
with the smallest number of assumptions, that is, in as model independent way
as possible; 
\vspace*{-2.5mm}
\item[$\bullet$]  ultimately, unravel the fundamental SUSY breaking  mechanism 
and shed light on the physics at the very high energy (GUT, Planck?) scale. 
\vspace*{-2.5mm}
\end{itemize}

Furthermore, the very precise knowledge of the properties of the lightest SUSY
particle and its interactions with the standard and other SUSY particles is
mandatory to predict the cosmological relic density of the DM, as well as its
rates in direct and indirect detection astroparticle experiments. Achieving
this goal would be the decisive test that a particular physics scenario is the
solution of the DM puzzle and would lay an additional bridge between collider
physics and the physics of the early universe.

\begin{figure}[ht!]
\begin{tabular}{ll}
\begin{minipage}{5.5cm}
\begin{center}
\vspace*{-5mm}
\psfig{file=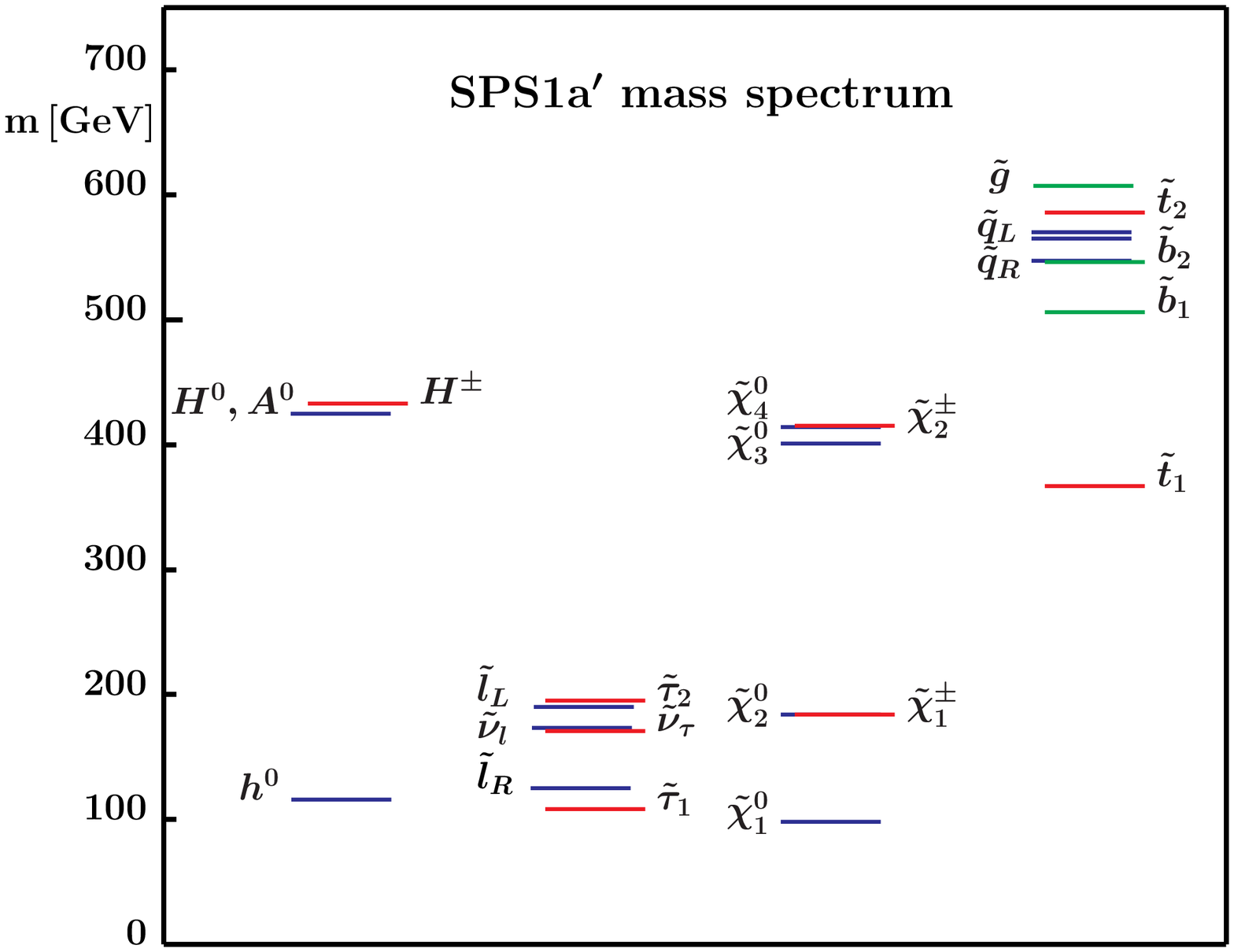,width=7.5cm}
\end{center}
\end{minipage} &\hspace*{2mm}
\begin{minipage}{10cm}
\begin{center}
\psfig{file=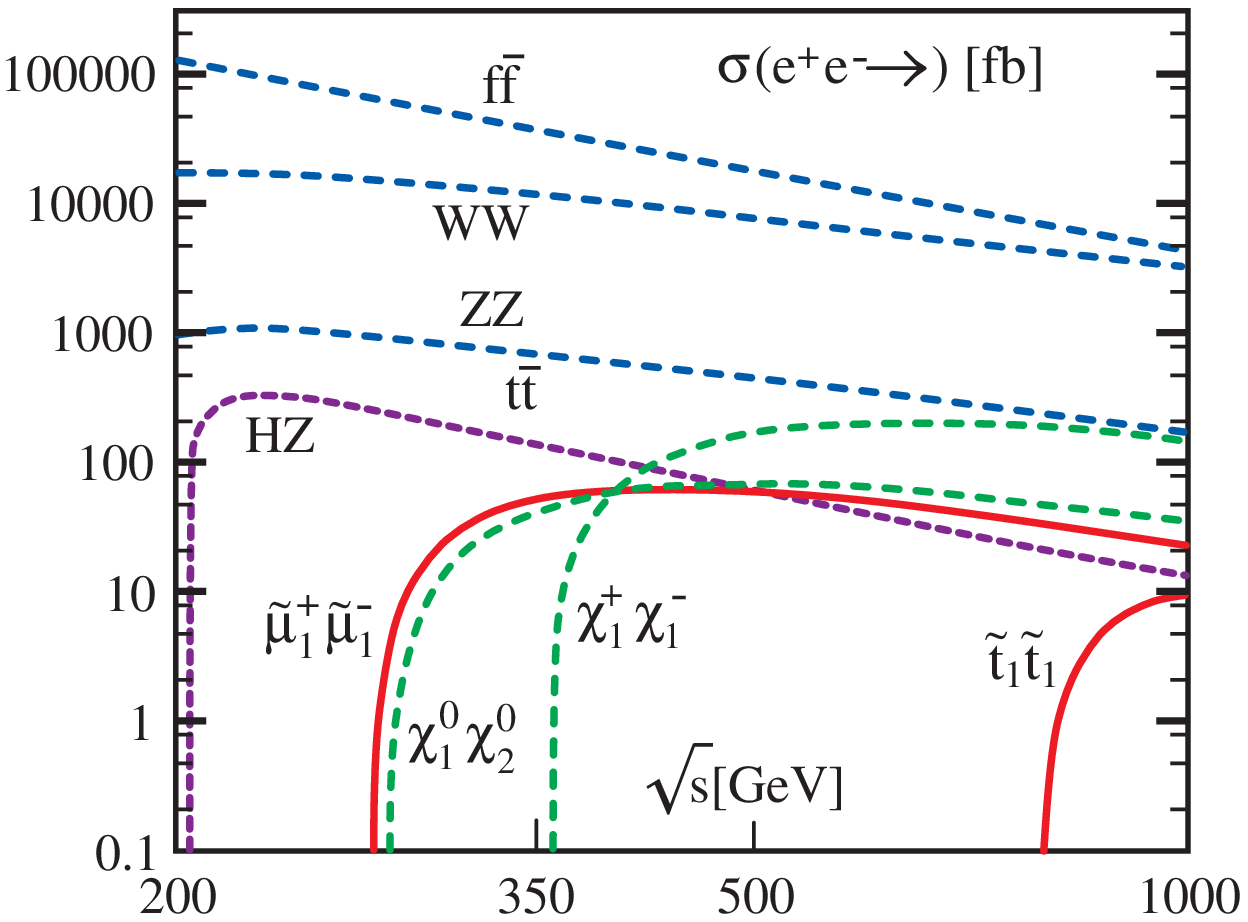,width=7.5cm,height=6.2cm}
\end{center}
\end{minipage}
\end{tabular}
\vspace*{-5mm}
\caption[SUSY spectrum in a benchmark point and production
cross sections at ILC]
{The spectrum of SUSY and Higgs particles in the benchmark SPS1a$^\prime$ 
cMSSM point \cite{SUSY-SPA} (left) and the production cross sections for 
various SM and SUSY processes in $e^+e^-$ collisions as a function of the c.m. 
energy in this scenario (right).} 
\label{fig:SPS}
\end{figure}

In most areas of the MSSM parameter space, in particular in cMSSM type scenarios
(except in the focus point scenario to be discussed later  in chapter
\ref{sec:cosmology}), the colored squarks and gluinos turn out to be much
heavier than the non--colored sparticles, the sleptons as well as the charginos
and neutralinos; see Fig.~\ref{fig:SPS}.  If the masses of the former sparticles
dot not significantly exceed the TeV scale, as required from naturalness
arguments, they can be copiously produced at the LHC either in pairs or in
association \cite{atlastdr,CMSTDR}. They will then decay in potentially long
chains which end in the LSP neutralino that signals its presence only via
missing energy. These decay chains will involve the other neutralinos and the
charginos, and possibly the sleptons, so that one can have access to these
weakly interacting particles as well.  Typically, one faces a situation in which
several SUSY particles are present in the same event, leading to rather
complicated final state topologies which are subject to very large backgrounds
from the SM and, more importantly, from SUSY itself. At the LHC, sparticle mass
differences can be determined by measuring the endpoints or edges of invariant
mass spectra (with some assumptions  on particle identification within the
chains) and this results in a strong correlation between the extracted masses;
in particular, the LSP mass can be constrained only weakly
\cite{Weiglein:2004hn}. Therefore, only in specific constrained scenarios with a
handful of input parameters, that some elements of SUSY can be reconstructed in
the complicated environment of the LHC.

On the other hand, the non--colored SUSY particles (and certainly the
lightest Higgs boson) would be accessible at the ILC with a c.m.
energy of $\sqrt s=500$ GeV, to be eventually upgraded to 1 TeV. This
is, for instance the case in a cMSSM typical scenario called SPS1a$^\prime$
\cite{SUSY-SPA} as shown in Fig.~\ref{fig:SPS}.  The cross sections
for chargino, neutralino and slepton pair production, when the states
are kinematically accessible, are at the level of 10--100 fb, which is
only a few orders of magnitude below the dominant SM background
processes; Fig.~\ref{fig:SPS}. Given the expected high--luminosity and
the very clean environment of the machine, large samples of events
will be available for physics analyses
\cite{Aguilar-Saavedra:2001rg,SUSY-Jan}. At the ILC, it will be thus
easy to directly observe and clearly identify the new states which appeared only through
cascade decays at the LHC. Most importantly, thanks to the unique
features of the ILC, tunable energy which allows threshold scans, the
availability of beam polarization to select given physics channels and
additional collider options such as $e^-e^-$ which allow for new
processes, very thorough tests of SUSY can be performed: masses and
cross sections can be measured precisely and couplings, mixing angles
and quantum numbers can be determined unambiguously.  Furthermore, the
ILC will provide crucial information which can be used as additional
input for the LHC analyses, as would be e.g. the case with the LSP
mass. The coherent analyses of data obtained at the LHC and the ILC
would allow for a better and model independent reconstruction of the
low energy SUSY parameters, connect weak--scale SUSY with the more
fundamental underlying physics at the GUT scale, and provide the
necessary input to predict the LSP relic density and the connection
with cosmology.

To highlight the unique abilities of the ILC to address these issues, we will 
often  use for illustration the cMSSM benchmark SPS1a$^\prime$ point with basic inputs
\cite{SUSY-SPA}: 

\centerline{ $m_{1/2}\!=\!250$ GeV, $m_0\!=\!70$ GeV, $A_0\!=\!-300$ GeV, 
$\tb\!=\!10$ and $\mu>0$,}

\noindent which, using one of the RGE codes ({\tt SPHENO}) of
Ref.~\cite{SUSY-ISAJET},  leads to the SUSY spectrum of Tab.~\ref{tab:SPS}. 
This testcase point is close to the point SPS1a \cite{SUSY-SPS} with 
$m_0=-A_0=100$ GeV and the same $m_{1/2}, \tb$ and $\mu$ values, which has been 
used for detailed LHC \cite{LHC-benchmark,SUSY-Desch} as well as ILC analyses,
but is not compatible anymore with all collider or cosmological constraints. 

\begin{table}[h!] 
\caption[Some superparticle masses for two minimal Supergravity benchmark points.]
{Some superparticle and their masses (in GeV) for the
cMSSM SPS1a$^\prime$ and SPS1a reference points.}  
\label{tab:SPS}
\begin{center}$
\begin{array}{|c||c|c|c||c|c|c|c|c|c||c|c|} \hline
\widetilde{p}/\mbox{mass} & \chi_1^0 & \chi_2^0 & \chi_1^\pm & \widetilde{e}_1/
\widetilde{\mu}_1 & \widetilde{e}_2/\widetilde{\mu}_2 & 
\widetilde{\nu}_e/\widetilde{\nu}_\mu & \widetilde{\tau}_1 & 
\widetilde{\tau}_2 & \widetilde{\nu}_\tau & \widetilde{t}_1 & 
\widetilde{b}_1 \\ \hline 
\mbox{SPS1a$^\prime$} & 97.7 & 183.9 & 183.7 & 125.3 & 189.9 & 172.5 & 107.9 & 
194.9 & 170.5 &  366.5 & 506.3\\ 
\mbox{SPS1a} & 96.1 & 176.8 & 176.4 & 143.0 & 202.1 & 186.0 & 133.2 & 
206.1 & 185.1 & 379.1 &  491.9\\ \hline
\end{array}$
\end{center}
\end{table}

\section{Precision SUSY measurements at the ILC}

\subsection{The chargino/neutralino sector}

The two charginos $\chi_{1,2}^\pm$ and the four neutralinos $\chi_{1,2,3,4}^0$
are obtained by diagonalyzing the mass matrices of the charged and neutral
gauginos and higgsinos. For charginos, the matrix depends on the wino and
higgsino mass parameters $M_2$ and $\mu$ and on $\tb$; for neutralinos, the
bino mass parameter $M_1$ enters in addition. These parameters determine to a
large extent the production and decay properties of the $\chi_i^0, \chi_i^\pm$
states that we will call ``inos" for short.  

Charginos are produced in pairs, $\epem \to \chi_i^+\chi_j^-$, through
$s$--channel $\gamma/Z$ boson and $t$--channel sneutrino exchanges; the latter
contribution can be suppressed with polarized $e^-_R/e^+_L$ beams. Neutralino
pair production, $\epem \to \chi_i^0\chi_j^0$, proceeds through $s$--channel $Z$
boson and $t$-- and $u$--channel $\tilde{e}_{L,R}$ exchanges. The ino states
decay into lighter charginos and neutralinos and (possibly virtual) gauge or
Higgs bosons as well as sfermion--fermion pairs; for the lighter inos, one would
then have the topologies $\chi_1^\pm \to f\bar f' \chi_1^0$ and $\chi_2^0 \to
f\bar f \chi_1^0$. These final states can be easily detected as the production
cross sections are sizable and the backgrounds involving a large amount of
missing energy are small.

\begin{figure}[htb]
\vspace*{-2.7cm}
\begin{center}
\mbox{\includegraphics*[width=69mm,height=59mm]{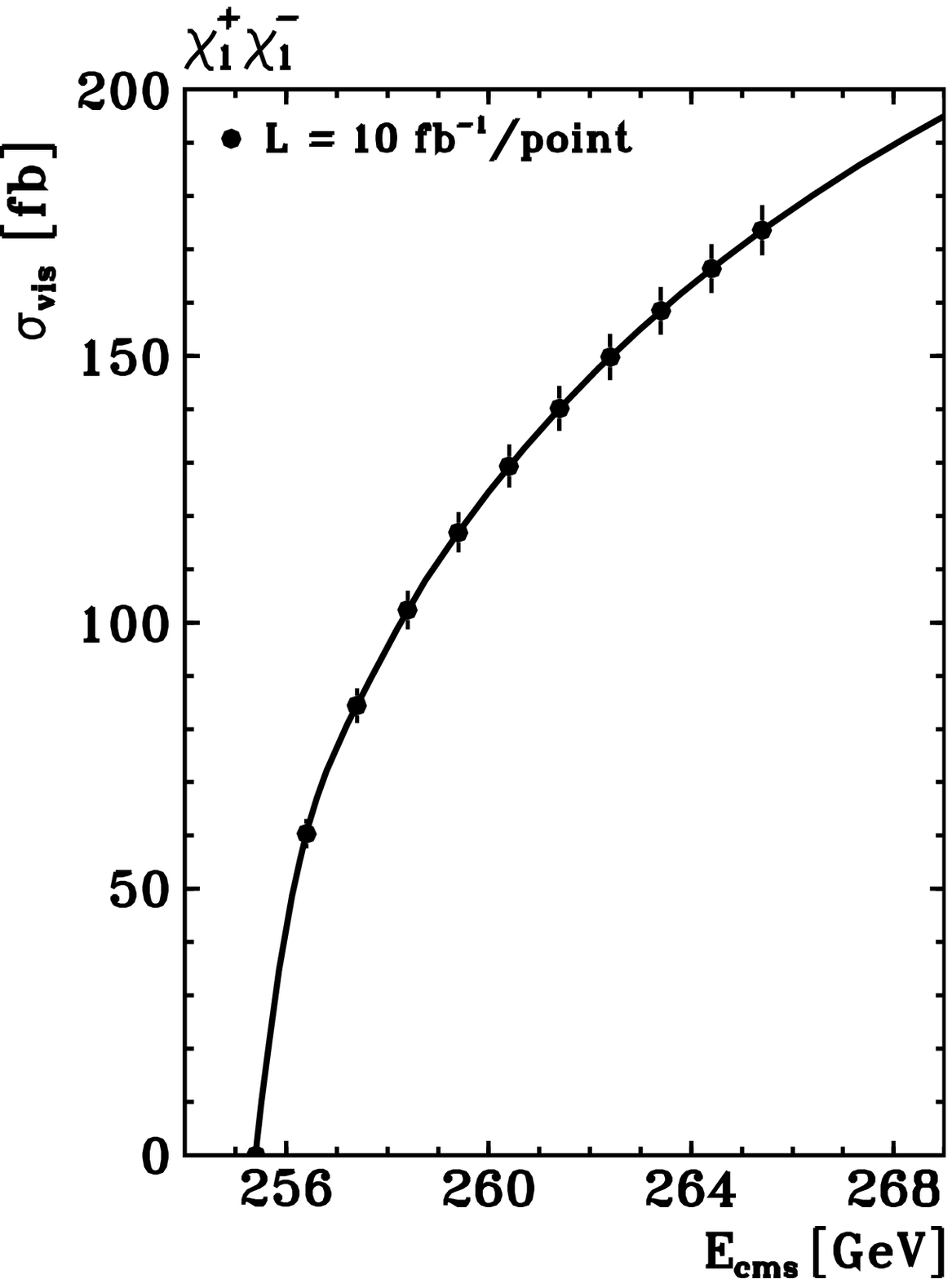}
\hspace*{-.9cm}
\includegraphics*[width=89mm,height=85mm]{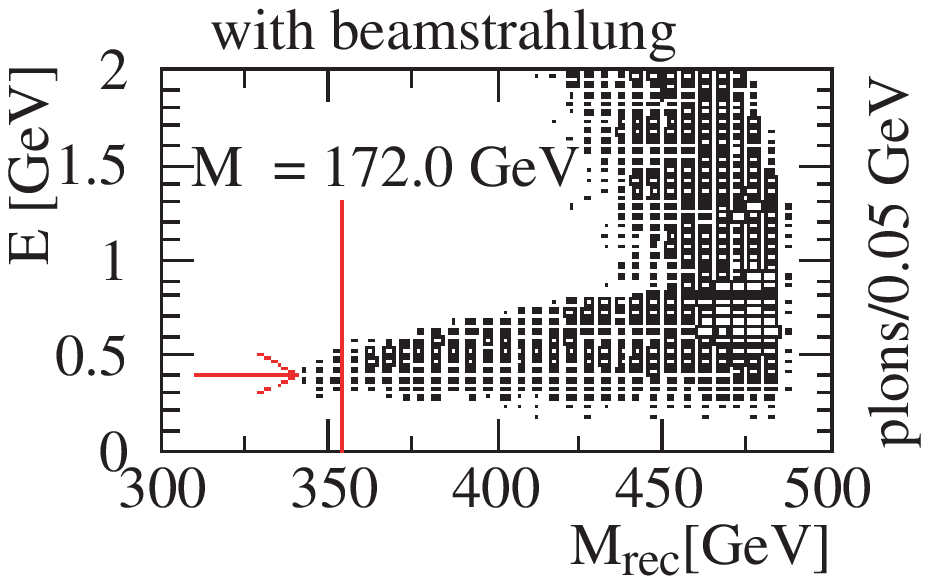} } 
\end{center}
\vspace*{-7mm}
\caption[Measurements of cross sections/distributions in chargino 
production at ILC]
{The cross section for $e^+_R e^-_L \to \chi^+_1 \chi^-_1 \to \ell^\pm
\nu_\ell \chi^0_1 q\bar q' \chi^0_1$ near threshold, with the error bars 
obtained with a luminosity of 10 fb$^{-1}$ per point \cite{SUSY-Martyn} (left). 
The initial state radiated photon  recoil mass for the process $e^+_R e^-_L \to
\chi^+_1 \chi^-_1 \gamma \to \pi^+\pi^-\gamma \ \slash \hspace*{-2mm}E$ (right)
\cite{SUSY-hensel}. }
\label{fig:ino} 
\end{figure}

The chargino masses can be determined in the continuum from the di--jet energy
distributions in the process $\epem \to \chi_1^+\chi_1^- \to \ell^\pm \nu q \bar
q' \chi_1^0 \chi_1^0$, which leads to a mass resolution $\Delta m_{\chi_1^\pm}
/m_{\chi_1^0}$ at the permille level. This can serve to optimize a scan around
threshold  which, because of the steep $\sigma  \propto \beta_\chi$ rise of the
excitation curve with the velocity, would lead to a mass resolution $\Delta
m_{\chi_1^\pm}\!=\!{\cal O}(50)$ MeV for $m_{\chi_1^\pm}\!\sim\! 170$ GeV; 
Fig.~\ref{fig:ino}.  The di--jet mass spectrum in $\chi_1^\pm$ decays allows
also to determine the chargino--neutralino mass difference with a high
precision, $\Delta (m_{\chi_1^\pm}\!-\!m_{\chi_1^0})\!=\!{\cal O}(50)$ MeV, from
which one can infer the mass of the escaping lightest neutralino. If the
chargino happens to be almost degenerate with the LSP neutralino, as is
typically the case in AMSB models, one can use ISR photons in the process $\epem
\to \chi_1^+\chi_1^- \gamma$ to measure both the $\chi_1^\pm$ and $\chi_1^\pm-
\chi_1^0$  masses from the spectra of, respectively, the photon recoil mass 
which peaks at $2m_{\chi_1^\pm}$ and the energy of the soft pions from
$\chi_1^+ \to \chi_1^0 + \pi_{\rm soft}$ which peaks at  $\Delta m_{\chi_1^\pm}
\!-\!m_{\chi_1^0}$; Fig.~\ref{fig:ino}. An uncertainty of a few percent is 
obtained in both cases.  

Similarly to the chargino case, the di--lepton mass and energy spectra in the
process $\epem \to \chi_2^0\chi_1^0 \to \ell^+ \ell^- \chi_1^0 \chi_1^0$, allow to
determine the mass difference of the two neutralinos at the permille level.  In
the case where the neutralino $\chi_2^0$ decays dominantly via a real or virtual
stau lepton, $\chi_2^0 \to \tilde \tau_1 \tau \to \tau^+ \tau^- \chi_1^0$, which
might occur at high $\tb$ values that lead to light tau sleptons, the resolution
on the $\chi_2^0$ mass deteriorates to the level of a few percent. The reason is
that the energy of the $\tau$'s cannot be reconstructed because of the missing
neutrinos and, in fact, this is also the  case for charginos in the decays 
$\chi_1^\pm \to \tilde \tau_1^\pm \nu_\tau \to \tau^\pm \nu_\tau \chi_1^0$.  A
better mass resolution, ${\cal O}(100)$ MeV, can be obtained with a threshold
scan in scenarios where sleptons are light, even in topologies involving
$\tau$'s.  For very heavy selectrons the error is larger since only the
$s$--channel $Z$ exchange contribution is present, leading to relatively smaller
cross sections and a less steep excitation curve, $\sigma \propto \beta^3_\chi$
because of the Majorana nature of the  neutralinos. An exception is when the 
neutralinos that are produced in mixed pairs have opposite CP parities, in 
which case the cross section increases steeply in S--waves. 

Note that for the verification of the spin--$\frac12$ character of the 
neutralinos and charginos, neither the onset of the excitation curves  near
threshold nor the angular distributions in the production processes provide
unique signals of the spin \cite{SUSY-spin-chi+}. However, decay angular
distributions of polarized neutralinos/charginos that are pair produced with
polarized beams provide an unambiguous determination of the spin--$\frac12$
character of the particles albeit at the expense of more involved experimental
analyses \cite{SUSY-spin-chi+}.

\begin{figure}[htb]
\vspace*{-2mm}
\setlength{\unitlength}{.7cm}
\begin{center}
{\epsfig{file=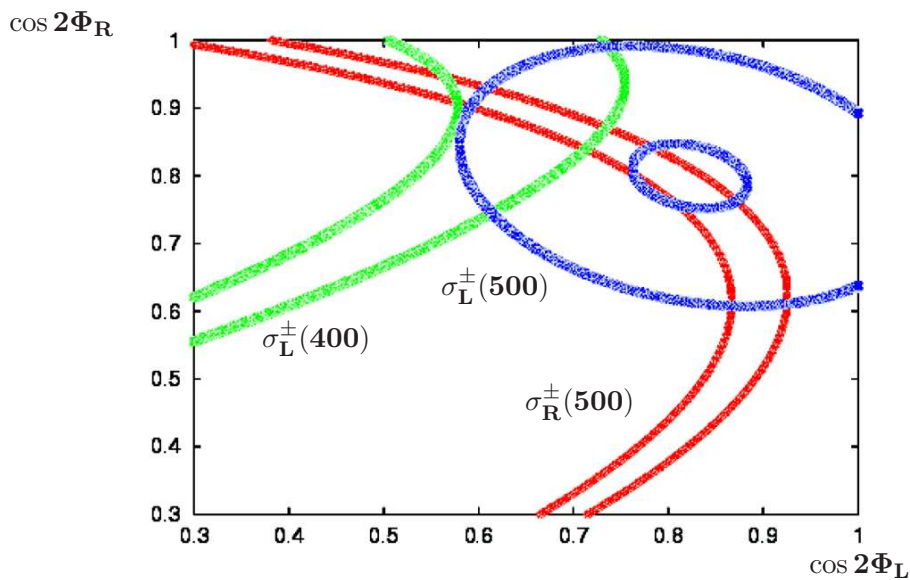,scale=.7}}
\put(-1.6,-.3){$\mathbf{\cos 2 \Phi_L}$}
\put(-16.8,10.){$\mathbf{\cos 2 \Phi_R}$}
\put(-8.6,5.){$\mathbf{\sigma_L^{\pm}(500)}$}
\put(-12.,4.){$\mathbf{\sigma_L^{\pm}(400)}$}
\put(-7,2.8){$\mathbf{\sigma_R^{\pm}(500)}$}
\end{center}
\vspace*{-5mm}
\caption[Contours for chargino production cross sections
with polarized e$^\pm$ beams]
{ Contours for the $\epem \to \chi_1^+ \chi_1^-$ production cross section
for polarized $e^\pm$ beams in the plane [$\cos2\phi_L,\cos2\phi_R$] at $\sqrt 
s=400$ and 500 GeV \cite{SUSY-Desch}.}
\label{fig:ino2}
\vspace*{-2mm}  
\end{figure}

The $\epem \to \chi_i^+\chi_j^-$ production cross sections are binomials in the
chargino mixing angles $\cos2\phi_{L,R}$ and the latter can be determined in a
model independent way using polarized beams. This is exemplified in the contours
shown in Fig.~\ref{fig:ino2} for two c.m. energies and assuming ${\cal
P}_{e^-}=0.8$ and ${\cal P}_{e^+}=0.5$. At $\sqrt s=500$ GeV, two regions of the
plane are selected, but one of them can be removed by moving to lower c.m. 
energies. For SPS1a, including the uncertainties in  the mass measurements, one
obtains the 95\% CL  limited range for the mixing angles
$\cos2\phi_{L}=[0.62,0.72]$ and $\cos2\phi_{R}=[0.87,0.91]$.  In the CP
conserving MSSM, the information obtained from chargino production and decay
processes would be sufficient to determine the basic parameters entering the
$\chi^\pm$--$\chi^0$ system with a very good accuracy.  Also, we recall that the
$t$--channel $\tilde \nu$  exchange can be suppressed using polarized beams and
$m_{\tilde \nu_e}$ can be measured from the cross section. If too heavy, one can
have an indirect sensitivity on multi--TeV sneutrinos and measure their masses
\cite{SUSY-ISR,Heavy-slep} unless the $e\chi_1^+ \tilde \nu$ coupling is small
[as for a higgsino $\chi_i^\pm$]. Thus, even if they are well beyond the
kinematical reach of the ILC, sleptons can be probed up to masses of ${\cal
O}(10$ TeV) thanks to the achievable high precision.

The neutralino mixing angles can also be determined in pair and mixed
production, leading to additional determinations of the basic SUSY parameters. 
By only using the processes $\epem \to  \chi_1^0 \chi_2^0$ and $\chi_2^0
\chi_2^0,$, the constraints on $M_2, \mu$ and $\tb$ can be improved and the
parameter $M_1$ can be determined from the production vertex.  This is
particularly true in models with CP violation, in which the parameters $\mu$ and
$M_{1,2}$ have complex phases that can be determined unambiguously in a fully
model independent way by combined information from $\chi^\pm$ and $\chi^0$
production. In fact, CP violation can be checked directly by measuring CP--odd
observables in neutralino production \cite{SUSY-CPV,CPV-neutralinos}.  

We note that in the SPS1a or SPS1a$^\prime$ scenarios, and in many SUSY cases, the
heavier neutralinos and chargino are not accessible in pair production unless
the ILC c.m. energy is upgraded to 1 TeV. However, mixed pair production $\epem
\to \chi_1^0 \chi_{3,4}^0$ for instance, might be accessible at energies only 
at or slightly above $\sqrt s=500$ GeV, but the production rates are small and
the backgrounds too large. A study at $\sqrt s=750$ GeV with 1 ab$^{-1}$
luminosity shows that the $Z/W$ boson energy spectra in the decays of these
heavier ino states allow their reconstruction with mass resolutions of a few
GeV. Note also that from the determination of the SUSY parameters in lighter
$\chi_{1,2}^0,\chi_1^\pm$ production and decays, one can predict the masses of
the heavier ino states  with a few percent accuracy.

\subsection{The slepton sector}

The sfermion system is described, in addition to $\tb$ and $\mu$, by three
parameters for each sfermion species: the left-- and right--handed soft--SUSY
breaking scalar masses $M_{\tilde{f}_L}$ and $M_{\tilde{f}_R}$ and the trilinear
couplings $A_f$. Sfermion mixing turns the current eigenstates $\tilde{f}_L$ and
$\tilde{f}_R$ into the mass eigenstates $\tilde{f}_1$ and $\tilde{f}_2$, but
only in the case of the third generation that this mixing, $\propto m_f$, is
important [for the first two sfermion generations, since $m_f \to 0$,
universality can be assumed in general as will be done here]. In the case of
$\tilde \tau$s,  it is significant at large $\tb$, leading to a $\tilde \tau_1$
that is much lighter than the other sleptons.  

The production of the second and third generation sleptons in $\epem$ collisions
is mediated by $s$--channel $\gamma/Z$ exchanges in P--waves with a
characteristic rise of the excitation curve, $\sigma\propto \beta^3_{\tilde
\ell}$. The production of selectrons and electronic sneutrinos proceeds, in
addition, through $t$--channel exchanges of neutralinos or charginos. The 
channels $\epem \to \tilde e^\pm_R e^\mp_L$ are generated in S--waves with a
steep threshold excitation curve, $\sigma \propto \beta_{\tilde \ell}$.
Selectrons can also be produced in $e^-e^-$ collisions through neutralino
exchange, with steep excitation curves for $ \tilde e^-_R \tilde e^-_R$ and
$\tilde e^-_L \tilde e^-_L$ final states. Thus, different states and their
quantum numbers can be disentangled by a proper choice of the beam energy and
the polarization. Since in many SUSY scenarios the sleptons are relatively
light, their decays are rather simple and involve in general only the light
chargino and neutralinos plus leptons. In SPS1a for instance, the decays of all
sleptons directly into the LSP, $\tilde \ell \to \ell \chi_1^0$, are the 
dominant ones.  

\begin{figure}[htb]
\vspace*{-7mm}
\begin{center}
\begin{tabular}{ll}
\begin{minipage}{7.cm}
\includegraphics*[width=70mm,height=60mm]{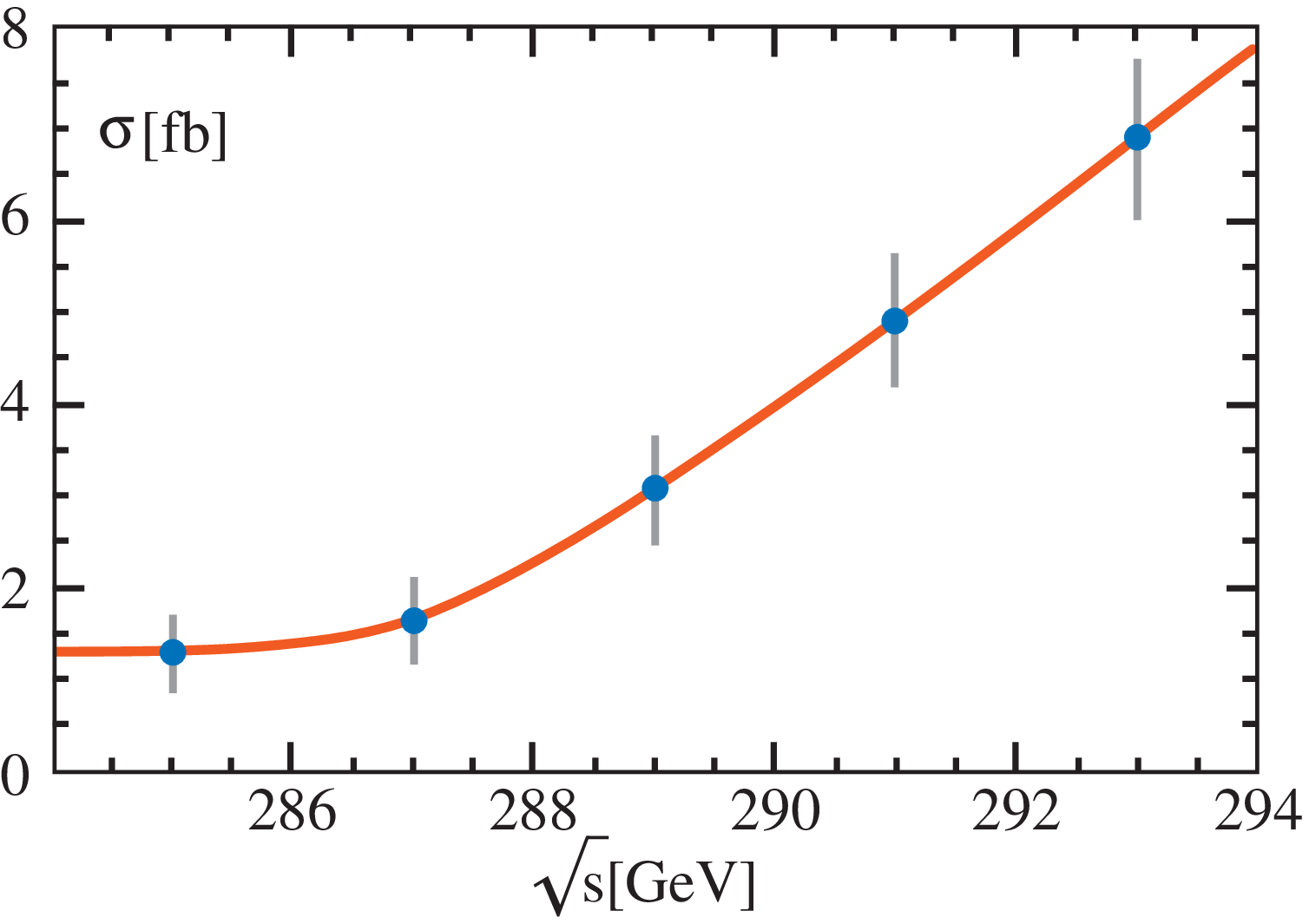}
\end{minipage} &  
\hspace*{-5mm}
\begin{minipage}{7.cm}
\vspace*{5mm}
\includegraphics*[width=65mm,height=75mm,angle=-90]{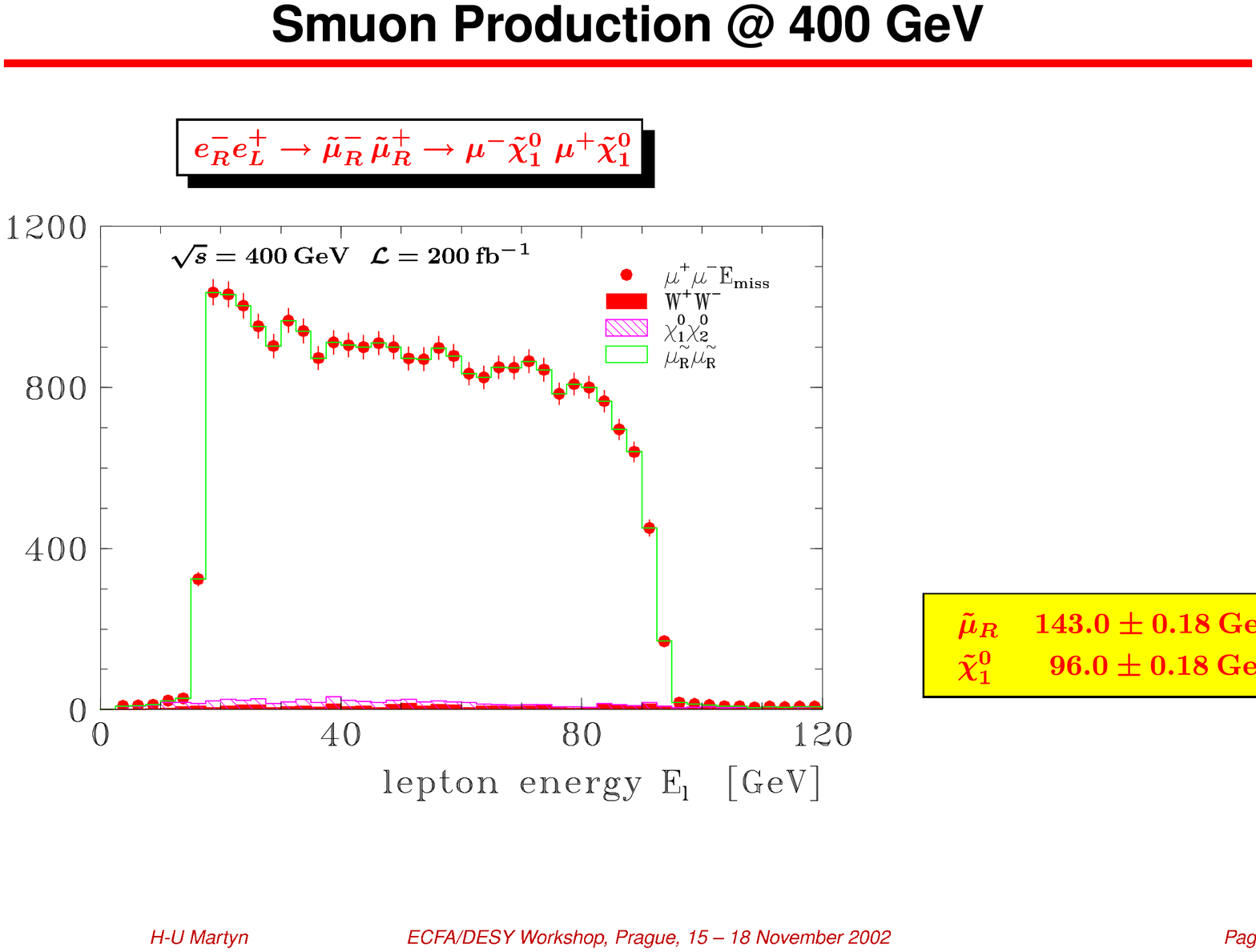} 
\end{minipage} 
\end{tabular}
\end{center}
\vspace*{-13mm}
\caption[Slepton mass measurements at the threshold and lepton energy spectra.]
{Slepton mass measurements in  SPS1a:   Cross sections at threshold
for $e^+_Le^-_R\to \tilde e^+_R \tilde e^-_R$ including background with 1
fb$^{-1}$ per point \cite{SUSY-Freitas} (left).    Lepton energy spectra in 
 $e^-_R e^+_L  \to \tilde \mu^-_R \tilde \mu^+_R \to \mu^-\chi_1^0 \mu^+
\chi_1^0$ at $\sqrt{s}\!=\!400$ GeV and  ${\cal L}\!=\!200$ fb$^{-1}$
\cite{SUSY-Martyn} (right).}
\label{fig:slep}
\vspace*{-3mm}
\end{figure}

Slepton masses can be measured in threshold scans or in the continuum.  At
threshold, $\tilde{\ell}^+_L \tilde{\ell}^-_L$ and  $\tilde{\ell}^+_R
\tilde{\ell}^-_R$ are excited in  a P--wave characterized by a slow rise of the
cross section. The experimental accuracy requires higher order corrections and
finite sfermion width effects to be included.  An example of a simulation for
the SPS1a point is shown in Fig.~\ref{fig:slep} for $\tilde \mu_R$.  Using
polarized $e^+e^-$ beams and ${\cal L}=50$ fb$^{-1}$, a highly correlated
2--parameter fit gives $\Delta m_{\tilde e_R} = 0.2$ GeV and
$\Delta\Gamma_{\tilde e_R}=0.25$ GeV; the resolution deteriorates by a factor
of $\sim2$ for $\tilde{\mu}^+_R \tilde{\mu}^-_R$ production.  For $e^-_R
e^-_R\to \tilde{e}^-_R \tilde{e}^-_R$, the gain in resolution is a  factor
$\sim 4$ with only a tenth of luminosity, compared to $\epem$ beams.

Above the threshold, slepton masses can be obtained from the endpoint energies
of leptons coming from slepton decays. In the case of two--body decays,  $\tilde
\ell^\pm  \to  \ell^\pm \chi_i^0$ and $\tilde \nu_\ell  \to  \ell^\pm 
\chi^\mp_i $, the lepton energy spectrum is flat with the minimum and maximum
energies providing an accurate determination of the masses of the primary
slepton and the secondary neutralino/chargino. A simulation of the $\mu$ energy
spectra of $\tilde{\mu}^+_R  \tilde{\mu}^-_R$ production, including
beamstrahlung, initial state radiation, selection criteria and detector
resolution, is shown in Fig.~\ref{fig:slep} for the point SPS1a
\cite{SUSY-Martyn}.  With a moderate luminosity of $ 200$ fb$^{-1}$ at
$\sqrt{s}=400$ GeV,  one obtains $m_{\tilde \mu_R}=143 \pm 0.10$ GeV and
$m_{\chi^0_1}=96 \pm 0.10$ GeV. If $m_{\chi^0_1}$ is known from
chargino/neutralino production, one can improve the slepton mass determination
by a factor of two  from reconstructed kinematically allowed slepton minima. 
Similar results are obtained in the case of selectron production in $\epem \to
\tilde{e}^-_R \tilde{e}^-_R$.  

\begin{figure}[htb]
\begin{center}
\psfig{figure=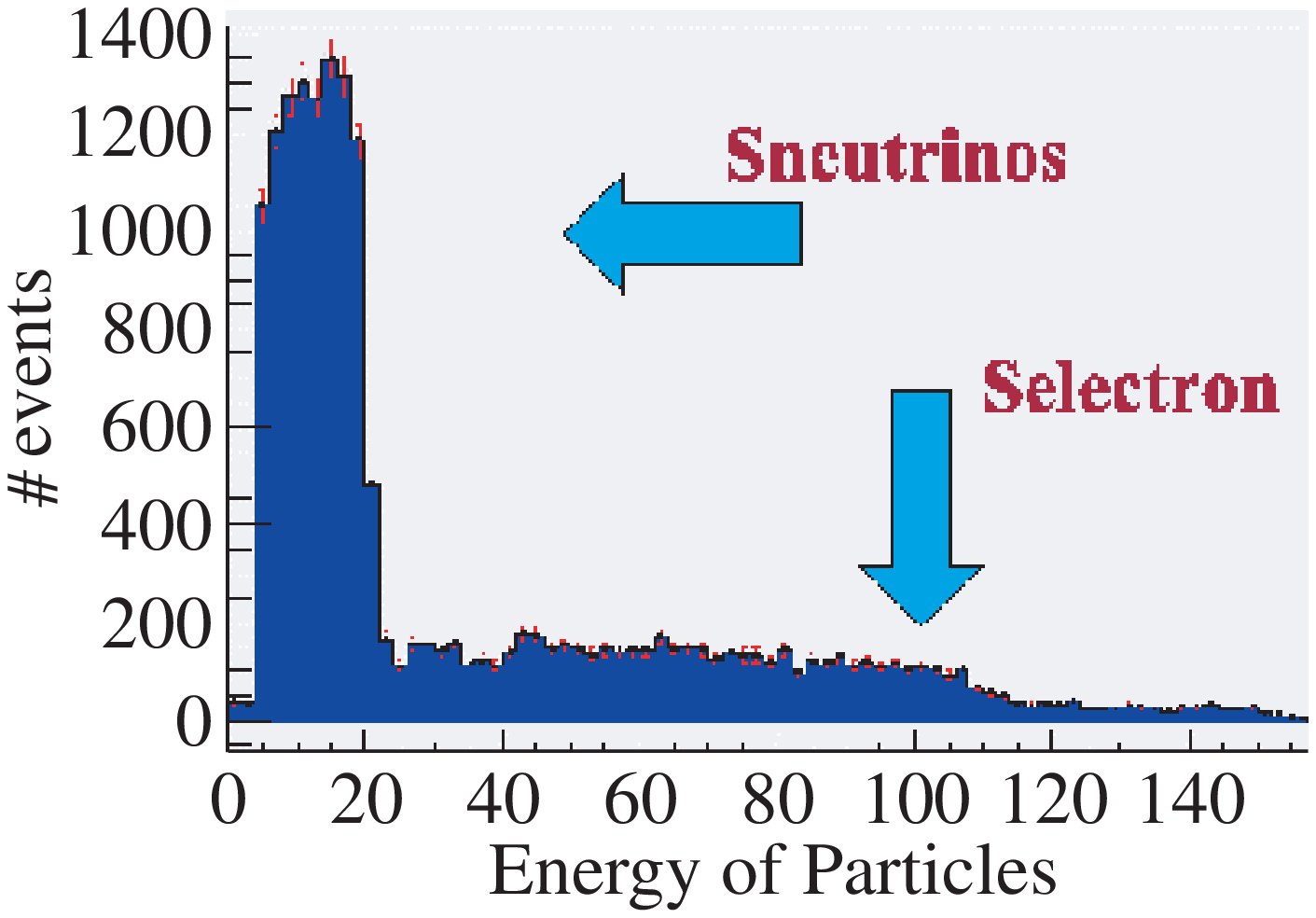, height=7cm,
clip=true}
\end{center}
\vspace*{-7mm}
\caption[Lepton  energy spectrum for sneutrino production and decay 
at the ILC]
{Lepton  energy spectrum for the sneutrino production and decay 
processes $\epem \to \tilde \nu_e \tilde \nu_e \to \nu_e \chi_1^0 e^\pm  
\chi_1^\mp \to e^\pm \mu^\mp + \slash  \hspace*{-2.5mm}E$ \cite{SUSY-Peter}.}
\label{fig:slep2}
\vspace*{-2mm}
\end{figure}

The sneutrino analysis is more involved in scenarios with light states which
decay dominantly into invisible channels, $\tilde \nu_\ell \to \nu_\ell
\chi_1^0$. The $\tilde \nu$ mass resolution could be optimized by looking at
the channel $\epem \to \tilde \nu_e\tilde \nu_e \to  \nu_e \chi_1^0 e^\pm
\chi_1^\mp$. This is exemplified in  Fig.~\ref{fig:slep2} for scenario
SPS1a, where the branching ratio for the $\tilde \nu_e \to \chi^\pm_1 e^\mp$
decay is about 10\%.  The sneutrino mass can be determined to the level $\Delta
m_{\tilde \nu}=  1.2$ GeV, which is comparable to the accuracy obtained from 
a  threshold scan.  

The $\sin^2\theta$ law for the angular distribution in the production of
sleptons (for selectrons close to threshold) is a unique signal of the
fundamental spin--zero character; the P--wave  onset of the excitation curve is
a necessary but not sufficient condition in this case \cite{SUSY-spin-chi+}.
Thus, the slepton spin determination is conceptually very simple at the ILC. 

As mentioned previously, large mixing effects are in general expected
in the stau sector, making as in SPS1a, $\tilde \tau_1$ the lightest
slepton.  The stau masses can be determined using the same methods as
described above and, for SPS1a, one obtains $\Delta m_{\tilde \tau_1}=
0.3$ GeV. Since in scenarios with $\tan\beta \gsim 10$, charginos and
neutralinos in the decay chain will dominantly lead to additional tau
leptons in the final state, it is difficult to disentangle the heavier
$\tilde \tau_2$ from the background of the lighter $\tilde \tau_1$ and
the $m_{\tilde \tau_2}$ measurement is still an open problem. Another
very difficult region is when $\tilde \tau_1$ is almost degenerate in
mass with the $\chi_1^0$ LSP, a possibility that is important as it
corresponds to the co--annihilation region in which the LSP has the
required cosmological relic density to make the DM. In this case, the
final state $\tau$ leptons are very soft and the two--photon processes
$\epem \to \tau \tau ee$ and $\epem \to ccee, bbee$ with the quarks
decaying semi-leptonically, besides $\epem \to WW \to \tau \tau \nu
\nu$, represent very large backgrounds. It has been nevertheless shown
in detailed simulations that the signal can be detected and accuracies
close to 1 GeV can be achieved on the $\tilde \tau$ mass for scenarios
where $m_{\tau_1} - m_{\chi_1^0} \gsim$ a few GeV; the uncertainty
drops by a factor of 2 if the c.m. energy is optimized.

In the case of $\tilde \tau$s, the mixing angle $\theta_{\tilde \tau}$ can be
extracted from two measurements of the cross section $\sigma(e^+e^- \to \tilde
\tau_1 \tilde \tau_1)$ with different beam polarizations
\cite{SUSY-taupol,SUSY-slep-tb}.  In the SPS1a scenario, one obtains a precision
at the percent level, $\cos 2\theta_{\tilde \tau} = -0.84 \pm 0.04$
\cite{SUSY-Martyn}.  The value of $\theta_{\tilde \tau}$ and the degree of
$\tau$ polarization in ${\tilde \tau}$ decays depend on the fundamental
 parameters $\mu$, $A_\tau$ and $\tan\beta$, which can therefore be
constrained by these measurements. In fact, the dominant decay mode
$\tilde{\tau}_1\to \chi^0_1 \tau$ can also be exploited  to determine
$\tan\beta$ if it is high enough, by using the polarization of $\tau$ leptons
which has been shown to be probed at the percent level
\cite{SUSY-taupol,SUSY-slep-tb}.  $\tau$ polarization would allow, for
instance, to discriminate between different GUT scenarios \cite{SUSY-staupol}.
Furthermore, since the trilinear $A_\tau$ coupling is enhanced by $\tb$ in the
couplings of the heavier scalar and  pseudoscalar Higgs bosons to $\tilde \tau$
states, this parameter can be measured in the Higgs decays $H,A \to \tilde
\tau_1 \tilde \tau_2$  \cite{SUSY-Hstau}. Finally, the important parameter $\tb$
can also be measured in $\tau \tau$ fusion to Higgs bosons at the $\gamma
\gamma$ option of the ILC \cite{SUSY-tautauH}.  

Note that in SUSY models which incorporate heavy right--handed neutrinos, 
spectacular flavor violating slepton decays such as $\tilde \tau_1 \to \mu
\chi^0_1$ may be observed at the ILC \cite{SUSY-LFV2}, in addition to 
lepton--number changing processes like $e^+ e^- \to \mu^\pm \tau^\mp$
\cite{SUSY-LFV1}.

\subsection{The squark sector}

\begin{figure}[!h]
{\small \phantom{} \hspace*{-1cm}
$\mathbf{cos\theta_{\tilde t}}$~~~~~ ~~~~~~~ ~~~~~~~~ ~~~~~~~~~
 ~~~~~~~~~~~~~~~~~~~~ $\mathbf{tan\beta}$}\\[1mm] ~~~~~~~~~~
\centering
\includegraphics*[width=75mm,height=50mm]{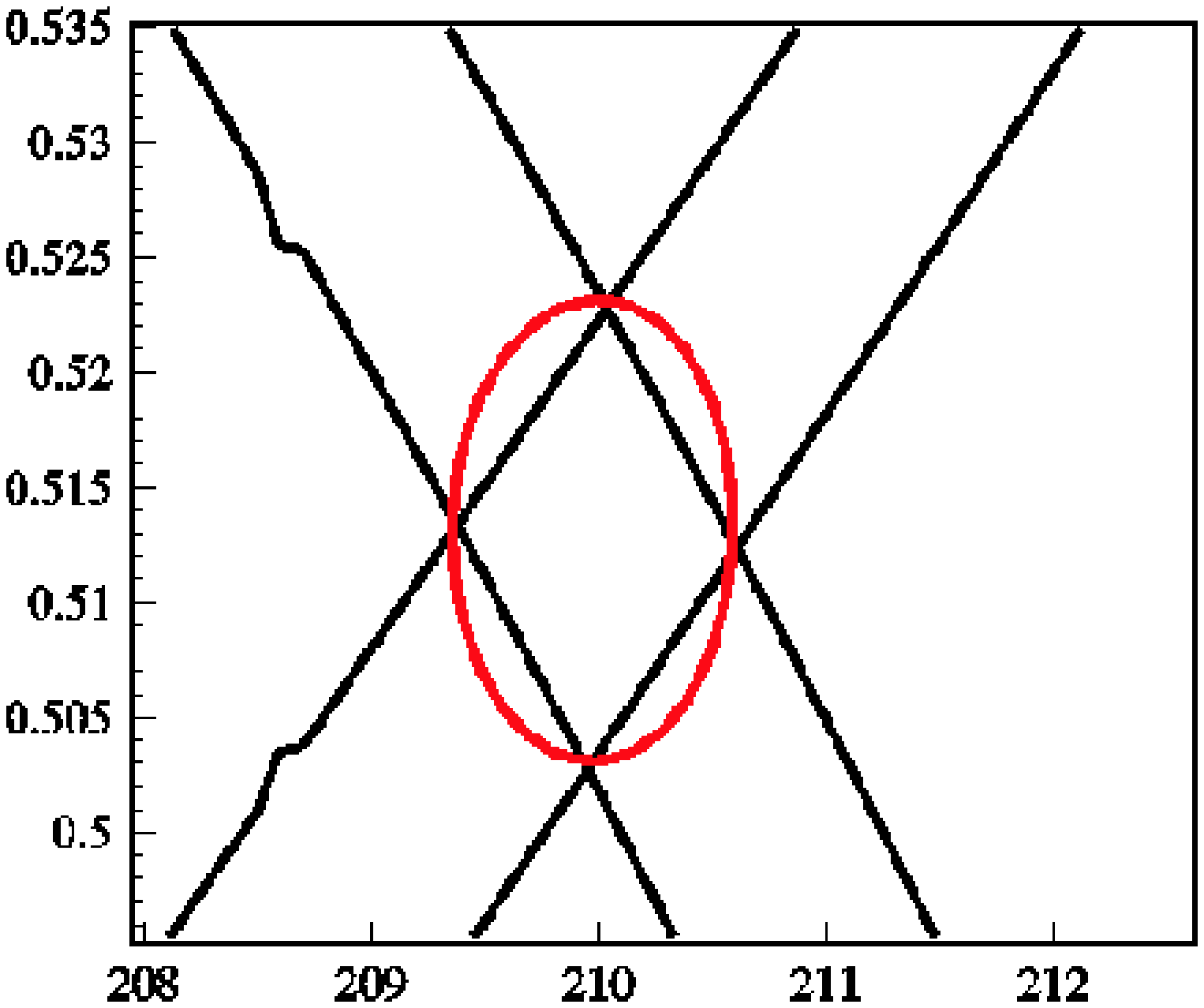} \hspace*{0.5cm}
\includegraphics*[width=62mm,height=50mm]{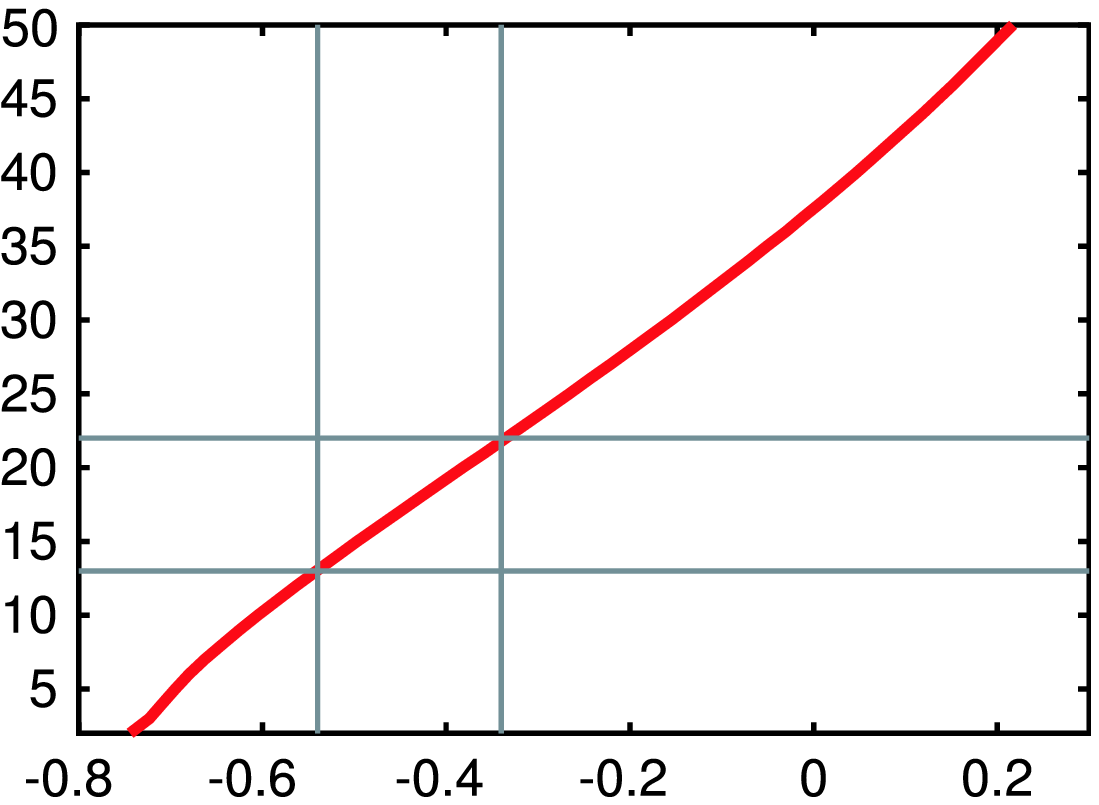}\\[-5mm] {\small
\phantom{} ~~~~~~~~ $\mathbf{m_{\tilde t}}$ {\bf [GeV]}~~~~~~~~~~~~~~ ~~~~~~~~~
~~~~~ ~~~~ ~~~~~~ $\mathbf{P_{\tilde b_1\to t \chi_1^\pm}}$} 
\vspace*{-5mm}
\caption[Determination of the mass and couplings of the top squark at the ILC]
{Left: Contours of $\sigma (e^-_R e^+_L \to \tilde{t}_1 \tilde{t}_1)$ 
and $\sigma (e^-_L e^+_R \to \tilde{t}_1 \tilde{t}_1)$ as a function of 
$m_{\tilde t_1}$ and $\cos\theta_{\tilde t_1}$ for $\sqrt{s} = 500$ GeV and 
${\cal L} = 2\cdot 500$ fb$^{-1}$ \cite{SUSY-FNS}. Right: $\tan\beta$ as a 
function of top polarization as obtained from a simulation in 
Ref.~\cite{SUSY-slep-tb}.  } 
\label{fig:sq}
\vspace*{-4mm}
\end{figure}

For the third generation squarks, $\tilde{t}$ and  $\tilde{b}$, the  mixing is
expected to be important and,  as a result of the large  top and bottom quark
Yukawa couplings, it is possible that the lightest top or bottom squarks are
much lighter than the other squarks and kinematically accessible at the ILC. 
This is for instance the case in SPS1a where $m_{\tilde t_1}=379.1$ GeV and
$m_{\tilde b_1}=491.9$ GeV in which case $\tilde t_1$, and to a lesser extent
${\tilde b_1}$, can be  produced at $\sqrt s=1$ TeV. In fact, to achieve 
electroweak baryogenesis in the MSSM (see chapter \ref{sec:cosmology}, the 
right--handed top squark must be lighter than the top quark in order that  a
strong first order transition is realized, while the other stop eigenstate is
very heavy. The $\tilde t_1$ state may escape detection at the LHC because of
the huge backgrounds, while it can easily  be observed at the ILC;
Fig.~\ref{fig:sq} \cite{SUSY-FNS}. Thus, there is a possibility that the stop
sector can be studied only at the ILC. 

The phenomenology of the $\tilde t$ and $\tilde b$ states is analogous to that
of the $\tilde \tau$ system. The masses and   mixing angles can be extracted
from production cross sections measured with polarized beams. For stop pair
production  with different beam polarizations, $\sigma (e^-_R e^+_L \to
\tilde{t}_1 \tilde{t}_1)$ and $\sigma (e^-_L e^+_R \to \tilde{t}_1 \tilde{t}_1)$
have been studied for $\tilde t_1\to b \chi^\pm_1$ and $\tilde t_1 \to
c\chi^0_1$ decay modes including full statistics SM  background. We mention here
a simulation using SIMDET in a dedicated ``light-stop'' scenario with $m_{\tilde
t_1}=210$~GeV and $m_{\chi^0_1}=121.2$~GeV \cite{SUSY-FNS} for which the decay
$\tilde t_1\to b \chi^\pm_1$ is not open and the SUSY background is thus small.
The charm tagging, helps to enhance the signal from the decay $\tilde t_1 \to
c\chi^0_1$.  The results, shown in the left panel of Fig.~\ref{fig:sq} provide
high accuracies on the $\tilde t_1$ mass $\Delta m_{\tilde t_1}\sim 0.7$ GeV and
mixing angle $\Delta\cos\theta_{\tilde t}\sim 0.01$.

Similarly to the $\tilde \tau$ case, the measurement of top quark polarization
in squark decays can provide information on $\tan\beta$.  For this purpose the
decay $\tilde{b}_1\to t \chi^\pm_1$  is far more useful than $\tilde{t}_1\to t
\chi^0_i$   since in the latter the top polarization is only weakly sensitive to
high  $\tan\beta$ values. A feasibility study of the reaction $e^+_Le^-_R \to 
\tilde   b_1 \tilde b_1 \to t \chi_1^- + \bar{t} \chi_1^+$ has been performed in
Ref.~\cite{SUSY-slep-tb} where a fit to the angular distribution with respect to
the angle between $\tilde b_1$ and a final quark in the top rest frame, allows
for a nice measurement of the polarization. One can then derive the value of
$\tb$ as illustrated in Fig.~\ref{fig:sq} where one obtains $\tan\beta\!=\!
17.5\! \pm\!4.5$ in the studied scenario with an input value of $\tb\!=\!20$. 
After fixing $\tan\beta$, measurements of the stop mass and mixing angle allows
to determine the trilinear coupling $A_{t}$ at the 10\% level.

Finally, first and second generation squarks, which will be produced copiously
and studied at LHC, might be accessible at ILC only at energies $\sqrt
s\gsim  1$ TeV. Compared to the LHC, $\tilde q$ pair production at the ILC if
kinematically possible  would allow for better mass measurements and a check of
their charge, spin and chirality numbers.

\subsection{Measurements in other scenarios/extensions} 

So far, we have only discussed the prominent features of the MSSM with gravity
mediated SUSY--breaking.  Interesting and important studies can also be
performed at the ILC in variants of the MSSM in which some underlying basic
assumptions are relaxed or in SUSY models with different breaking patterns. In
the following, we will briefly summarize some of the studies which can be made
at the ILC. 

\underline{In Gauge mediated SUSY breaking models} \cite{SUSY-GMSB}, the LSP is
the lightest gravitino $\tilde G$ which has a very small mass, leading to NLSP
decay lengths ranging from micro--meters to tens of meters. This NLSP is in
general either the lightest neutralino which decays into a gravitino and a
photon, $\chi_1^0 \to \tilde G \gamma$, and produces displaced photons not
pointing to the interaction vertex, or the $\tilde \tau_1$ with decays $\tilde
\tau_1 \to \tilde G \tau$.  The phenomenology of the other SUSY particles, and
even that of the NLSP if its lifetime is large and decays outside the detector,
is the same as in gravity mediated models but with different spectra.  Detailed
simulations \cite{Aguilar-Saavedra:2001rg} show that a signal with displaced
photons can be observed for NLSP masses close to the production kinematical
limit and that various techniques [such as tracking, pointing calorimetry and
photon counting] allow to measure the decay length over a large range and
determine the SUSY scale.  From the rest of the SUSY spectrum, a precise
determination of the GMSB parameters is possible. The scenario with $\tilde 
\tau_1$ NLSP has also been studied \cite{DM-Martyn} and it has been shown that in many
cases that the long $\tilde \tau$ lifetime allows a precise determination of
$m_{\tilde G}$. 

\underline{In Anomaly mediated SUSY breaking models} \cite{SUSY-AMSB}, the most
characteristic feature is that the LSP neutralino is wino like and is  nearly
mass degenerate with the lightest chargino $\chi_1^\pm$. As mentioned
previously, chargino $\epem \to \chi_1^+ \chi_1^-$ production will be then a
difficult process and one should rely on new search strategies
\cite{SUSY-hensel,SUSY-ISR}, depending on the $\chi_1^\pm$ lifetime and decay
modes which are related to the small mass difference $
m_{\chi_1^+}-m_{\chi_1^0}$. Signatures like ISR photons, heavy ionizing
particle, terminating tracks decaying to pions, etc.., can be used for
detection. Chargino with masses very close to the beam energy can be observed. 
Another interesting feature of AMSB models is the near mass degeneracy of 
$\tilde \ell_L$ and $\tilde \ell_R$ which can be tested precisely at the ILC.  

\underline{The MSSM with R--parity breaking} \cite{SUSY-RPV} is an interesting
scenario as it provides a nice framework to describe  \cite{SUSY-RPV-neutrinos}
the mass and the mixing patterns of the SM light neutrinos. The LSP is not
anymore stable and does not provide a DM candidate and, since astrophysical
constraints do not apply, it can be a priori any SUSY particle. Nevertheless,
the LSP is generally again the $\chi_1^0$ or the $\tilde \tau_1$ and, depending
on whether $\not \hspace*{-1.5mm}R_p$ couplings are lepton or baryon number
violating, it will decay either into leptons or jets.  For small $\not
\hspace*{-1.5mm}R_p$ couplings, as required by data in the leptonic and light
quark sectors, the production and decay characteristics of the SUSY particles
are identical to the usual MSSM, except for the LSP decays which lead to visible
particles and not missing energy.  The signatures with multi--lepton or/and
multi--jet final states have been shown to be straightforwardly observable using
the over-constrained kinematics of the final states, and easily recognizable
from the SM and usual MSSM expectations \cite{Aguilar-Saavedra:2001rg}.  For
large $\not \hspace*{-1.5mm}R_p$ couplings, interesting new signals,  such as
single production of sneutrinos $\epem \to \tilde \nu \to \ell^+
\ell^-, \nu \chi_1^0, \ell^\pm \chi_1^\pm$ might occur and extend significantly
the accessible mass reach of the ILC. Significant $\not \hspace*{-1.5mm}R_p$
couplings can be present in the third generation sfermion sector, in particular
for $\tilde t_1$, leading to an interesting phenomenology and new signatures
which can be also precisely probed at the ILC.  

\underline{The next--to minimal SSM}, is a very interesting extension of the
MSSM as it solves the problem of the $\mu$ parameter, which is a SUSY parameter
but with values of the order of the SUSY--breaking scale. By adding a singlet
superfield $S$ in the superpotential, $W \supset \lambda H_1 H_2 S -\frac{1}{3}
\kappa S^3$ \cite{SUSY-NMSSM}. The scalar component of $S$ develops a vev
$x=\langle S\rangle $ which generates an effective $\mu$--term, $\mu=\lambda
x$.  The fermionic component of the extra superfield, the singlino, will mix
with the neutral gauginos and higgsinos, leading to a $5\times 5$ neutralino
mass matrix which will depend on $M_1$, $M_2$, $\tan\beta$, $x$ and the
trilinear couplings $\lambda$ and $\kappa$. In some regions of the parameter
space, the singlino $\chi_S^0$ may be the LSP and can be searched for in
associated production with the usual neutralinos, $\epem \to \chi^0_S \chi_i^0$. 
If the singlino dominated LSP has small couplings to the other neutralinos, the
usual SUSY production processes  will lead to signatures involving displaced
vertices due to the decay of the NLSP neutralino into the singlino which would
signal the extended structure \cite{SUSY-NMSSMp}. Another possibility of
discriminating the MSSM from the NMSSM when the spectra look identical but the
neutralino--singlino mixing is substantial, would be to study the summed up
production cross sections  for the four neutralinos, $\sum \sigma (\epem \to
\chi^0_i \chi_i^0)$, if they are all kinematically accessible 
\cite{SUSY-TH-neut}. 

\underline{The CP violating MSSM} \cite{SUSY-CPV} has been already
mentioned previously. In the chargino and neutralino sectors, the phases of
$\mu, M_1$ and $M_2$ can be determined from the precise measurement of the
$\chi^0, \chi^\pm$ masses and mixing angles, even if only the light states are
accessible kinematically; the availability of beam polarization \cite{gudi}
is crucial here. In the sfermion sector, the phases of the trilinear couplings
$A_f$ and $\mu$ can be studied in the production and decays of the third
generation $\tilde t, \tilde b$ and $\tilde \tau$ states. 

Other scenarios, such as those inspired by superstring models or incorporating
right--handed sneutrinos or heavy right--handed neutrinos, have been also
discussed. 

\section{Determining the SUSY Lagrangian}

\subsection{A summary of measurements and tests at the ILC} 

Let us first summarize the results of the SPS1a sparticle mass measurements to
highlight the high precision that can be achieved at the ILC. These are
displayed in Tab.~\ref{sps1_results} from Ref.~\cite{SUSY-Martyn}, where  quoted
are the best values expected from either production in the continuum or in
threshold scans. In most cases, they are based on realistic Monte Carlo and
detector simulations with reasonable assumptions on the ILC performance. Only
for the heavy $\chi^0$, $\chi^\pm$ and $\tilde t_1$ states some plausible
estimates are made. Typical accuracies in the percent to the permille range are
expected.  

It should be pointed out once more that the ILC provides much more valuable
information than sparticle masses. Accurate values on sparticle mixing angles
and couplings can also be obtained and the spin--quantum numbers can be easily
determined. Other aspects, such as the chirality of the sleptons, the Majorana
nature of the neutralinos, the presence of CP--violation, etc.., can be directly
verified. All these precision measurements serve as a valuable input to explore
SUSY scenarios in a model independent way. For some of these studies, the
polarization of both electron and positron beams is very important \cite{gudi}.

\begin{table}[!h] 
\vspace*{-2mm}
\renewcommand{\arraystretch}{.95}
\caption[Expected accuracy on some sparticle masses at ILC in a benchmark point]
{Sparticle masses and their expected accuracies at the ILC 
in SPS1a$^\prime$ \cite{SUSY-Martyn,SUSY-SPA}.}
\label{sps1_results}
\centering
\begin{tabular}{|c|c|c|l|}
\hline
               & $m~[{\rm GeV}]$ & $\Delta m~[{\rm GeV}]$ & Comments\\ \hline
$\chi^\pm_1$ & 183.7           & 0.55      & simulation threshold scan,
                                                     100 fb$^{-1}$ \\
$\chi^\pm_2$ & 415.4           & 3         & estimate
    $\chi^\pm_1\chi^\mp_2$, spectra $\chi^\pm_2 \to Z \chi^\pm_1,\, W \chi^0_1$
               \\ 
$\chi^0_1$   &  97.7           & 0.05      & combination of all methods \\
$\chi^0_2$   & 183.9           & 1.2       & simulation threshold scan 
         $\chi^0_2\chi^0_2$,             100 fb$^{-1}$ \\
$\chi^0_3$   & 400.5           & 3--5    & spectra
       $\chi^0_3\to Z \chi^0_{1,2}$, $\chi^0_2\chi^0_3, \chi^0_3\chi^0_4$, 750 GeV, $\gsim 1~$ab$^{-1}$ \\
$\chi^0_4$   & 413.9           & 3--5    & spectra
      $\chi^0_4\to W \chi^\pm_1$, $\chi^0_2\chi^0_4, \chi^0_3\chi^0_4$,  750 GeV, $\gsim 1~$ab$^{-1}$ \\
\hline
$\tilde{e}_R$        & 125.3           & 0.05      & $e^-e^-$ threshold scan,
                                                     10 fb$^{-1}$ \\
$\tilde{e}_L$        & 189.9           & 0.18       & $e^-e^-$ threshold scan 
                                                     20 fb$^{-1}$ \\
$\tilde{\nu}_e$      & 172.5           & 1.2       & simulation 
                                                     energy spectrum, 500 GeV,
                                                     500 fb$^{-1}$ \\
$\tilde{\mu}_R$      & 125.3           & 0.2       & simulation
						     energy spectrum, 400 GeV,
                                                     200 fb$^{-1}$ \\
$\tilde{\mu}_L$      & 189.9           & 0.5       & estimate threshold scan,
                                                  100 fb$^{-1}$  \\
$\tilde{\tau}_1$     & 107.9           & 0.24       & simulation 
                                        energy spectra, 400 GeV, 200 fb$^{-1}$ \\
$\tilde{\tau}_2$     & 194.9           & 1.1       & estimate threshold scan,
                                            60 fb$^{-1}$  \\ 
\hline
$\tilde{t}_1$        & 366.5
           & 1.9         & estimate 
                    $b$-jet spectrum, $m_{\rm min}(\tilde t_1)$, 1TeV, 1000 fb$^{-1}$ \\
\hline
\end{tabular} 
\vspace*{-.3cm}
\end{table}

A very important test to be performed at the ILC is the fundamental SUSY
identity between the gauge couplings $g$ and the corresponding gaugino Yukawa
couplings $\hat{g}$ in the electroweak and strong sectors.  The cross
sections of the first generation sleptons are sensitive to the SUSY Yukawa
couplings $\hat{g}(e\tilde e \chi^0)$ and $\hat{g}(e\tilde \nu \chi^\pm)$ and,
from the measurement of $\tilde e_R$, $\tilde e_L$ and $\tilde \nu$ production
rates, one can test the SUSY identity in the electroweak sector
\cite{SUSY-Freitas,SUSY-Peter}. For $\tilde e$ production, beam polarization is
crucial for disentangling  the SU(2) and U(1) couplings: taking into account
uncertainties from the selectron mass and the neutralino parameters,  the
couplings $\hat{g}$ and $\hat{g}'$, can be extracted with a precision of 0.7\%
and 0.2\%, respectively, at a 500 GeV collider with 500 fb$^{-1}$ in the SPS1a
scenario \cite{SUSY-Peter}. Sneutrino production is only sensitive to the SU(2)
coupling $\hat{g}$, but here,  the dominantly invisible $\tilde \nu_e$ decay
limits the expected precision to 5\% \cite{SUSY-Peter}.  The equality of the
gauge and SUSY Yukawa couplings in the SU(3) sector can be checked only if the
squarks and gluinos are also relatively light, in which case the associated
production of squarks and gluinos, $\epem \to q \tilde q \tilde g$ can be used
\cite{SUSY-Peter}.  

Note that the identity between the Yukawa and the electroweak gauge couplings
can also be tested in chargino/neutralino pair production \cite{SUSY-TH-neut};
this is worth noting as this method works also in the case where the sleptons
are too heavy to be directly accessible. 

\subsection{Determination of the low energy SUSY parameters} 

Once masses and mixing angles of superparticles have been measured, the 
Lagrangian SUSY breaking parameters can be then determined. We briefly 
summarize below the procedure, ignoring higher order effects to simplify the 
picture. 

\, From chargino--neutralino measurements, one obtains $M_{1,2},\mu$ and $\tb$
\cite{SUSY-TH-neut,SUSY-TH-char}: 
\begin{eqnarray}
&M_1=[\Sigma_i m_{\chi_i^0}^2 - M_2^2 -\mu^2 -2 M_Z^2]^{1/2}\, , \
&M_2=M_W[\Sigma - \Delta[\cos2\phi_R+\cos2\phi_L]]^{1/2}\nonumber \\
&\left|\mu\right|=M_W[\Sigma + \Delta[\cos2\phi_R+\cos2\phi_L]]^{1/2}
\, , \ 
&\tan\beta=\left[ (1+\Delta')/ (1-\Delta')\right]^{1/2} \nonumber
\end{eqnarray}
with $\Delta\!=\!(m^2_{\tilde{\chi}^\pm_2}-m^2_{\tilde{\chi}^\pm_1})/4M^2_W$,
$\Delta'\!=\!\Delta (\cos 2\phi_R-\cos 2\phi_L)$
and $\Sigma\!=\!(m^2_{\tilde{\chi}^\pm_2}+m^2_{\tilde{\chi}^\pm_1})/2M^2_W -1$.
 
It has been demonstrated in detail \cite{SUSY-TH-neut} that using the
chargino/neutralino sector, the four parameters  can be determined from the
measurement of the ino masses and mixing angle even if only the light states are
accessible kinematically.

The sfermion mass parameters and trilinear couplings are obtained through
\begin{eqnarray}
&&m^2_{\tilde{f}_{L,R}}={ M^2_{\tilde{f}_{L,R}}} +M^2_Z\cos2\beta\,(I^3_{L,R}
-Q_f \sin^2\theta_W) +m_f^2\nonumber \\  
&& A_f -\mu (\tb)^{-2I_3^f}= (m^2_{\tilde f_1}-m^2_{\tilde f_2})/(2m_f) 
\cdot \sin2\theta_{
\tilde f} \nonumber
\end{eqnarray}

Parameter determination from the Higgs sector is more involved as one needs to
include the large radiative corrections that are present.  In any case, the
expected precise measurement of the lightest $h$ boson mass at the ILC, $\Delta
M_h \sim 50$ MeV, allows to severely constrain and with some assumptions to
determine some parameters in the stop sector, such as the trilinear coupling
$A_t$ and  the heavier stop mass $m_{\tilde t_2}$ (which are difficult to measure
at the LHC), if they cannot be accessed directly at ILC
\cite{Heinemeyer:2003ud}. 

In view of the high accuracy that is achievable at the ILC an even more involved
approach is required and the radiative corrections to the previous relations 
need to be implemented. This leads to a highly non--linear system of relations
which has to be solved numerically; several codes which do this job
\cite{SUSY-Sfitter,SUSY-fitino}  are available.   In Tab.~\ref{tab:mssm}, we
display values of SUSY parameters that can be derived for the general MSSM in
SPS1a \cite{SUSY-Sfitter} using mass measurements at the ILC given previously
and the LHC \cite{SUSY-Desch} after a global fit.  As expected, a very high
precision is achieved in the gaugino and slepton sectors, while the gluino and
squark (except for $\tilde t_1$) sectors are the territory of LHC.   However,
the precision measurements at the ILC also allow for mass  predictions for
heavier sparticles. Providing such mass predictions lead to an  increase in
statistical sensitivity for observing these heavier particles  in the decay
chains at the LHC. Verifying subsequently the predicted  particle masses at the
LHC leads to a powerful test of the underlying model. On the other hand, fitting
this information back to the ILC analyses enhances  the accuracy of the
parameter determination \cite{SUSY-Desch}.

\begin{table}[htb]
\vspace*{-2mm}
\caption[Determination of the low energy MSSM parameters at the ILC and LHC.] 
{Results for the  MSSM parameter determination in 
SPS1a \cite{SUSY-Sfitter} and SPS1a$^\prime$ \cite{SUSY-SPA} using the mass measurements
at the ILC and the LHC \cite{SUSY-Desch} after a global fit; the central values
are  approximately reproduced. Not that the two analyses use different sets
of measurements, assume slightly different accuracies and treat differently 
the theoretical errors; this explains the slight discrepancies in the outputs.}
\label{tab:mssm}
\begin{center} \begin{small}
\begin{tabular}{|l|rrrr|rr|}
\hline
       & $\Delta$LHC & $\Delta$ILC & $\Delta$LHC+ILC    & SPS1a & 
       $\Delta$LHC+ILC & SPS1a$^\prime$ \\
\hline
$\tan\beta$          &     $\pm$9.1   &   $\pm$0.3  &  $\pm$0.2   &       10 
& $\pm 0.3$ & 10  \\
$\mu$                &     $\pm$7.3   &   $\pm$2.3  &  $\pm$1.0  &    344.3 
& $\pm 1.1$ & 396 \\
$M_A$                &     fixed 500  &   $\pm$0.9  &  $\pm$0.8   &    399.1 
& $\pm 0.8$ & 372\\
$A_t$                &     $\pm$91    &   $\pm$2.7  &  $\pm$3.3   &   $-504.9$ 
& $\pm 24.6$ & $-565.1$ \\
$M_1$                &     $\pm$5.3   &   $\pm$0.1  &  $\pm$0.1   &    102.2 
& $\pm$0.1 & 103.3 \\
$M_2$                &     $\pm$7.3   &   $\pm$0.7  &  $\pm$0.2   &    191.8 
& $\pm$0.1 & 193.2 \\
$M_3$                &     $\pm$15    &   fixed 500 &  $\pm$11    &    589.4 
& $\pm$7.8 & 571.7 \\
$M_{\tilde{\tau}_L}$ &     fixed 500  &   $\pm$1.2  &  $\pm$1.1   &    197.8 
& $\pm$1.2 & 179.3 \\
$M_{\tilde{e}_L}$    &     $\pm$5.1   &   $\pm$0.2  &  $\pm$0.2   &    198.7 
& $\pm$0.18 & 181.0 \\
$M_{\tilde{e}_R}$    &     $\pm$5.0   &   $\pm$0.05 &  $\pm$0.05  &    138.2 
& $\pm$0.2 & 115.7 \\
$M_{\tilde{Q}3_L}$   &     $\pm$110   &   $\pm$4.4  &  $\pm$39    &    501.3 
& $\pm$4.9 & 471.4 \\
$M_{\tilde{Q}1_L}$   &     $\pm$13    &   fixed 500 &  $\pm$6.5   &    553.7 
& $\pm$5.2 & 525.8 \\
$M_{\tilde{d}_R}$    &     $\pm$20    &   fixed 500 &  $\pm$15    &    529.3 
& $\pm$17.3 & 505.7 \\
\hline
\end{tabular}
\end{small} \end{center} 
\vspace*{-3mm}
\end{table}

\subsection{Reconstructing the fundamental SUSY parameters}

Although low energy SUSY is characterized by energy scales of ${\cal O}(1$ TeV),
the roots for all the phenomena we will  observe experimentally in this range 
may go  to energies near the GUT or Planck scales.  Fortunately, SUSY provides
us with a stable bridge  between these two vastly different energy regions: RGEs
by which parameters from low to high scales are evolved based on nothing but
experimentally measured quantities. This procedure, which has very successfully
been pursued for the three gauge couplings, can be expanded to the soft--SUSY
breaking parameters: gaugino and scalar masses and trilinear couplings. 
This bottom--up approach makes use of the low-energy measurements to the maximum
extent possible and allows to reconstruct the fundamental theory at the high
scale in a transparent way.

In this approach, the combination of measurements performed at both the LHC and
the ILC will be crucial. As a matter of fact, most of the strongly interacting
particles are too heavy and will not be accessible at the ILC, while they will
be copiously produced and their masses measured at the LHC. In turn, the
precision of the LHC measurements alone will not be sufficient for a
comprehensive and high--precision picture of SUSY at the weak scale; in fact,
some of the low energy SUSY--breaking parameters cannot be constrained at all. 
Thus, only the LHC--ILC tandem can provide us with such a picture and allows the
reconstruction of the fundamental SUSY theory at the high scale.  

\begin{figure}[h!]
\vspace*{-3mm}
\hspace*{1cm}
\begin{minipage}{0.42\linewidth}
\begin{center}
{\small $\mathbf{1/M_i\,[GeV^{-1}]}$}\\
\includegraphics*[width=\linewidth]{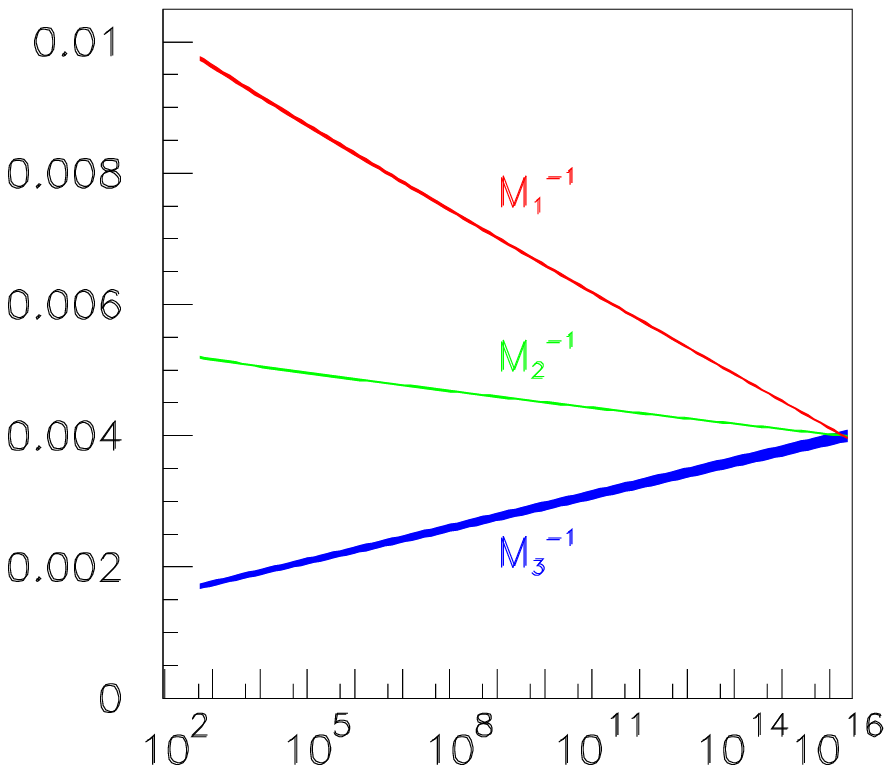} \\
{\small $\mathbf{Q~[GeV]}$}
\end{center}
\end{minipage}
\begin{minipage}{0.42\linewidth}
\begin{center}
{\small $\mathbf{M^2_{\tilde j}\,[10^3\,GeV^2]}$}\\
\includegraphics*[width=\linewidth]{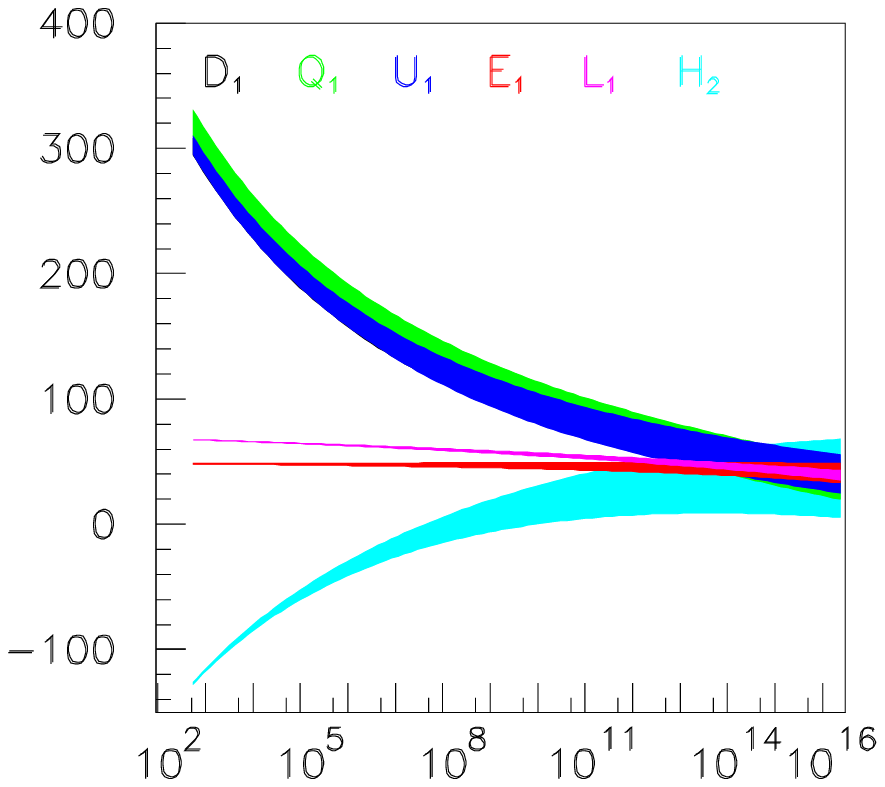}\\
{\small $\mathbf{Q~[GeV]}$}
\end{center}
\end{minipage}
\vspace*{-7mm}
\caption[Evolution from low to high scales of SUSY mass parameters
in the cMSSM]
{Evolution from low to high scales of gaugino and scalar mass 
parameters in the cMSSM point SPS1a$^\prime$; the widths of the bands indicate 
the 1$\sigma$ CL \cite{SUSY-bpz-new}. }
\label{fig:GUT}
\vspace*{-3mm}
\end{figure}

This discussion will be again illustrated using a cMSSM scenario. Adding the
measurements of the masses of the heavy states [the colored $\tilde{q}_L,
\tilde{q}_R, \tilde{b}_1$ and $\tilde{g}$ and the heavy electroweak
$\chi_{3,4}^0, \chi_2^\pm$ states] which can be performed at the LHC at the
percent level provided a very high luminosity is collected, and the ILC measurements
discussed previously, one can determine to a high precision the soft
SUSY--breaking gaugino mass parameters $M_{1,2,3}$ and the sfermion mass
parameters $m_{ \tilde{f}_{L,R} }$. One can then evolve these parameters using
standard RGEs up to the GUT scale, the value of which is derived from the
measurement of the gauge coupling constants at the Giga--Z option of the ILC. In
SPS1a$^\prime$, one obtains (ignoring threshold effects)  $M_{\rm GUT}= (2.47 \pm 0.02)
\cdot 10^{16}$ GeV, which leads to a common value of $\alpha_{\rm GUT}^{-1}=
24.17 \pm 0.06$. This is shown in Fig.~\ref{fig:GUT}, where the thickness of the
curves reflect the $1\sigma$ errors. 

Note that while the parameters are determined accurately in  the gaugino and
slepton sectors, the errors are larger for squarks. Nevertheless, one can see
that the two sets unify nicely, providing a strong confidence that we are indeed
in a cMSSM--type scenario.  

One can then derive the basic parameters of the model at the scale $M_{\rm
GUT}$. A global fit of all the SUSY parameters obtained from measurements at the
LHC and the ILC as given in Tab.~\ref{tab:mssm}, can be used to determine the
GUT values of the common gaugino and scalar masses $m_0$ and $m_{1/2}$, the
universal trilinear coupling $A_0$ as well as the value of $\tb$.  The result of
a fit performed in Ref.~\cite{SUSY-Sfitter} for the SPS1a scenario is shown in
Tab.~\ref{tab:msugra_fit}, with the sign of $\mu$ fixed to its true value, i.e. 
$\mu>0$; for further analyses, see e.g. Ref.~\cite{SUSY-fitino}.  At the LHC,
these fundamental parameters can be determined at the percent level but the ILC
improves the determination by an order of magnitude; a very accurate picture is
achieved  when the LHC and ILC data are combined. 

\begin{table}[!h]
\vspace*{-2mm}
\caption[Determination of the fundamental SUSY parameters at the LHC and ILC.]
{Summary of the cMSSM fit in SPS1a (with $\mu>0$ fixed) and SPS1a$^\prime$
based on the parameter values of Tab.~\ref{tab:mssm}
at the LHC, ILC and their combination. The same warnings on the differences
between the two analyses as in the caption of Table \ref{tab:mssm} hold also 
in this context.}
\label{tab:msugra_fit}
\begin{center} \begin{small}
\begin{tabular}{|l|rrrr|rr|}
\hline
& SPS1a & LHC & ILC & LHC+ILC & SPS1a$^\prime$ & $\Delta_{\rm LHC+ILC}$ \\ \hline
$m_0$   & 100 & $100.03 \pm 4.0$ & $100.03 \pm 0.09$ & $100.04 \pm 0.08$ 
& 70 & 0.2 \\
$m_{1/2}$ & 250 & $249.95 \pm 1.8$  & $250.02 \pm 0.13$ & $250.01 \pm 0.11$
& 250 & 0.2 \\
$\tan\beta$ &  10 & $9.87 \pm 1.3$  & $9.98 \pm 0.14$ &   $9.98 \pm 0.14$ 
& 10 & 0.3 \\
$A_0$       &$-100$ & $-99.29 \pm 31.8$ & $-98.26 \pm 4.43$ &$ -98.25 \pm 4.13$ 
& $-300$ & 13\\
\hline
\end{tabular}
\vspace*{-7mm}
\end{small} \end{center} 
\end{table}

\subsection{Analyses in other GUT scenarios}

The case of the cMSSM discussed previously demonstrates that high-precision
measurements allow us to reconstruct physical scenarios near the Planck scale. 
This can be done in many other GUT scenarios and the example of string effective
theories is briefly discussed below. another example, left--right symmetric
models which  incorporate the seesaw mechanism to generate the small neutrino
masses will be discussed in chapter \ref{sec:cosmology}. 

Heterotic string theories give rise to a set of 4-dimensional dilaton $S$ and
moduli $T$ superfields after compactification.  The vacuum expectation values of
$S$ and $T$, generated by non--perturbative effects, determine the soft
supersymmetry breaking parameters.  The properties of the  theories are quite
different for dilaton and moduli dominated scenarios, quantified by the mixing
angle $\theta$. This angle $\theta$ characterizes the $\tilde S$ and $\tilde T$
components of the wave function of the Goldstino, which is associated with the
breaking of supersymmetry.  The mass scale is set by the second parameter of the
theory, the gravitino mass $m_{3/2}$.

In leading order, the masses \cite{SUSY-stringsT} are given by, 
$M_i\!\propto\!- g_i^2 m_{3/2} \langle S \rangle {\sqrt{3} \sin \theta}$ and 
$M_{\tilde{j}}^2\!  \propto\! m^2_{3/2} \left(  1 + n_j \cos^2 \theta \right)$
for the gaugino and scalar sectors, respectively.  A dilaton dominated scenario,
$\sin \theta \to 1$, leads to universal boundary conditions of the soft--SUSY
breaking parameters while in moduli dominated scenarios, $\cos\theta \to 1$, the
gaugino masses are universal but not the scalar masses. The breaking is
characterized by integer modular weights $n_j$ which quantify the couplings
between  matter and moduli fields. Within one generation, significant
differences between left and  right sfermions and between sleptons and squarks
can occur.  

\begin{figure}[h]
\vspace*{-9mm}
\begin{center}
\begin{minipage}{7cm}
\hspace*{-1.2cm}
\includegraphics[width=7.5cm,angle=270]{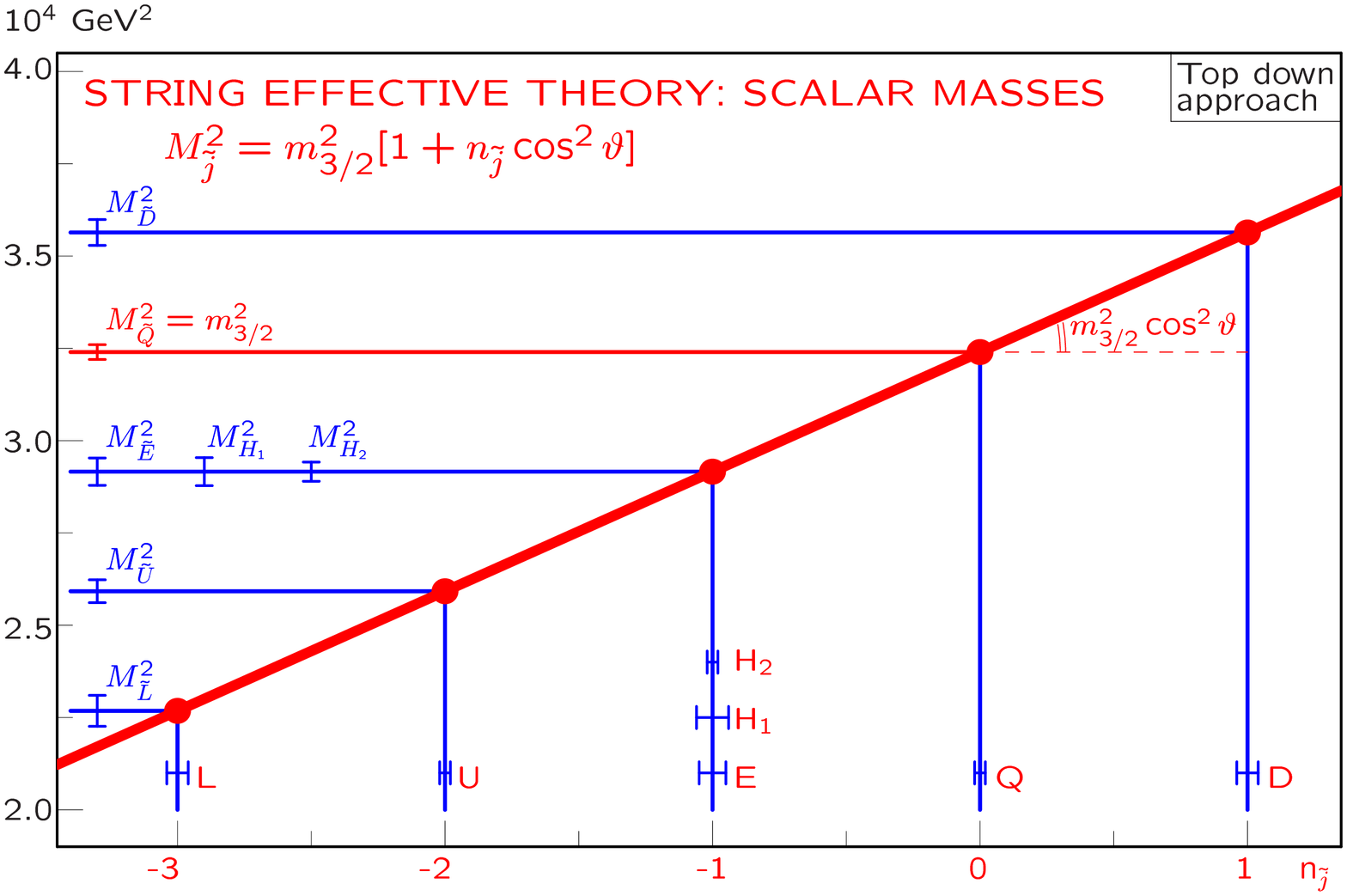}
\end{minipage}
\begin{minipage}{7cm}
\hspace*{2.cm}
\includegraphics[width=5.5cm]{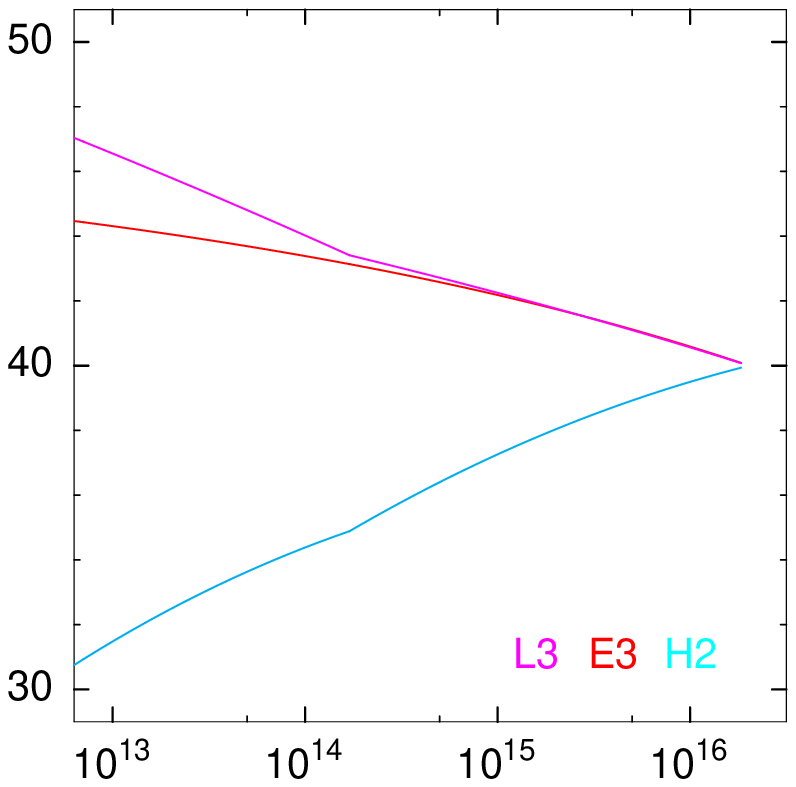}
\end{minipage}
\vspace{-12mm}
\caption[The evolution of SUSY parameters in effective string and left--right models]
{Left: the linear relation between integer modular weights and scalar 
mass parameters in string effective theories \cite{SUSY-bpz-new}.
Right: impact of the heavy right--handed neutrino mass on the evolution of 
the scalar mass parameters in left--right symmetric theories \cite{SUSY-FPZ}.}
\label{fig:stringLR}
\end{center}
\vspace*{-4mm}
\end{figure}

The results \cite{SUSY-bpz-new} for the analysis of a mixed dilaton/moduli
superstring scenario with dominating dilaton component, $\sin^2 \theta \!=\!
0.9$, and with different couplings of the moduli field to the (L,R) sleptons,
the (L,R) squarks and to the Higgs fields corresponding to the O--I
representation $n_{L_i} \!=\! -3$, $n_{E_i} \!=\! -1$, $n_{H_1} \!=\!n_{H_2}
\!=\!-1$, $n_{Q_i} \!=\! 0$, $n_{D_i} \!=\! 1$ and $n_{U_i} \!=\!  -2$, are
presented in Fig.~\ref{fig:stringLR}.  The gravitino mass is set to 180~GeV in
this analysis. Given this set of superstring induced parameters, the evolution
of the gaugino and scalar mass parameters can be exploited to determine the
modular weights $n$. Fig.~\ref{fig:stringLR} demonstrates how stringently this
theory can be tested by analyzing the integer character of the entire set of
weights. 

Thus, high-precision measurements at the ILC may provide access to crucial  low
and high--scale parameters which allow to discriminate between various theories 
beyond the SM. 

Another example of model parameterization at the very high scale is provided  by
left--right symmetric extensions of the SM. The complex structure observed in
the neutrino sector requires the extension of the MSSM  by a superfield
including the right--handed neutrino field and its scalar partner.  If the small
neutrino masses are generated by the seesaw mechanism \cite{ SUSY-seesaw}, a
similar type of spectrum is induced in the scalar sneutrino sector, splitting
into light TeV scale and very heavy masses.  The intermediate seesaw scales will
affect the evolution of the soft mass terms which break the supersymmetry at the
high (GUT) scale, particularly in the third generation with large Yukawa
couplings.  

If sneutrinos are lighter than charginos and the second lightest neutralino, as
encoded in SPS1a$^\prime$, they decay only to invisible $\nu \tilde{\chi}_1^0$ final
states, but sneutrino masses can be measured in chargino decays to sneutrinos
and leptons.  These decays develop sharp edges at the endpoints of the lepton
energy spectrum for charginos produced in $e^+e^-$ annihilation.  Sneutrinos of
all three generations can be explored this way \cite{SUSY-FPZ}. As seen before, 
the errors for the first and second generation sneutrinos are expected at the
level of 400 MeV, doubling for the more involved analysis of the third
generation.  

This will provide us with the opportunity to measure, indirectly, the
intermediate seesaw scale of the third generation \cite{SUSY-FPZ}.  This can be
illustrated in an SO(10) model  in which the Yukawa couplings in the neutrino
sector are proportional to the up--type quark mass matrix. The masses of the
right--handed Majorana neutrinos are hierarchical,  $\propto m^2_{\rm up}$, and
the mass of the heaviest neutrino is given by $M_{R_3} \sim m^2_t / m_{\nu_3}$
which, for $m_{\nu_3} \sim 5 \times 10^{-2}$ eV, amounts to $\sim 6 \times
10^{14}$ GeV, i.e., a value close to the GUT scale.

Since the $\nu_{ R}$ is unfrozen only beyond $Q=M_{\nu_{R}}$ the  impact of
the left--right  extension will be visible in the evolution of the scalar  mass
parameters only at very high scales.  The effect of $\nu_{ R}$ can  be
manifest only in the third generation where the Yukawa coupling is large enough;
the evolution in the first two generations can thus be used to calibrate the
assumption of universality for the scalar mass parameters at the unification
scale.  In Fig.~\ref{fig:stringLR} the evolution of the scalar mass parameters in
the third generation and the Higgs mass parameter are displayed. The lines
include the effects of the right--handed neutrino which induce the kinks. Only
the picture including $\nu_{R}$, $\tilde{\nu}_{ R}$ is compatible with
the assumption of unification.

The kinks in the evolution of $M^2_{{\tilde L}_3}$ shift the physical masses
[squared] of the ${\tilde{\tau}_{L}}$ and ${\tilde{\nu}_{\tau { L}}}$
particles of the third generation by the amount $\Delta_{\nu}[M_R]$ compared
with the slepton masses of the first two generations. The measurement of
$\Delta_{\nu}[M_{{R}_3}] \propto  m_{\nu_3} M_{R_3} \log(M^2_{
GUT}/M^2_{R_3})$ can be exploited to  determine the neutrino seesaw scale of the
third generation, $M_{R_3} = 3.7$--$6.9 \times 10^{14}$ GeV \cite{SUSY-FPZ},  in
the LR extended SPS1a$^\prime$ scenario with an initial value of $6 \times 10^{14}$
GeV. 

Thus, this analysis provides us with a unique possibility of indirectly 
verifying the seesaw mechanism and estimate of the high-scale $\nu_{R}$ seesaw
mass parameter $M_{R_3}$. This would have an impact in explaining the baryon
asymmetry of the universe if it is triggered by leptogenesis as will be discussed
in chapter \ref{sec:cosmology}. 

%
\chapter{Alternative scenarios}
\label{sec:alternatives}

\section{General motivation and scenarios}

Besides supersymmetric models, there are many proposals for physics scenarios
beyond the Standard Model. These alternative scenarios involve new dynamics on
the electroweak symmetry breaking and/or new concepts on space--time and their
main motivation is, in most cases,  to provide a solution to the naturalness
problem. Since this problem is connected with the stability of the electroweak
symmetry breaking scale, and the new ingredients are closely related to the
physics of the Higgs sector, its solution necessarily involves new particles
and/or new interactions at the Terascale. Furthermore, these models need to
address the question of the dark matter  which calls for a new stable particle
with a mass near the EWSB scale. Among the plethora of scenarios which have been
proposed, some examples are as follows: 

\underline{Models with large extra dimensions} \cite{LED}:  If there is an extra
dimensional space where only gravitons can propagate, the weakness of the
gravitational interaction can be explained. In this case, the  four--dimensional
Planck mass is a fictitious mass scale, and the fundamental gravity mass scale
in the higher dimension can be close to the TeV scale. A characteristic collider
signal is Kaluza--Klein (KK) graviton emission  where topologies with missing
energy are expected at the LHC and ILC. KK graviton exchange in fermion  pair
production will play an important role  to confirm the gravitational nature of
the new particles.

\underline{Warped extra-dimension models} \cite{WED}:  In the setup proposed by
Randall and Sundrum (RS), two three--dimensional branes are placed at different
points in the fifth dimensional direction, and the space--time between two
branes is part of a five--dimensional anti--de Sitter space. In this case, the
mass scale on the SM brane is exponentially suppressed compared to that on the
Planck brane. The weakness of gravitation is explained by the  suppression of
the graviton wave function at the SM brane. The KK modes of the graviton,
however, can couple strongly to the SM particles, and these may be produced as
spin--two resonances at the LHC and ILC. Their effects may also appear
indirectly in SM particle production processes. Note that five--dimensional
RS models are dual to strongly coupled four--dimensional models. 

\underline{Universal extra dimension (UED) models} \cite{Appelquist:2000nn}: In
these models, all SM particles are assumed to propagate in a flat 
extra--dimensional space. With a suitable orbifold compactification, one can
construct a phenomenologically viable model. These models look like a bosonic
supersymmetric theory since the first KK modes play the role of superpartners in
SUSY models but with the wrong spins. One can introduce a KK parity which makes
the lightest first KK particle absolutely stable and a potential dark matter
candidate.

\underline{Strong interaction models}: Within the SM and its supersymmetric
extensions, the Higgs field is introduced as a fundamental degree of freedom. 
Dynamical electroweak symmetry breaking is rooted in new strong interactions,
not necessarily involving a Higgs boson \cite{Strong-review}.  If global
symmetries of these interactions are broken spontaneously, a set of Goldstone
bosons will be generated, such as pions after breaking chiral symmetries in QCD.
By absorbing these Goldstones, longitudinal degrees of freedom and masses are
generated for gauge bosons. Several scenarios have been developed along this
path quite early \cite{H-technicolor,Technicolor} as an alternative to the
standard Higgs mechanism and more recently \cite{LHM}  in a variant responding
to the success of the light Higgs picture in accounting for the high--precision
electroweak data.

\underline{Little Higgs models} \cite{LHM}: These are models with a composite
Higgs boson but, unlike traditional Technicolor models \cite{Technicolor}, the
dynamical scale is around 10 TeV  and the physical Higgs boson is considered to
be a part of composite field. The quadratic divergence of the Higgs boson mass
renormalization is canceled at the one--loop level by extra gauge bosons and top
partners with a carefully chosen global and gauge symmetry structure. An
interesting class of little Higgs models are those with T parity \cite{DM-LTP1}
in which the new particles can be  much lighter than 1 TeV without conflict with
the precision electroweak data. In particular, the lightest T-odd particle, a
heavy photon, can be even lighter than  a few hundred GeV.

There is a variety of possibilities in each of the above scenarios. In models
with extra dimensions, phenomenological implications depend on which particles
are allowed to propagate in the extra dimensions. The Higgsless model proposed 
in Ref.~\cite{Hless} is one type of a five--dimensional model. There are also
proposals where the idea of extra space dimensions is combined with  low energy
supersymmetry. Some models in warped extra  dimensions can be considered to be
the dual description of strongly  coupled conformal field theories \cite{SCFT}
and composite Higgs scenarios have been proposed based on this duality
\cite{CompositeH}.  

The above alternative models introduce new particles and interactions at the TeV
scale and new signals are expected at the LHC experiment. If some signals are
indeed observed, the nature of the new physics could be determined by various
precise measurements at the ILC. In this respect, indirect searches for new
physics effects in SM and Higgs processes are also important at the ILC. In the
following, typical examples of ILC studies are presented.

\section{Extra dimensional models}

\subsection{Large extra dimensions}

In the models with large extra dimensions, the effective four--dimensional
Planck mass $M_{P}$ is related to the fundamental gravity mass scale $M_D$ in
the $4+\delta$ dimensional space--time by $M_{P}^2=V_{\delta}M _D^{2+\delta}$
where $V_{\delta}$ is the volume of the extra--dimensional space. For example,
taking $M_D= 1$ TeV, the size of the extra dimension is 0.1 mm to 1 fm for
$\delta = 2$ to $6$. The KK modes of the graviton have, therefore, an almost
continuous spectrum. 

At the ILC, the observation of a single photon with missing energy due to the 
emission of a KK graviton in the reaction $e^+e^- \to G_{KK} \gamma$ is a robust
signal of the model. The sensitivity to the scale $M_D$ in this channel is shown
in Table~\ref{tab-xdim} for polarized and unpolarized $e^\pm$ beams. Beam
polarization is very effective in this case as the main background process,
$e^+e^- \to \nu \bar{\nu} \gamma$, can be suppressed significantly. The search
limit for the scale $M_D$ is similar to that obtained in gluon and KK graviton
emission  at the LHC. Note that there are severe cosmological and astrophysical
\cite{LED-cosmo1} bounds on the mass $M_D$ in this scenario; a recent analysis
\cite{LED-cosmo2} of astrophysical data sets a lower limit of several hundred
TeV in the case of two extra dimensions. The limit is weaker for a larger number
of extra dimensions and the constraints are not strong for $\delta \geq 4$. 

\begin{table}[hbt]
\vspace*{-2mm}
\caption[ILC sensitivity to the effective gravity scale for large extra dimensions]
{The sensitivity at the 95\% CL in the mass scale $M_D$ (in TeV) for 
direct graviton production in the polarized and unpolarized $e^+e^- \to
\gamma \mathrm{G_{KK}}$ process for various $\delta$ values assuming a 0.3\%
normalization error \cite{Aguilar-Saavedra:2001rg}.}
\label{tab-xdim}
\centering
 \renewcommand{\arraystretch}{1.3}
\begin{tabular}{|l|c|c|c|c|c|}
\hline
$\delta$                               &  3  &  4  &  5  &  6  \\ \hline
$M_D ( {\cal P}_{e^-}={\cal P}_{e^+}=0)$  & 4.4 & 3.5 & 2.9 & 2.5 \\
$M_D ({\cal P}_{e^-}=0.8)$             & 5.8 & 4.4 & 3.5 & 2.9 \\
$M_D ({\cal P}_{e^-}=0.8,{\cal  P}_{e^+}=0.6)$ & 6.9 & 5.1 & 4.0 & 3.3\\ \hline
\end{tabular}
\vspace*{-2mm}
\end{table}

\begin{figure}[h!] 
\vspace*{-5mm}
\centering
    \includegraphics[width=7.5cm,height=6cm,angle=0]{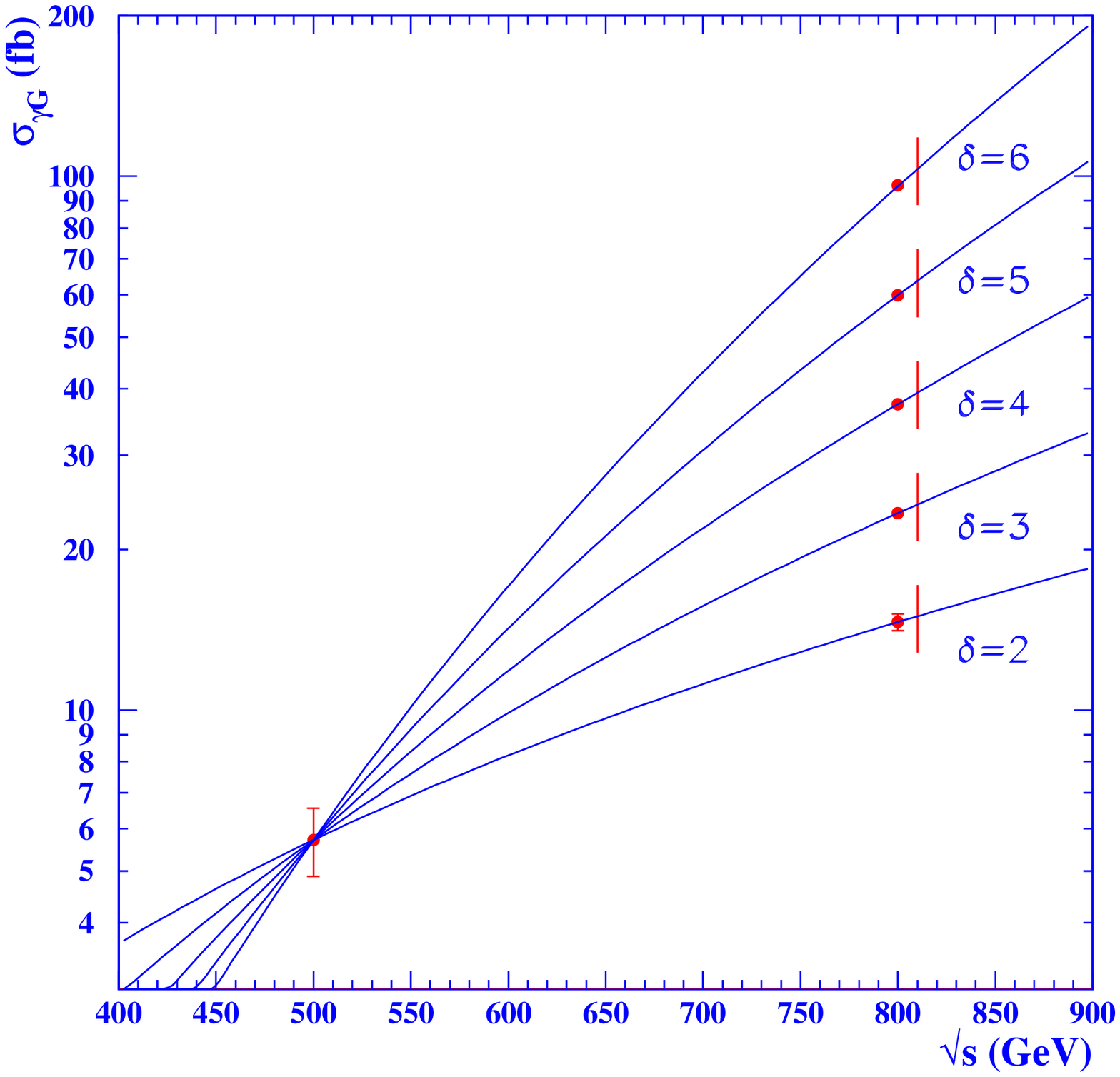}
    \includegraphics[width=5.5cm,height=7.5cm,angle=90]{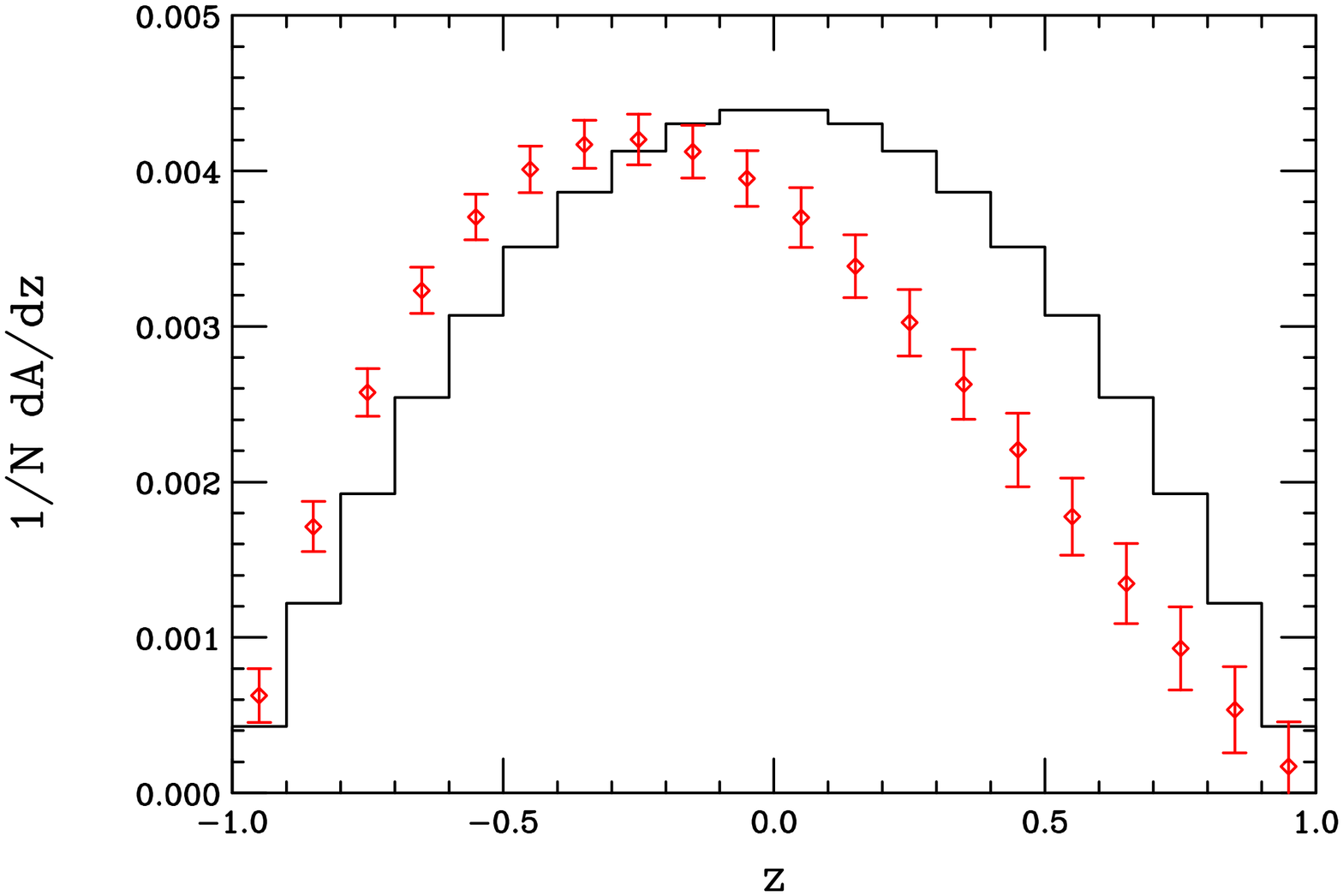}
\vspace*{-7mm}
\caption[Determination of the number of large extra-dimensions at the ILC]
{Left: determination of the number of extra-dimensions at the ILC
at two center of mass energies $\sqrt s=500$ and 800 GeV~\cite{wilson}. Right:
the differential azimuthal asymmetry distribution for $e^+e^- \to \ell^+ 
\ell^-$  at 500 GeV ILC with 500~fb$^{-1}$ data in the SM (histogram) and in 
the LED model with a cut--off of 1.5 TeV (data points); $e^\pm$  are assumed 
to be 80\% and 60\% polarized, respectively \cite{Rizzo:2002ww}.}
\label{fig:LED}
\vspace*{-3mm}
\end{figure}

Once the missing energy signal is observed, the next step would be to confirm
its gravitational nature and determine the number of extra dimensions. The ILC
will play an essential role here. The number of extra dimensions can be
determined from the energy dependence of the production cross section. In the
left--hand side of Fig.~\ref{fig:LED}, it is shown that its measurement at two
collider energies, $\sqrt s= 500$ GeV and 800 GeV,  can discriminate between
scenarios with different numbers of extra dimensions. Additional information on
the number of extra dimensions can also be obtained from the missing mass
distribution.

An alternative signal for the presence of extra dimensions is provided by
KK--graviton exchange in processes such as $e^+e^- \to f\bar{f}$. The mass
reach  in this channel is similar to that obtained in KK--graviton emission.
Since many new physics models can generate deviations in this reaction, it is
important to discriminate the extra--dimensional model from other scenarios.
$s$--channel KK--graviton exchange has the characteristic signature of spin--two
particle in the angular distributions of the  $e^+e^- \to f\bar{f},  WW$ and $
HH$ production processes \cite{A-graviton}. Furthermore, if both electron and
positron are transversely polarized, the azimuthal asymmetry distribution
provides a powerful tool to identify the spin--two nature of the virtually
exchanged particle ~\cite{gudi,Rizzo:2002ww} as shown in the right--hand side of
Fig.~\ref{fig:LED}.

\subsection{Warped extra dimensions}

In the original proposal of Randall and Sundrum \cite{WED}, only the graviton
was assumed to propagate in the extra--dimensional space and the SM fields were
confined on the TeV brane.  In this model, the mass scales of the dimensionful
parameters  in the action are set by the Planck scale, but the physical mass
scales  on the TeV (or SM) brane are reduced by the warp factor  of $e^{-\pi k
r_c}$ where $kr_c \sim 11$ to explain the hierarchy between the weak and Planck
scales.  A characteristic signal of this extension is the presence of the
graviton  KK modes near the TeV scale. In fact, the model is specified  by two
parameters, for instance, the mass of the first KK mode and  $k/M_{*}$ where 
$M_{*}$  is the four--dimensional reduced Planck mass. KK graviton resonances
can be searched for through the Drell--Yan process at the LHC and the  mass
reach can be 3--4 TeV, covering most of the interesting parameter space of the
model~\cite{Davoudiasl:2000wi}. 

If such resonances are indeed observed at the LHC, one  needs to establish their
gravitational nature. The spin of the resonance can be  determined from the
angular distribution of the final lepton pairs at the LHC and ILC
\cite{Davoudiasl:2000wi,Allanach:2000nr}.  The search reach through contact
interactions at the ILC with a c.m. energy of 500 GeV is similar to the LHC
direct search reach and a 1 TeV ILC can significantly extend the discovery
limit~\cite{Weiglein:2004hn}. 

\begin{figure}[h]
\vspace*{-2mm}
\begin{center}
\mbox{
\includegraphics[width=6.5cm,height=10.cm,angle=90]{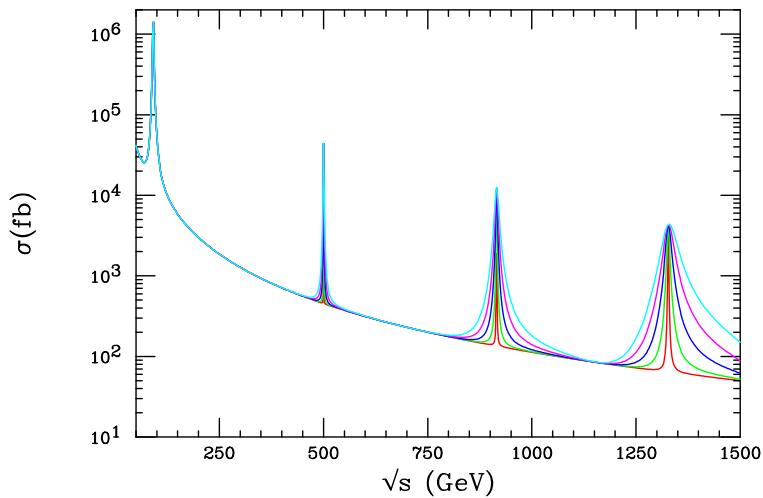}
}
\vspace*{-5mm}
\caption[Graviton resonance peaks in  ${\rm e^+ e^- \to\! \mu^+\mu^-}$ in the Randall--Sundrum  model]
{Graviton resonance production at the ILC in  $\eei\!\to\!\mu^+\mu^-$
in the RS model with the mass of the first KK mode taken to 
be 500 GeV;  the exchange of a KK tower is included and the ever 
widening resonances correspond to increasing the value of  $k/M_*$ 
in the range of 0.01--0.1. From Ref.~\cite{Davoudiasl:2000wi}. }
\label{fig:A-KKRS}
\end{center}
\vspace*{-5mm}
\end{figure}

Another important property which has to be verified is the universal structure
of the graviton couplings to other particles. For this purpose, the branching
ratios of the resonances have to be determined precisely. An ultimate
confirmation of the model would be  provided by the $s$--channel production of
the KK graviton state at the ILC as shown in  Fig.~\ref{fig:A-KKRS}. From 
line--shape analyses, the two independent parameters, the first KK mode mass and
the ratio $k/M_{*}$, can be precisely determined along with  the various decay 
branching ratios.    

\begin{figure}[!h]
\begin{center}
\vspace*{-.5cm}
\includegraphics[width=5.7cm,height=7cm,angle=90]{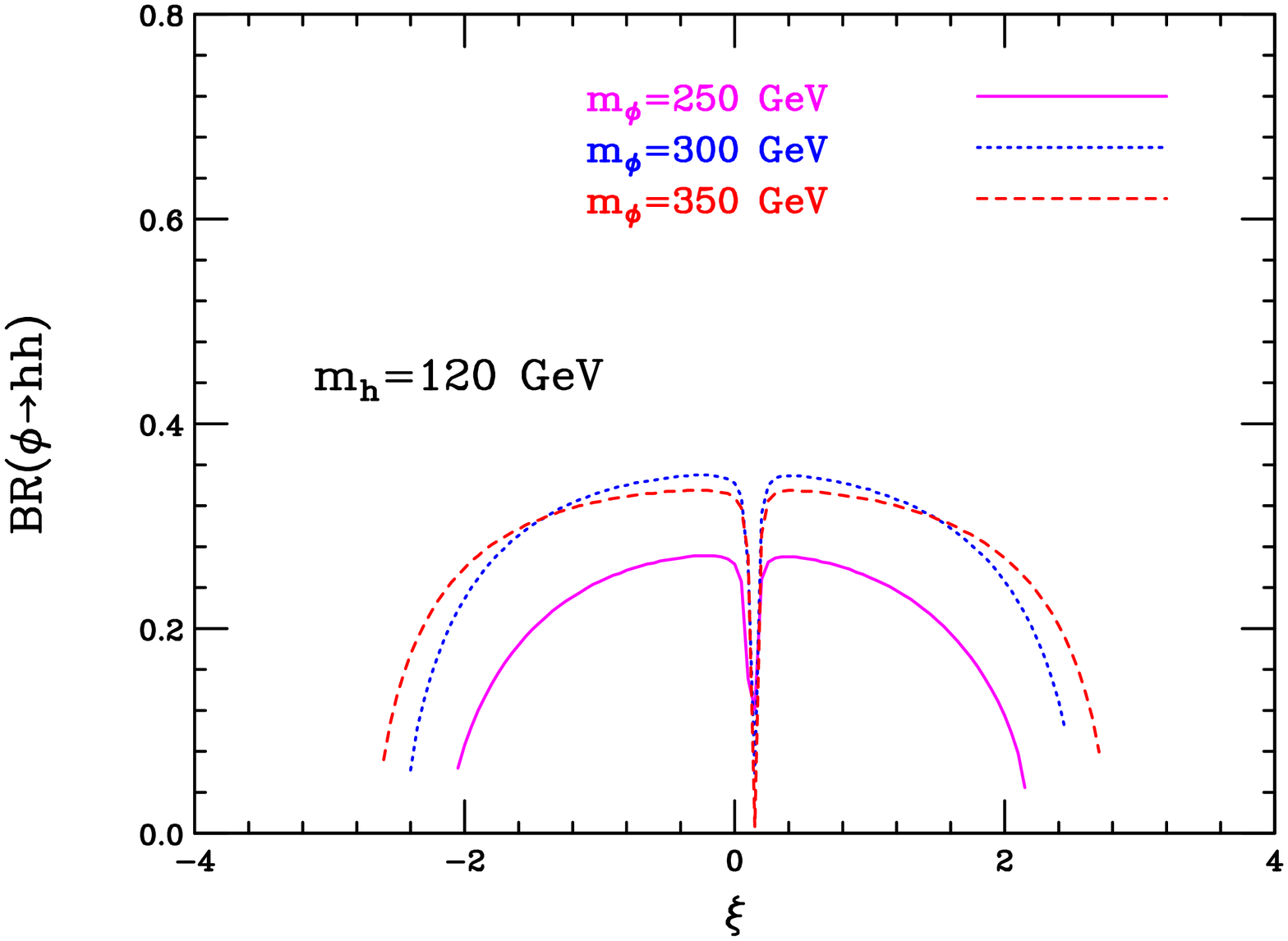}\hspace*{1cm}
\epsfig{file=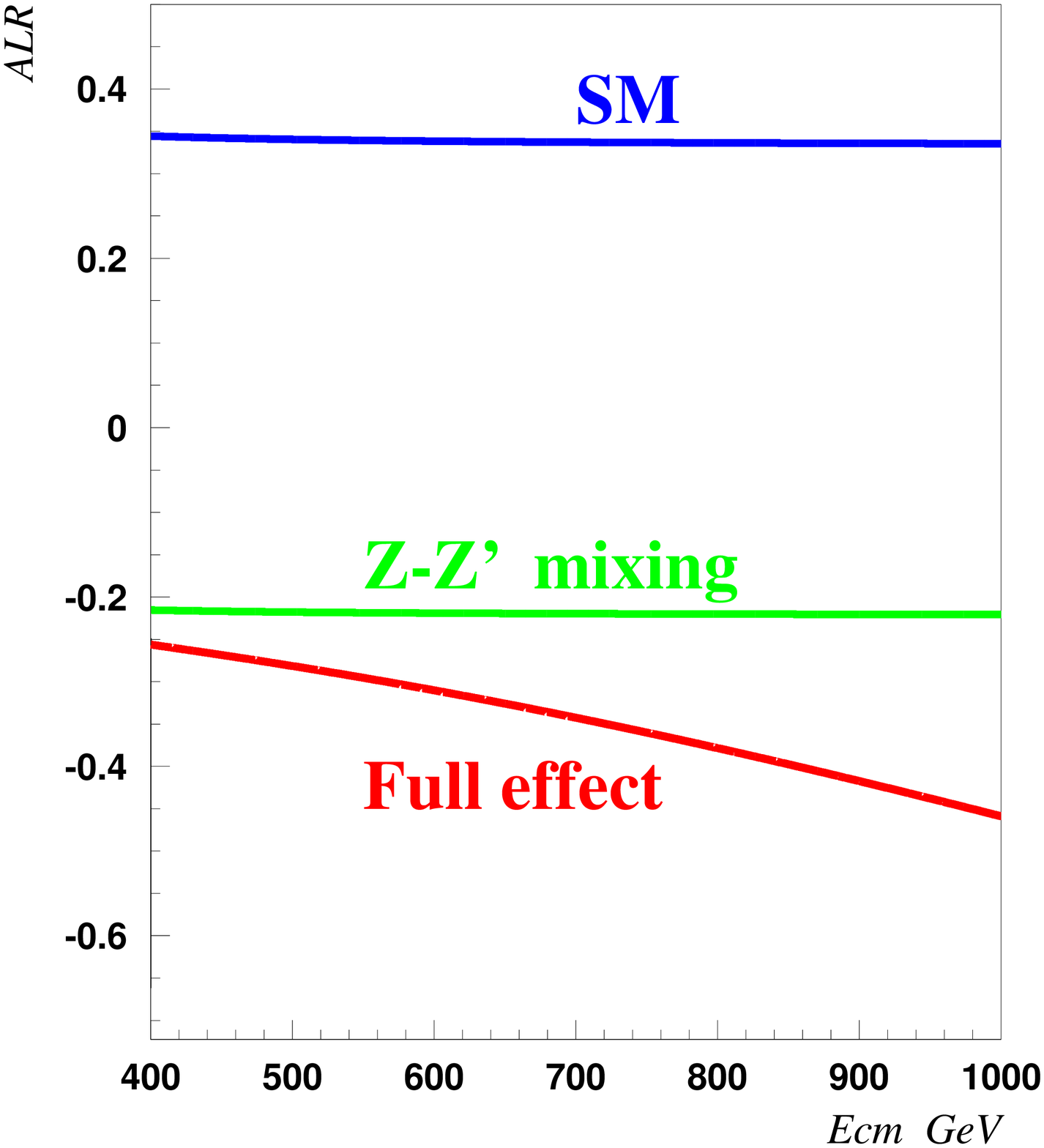,width=7.cm,height=6.cm}
\end{center}
\vspace*{-.9cm}
\caption[Effects of radions and Kaluza--Klein excitations of gauge bosons at
the ILC] 
{Left: the $\phi\to HH$ branching ratios as functions of the parameter
$\xi$ for $M_H=120$ GeV and $\Lambda=5$ TeV for several values of $M_\phi$
\cite{Dominici:2002jv}. Right: the energy dependence of the left--right 
polarization asymmetry for $t \bar t$  production at the ILC in the SM 
and in the RS scenario in the pure $Z$--$V_{KK}$ mixing case
and taking also into account the virtual KK exchange 
\cite{Djouadi:2006rk}.}
\label{fig:Aqq}
\vspace*{-.5cm}
\end{figure}

In RS models, one would expect the presence of a radion which  will mix with 
the Higgs boson whose properties could be significantly  altered. The Higgs
couplings to various particles, for instance, could be reduced at the level of a
few 10\%.  These effects can be easily identified with the  precision ILC
measurements as discussed in chapter \ref{sec:higgs}. The radion has substantial
couplings to the $W/Z$ bosons and can  be produced  in the Higgs--strahlung
$\eei \to \phi Z$ or  $WW$ fusion $\eei \to \phi \nu_e \bar \nu_e$ processes. If
it is relatively heavy, $M_\phi \gsim 2M_H$, it could decay into two Higgs
bosons with large  rates. This is illustrated in the left--hand side of
Fig.~\ref{fig:Aqq} where BR($\phi \to HH)$ is displayed as a function of the
Higgs--radion mixing parameter $\xi$. Besides the dominating $\phi \to WW,ZZ$
decay modes, the channels $\phi \to HH$ can reach branching fractions of ${\cal
O}(30\%)$ leading to a significant excess of Higgs pairs compared to the SM.
Other decay channels of the radion, such  as $\phi \to t\bar t$ and $gg$,
besides $WW$ and $ZZ$ decays, can reach the level of few ten percent when
kinematically accessible. These decays could also be probed at the ILC and the
branching fractions measured very precisely.

The version of the RS model with bulk matter offers the possibility of
generating the large mass hierarchies prevailing among SM fermions   if  they
are placed differently along the extra dimension \cite{RSloc}.  An interesting
aspect of this scenario is related to the KK excitations of gauge bosons. If the
SM symmetry is enhanced to  ${\rm  SU(2)_L\! \times\! SU(2)_R\! \times\!
U(1)}$,  the  high--precision data can be fitted while keeping the KK masses
down to values as low as $3$ TeV. Since the third generation fermions should be
localized closer to the  TeV--brane to get higher masses, their couplings to the
KK gauge  bosons are larger and generate more important effects in the $t$ and
$b$ sectors. In particular, the stronger $b$  couplings induce  a large  mixing
between the $Z$ and KK bosons which allows to resolve the LEP anomaly on the 
asymmetry $A_{FB}^b$ \cite{Z-Pole}.  With the high precision in the measurement
of the production rates and polarization/angular asymmetries in the $\eei \to t
\bar t$ and $b \bar b$ processes,  KK  excitations exchanged in the $s$--channel
can be probed even for masses up to $\sim 20$ TeV \cite{Djouadi:2006rk}. This is
exemplified in Fig.~\ref{fig:Aqq} (right) where the deviations in the
left--right asymmetry $A_{LR}^{t}$ in $\eei \to t\bar t$  are displayed as a
function of $\sqrt s$ for fermion localizations and couplings which resolve the
$A_{FB}^b$ anomaly with a KK mass of  3 TeV. With the ILC accuracy, a
measurement of 10\% of the KK mass can be achieved.  Additional information on
the  KK couplings can be obtained from a more precise measurement of  $A_{FB}^b,
A_{LR}^b$ and $\Gamma( Z\to b\bar b)$ at the GigaZ option of the ILC.

Note that in such models, there may be also new fermions with not too large
masses. For instance, the ${\rm SU(2)_R}$ partner of $t_R$, $b_R^\prime$, 
typically reaches KK masses as low as a few hundred GeV and can be thus produced
and studied in detail at the ILC. This new quark might affect dramatically the
production rates of the Higgs boson at the LHC as discussed earlier.

\subsection{Universal extra dimensions}

Universal extra dimensions (UED) \cite{Appelquist:2000nn} is the model which
resembles the  most to the original Nordstr\"om--Kaluza--Klein scenario. All SM
particles are assumed to propagate in a flat extra--dimensional space which is
compactified to an orbifold. In the minimal version, the extra one--dimensional
space is compactified in the form of an $S_1/Z_2$ orbifold, where a circle $S_1$
is divided in half by $Z_2$ projection. Viewed as a four dimensional theory, the
UED model introduces a Kaluza--Klein tower for each SM particle. The common mass
of the $n$th KK states is roughly given by $n/R$ where $R$ is the
compactification radius, but radiative corrections and  boundary terms lift the
initial mass degeneracy of the $n$th KK states. 

In UED models, momentum conservation in the fifth dimension is replaced by a
conserved parity, called KK--parity  \cite{Hooper:2007qk,LKP-heavy}. The zero
modes, i.e. the SM particles, are even under this parity but the lightest
massive modes are odd. This has the major consequence that the lightest KK
particle (LKP), which in general corresponds to the KK hypercharge gauge boson,
is absolutely stable. It gives missing transverse energy signals at colliders
and is a good dark matter candidate as will be discussed in chapter
\ref{sec:cosmology}. Another important consequence of this parity is that  $n=1$
KK particles are only produced in pairs. This suppresses their virtual
corrections to SM  processes, allowing the UED scale $1/R$ to be as low as 300
GeV without conflicting with  high--precision electroweak data.

\begin{figure}[ht!]
\vspace{-0.1cm} 
\begin{center}
\includegraphics[height=5.cm,width=7.cm,angle=0]{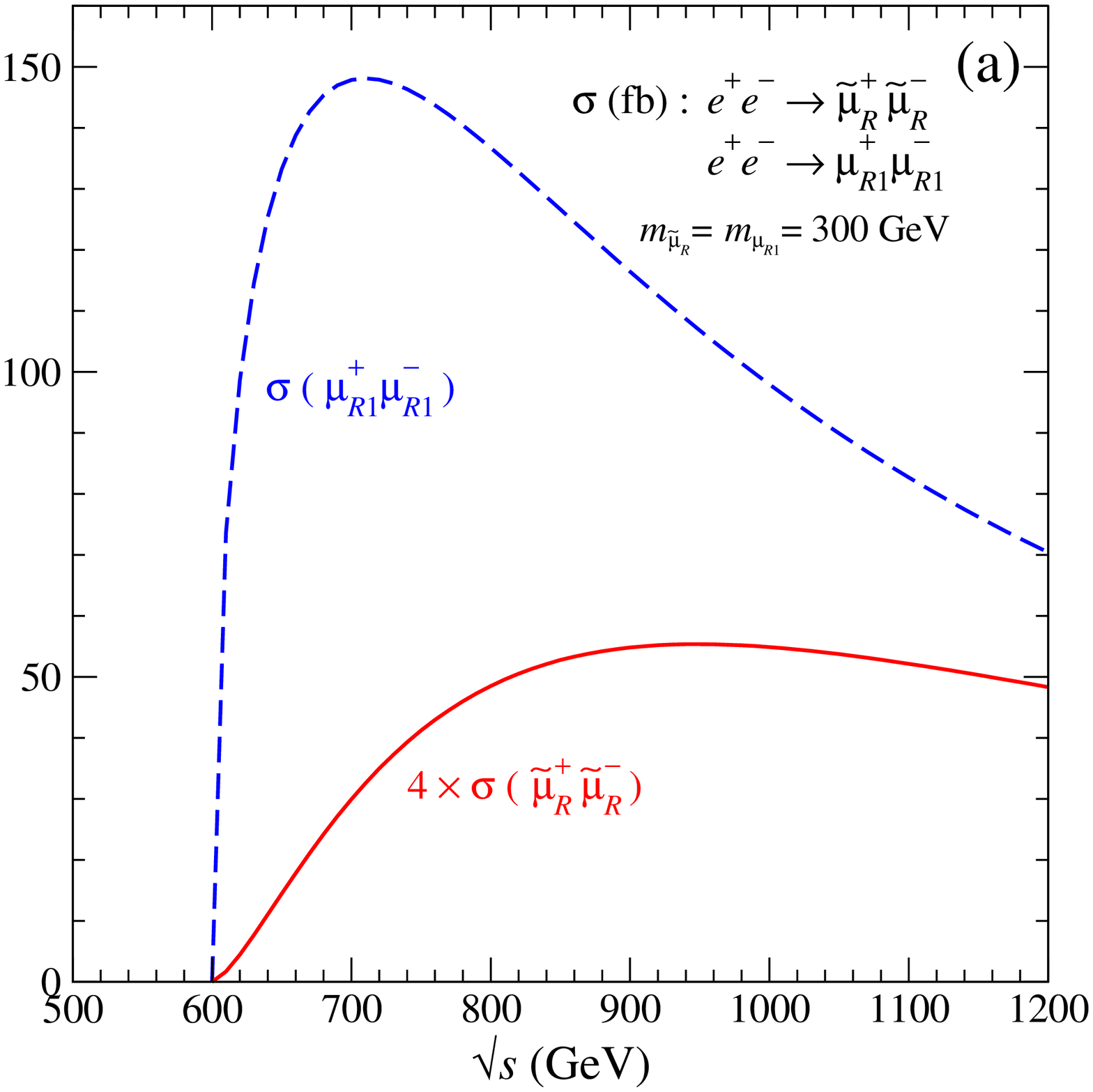}
\hskip 1.cm
\includegraphics[height=5.cm,width=7.cm,angle=0]{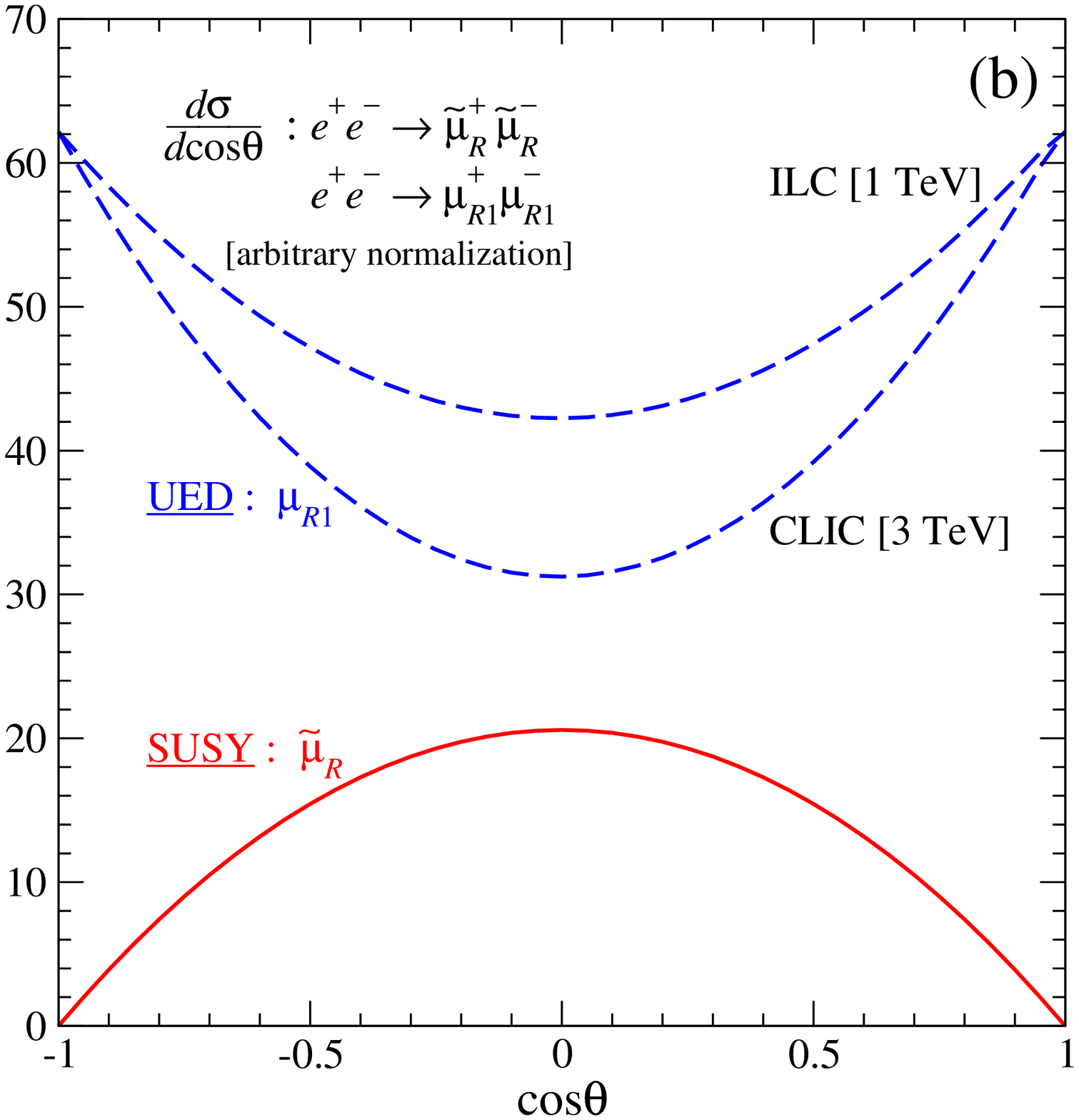}
\end{center}
\vspace{-0.7cm}  
\caption[Discrimination between SUSY and universal extra dimension models 
at ILC]  
{The threshold excitation for smuons (a) and the angular distribution 
(b) in the case of smuons in the MSSM and the first KK excitation 
$\mu_{R1}^\pm$ in UED in pair production at the ILC; from 
Ref.~\cite{SUSY-spin-chi+}.} 
\label{fig:smuon_muon1}
\vspace{-0.3cm} 
\end{figure}

From the previous discussion, one concludes that the situation in UED models  is
quite analogous to the minimal supersymmetric SM extension with conserved
R--parity, except that here, the lightest particle is a spin--one particle, a
heavy photon.  Thus, if only  the first massive KK modes are produced, UED
models would look very much like a subset of SUSY models in terms of their
collider signatures. Even if one detects a few of the second level KK modes, it
is not obvious that this will discriminate the signatures from an extended SUSY
model. The crucial discriminators, of course, are the spins of the heavy partner
particles. At the LHC, distinguishing these spins is a significant experimental
challenge. The ILC will play an important role in this context as the  spin
difference between  superpartners and KK excitations can be determined in
detailed angular distribution studies and threshold scans. This is exemplified
in Fig.~\ref{fig:smuon_muon1} where the threshold excitation curve and the
angular distribution in  the case of $e^+e^-\to {\mu}^+_{R1} {\mu}^-_{R1}$ for
the first muon KK excitation in UED models is compared to smuon pair production
in the MSSM,  $e^+ e^- \to \tilde  \mu^+_R \tilde \mu^-_R$ \cite{SUSY-spin-chi+}.

\section{Strong interaction models}

\subsection{Little Higgs models}

To interpret the Higgs boson as a (pseudo-)Goldstone boson has been a very 
attractive idea for a long time. The interest in this picture has been renewed 
within the little Higgs scenarios  \cite{LHM}, that have recently been
developed  to generate the electroweak symmetry breaking dynamically by new
strong  interactions. Little Higgs models (LHMs) are based on a complex system
of symmetries and symmetry breaking mechanisms.  Three points are central in
realizing the idea: (i)  the Higgs field is a Goldstone field associated with
the breaking  of a global symmetry G at an energy scale of order $\Lambda_s \sim
4 \pi f  \sim$ 10 to 30 TeV, with $f$ characterizing the scale of the symmetry 
breaking parameter;  (ii) in the same step, the gauge symmetry ${\rm G_0 \subset
G}$ is broken  down to the SM gauge group, generating  masses for heavy vector
bosons and fermions which cancel the standard  quadratic divergencies in the
radiative corrections to the light Higgs mass; since the masses of these new
particles are generated by the breaking of  the gauge symmetry ${\rm G_0}$ they
are of the intermediate size $M \sim g f \sim 1$ to 3 TeV;  (iii) the Higgs
bosons acquires a mass finally through radiative  corrections at the standard
electroweak scale of order  $v \sim g^2 f / 4 \pi \sim$ 100 to 300 GeV.

Thus, three characteristic scales are encountered in these models: the strong 
interaction scale $\Lambda_s$, the new mass scale $M$ and the electroweak 
breaking scale $v$, ordered in the hierarchical chain $\Lambda_s \gg M \gg v$. 
The light Higgs boson mass is protected at small value by requiring the 
collective breaking of two symmetries. In contrast to the boson--fermion 
symmetry that cancels quadratic divergencies in supersymmetry,  the
cancellation  in LHMs operates in the bosonic and fermionic sectors
individually,  the cancellation ensured by the symmetries among the couplings of
the SM  fields and new fields to the Higgs field.

A generic feature of LHMs is the existence of extra gauge bosons, Higgs
particles  and partners of the top quark. The masses of these new particles are
constrained by electroweak precision measurements. Although the precise values
depend on the specific model under consideration, these are usually beyond a few
TeV, so that their direct production is kinematically not accessible at the ILC.
If one introduce T--parity, these masses can be below the TeV scale, but T--odd
particles should be pair produced. Even if the new particles are beyond the
kinematical reach of the ILC, indirect searches for effects of  LHMs is possible
in SM processes such as $e^+e^- \to f\bar{f},  t\bar{t}, ZH$ and $\gamma \gamma
\to H$.

\begin{figure}[h] \centering
\vspace*{-3mm}
\includegraphics[width=7.5cm,height=7.5cm,angle=0]{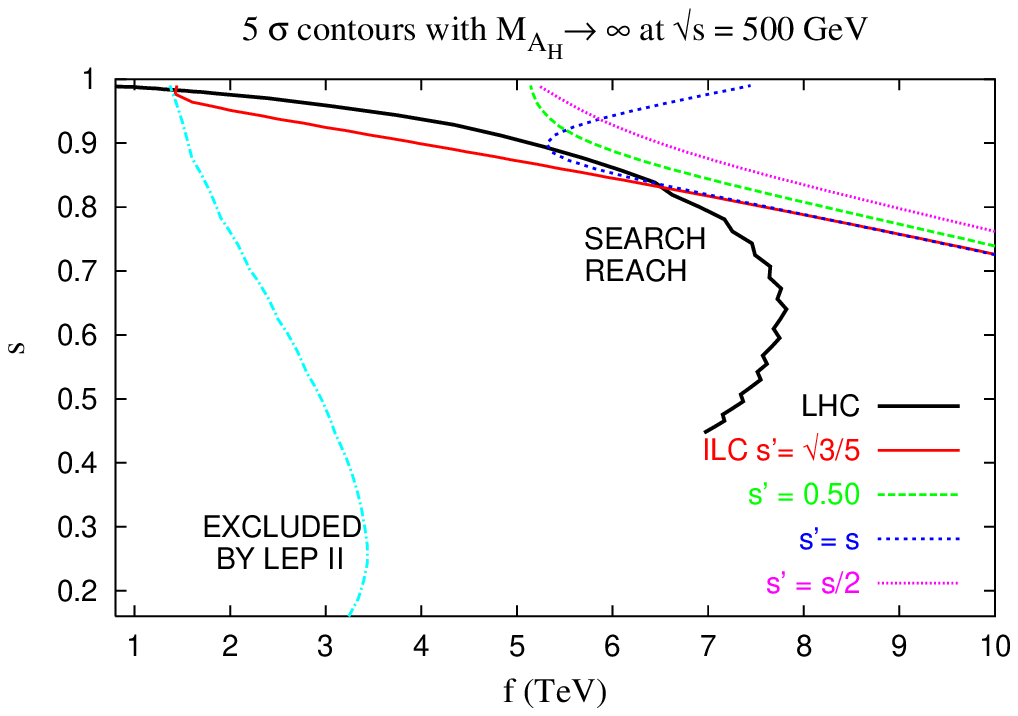}
\includegraphics[width=7.5cm,height=7.5cm,angle=90]{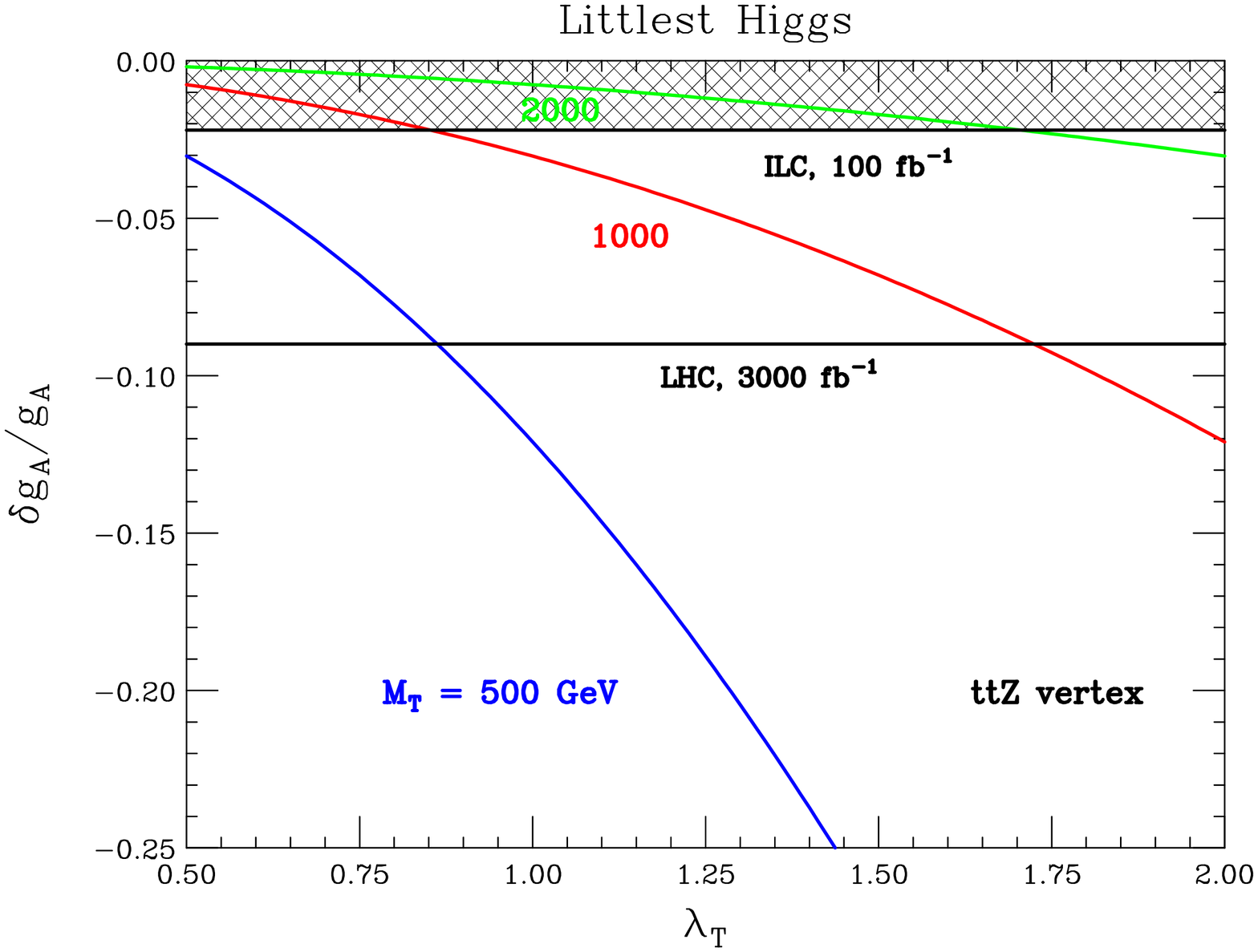}
\vspace*{-9mm}
\caption[The ILC search reach in little Higgs models and comparison with the LHC]
{Left: the ILC search reach in LHMs, as derived from the process 
$e^+e^- \to f \bar{f}$, is compared to the LHC reach in heavy $Z^\prime$ boson 
searches;  the decoupling limit of the heavy photon is taken 
\cite{Conley:2005et}. Right: the  corrections to the $t\bar{t}Z$  coupling in 
LHMs  with conserved T--parity for two values of the heavy top quark 
partner compared to the (super)LHC and ILC sensitivities \cite{Berger:2005ht}.}
\label{fig:A-LHM}
\vspace*{-5mm}
\end{figure}

An example of indirect search of the new states at the ILC  is shown in the
left--hand side of  Fig.~\ref{fig:A-LHM}. The figure displays the limit on the
vev  $f$ associated with ${\rm SU(5) \to SO(5)}$ symmetry breaking in LHMs as
derived from the $e^+e^- \to f\bar{f}$ processes with a center of mass energy
$\sqrt s=500$ GeV and an integrated luminosity of 500 fb$^{-1}$. Two new mixing
angles $s$ and $s^\prime$ specify the gauge symmetry breaking of ${\rm [SU(2) \times
U(1)]^2 \to SU(2)_L \times U(1)_Y}$. For comparison, the LHC search reach for
the heavy gauge boson $Z_H$ is also shown. As can be seen, the indirect searches
at the ILC  can extend the LHC search limit substantially.  A similar search can
be performed in the $e^+e^- \to ZH$ process but with less sensitivity.

In order to cancel the quadratic divergence in the Higgs mass in LHMS, the top 
quark sector has to be extended. The ordinary top quark is identified as one
light combination of the extended top sector so that there could be sizable
deviations in the top coupling to $W/Z$ bosons. In Fig.~\ref{fig:A-LHM} (right),
the correction to the $t\bar{t}Z$ coupling is shown in the case of LHMs with
T--parity. The displayed ILC search limit indicates that most of the interesting
parameter region is covered by future high--precision top quark measurements.

\begin{figure}[h] \centering
\vspace*{-1mm}
    \includegraphics[width=7.cm,]{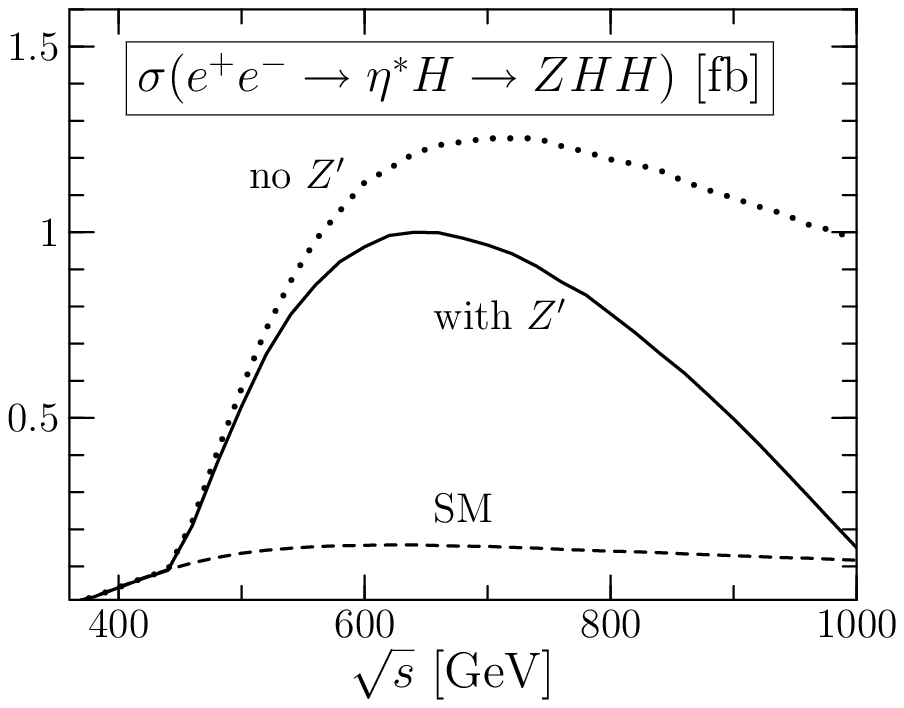}\hspace*{0.7cm}
    \includegraphics[width=7.cm]{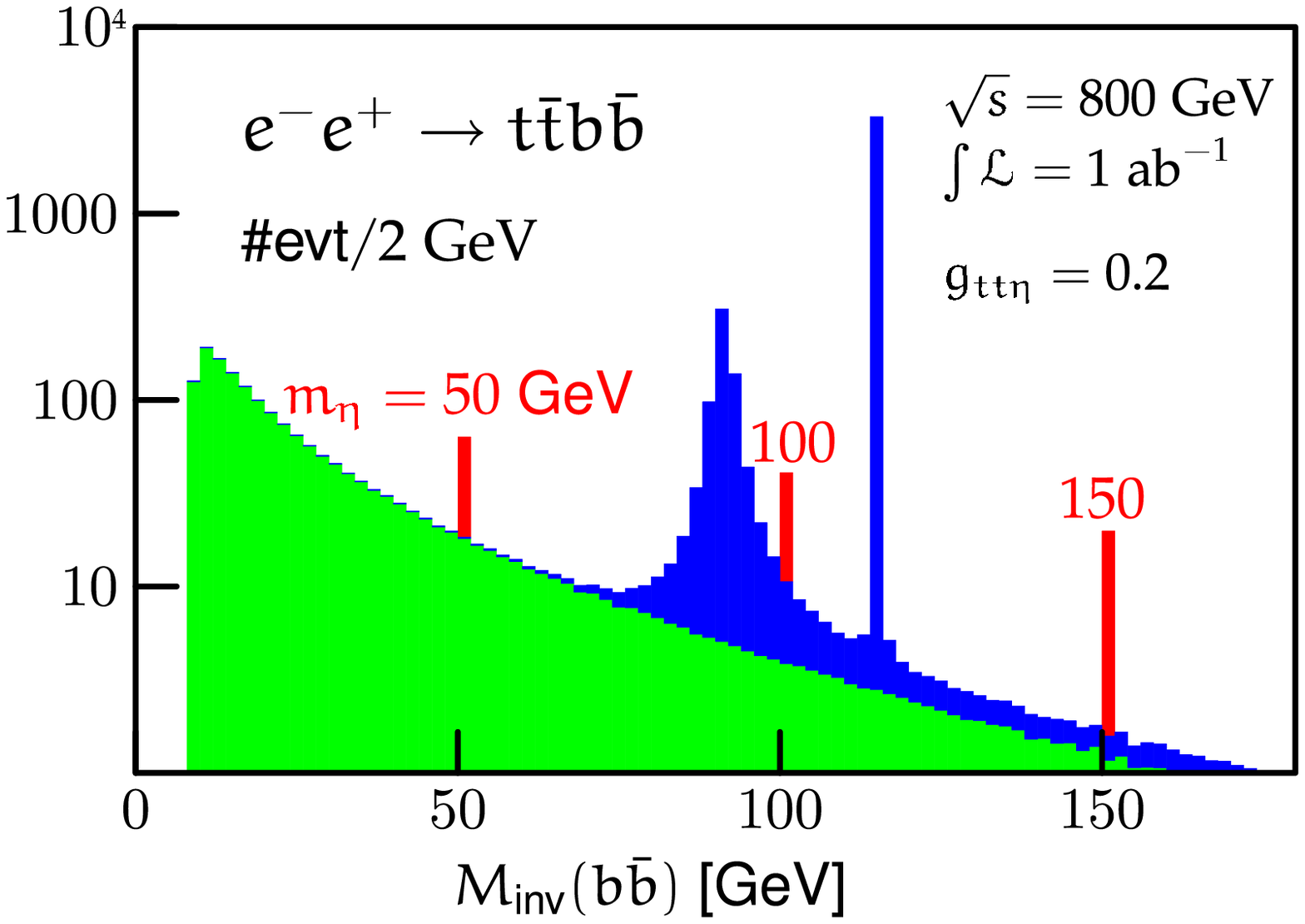}
\vspace*{-5mm}
    \caption[Cross sections at ILC for pseudo--axion $\eta$ production in little
    Higgs models] 
    {Left: the cross section of double Higgs production with and 
    without $Z^\prime$ exchange compared with the SM prediction in the simplest 
    LHM for $M_H \sim 130$ GeV and $M_\eta \sim 300$ GeV. Right: the
    reconstructed $b\bar b$ invariant mass in the process $\ee \to 
    t\bar t\eta \to t\bar t b\bar b$ compared to the SM background; 
    the peaks correspond to $Z,H$ production and to the $\eta$ resonance
    for several $M_\eta$ values. From Ref.~\cite{Kilian:2006eh}.}
\label{fig-Azhh}
\vspace*{-3mm}
\end{figure}

Even if T--parity is not imposed, a pseudo--axion might be light enough to
be accessible at the ILC in the case where LHMs  possess a spontaneously broken
approximate U(1) symmetry as in the simplest model  \cite{Schmaltz:2004de}.   In
such a case the pseudo--axion $\eta$ could  be produced in association with the
Higgs boson, $\eei \to H\eta$ and would decay via $\eta \to HZ$.  This is
exemplified in Fig.~\ref{fig-Azhh} (left) where the cross section for the $\eei
\to \eta H \to HHZ$ process is shown as a function of $\sqrt s$ for scenarios
with and without the contribution of a $Z^\prime$ boson \cite{Kilian:2006eh}. The new 
contributions increase the $ZHH$ rate by an order of magnitude compared to the
SM. A relatively light $\eta$ boson could also be produced in association with 
top quark  pairs, $\ee \to t\bar t \eta$ and the signal in which the $\eta$
resonance dominantly decays into $b\bar b$ pairs could be easily observed at the
ILC as shown in Fig.~\ref{fig-Azhh} (right) for several $M_\eta$ values.

\subsection{Strong electroweak symmetry breaking}     

If no Higgs boson will be observed with mass below 1 TeV, quantum--mechanical
unitarity demands strong interactions between the electroweak gauge bosons,
becoming effective at energies $(8\pi/\sqrt2 G_F)^{1/2}\simeq 1.2$ TeV, to damp
the growth of the amplitudes for (quasi--)elastic massive gauge boson scattering
processes \cite{H-LQT}. 

As discussed in chapter \ref{sec:couplings}, the new interactions between the
weak bosons can be expanded in a series of effective interaction terms with
rising dimensions~\cite{Appelquist}. Scattering amplitudes are expanded 
correspondingly in a series characterized by the energy coefficients
$s/\Lambda_\ast^2$. Demanding CP and isospin invariance, for instance,  only two
new dimension--four interaction terms (out of the 10 terms present in the
general case) must be included in the expansion, $L_4$ and $L_5$, with
coefficients $\alpha_{4,5} = v^2/\Lambda_{\ast 4,5}^2$ with scale parameters
bounded from above by the value $4\pi v \sim 3$ TeV.  The parameters  $\alpha_i$
can be measured in the quasi--elastic $VV$ scattering processes  $e^+e^- \to
\ell \ell  VV$ and triple gauge boson production $\ee \to VVV$, as  the new
interaction terms affect the total cross sections and the final state
distributions.

\begin{figure}[!h]
\begin{minipage}{7cm}
\includegraphics[width=70mm]{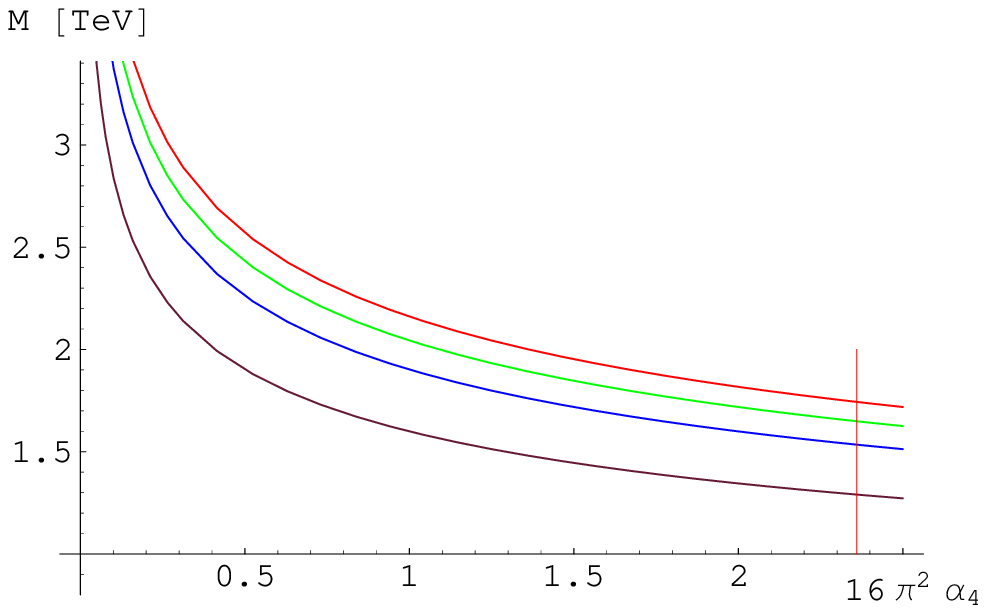}
\end{minipage}\hspace*{1cm}
\begin{minipage}{7cm}
\includegraphics[width=6cm,angle=270]{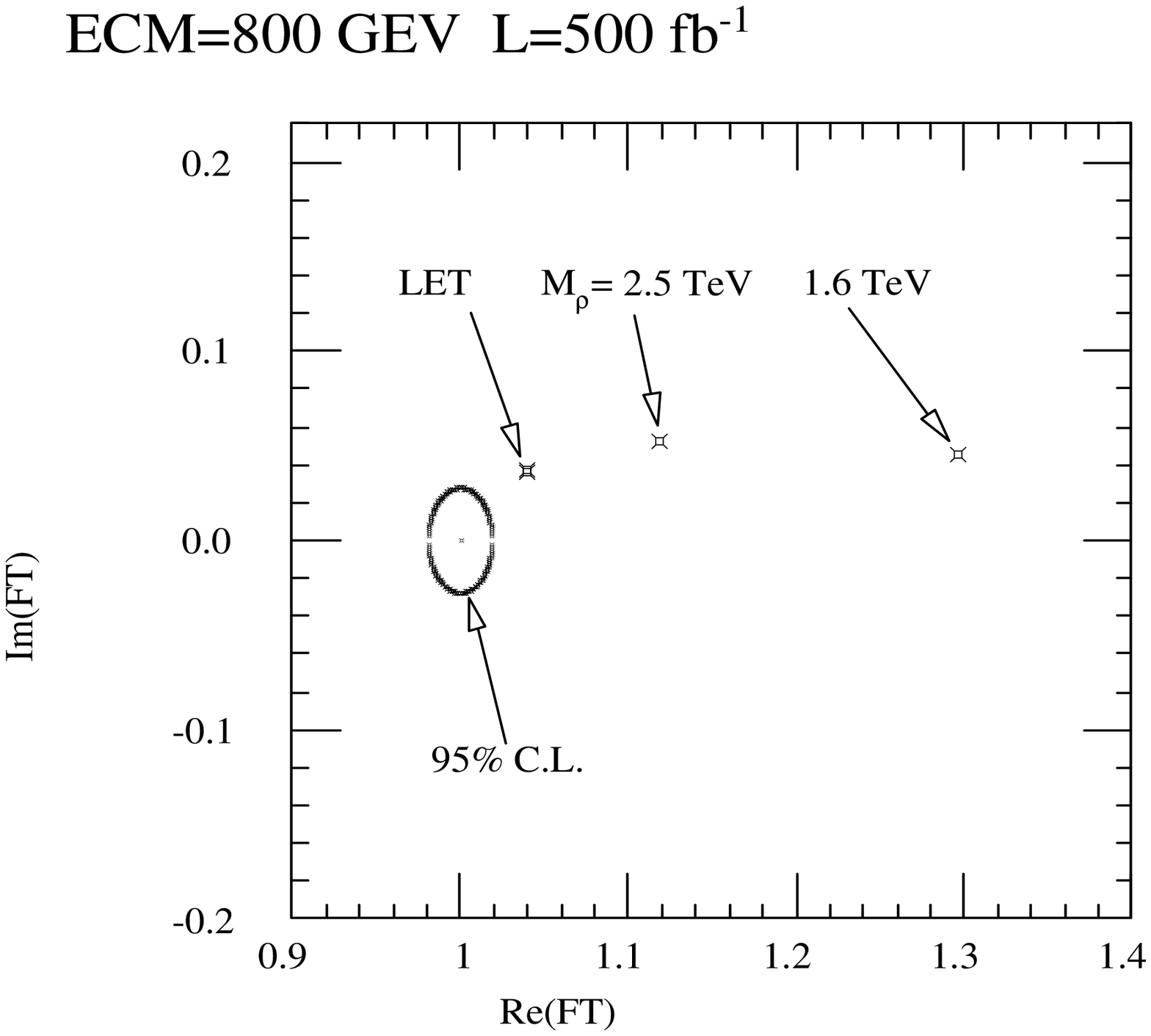} 
\end{minipage}
\vspace*{-5mm}
\caption[ILC sensitivity to new resonances for strong electroweak symmetry
breaking]
{Left: dependence of the mass of a singlet vector resonance on
$\alpha_4$ for different values of $\Gamma_\pi /M_\pi = 1.0$ (red), $0.8$ (blue),   $0.3$
(brown) \cite{predrag}. Right:  sensitivity for a resonance form factor at a 800
GeV ILC with 500 fb$^{-1}$ data assuming perfect charm tagging
~\cite{LET:Barklow}.} \label{A-strongWZ} 
\vspace*{-4mm}
\end{figure}

As can be seen from Figs.~\ref{fig:VVVfit} and \ref{fig:predrag} of chapter
\ref{sec:couplings}, at $\sqrt s= 1$ TeV  with 1 ab$^{-1}$ data, the entire
range of $\Lambda_\ast$ values can be covered,  $\Lambda_\ast \leq 4 \pi v
\simeq 3$ TeV.  These values can be conveniently re--expressed  in terms of the
maximal mass of the heavy resonances associated with the new interactions the
measurement can be sensitive to, under the most favorable conditions;
Fig.~\ref{A-strongWZ} (left).  In Table~\ref{tab:resonance}, displayed are the
combined  results obtained in the full analysis of Ref.~\cite{predrag} for the
sensitivity on the scale $\Lambda_\ast$  for all possible spin/isospin
channels.  In the left--hand side of the table, a conserved ${\rm SU(2)_c}$ is
assumed and in this case,  only the channels with even $I\!+\!J$ couple to  weak
boson pairs; in the right--hand side, shown are the results without this
constraint.  In each case, a single resonance with maximal coupling  was assumed
to be present.  As one can see, scales from $\sim  1.5$ to $\sim 6$ TeV can be
probed.  

\begin{table}[h]
\vspace*{-4mm}
\caption[Accessible scales for new heavy resonances at the ILC in no--Higgs scenarios.]
{Accessible scales $\Lambda_\ast$ in $\TeV$ for all possible
spin/isospin channels from a complete analysis of vector boson scattering
processes at 1 TeV the ILC, assuming a single resonance with optimal 
properties \cite{predrag}. The numbers in the left--(right--)hand side are 
with (without) assuming the custodial symmetry.}
\label{tab:resonance}
\centering
\renewcommand{\arraystretch}{1}
\begin{tabular}{|c||c|c|c||c|c|c|} \hline
Spin & I=0 & I=1 & I=2 & I=0 & I=1 & I=2 \\ \hline \hline
0 & 1.55 & -- & 1.95 & 1.39 & 1.55 & 1.95 \\
1 & -- & 2.49 & --  &   1.74 & 2.67 & -- \\
2 & 3.29 & -- & 4.30 & 3.00 & 3.01 & 5.84 \\ \hline
\end{tabular}
\vspace*{-6mm}
\end{table}

\hspace*{1.cm}  


Alternatively, when resonances below the scale $\Lambda_\ast$ are present, the
vector boson pair production amplitude can be unitarised by a Omn\`es
rescattering factor  with one contribution reproducing the low energy theorem
$\delta_{\mathrm{LET}}(s)=s/(8\Lambda_{\mathrm{EWSB}}^2)$ for Goldstone boson
scattering at threshold far below any resonance and a second contribution from a
resonance $\delta_\rho(s)=3\pi/8\cdot(\tanh(s-M_\rho^2)/(M_\rho\Gamma_\rho)+1)$.
A  study performed in Ref.~~\cite{LET:Barklow} has shown that $W^+ W^-$
production at the ILC with $\sqrt s=800$ GeV and ${\cal L}=500\,
\mathrm{fb}^{-1}$ is competitive with the LHC.  As shown in the right--hand side
of Fig.~\ref{A-strongWZ}, there is a~$6\sigma$ exclusion limit for the LET and
one can exclude a $\rho$--like resonance of $2.5\; (1.6)$ TeV at the $16\; (33)
\sigma$ level.

\begin{figure}
\vspace*{-3mm}
  \begin{center}
\mbox{ 
\includegraphics[width=0.4\textwidth]{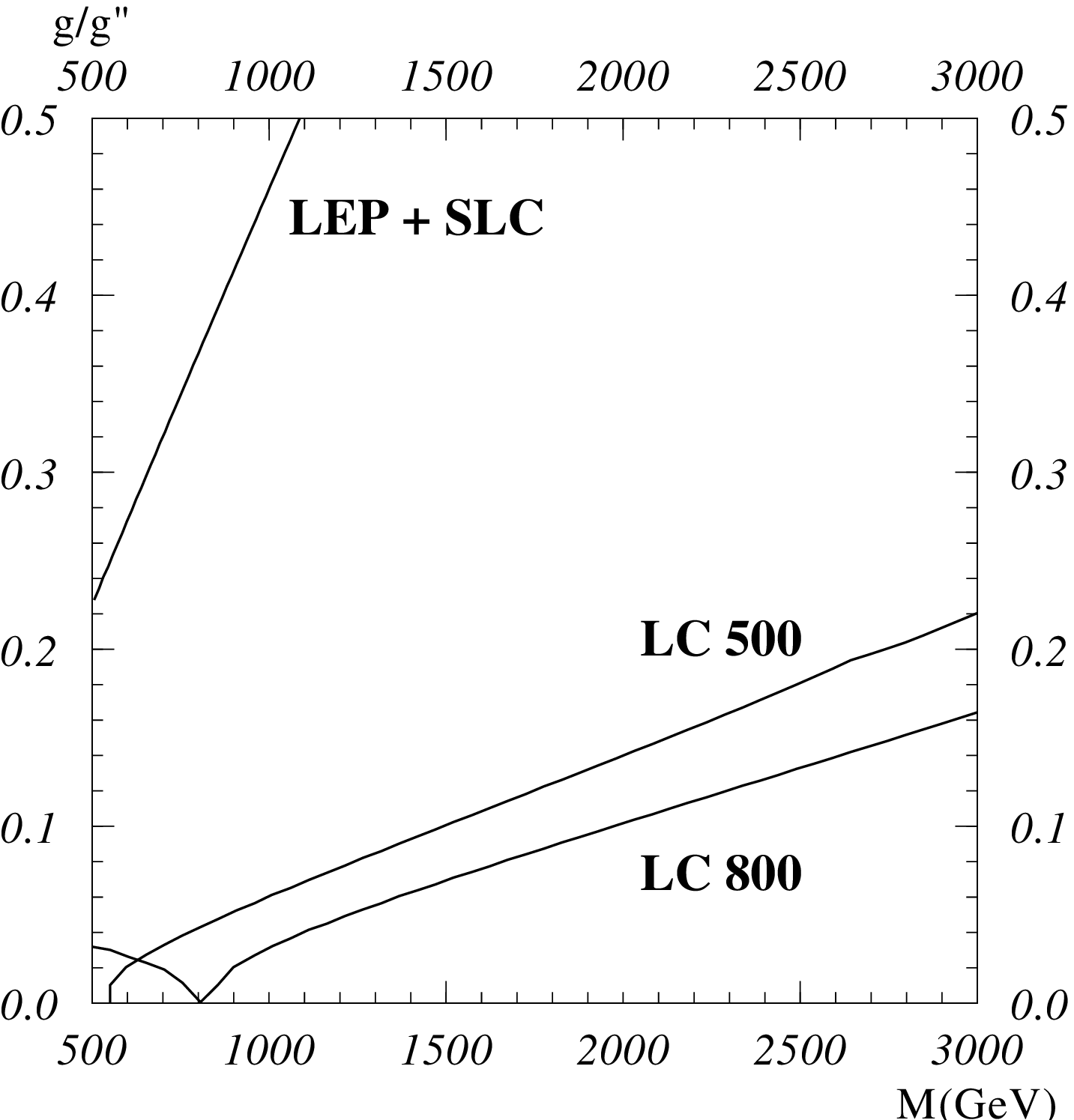}\hspace*{6mm}
\includegraphics[width=0.46\textwidth]{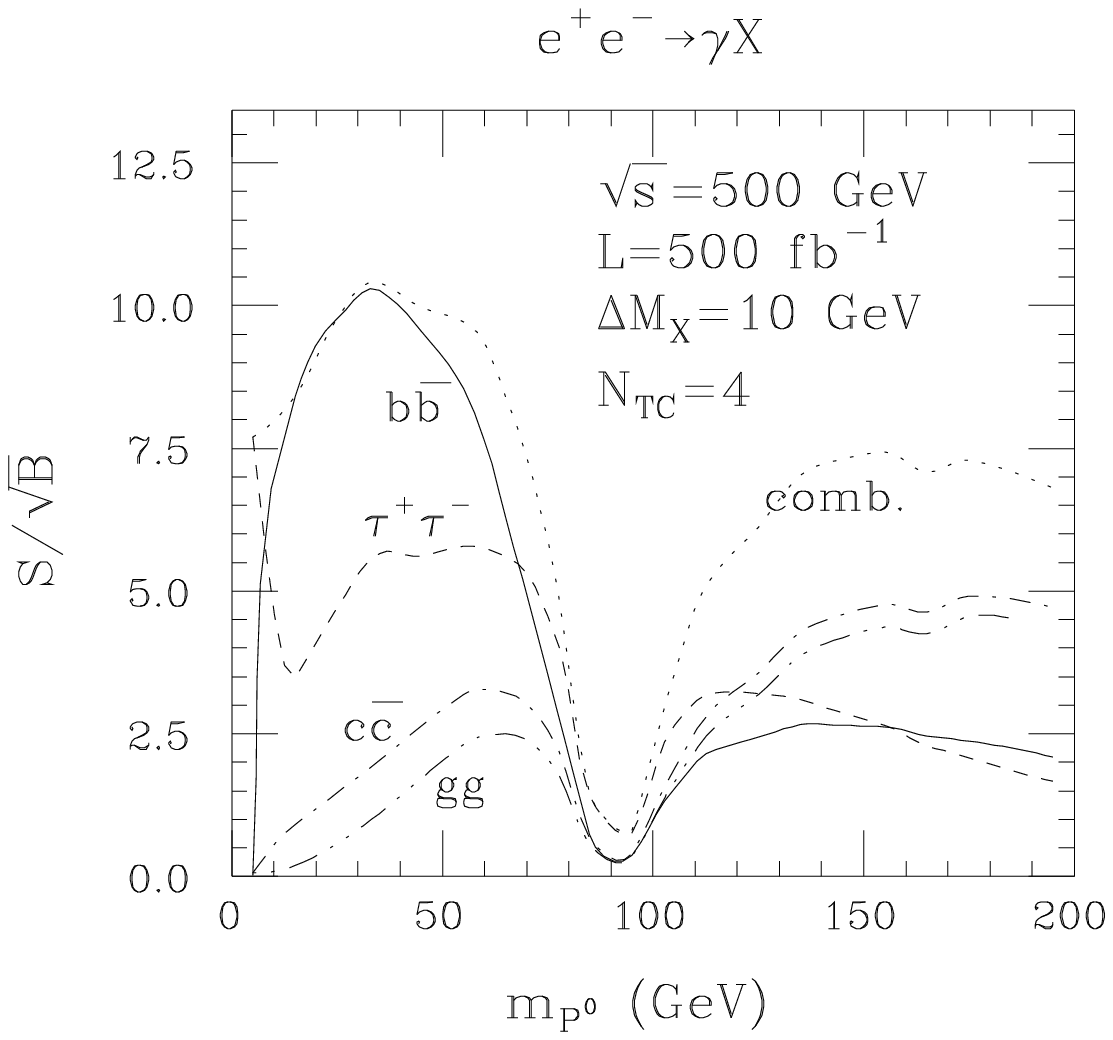}
}
\vspace*{-7mm}
\caption[Sensitivity of the ILC on the parameters and particles of the BESS model] 
{\label{fig:bess-limits}%
Left:  the 95\,\%\,C.L. contours for the BESS model parameters from the ILC  
at $\sqrt{s}=500$ and 800 GeV compared to present constraints. 
Right: statistical significance for a $P^0$ signal in various tagged channels 
as a function of $m_{P^0}$ at  $\sqrt s=500$ GeV with 500 fb$^{-1}$ data.
From Ref.~\cite{Casalbuoni:1999mm}. 
}
\label{Fig:bess}
\end{center}
\vspace*{-7mm}
\end{figure}

A concrete example of models with a strong EWSB sector is the BESS
model~\cite{A-BESS}, which includes most Technicolor models \cite{Technicolor}.
It  assumes a triplet of new vector resonances~$V^{\pm,0}$, similar to the
$\rho$ or techni--$\rho$, which  mix with the $W/Z$ bosons with  a mixing
$\propto g/g''$, where $g''$ is the self--coupling of the~$V^{\pm,0}$ state.  
The $f\bar fV^{\pm,0}$ couplings are determined by a second parameter $b$. A
variant of the model, the degenerate BESS, is when the axial and vector
resonances are almost degenerate in mass. As many scenarios of dynamical EWSB,
it predicts the presence of pseudo Nambu Goldstone bosons (PNGBs).

The vector resonances of the BESS model can be observed in  $\ee\! \to\! W^+W^-$
in the general or in $\ee\! \to\! f\bar f$ in the degenerate cases. Combining
all  possible observables in these two channels and using beam polarization, 
the sensitivity of the ILC on the parameters of the general model is larger than
the one  expected at the LHC. In the degenerate case, the ILC sensitivity is
shown in Fig.~\ref{Fig:bess} (left)  and if a resonance below 1\,TeV is observed
at the LHC, one can study it in detail and attempt to split the two nearly
degenerate resonances and measure their widths~\cite{Casalbuoni:1999mm}. In
addition, the lightest PNGB $P^0$ can be  produced at the ILC e.g. in the
reaction $\ee \to \gamma P^0$ as shown in  Fig.~\ref{Fig:bess} (right); unlike
at the LHC, low $P^0$ masses can be probed and rates for interesting decay modes
can be  measured \cite{Casalbuoni:1999mm}.

\subsection{Higgsless scenarios in extra dimensions}

Also in theories with extra space dimensions, the electroweak symmetry can  be
broken without introducing a fundamental scalar field, leading to Higgsless
theories \cite{Hless}. Since in five--dimensional theories the wave functions 
are expanded by a fifth component, the electroweak symmetry can be broken by
applying  appropriately chosen boundary conditions to this field component. 
This scalar component of the original five--dimensional gauge field is  absorbed
to generate the massive KK towers of the gauge fields in  four dimensions.  The
additional exchange of these towers in $WW$ scattering  damps the scattering
amplitude of the SM and allows, in principle,  to extend the theory to energies
beyond the ${\cal O}(1)$ TeV unitarity bound of  Higgsless scenarios. However,
it is presently unclear whether realistic models  of this type can be
constructed that give rise to small enough elastic $WW$  scattering amplitudes
compatible with perturbative unitarity \cite{ED-review}.

\begin{figure}[!h]
\vspace*{-2mm}
\centering
\includegraphics[width=70mm]{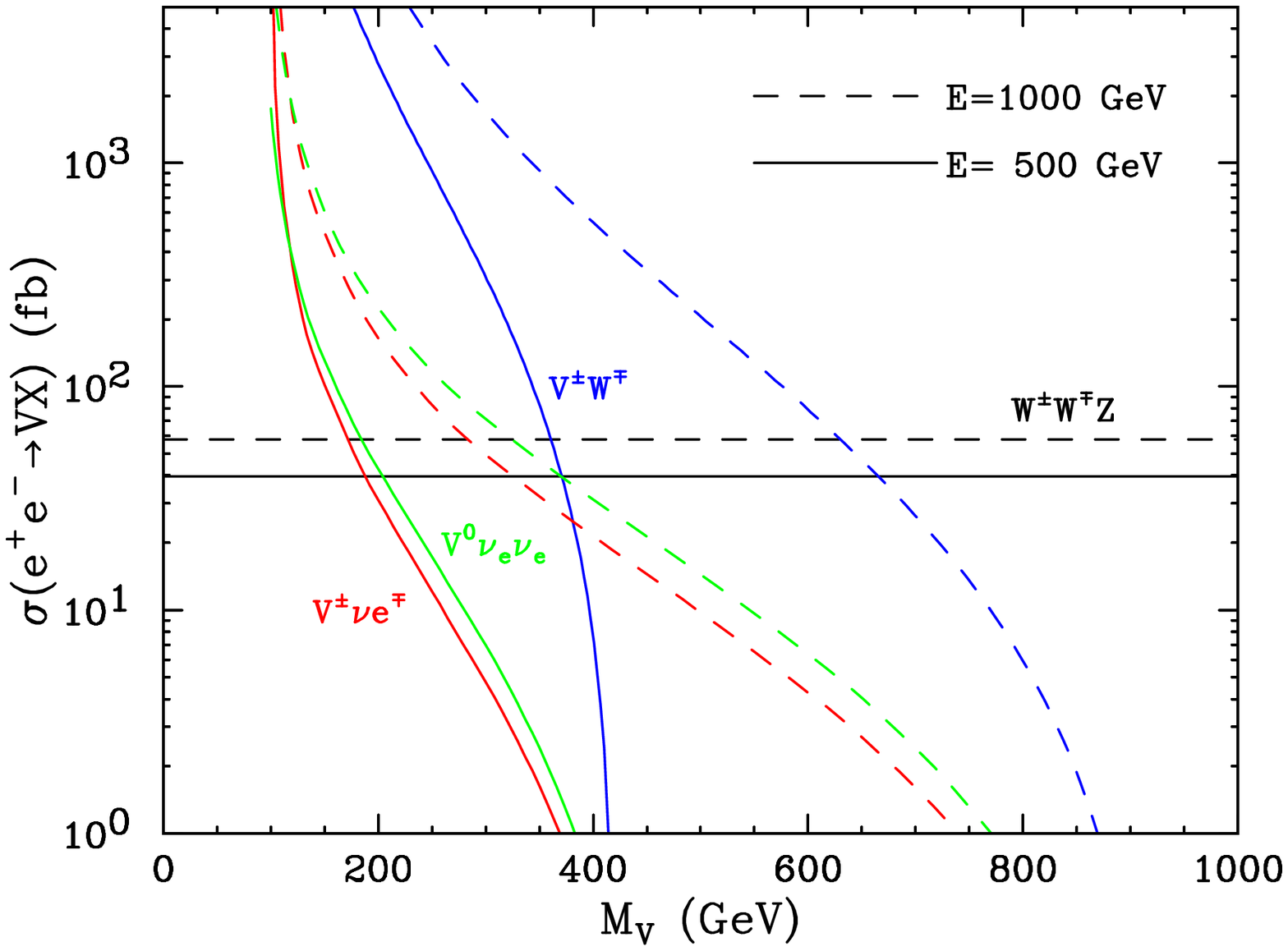}
\includegraphics[width=70mm]{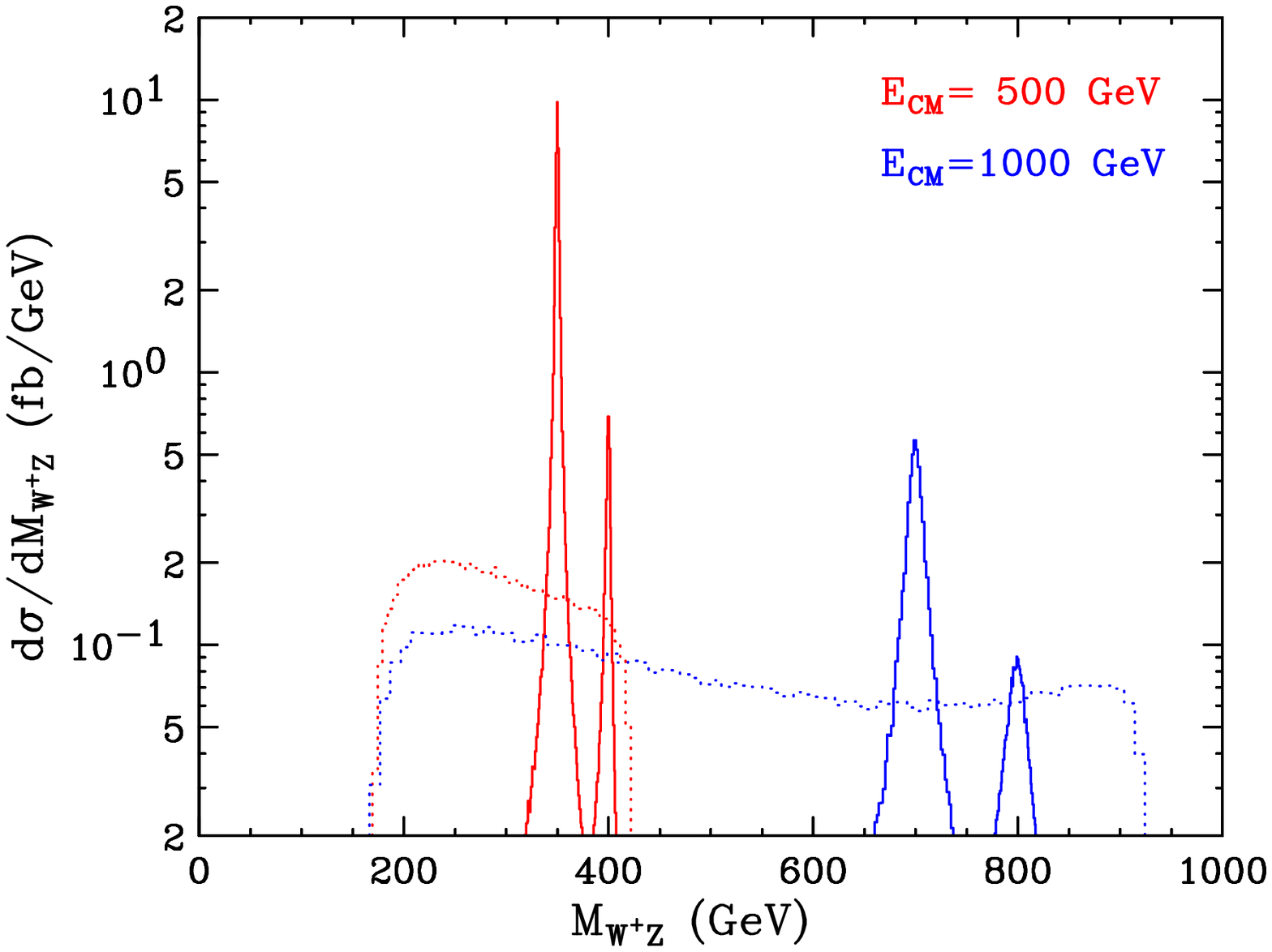}
\vspace*{-6mm}
\caption[The production of new vector bosons at the ILC in Higgsless scenarios]
{Left: the production cross sections for the new gauge bosons $V_1$ 
and the continuum SM background at the ILC. Right: the $WZ$ invariant mass 
distribution for the signal in Higgsless models and the SM background. In  
both cases, the c.m. energy is $\sqrt s= 500$ GeV and 1 TeV. From 
Ref.~\cite{Higgsles-plots}. }
\label{fig:Higgsless}
\vspace*{-10mm}
\end{figure}

Higgsless models can be best tested at the ILC if the energy is pushed to its
maximum. Unlike for Technicolor models, one expects  that the masses of the new
vector bosons, collectively called  $V_1$, are below the TeV scale and thus
kinematically accessible.  In this case, they can be produced in the $W/Z$
fusion processes $e^+e^-\to V^\pm_1 e^\mp \nu_e$ and $e^+e^-\to V^0_1 \nu_e
\bar{\nu}_e$  for the charged and neutral states, respectively. The cross
sections for these processes, as well as the one for the associated production
process $e^+e^-\to V^\pm  W^\mp$, are shown as a function of the $V_1$ mass in 
the left--hand side of Fig.~\ref{fig:Higgsless} for c.m. energies of $\sqrt s=
500$ GeV and $\sqrt s=1$ TeV  and compared to the SM $W^\pm W^\mp Z$   continuum
background \cite{Higgsles-plots}. One can see that  the rates are rather large,
exceeding the femtobarn level for $V_1$ masses close to $M_{V_1}=800$ GeV at a 1
TeV c.m. energy, before experimental cuts and   efficiencies are applied. Thanks
to the clean environment, the dominant hadronic decays of the $W/Z$ bosons can
be used and the invariant masses of the $V_1$ resonances can be easily
reconstructed. This provides an extra handle for suppressing the SM background
as shown in the right--hand side of  Fig.~\ref{fig:Higgsless} where the $WZ$
invariant mass distribution for the  signal of Higgsless models and the SM
background are compared for the same two  c.m. energies  and several values of
the  resonance masses. Thus, the ILC has a real potential to test some of the 
generic predictions of Higgsless models.

\section{New particles and interactions}

New gauge and/or matter particles, not necessarily related to the electroweak 
symmetry breaking mechanism, are predicted in many extensions of the Standard
Model. If any signals for these new particles are seen, it will be crucial to
distinguish among the variety of possible new states.  Total cross sections,
angular distributions and the final polarization  can be used to discriminate
among the different possibilities; longitudinally polarized beams allow for
additional methods to unravel the helicity structure of the new underlying
interactions. If new states are directly or indirectly accessible, the ILC will
be the ideal instrument to determine their characteristics as will be briefly
illustrated below.  

\subsection{New gauge bosons}

Gauge bosons in the intermediate TeV scale are motivated by many theoretical
approaches \cite{A-ZPreviews}.  For instance, the breaking of GUTs 
based on SO(10) or E$_6$ symmetries, may leave one or several U(1)
remnants unbroken down to TeV energies, before the symmetry  reduces
to the SM symmetry.  In the case of the ${\rm E_6}$ model, one has the possible
breaking pattern: 

\centerline{
${\rm E_6   \to SO(10) \times U(1)_\psi \to SU(5) \times U(1)_\chi
\times U(1)_\psi \to  SM \times U(1)^\prime}$} 

\noindent and the new $Z^\prime$ corresponding to the final ${\rm U(1)^\prime}$
remnant, is a  linear combination of the gauge bosons of the U(1)$^\prime$s
generated in the two--step symmetry breaking, $Z^\prime\!=\! Z_\chi \cos\beta\!
+\!Z_\psi \sin\beta$. The  value $\beta= {\rm arctan}(-\sqrt{5/3})$ would
correspond to a $Z^\prime_\eta$  originating from the direct breaking of ${\rm E_6}$
to a rank--5 group in superstrings inspired models. Another interesting option
is left--right (LR)  models, based on the group  ${\rm SU(2)_R \times SU(2)_L
\times U(1)_{B-L}}$ in which  the new  $Z^\prime_{LR}$ will couple to a linear 
combination of the right-handed and B--L currents with a parameter
$\alpha_{LR}^2 \sim  3 g_R^2/ g_L^2-1$. The value  $\alpha_{LR}\!\sim\!\sqrt 2$
corresponds  to a LR symmetric model with equal ${\rm SU(2)_R}$ and ${\rm
SU(2)_L}$ couplings, $g_R\!=\!g_L$.   As has been discussed previously, new
gauge bosons also appear in little Higgs models and, in extra--dimensional
models, the Kaluza--Klein excitations of the electroweak gauge bosons can have
masses in the range of a few TeV.

Such intermediate gauge bosons can be searched for at the LHC in the Drell--Yan 
process, $q\bar q \to Z^\prime \to \ell^+ \ell^-$ with $\ell =e,\mu$, and masses up to
about 5 TeV can be reached in general \cite{atlastdr,CMSTDR}. If $Z^\prime$ bosons are
found at the LHC, the role of the ILC will be twofold . First, by analyzing the
effect of virtual $Z^\prime$ $s$--channel exchange on the cross sections and angular
distributions of fermion pair production, $e^+e^- \to f \bar{f}$, the
sensitivity to new gauge boson scales can be extended significantly. Second, the
couplings of the new $Z^\prime$ boson to SM fermions can be determined very precisely
using  forward--backward asymmetries and the polarization dependence of the
cross sections. The various models could be then clearly  discriminated and the
nature of the underlying gauge symmetry or model could be identified.

\begin{figure}[h]
 \vspace*{-6mm} 
\centering
\mbox{\hspace*{-1cm}
\includegraphics[width=9.cm,angle=0]{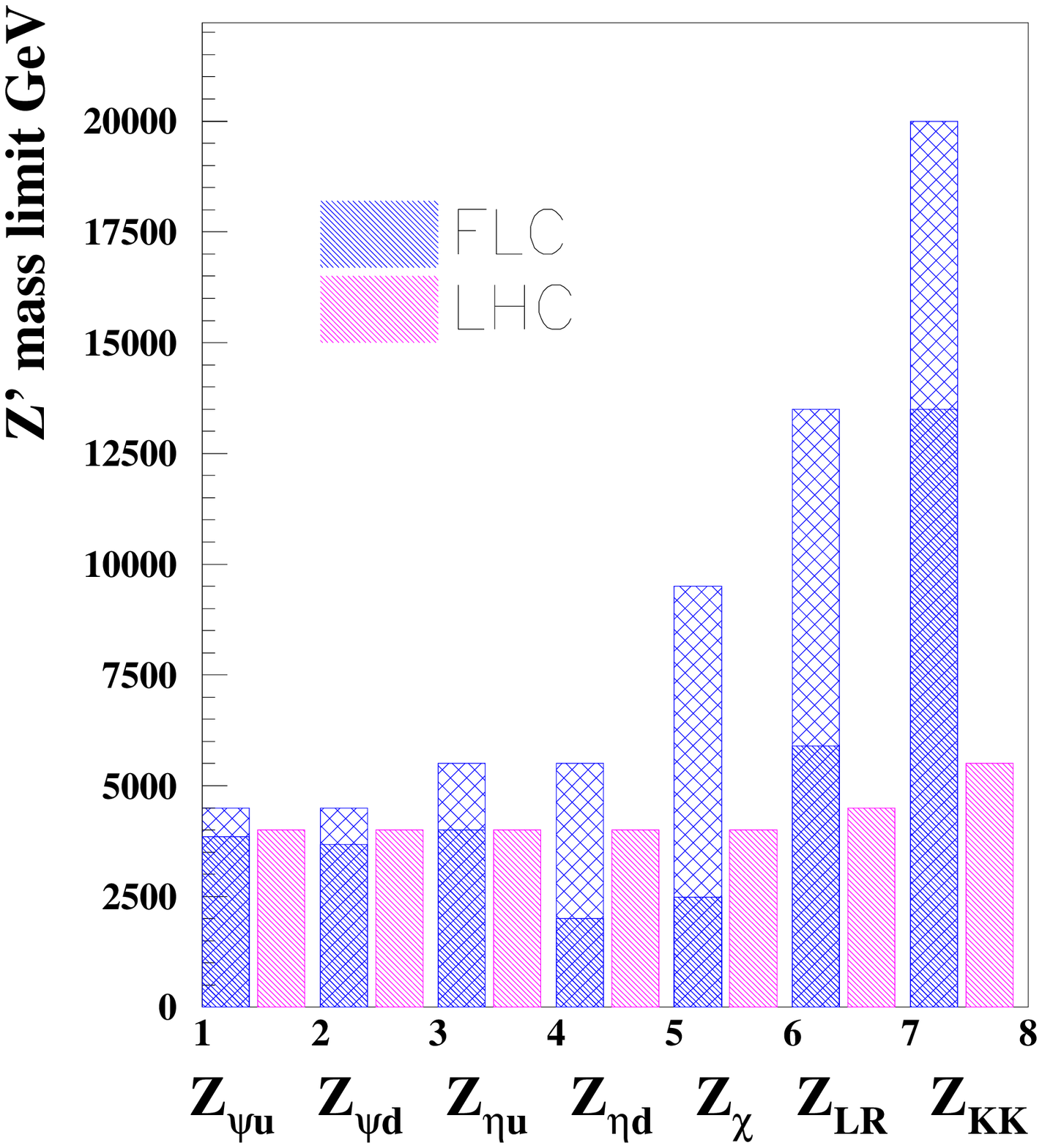}\hspace*{-3mm}
 \includegraphics[width=7.cm,angle=0]{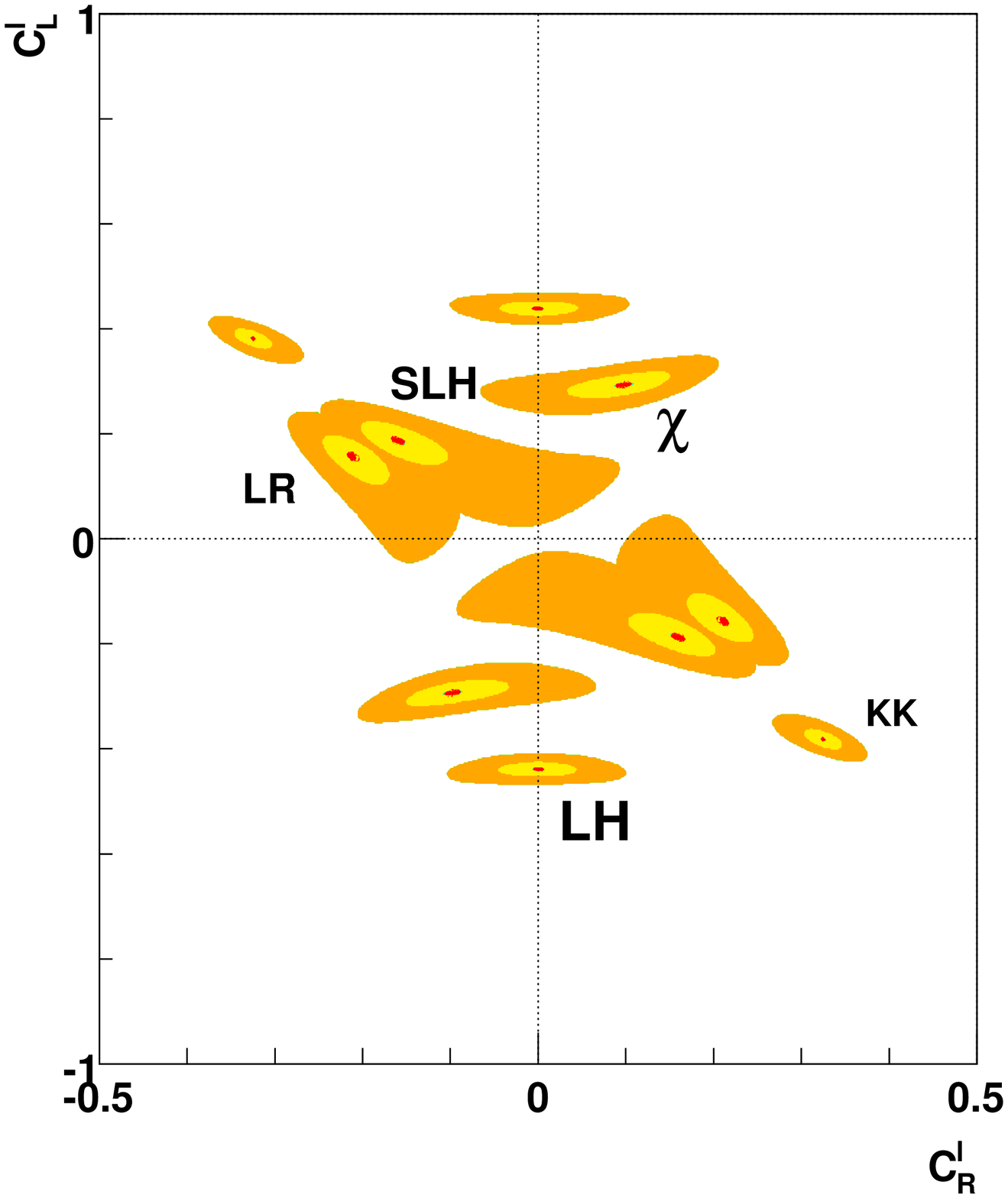} 
}
 \vspace*{-10mm}
\caption[Mass reach and coupling measurements for a heavy Z$^\prime$ boson at
the ILC]{Left: the mass range covered by the LHC and the ILC (FLC) for a 
$Z^\prime$ boson in various scenarios; for the ILC the heavy hatched region is 
covered by exploiting the GigaZ option (sensitive to the $Z$--$Z^\prime$ mixing)
and the high energy region (sensitive to the $\gamma,Z$--$Z^\prime$
interference) \cite{Weiglein:2004hn, Richard:2003vc}. Right: the ILC resolving
power (95\% CL) for  $M_{Z^\prime}=1, 2$ and 3 TeV  for left-- and right--handed
leptonic couplings ($c^l_L$ and $c^l_R$) based on the leptonic observables
$\sigma^\mu_{\rm pol}$,  $A_{LR}^\mu$  and $A_{FB}^\mu$; the smallest (largest) 
regions correspond to $M_{Z^\prime}\!=\!1$~TeV (3 TeV) \cite{Godfrey:2005pm}.
In  both figures $\sqrt{s} \!=\!500$ GeV and ${\cal L}\!=\!1$ ab$^{-1}$ are
assumed.}
\vspace*{-4mm}
\label{fig-newZ}
\end{figure}

By studying the interference between the $\gamma,Z$ and the $Z^\prime$ boson exchange
contributions in the process $\eei \to f\bar f$, the effects of the new gauge
boson can be probed for masses in the multi--TeV range \cite{A-Zpee}. Already at
a $\sqrt s=500$ GeV ILC, the mass reach is comparable to that of the LHC as
exemplified in the left--hand side of Fig.~\ref{fig-newZ} for several models.
This is particularly the case for $Z_{LR}^\prime$  boson and the KK excitations where
the mass reach exceeds 5 TeV and 10 TeV, respectively. The sensitivity will be
significantly  increased  when the ILC will be upgraded to $\sqrt s=$ 1 TeV if
the same integrated luminosity is collected.  

The $Z^\prime$ mass reach can also be further extended using the GigaZ option of
the ILC. Precision electroweak measurements at the $Z$ pole provide a
complementary information as they are sensitive to the mixing between the $Z$
and the $Z^\prime$ bosons which is expected to be proportional to  the
$Z/Z^\prime$ mass ratio.  With precisely determined top and Higgs boson masses
at the ILC, the $Z^\prime$ mass reach can be significantly larger than  the LHC
direct $Z^\prime$ search limit for some models, as also illustrated in the
left--hand side of Fig.~\ref{fig-newZ}.

In a second step, the couplings of the $Z^\prime$ boson need to be probed and the
model origin determined. An example of chiral coupling determination in several
extended models is shown in the right--hand side of Fig.~\ref{fig-newZ}. Here,
$Z^\prime$ bosons originating from the $E_6$ $\chi$ model ($\chi$), a left--right
symmetric model (LR), the littlest Higgs model (LH), the simplest little Higgs
model (SLH), and KK excitations originating from theories of extra dimensions
(KK) are considered. Only  leptonic observables have been taken into accounted
and electron and positron beam polarizations are assumed to be 80\% and 60\%,
respectively. As can be seen, for $M_{Z^\prime} = 2$ TeV, the various models can be
clearly distinguished. This is a very important step to identify the underlying
theory if a new vector resonance is observed at the LHC.

Finally, new charged gauge bosons $W^\prime$ also appear in extensions of the SM such
as left--right symmetric models. These particles can be produced at the LHC up
to masses of the order of 5 TeV in some cases. Complementing the LHC detection
of these states, the ILC could allow to reconstruct the $W^\prime$ couplings. A
detailed simulation \cite{A-Arnd} shows that  $W^\prime$ bosons can be observed via 
their virtual  effects in the process $\ee \to \nu \bar \nu \gamma$  and, at
$\sqrt s=500$ GeV with 1 ab$^{-1}$ data, masses up to  $M_{W^\prime} \sim 1.3$ TeV in
left--right models and up to  $M_{W^\prime} \sim 5$ TeV for a SM--like heavy $W^\prime$ and
the KK excitation of the $W$ boson, can be probed if the systematical errors are
assumed to be smaller than 0.1\%. The sensitivity can be slightly improved  by
considering the $e\gamma \to \nu q+X$  process in the $e\gamma$ option of the
ILC.  In the case where a heavy SM--like $W^\prime$ boson with a mass of 1.5 TeV is
observed, its couplings to quarks and leptons could be measured with an accuracy
of a few percent in some cases \cite{A-Arnd}. 

\subsection{Exotic fermions}

Many theories beyond the SM such as GUTs or extra--dimensional require the
existence of new matter particles with the possibility of new interactions not
contained in the SM; for a review, see e.g.  Ref.~\cite{A-NFreviews}.  Examples
of new elementary fermions include sequential fourth generation fermions,
vector--like fermions with both left-- and right--handed components in weak
isodoublets, mirror fermions which have the opposite chiral properties as the SM
fermions and isosinglet fermions such  as the SO(10) Majorana neutrino. Exotic
fermions, i.e. fermions that have the usual  lepton/baryon  but  non-canonical
${\rm SU(2)_L\times U(1)_Y}$  quantum numbers,  occur naturally in GUT models
that contain a single representation into which a complete generation of SM
quarks and leptons  can be embedded. For instance, in the  E$_6$ group, each
fermion generation lies in the {\bf 27} representation, which contains 12 new
fermions in addition to the 15 chiral fermions of the SM. It is conceivable that
these new fermions acquire masses not much larger than the EWSB scale, if these
masses are protected by some symmetry.  In fact, this is necessary if the
associated new gauge bosons are relatively light.

\begin{figure}[hbt]
\begin{center}
\vspace*{-5mm}
\mbox{
\includegraphics[height=5.99cm,angle=-90, bb= 29 175 546 695]{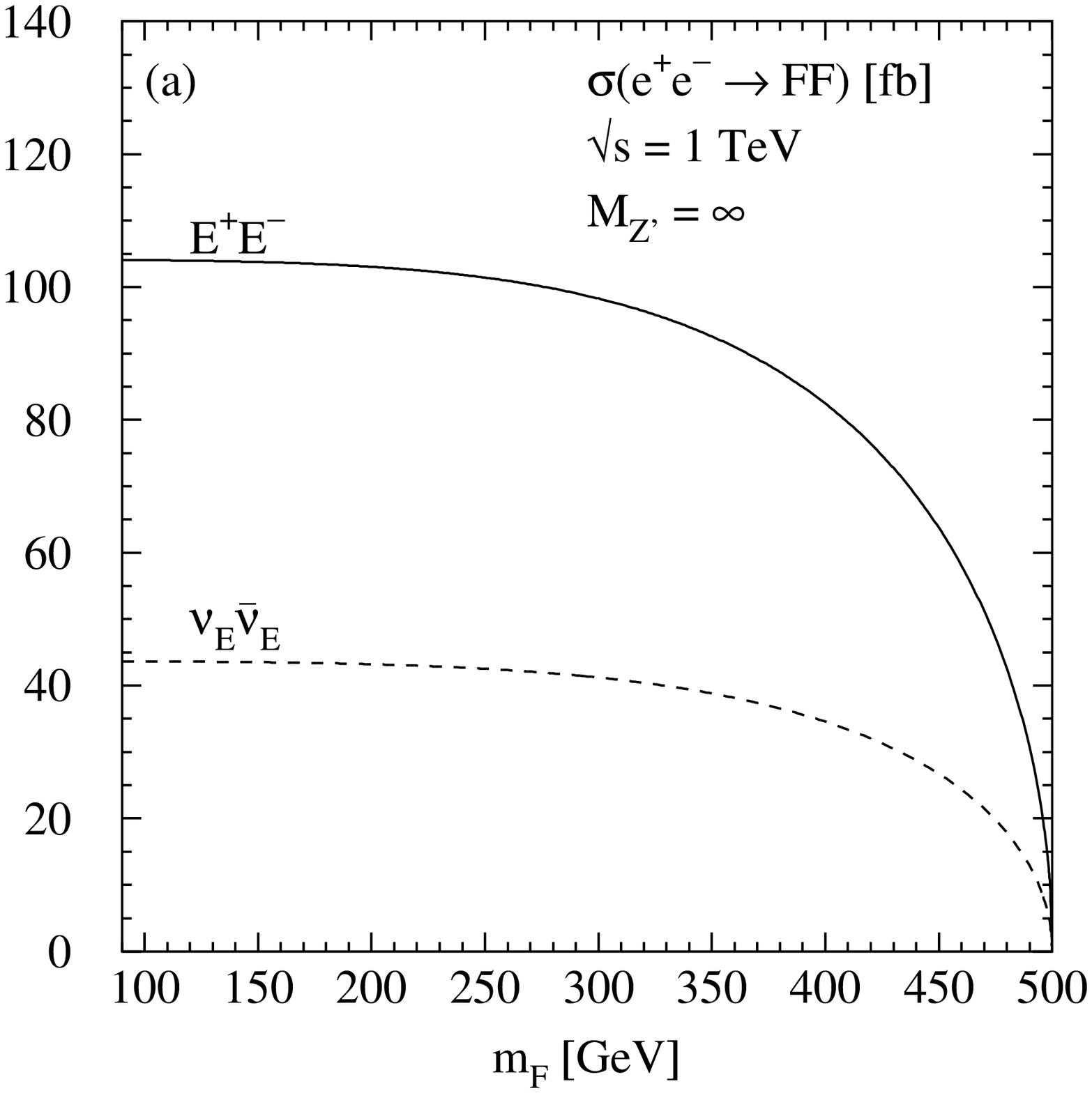}
\hspace*{1cm}
\includegraphics[height=5.99cm,angle=-90, bb= 29 175 546 695]{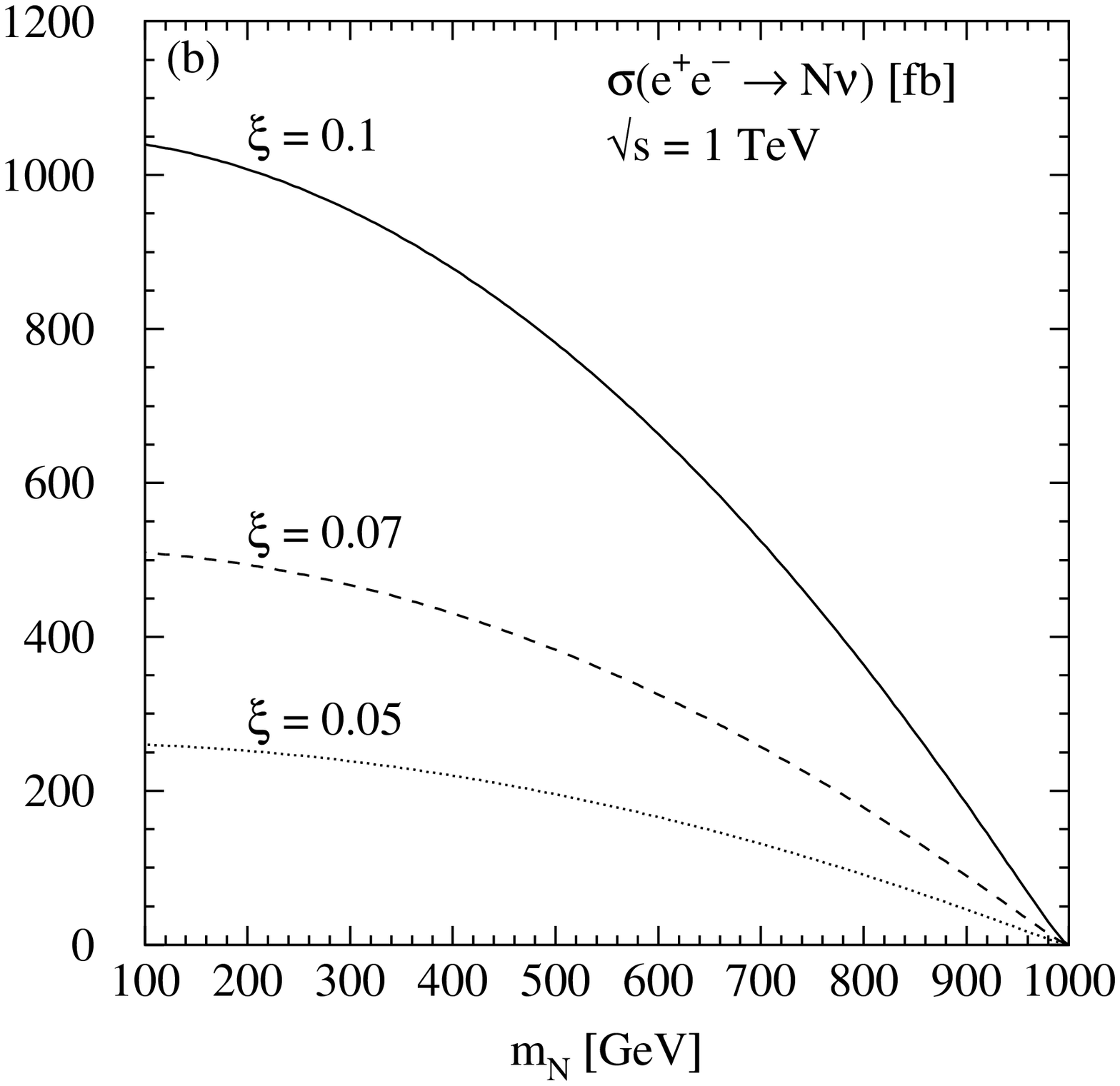}
}
\end{center}
\vspace*{-9mm}
\caption[Cross sections for pair and single production of new leptons at the ILC]
{The production  cross sections for new heavy leptons at $\sqrt s=1$ TeV:  
pair production (left) and single neutrino production for various mixing 
angles (right). From Ref.~\cite{A-exotics-fig}.   }
\label{fig:E6fermion}
\vspace*{-2mm}
\end{figure}

Except for singlet neutrinos, the new fermions couple to the photon and/or to
the weak gauge bosons $W/Z$ with full strength; these couplings allow for pair
production, $\ee \to F \bar F$, with practically unambiguous cross sections and,
masses very close to the kinematical limit, $m_F \sim \frac12 \sqrt s$,  can be
probed; see Fig.~\ref{fig:E6fermion} (left). In general, the new fermions will
mix with their SM light partners  which have the same conserved quantum numbers.
This mixing, which is expected to be small $\xi \lsim 0.1$ from LEP constraints,
gives rise to new currents which determine the decay  pattern of the heavy
fermions, $F \to fZ/ f^\prime W$. 

The  mixing also allows for the single production of the new fermions, $\ee \to
F\bar f$. In the case of  quarks and second/third generation leptons, single
production proceeds only via $s$--channel $Z$ exchange and the rates are
moderate. For the first generation neutral and charged leptons, one has
additional $t$--channel exchanges which  significantly increase the production
cross sections; see Fig.~\ref{fig:E6fermion} (right).  For not too small mixing,
lepton masses close to the center of mass energy can be  produced. A full
simulation  \cite{A-exotics} of the  processes $\ee\!\to\!  N\nu_e\!\to\! e^\pm
W^\mp \nu_e$ and $\ee\!\to\! E^\pm e^\mp\! \to\!  e^\pm  e^\mp Z$,  taking into
account the dominant backgrounds and detector efficiencies, shows that for
$M_{N,E}=350$ GeV , mixing angles down to $\xi  \sim 0.002$ and $0.01$ can be
probed at a 500 GeV ILC with 500 fb$^{-1}$ data in, respectively, the neutral
and charged lepton case.

\subsection{Difermions}

Difermions are new spin--zero or spin--one bosons that have unusual baryon
and/or lepton quantum numbers \cite{A-NFreviews}. Examples are leptoquarks  with
B$=\pm 1/3$ and L$=\pm 1$, diquarks with B$=\pm 2/3$ and L$=0$ and dileptons
with B$=0$ and L$=\pm2$. They occur  in models of  fermion compositeness as well
as in some GUT models such as $E_6$ where a  colored weak isosinglet new
particle can be either a leptoquark or a diquark. In the case of leptoquarks,
starting from an effective Lagrangian with general
SU(3)$\times$SU(2)$\times$U(1) invariant couplings and conserved  B and L
numbers, one obtains 5 scalar and 5  vector states with distinct SM
transformation properties. In addition to the usual couplings to gauge bosons,
difermions have couplings to fermion pairs which determine their decays. In
supersymmetric models with R--parity violation, the scalar partners of sfermions
may be coupled to two fermions giving rise to production and  decay mechanisms
that are analogous to those of difermions.

\begin{figure}[!h] 
\vspace*{-2.4cm}
\begin{center}
\begin{minipage}{7cm}
\hspace*{-1.5cm}
\includegraphics[height=12.5cm]{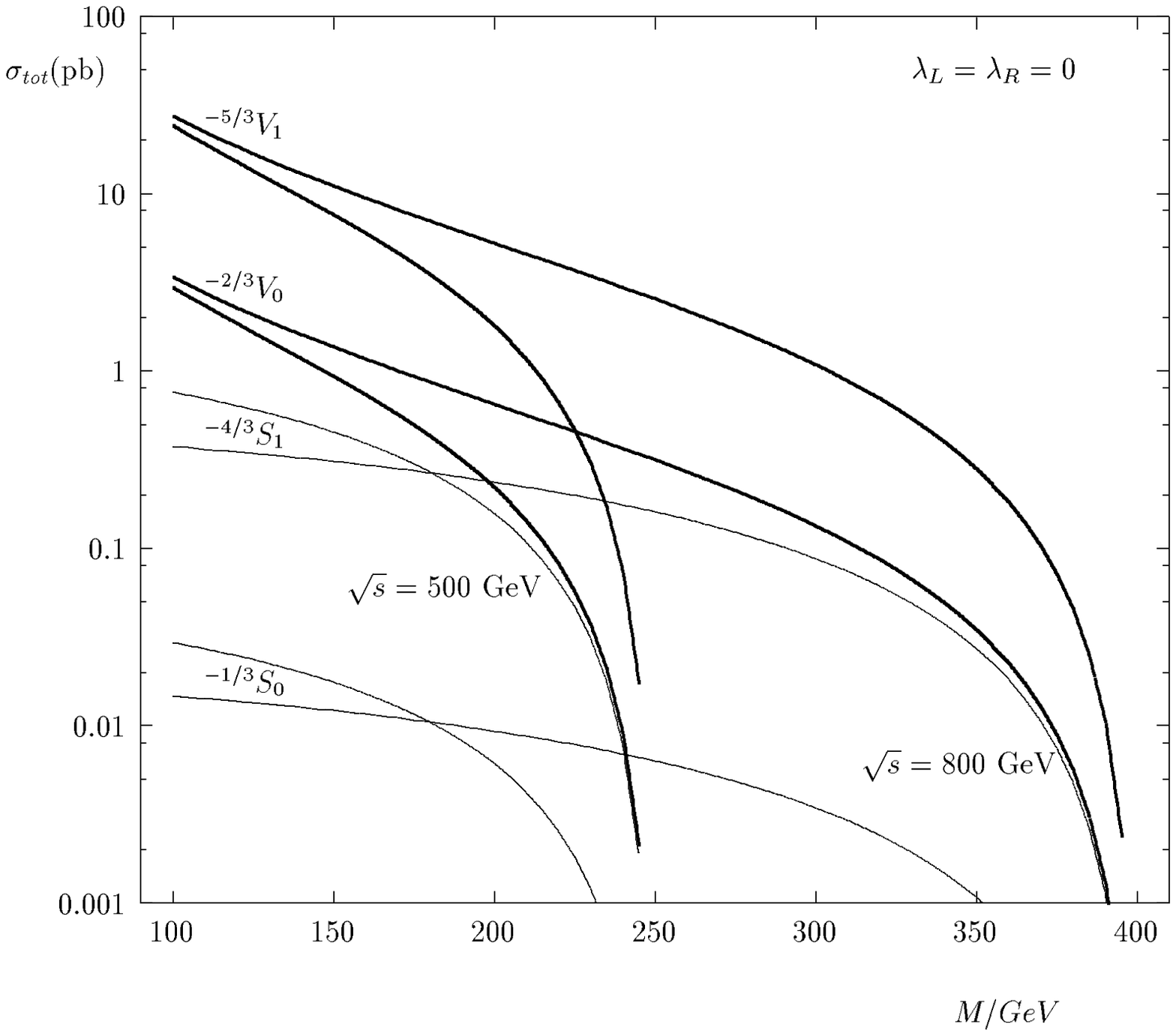}
\end{minipage}
\begin{minipage}{7cm}
\hspace*{-.9cm}
\includegraphics[height=9cm,angle=-90]{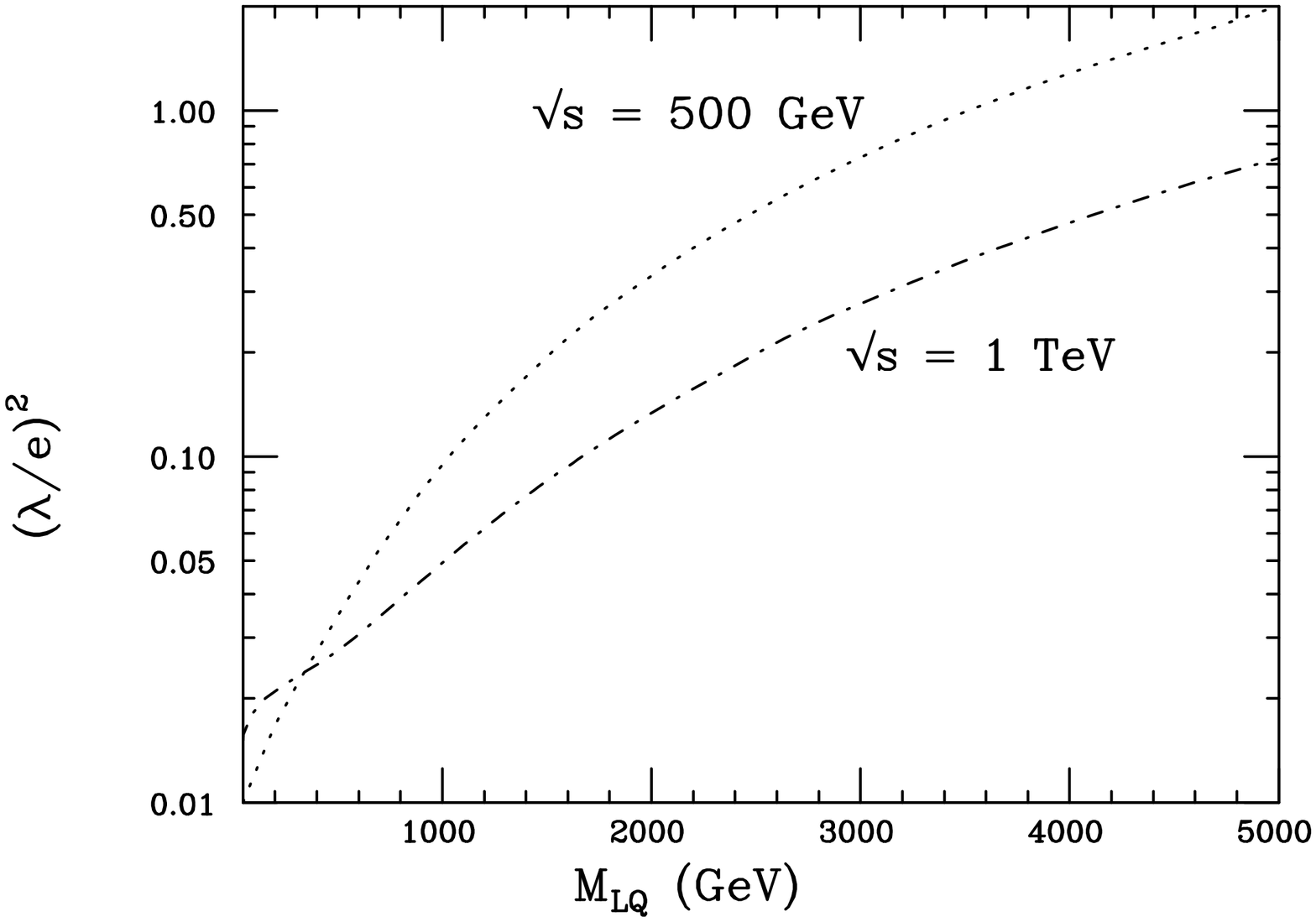}
\vspace*{2cm}
\end{minipage}
\end{center}
\vspace*{-5.3cm}
\caption[Direct production and indirect effects of heavy leptoquarks at the ILC]
{Left: total cross sections for various leptoquark pair production at 
the ILC with $\sqrt s=500$ and 800 GeV with vanishing Yukawa couplings and 
including the  corrections due to beamstrahlung and ISR \cite{LQ-ruckl}.
Right: $95\%$ CL indirect $^{-1/3}S_0$ leptoquark discovery regions (to the 
left of the curves) at $\sqrt=500$ GeV and 1 TeV with 50 and 100 fb$^{-1}$ 
data \cite{LQ-joanne}. 
}
\label{Fig:LQs}
\vspace*{-.3cm}
\end{figure}

Leptoquarks can be produced in pairs at $e^+e^-$ colliders
\cite{LQ-rizzo,LQ-ruckl} through gauge boson $s$--channel exchange; significant
$t$-channel quark exchange can also be present in some channels if the
quark--lepton--leptoquark coupling squared  $\lambda^2/e^2$ are not too small.
Depending on  the charge, the spin and isospin of the leptoquark, the cross
sections  can vary  widely as shown in the left--hand side of Fig.~\ref{Fig:LQs}
for $\sqrt s= 500$ and 800  GeV.  In a detailed simulation, it has been shown
that scalar and vector leptoquark masses very close to the beam energy can be
detected with the exception of the $^{-1/3}S_0$ state which can be probed only
for masses $\sim 40\%  \sqrt s$ because of the lower cross section
\cite{LQ-ruckl}. Once the leptoquarks have been observed, besides the total
cross sections, the study of the angular distribution gives an additional handle
on the spin and the relative size of the couplings to gauge bosons and fermions.

Single production of scalar and vector leptoquarks  can also occur
\cite{LQ-rizzo}, in particular for those states coupling to first generation
leptons which can be produced with large rates in $e \gamma$ initiated 
subprocesses. Though suppressed  by the unknown Yukawa coupling to quark--lepton
pairs $\lambda/e$,  these processes could allow to extend the kinematical reach
to masses up to $\sim \sqrt s$.  First generation leptoquarks can also be
observed at the ILC in $e \gamma$  option: the rates are much larger than in the
$\ee$ option but the mass reach is slightly lower due to the reduced energy.

One can also indirectly probe the existence of very heavy leptoquarks that are
not kinematically accessible at a given c.m. energy in the $e^+e^-\!  \to\!
q\bar q$ process as $t$--channel leptoquark  exchange can contribute 
significantly to the cross section, provided the Yukawa coupling is sufficiently
large. From the total cross section and angular distribution measurements at
$\sqrt s=500$ GeV, one can probe the E$_6$ leptoquark $^{-1/3}S_0$ for  $M_S
\sim 4$ TeV and $\lambda/e \sim 1$ with only 50 fb$^{-1}$ data as shown in  
Fig.~\ref{Fig:LQs} (right) \cite{LQ-joanne}. The effects of a  2 TeV state with
couplings as low as  $\lambda/e \sim 0.1$, can be probed at $\sqrt s=1$ TeV and 
${\cal L}=100$ fb$^{-1}$.

Dileptons, like doubly charged Higgs bosons, would lead to the spectacular  four
lepton signature if they are pair produced, $\ee \to L^{++} L^{- -} \to
4\ell^\pm$. Because of the large electric charge $Q_{\ell \ell}=2$, the rates
are significant in the $\eei$  mode of the ILC and even more in the $\gamma
\gamma$ mode as $\sigma \propto Q_{\ell \ell}^4$. They can also be singly
produced and, in particular, they could appear as $s$--channel resonances in
$e^- e^-$ collisions for mass close to the c.m. energy. Diquarks can be pair
produced in $e^+e^-$  collisions for masses smaller than $\frac 12 \sqrt{s}$ and
lead to an excess of four--jet events which could be easily searched for in
contrast to the LHC.  

\subsection{Compositeness}

As a possible physical scenario, strongly interacting electroweak bosons at
energies of order 1 TeV  could be interpreted as a signal of composite
substructures of these particles at a scale of $10^{-17} $~cm.  Moreover, the
proliferation of quarks and leptons could be taken as evidence for possible
substructures in the fermionic sector.  In this picture, masses and mixing
angles are a consequence of the interactions between a small number of
elementary constituents, in analogy to the quark/gluon picture of hadrons.
Although no satisfactory theoretical formalism has been set up so far, one can
describe  this scenario in a  purely phenomenological way. 

Compositeness in the fermion sector can be tested at the ILC through the
measurement of the $\ee \to q\bar q$ and $\ell^+  \ell^-$ cross sections and
asymmetries and the search for four--fermion contact interactions generated  by
the exchange of the fermion subconstituents. As discussed in chapter
\ref{sec:couplings},  compositeness scales $\Lambda$ up to 100 TeV can be probed
at the ILC; Fig~\ref{fig:contact}. 

The existence of excited fermions is a characteristic signal of substructure in
the fermionic sector: if the known fermions are composite objects, they should
be the ground state of a rich spectrum of excited states which decay down to the
former states via a magnetic dipole type de--excitation. In this case, decays to
a light partner fermion and a photon  with branching ratios of the order of 30\%
is possible. These decays constitute a characteristic signature of excited
fermions and  discriminate them from the exotic fermions discussed above. 

The pair production of excited fermions \cite{A-excited} follows the same
pattern  as for the exotic fermions and, for excited leptons, the cross sections
are similar to those shown in Fig.~\ref{fig:E6fermion} (left) generating event
samples that allow for an easy discovery of these states for masses smaller than
the beam energy.  Single production of excited fermions at the ILC
\cite{A-excited} is also  similar to that of exotic fermions,  with the notable 
exception  of single production of excited electrons which, in  $\eei$
collisions,  is strongly enhanced  by $t$--channel photon exchange. This state
can also be produced as an $s$--channel resonance in $e \gamma$ collisions. The
single production of  excited electronic neutrinos in $\eei$ collisions is also
enhanced by $t$--channel $W$ exchange and leads to the interesting signature of
an isolated monochromatic photon and missing energy.

%

\chapter{Connections to cosmology}
\label{sec:cosmology}

Dark matter has  been established as a major component of the universe. We know
from several independent observations, including the cosmic microwave 
background, supernovas  and galaxy clusters, that DM is responsible  for $\sim
20\%$ of the energy density of the universe. Yet, none of the  SM particles can
be responsible for it and the observation of DM, together  with neutrino
masses,  is likely the  first direct signal of new physics beyond the SM.
Several particles and objects have been nominated as candidates for DM. They
span a wide range of  masses, from 10$^{-5}$~eV, in the case of axions, to
10$^{-5}$~solar masses, for  primordial black holes. Cosmology tells us that a
significant fraction of the  universe mass consists of DM, but does not provide
clues on its nature. Particle  physics tells us that new physics must exist at,
or just beyond, the electroweak scale  and new symmetries may result in new,
stable particles. Establishing the  inter--relations between physics at the
microscopic scale and phenomena at  cosmological scale will represent a major
theme for physics in the next decades.  

The ILC will be able to play a key role in elucidating these inter--relations.
Out of these many possibilities, there is a class of models which is especially
attractive since its existence is independently  motivated and DM, at about the
observed density, arises naturally. These are  extensions of the SM which
include an extra symmetry protecting the lightest  particle in the new sector
from decaying into ordinary SM states.  The lightest particle becomes stable and
can be chosen to be  neutral. Such a particle is  called a weakly interacting
massive particle (WIMP) and arises in theories beyond the SM, such as
supersymmetry with conserved R---parity  but also in extra dimensional models 
with KK--parity. 

Current cosmological data, mostly  through the WMAP satellite measurements of
the CMB, determine the DM density  in the universe to be

\centerline{$\Omega_{\rm DM}\, h^2  = 0.111 \pm 0.006 \, , $}

\noindent  which is already a determination to 6\% accuracy. The accuracy is
expected to be improved to the percent level by future measurements by the
Planck satellite \cite{Planck}.  The next decades promise to  be a time when
accelerator experiments will provide new breakthroughs and highly  accurate data
to gain new insights, not only on fundamental questions in particle physics, but
also in cosmology, when studied alongside the observations from  satellites and
other experiments. The questions on the nature and the origin of  DM offer a
prime example of the synergies of new experiments at hadron and lepton 
colliders, at satellites and ground--based DM experiments. In this context, the
ILC will play  a major role as will be discussed here. 

Explaining the baryon asymmetry of the universe is another outstanding problem
in cosmology. Both the WMAP experiment and the theory of primordial
nucleosynthesis indicate that the baryon-to-entropy ratio of the present
universe is $\sim 10^{-10}$. This asymmetry has to be created after the
inflationary period which likely occurred in the evolution of the universe. In
order to generate the baryon asymmetry after inflation, the three Sakharov
conditions are required, namely, baryon number violation, C and CP violation and
a deviation from thermal equilibrium \cite{Sakharov:1967dj}.  Two main
approaches for generating the baryon asymmetry in our universe have been
proposed: baryogenesis mediated by leptogenesis and electroweak baryogenesis.
Both options need the introduction of new physics beyond the SM and can be
formulated in the context of supersymmetric models. This is, therefore,  another
aspect that is highlighting an interface between collider particle physics and
cosmology. Also in  this fundamental issue,  the ILC might play a key role.  

\section{Dark matter}

\subsection{DM and new physics}

Since there is no WIMP candidate within the SM, cold DM is a clear evidence for
physics beyond the SM and  in chapters \ref{sec:susy} and
\ref{sec:alternatives}, we discussed SM extensions in which appropriate DM
candidates exist. These particles are in  general  electrically neutral,
relatively massive and absolutely stable;   in addition, they have rather weak
interactions in such a way that  their cosmological relic density, which is
inversely proportional to their annihilation cross section $\sigma_{\rm ann}
\equiv \, \sigma( {\rm WIMP+ WIMP}  \rightarrow {\rm SM\, particles})$, falls in
the range required by WMAP.

\underline{Supersymmetry:} a standard way to suppress unwanted interactions
leading to unreasonable proton decay rates in SUSY models is to impose R-parity.
By virtue of this symmetry, the lightest supersymmetric particle (LSP) is
absolutely  stable and represents a good candidate for cold DM
\cite{DM-Ellis,DM-review}. In particular, the lightest neutralino is considered
to be the prime candidate, but other interesting possibilities  are the axino
and the gravitino. A detailed description of SUSY dark matter is given in the
next two sections.

\underline{Models of extra dimensions:} which introduce a KK tower for each SM
particle. In universal extra--dimensional (UED) models, a discrete quantity
called KK--parity is conserved so that the lightest KK particle (LKP), generally
corresponding to the KK  hypercharge gauge boson, is stable and is a DM
candidate \cite{Hooper:2007qk,LKP-heavy}. In warped Randall--Sundrum (RS) models
embedded in GUTs, a $Z_3$ symmetry ensures also that the lightest KK state
(LZP), the excitation of a Dirac right--handed neutrino, could be stable and a 
good DM candidate \cite{DM-LZP} as a result of  a baryon number symmetry. These
two options will be briefly discussed here.

\underline{Little Higgs models:} in a class of which, a discrete symmetry called
T--parity can be introduced \cite{DM-LTP1} which  forbids direct interactions
between new heavy gauge bosons and ordinary fermions. The lightest T--odd
particle (LTP) is a heavy partner of a U(1) gauge boson and  is a good DM
candidate \cite{DM-LTP2}; in this respect, these models are four--dimensional
reminiscent of UED models mentioned above. Note, however, that it has been
recently pointed out that T--parity might be broken by anomalies in some cases
\cite{DM-anomaly}.   

As in these examples, a new continuous or discrete symmetry has to be introduced
in order that a new physics model incorporates an electrically neutral particle
that is absolutely stable to be an appropriate DM candidate. If thermal
production of these particles is assumed in the early universe, their mass  and
their interactions, which enter in the annihilation cross section, are
constrained by the relic density.  In most cases, the resulting mass range turns
out to be roughly in the vicinity of the electroweak symmetry breaking  scale.
It is therefore generally expected that such DM particles can be detected at the
LHC in the decay products of the new colored particles that are also present in
the new physics model and which can be copiously produced \cite{DM-LHC}. A
characteristic signal of DM particle production is, thus, cascade decays with
large missing transverse energy due to the escaping WIMPS, just as in the SUSY
case. In order to distinguish between different possibilities and identify 
unambiguously the DM particle, one needs to determine its mass, spin and other
quantum numbers  as well as the model parameters that are relevant in the
calculation of its thermal relic abundance and its detection rates in
astrophysical experiments. In fact, there are four main steps in the physics
program which allows  for a complete understanding of the nature of the DM
candidate:

\begin{itemize} 
\vspace*{-2.5mm} 
\item[$\bullet$] discover the WIMP candidate in collider physics experiments in 
missing energy events (and in direct detection experiments) and measure
precisely their mass, 

\vspace*{-2.5mm}  
\item[$\bullet$]  determine the physics of the new  model that leads to  the
WIMP, 

\vspace*{-2.5mm} 
\item[$\bullet$] determine precisely the parameters of this model  and predict
the relic density as well as the direct and indirect detection cross sections in
astrophysical experiments,

\vspace*{-2.5mm}
\item[$\bullet$] observe the DM particle in astroparticle physics experiments
and measure products of cross sections and densities to reconstruct the density
distribution of DM. 
\vspace*{-2.5mm}
\end{itemize}

This ambitious program of precision measurements should reveal what the DM
particle is and how it is distributed in the universe. If the determination  of
the properties of the DM particle matches cosmological observations to high
precision, then (and only then) we will be able to claim to have determined what
DM is.  Such an achievement would be a great success of the particle
physics/cosmology connection and would give us confidence in our understanding
of the universe.

The high  precision measurements to be performed at the ILC  will play a
significant role in this context. This is demonstrated for SUSY dark matter in
the following sections.

\subsection{SUSY dark matter}

In the MSSM, the LSP neutralino  is an ideal cold DM candidate
\cite{DM-Ellis,DM-review}.  In some areas of the SUSY parameter space, the
$\chi_1^0$ cosmological relic density falls in the range required by WMAP. In
particular, in the constrained MSSM, there are generally four  regions in which
this constraint (together with the constraints from collider physics) is
satisfied \cite{DM-review}:  

$1)$ Scenarios where both $m_0$ and $m_{1/2}$ are rather small, the ``bulk
region'', are most natural from the point of view of EWSB but are severely
squeezed by bounds from colliders searches.  

$2)$ The ``focus point''  region occurs at $m_0 \gg m_{1/2}$, and allows
$\chi_1^0$ to have a significant higgsino component, enhancing its annihilation
cross sections into final states containing gauge and/or Higgs bosons; this
solution generally requires  multi--TeV scalar masses. 

$3)$ In the ``co--annihilation'' region, one has near mass degeneracy between
the LSP and the lightest stau $m_{\chi_1^0} \simeq m_{\tilde  \tau_1}$, leading
to enhanced destruction of sparticles since the $\tilde  \tau_1$ annihilation
cross section is much larger than that of the  LSP; this requires $m_{1/2} \gg
m_0$.  

$4)$ If $\tan\beta$ is large, the $s-$channel exchange of the CP--odd Higgs
boson $A$ can become nearly resonant, the ``$A$--funnel'' region,  again leading
to an acceptable relic density. 

\begin{figure}[h!]
\vspace*{-.3cm}
\begin{center}
\hspace*{.4cm}
\mbox{\includegraphics[height=2.3in,width=2.3in]{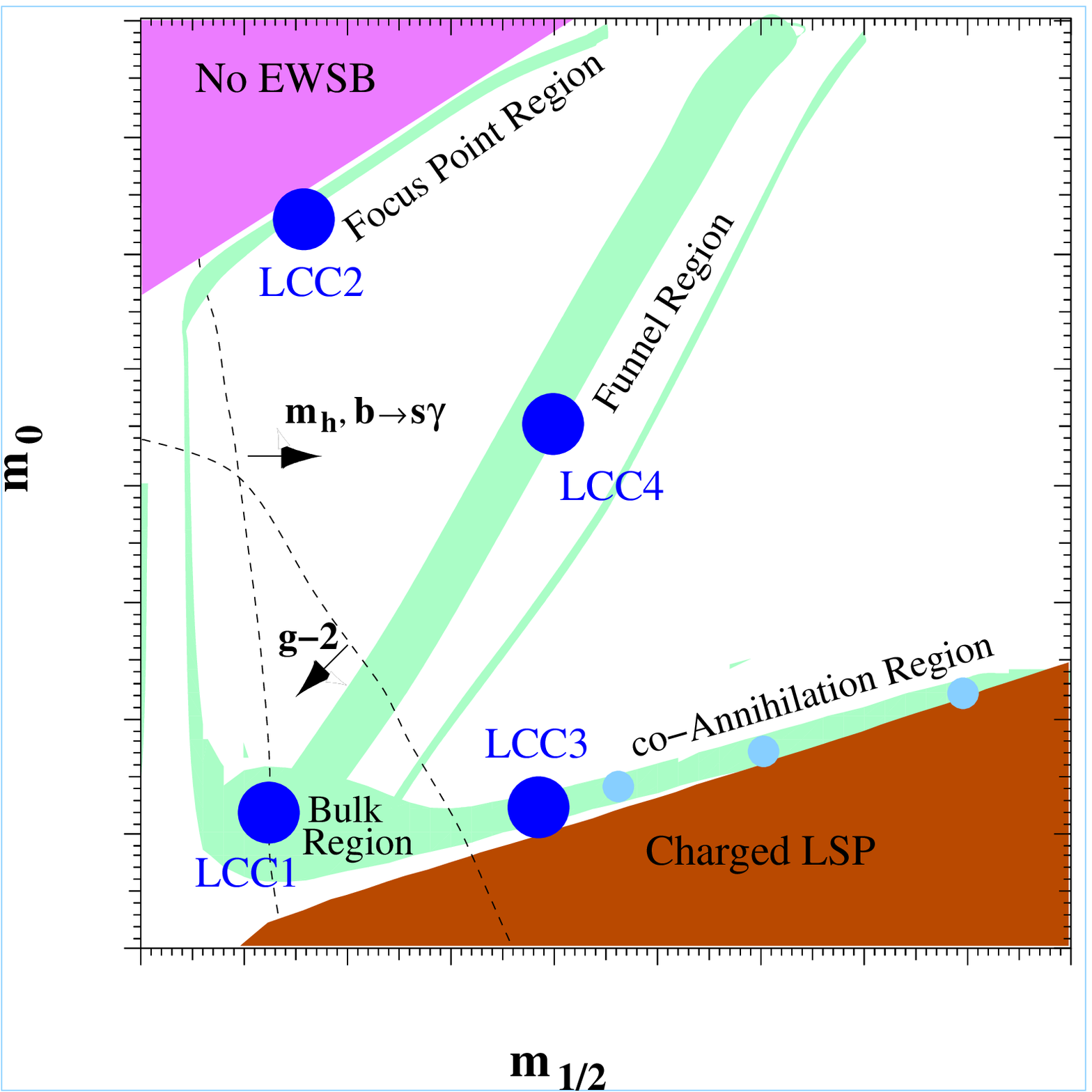}
\hspace*{1cm}
\includegraphics[width=0.5\textwidth]{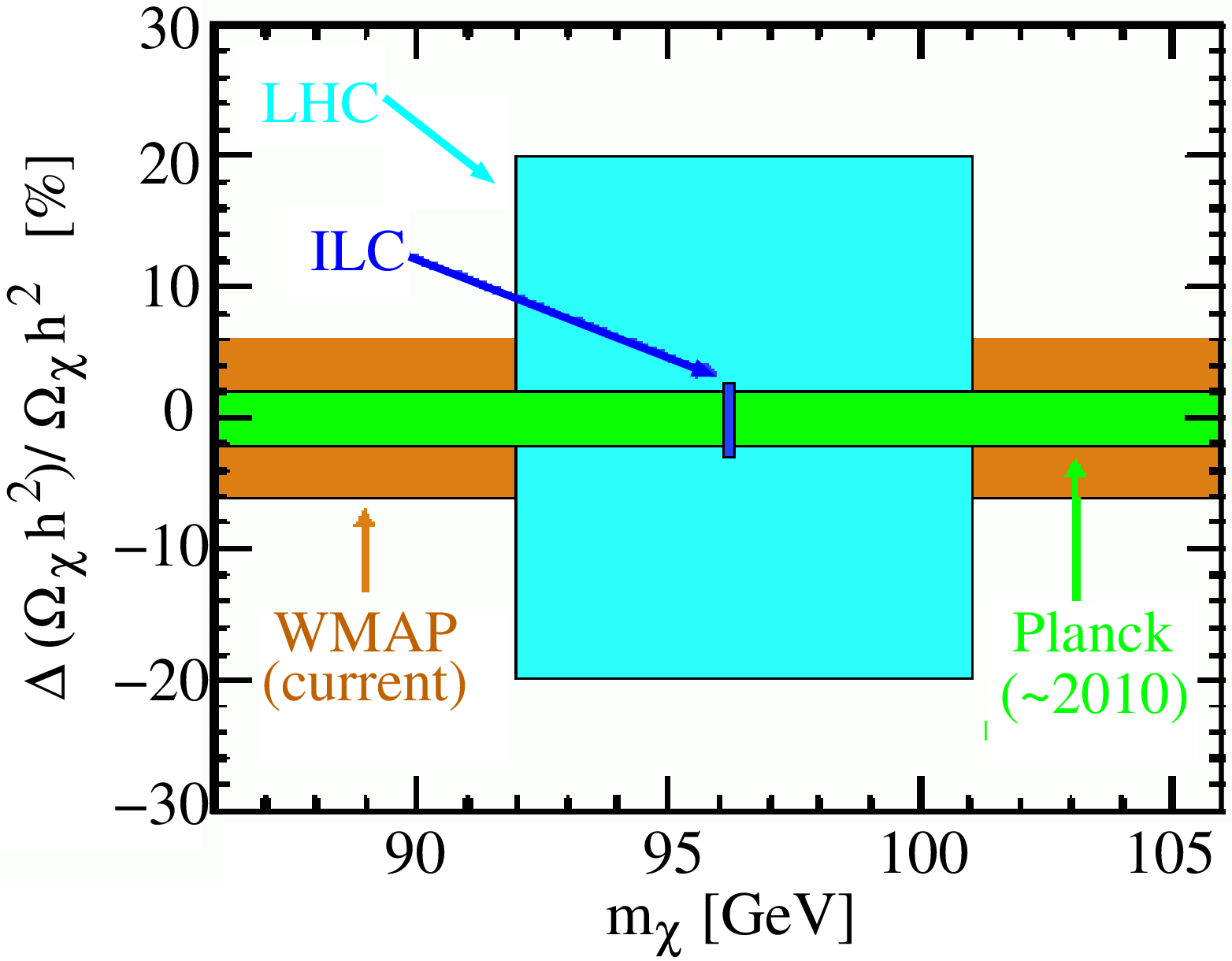} }
\vspace*{-.5cm}
\end{center}
\vspace*{-.7cm}
\caption[Favored regions for dark matter in mSUGRA and relic density
determination]
{Left: the DM--favored regions in the cMSSM $[m_{1/2},m_0]$ parameter space 
with all experimental and theoretical constraints imposed
\cite{DM-Olive,DM-Peskin}. Right: accuracy of WMAP and expected accuracy of
Planck compared to the LHC and ILC accuracies in the determination of the LSP
neutralino mass and the cosmological relic density in  the cMSSM  point SPS1a$^\prime$
\cite{DM-Feng}.}
\label{fig:DM1}
\vspace*{-.3cm}
\end{figure}

Fig.~\ref{fig:DM1} (left) summarizes the areas in the $[m_0, m_{1/2}$] cMSSM
parameter space for $A_0=0$ and $\mu>0$ in which all constraints from collider 
searches and high--precision measurements are imposed and the LSP abundance
matches the WMAP constraint \cite{DM-Olive,DM-Peskin};  their precise locations
vary with $\tan \beta$  and thus the $m_0, m_{1/2}$ axes are given without
units.  Note that a fifth possible region is when $2 m_{\chi_1^0} \sim M_h$ and
the $s-$channel $h$ exchange is nearly resonant allowing the neutralinos to
annihilate efficiently \cite{DM-hpole}; this ``$h-$pole'' region,  in which the
inos are very light and can be studied in detail at the ILC,  is however squeezed
by the LEP2 lower limit on $M_h$ \cite{Barate:2003sz}. Another possibility in
the unconstrained MSSM is the stop co--annihilation region  \cite{DM-stop}, with
a small $\tilde t_1$--$\chi_1^0$ mass difference, which is important for 
scenarios of electroweak baryogenesis in the MSSM \cite{Baryo-Carena}; it will
be discussed later in this chapter.

As seen previously, SUSY particles can be produced abundantly at the LHC and the
ILC. However, to determine the predicted WIMP relic density, one must
experimentally constrain all processes contributing  to the LSP pair
annihilation cross section. This requires detailed knowledge, not only of the
LSP properties, but also of all other particles contributing to their
annihilation.  This is not a simple task and all unknown parameters entering the
determination of $\Omega_ \chi h^2$ need to be experimentally measured or shown
to have marginal effects. The very high precision that can be achieved at the
ILC, eventually combined with measurement for squarks, gluinos and the heavy
Higgs bosons at the LHC, will allow  to achieve this goal. The results of a 
study  in the cMSSM SPS1a' scenario are summarized in Fig.~\ref{fig:DM1}
(right), where the expected precision at ILC and LHC are compared with the
satellite determination of $\Omega_ \chi h^2$.  The figure shows that the ILC
will provide a percent determination of $\Omega_\chi h^2$ in the case under
study, matching WMAP and even the very high accuracy expected from Planck.

Other SUSY WIMP candidates such as the axino \cite{DM-axinos} or the gravitino
\cite{DM-grav0} are also possible.  If DM is composed of the lightest SUSY
particle, the ILC, in some cases when some information from the LHC is added,
will be able to determine the mass and the properties of the LSP and  pin down
its relic density.

\subsubsection{Neutralino DM scenarios at the ILC}

To quantify the prospects for determining the neutralino DM relic density at the
ILC and the connection of the ILC with cosmology (LCC), four benchmark cMSSM
scenarios which correspond to the four areas discussed above and in which the
model is compatible with WMAP data (for the first scenario, see the next
footnote  however), Fig.~\ref{fig:DM1} (left) with their basic  input parameters
given in Tab.~\ref{tab:LCCpoints},  have been selected:

\begin{description}
\vspace*{-2mm}

\item[LCC1:] this is simply the SPS1a point with light sleptons with masses just
above the LSP mass\footnote{As discussed earlier, this point is ruled  out as it
gives a relic density that is outside the WMAP range, $\Omega_\chi  h^2 =0.19$.
However, since the corresponding phenomenology is rather close to that of the
SPS1a' point (see for instance Tabs.~\ref{tab:mssm} and \ref{tab:msugra_fit})
which has the correct relic density, $\Omega_\chi h^2 =0.115$, we will keep this
problematic point for illustration. The accuracy in the determination of the
relic density  is different in the two scenarios, though, and in SPS1a' one
obtains $\Omega h^2$ at the percent level only.}. The
important DM annihilation process is through $t$--channel $\tilde \ell= \tilde 
e,\tilde \mu, \tilde  \tau$ exchange, so that the masses $m_{\tilde \ell}$ need
to be very accurately measured. This is indeed the case at a 500 GeV ILC as
shown previously.  \vspace*{-3mm}

\item[LCC2:] in which all sfermions are too heavy to be observed either at the
ILC or at the LHC while all charginos and neutralinos can be produced at the LHC
and then measured at the ILC. The main contribution to DM is when these states
are exchanged in the $t$--channel of LSP annihilation into gauge and Higgs
bosons and thus, $\Omega_\chi h^2$ strongly depends on the gaugino--higgsino
mixing which needs to be measured accurately. \vspace*{-3mm}

\item[LCC3:] in this scenario the $\tilde \tau_1$ and the $\chi_1^0$ LSP are
very close in mass, $m_{\tilde \tau_1}-m_{\chi_1^0}=10.8$ GeV, so that
co--annihilation dominates annihilation of SUSY particles in the early universe.
Here, only these two particles (and $\chi_2^0$) are light enough to be
accessible at the 500 GeV ILC, but their important mass difference can be
measured with an error of 1 GeV. \vspace*{-3mm}

\item[LCC4:] here, LSP annihilation occurs mainly through the exchange of the
$A$ boson which has a mass $M_A\!=\!419$ GeV; the measurements of $M_A$ and the
total width $\Gamma_A$ are crucial and, at the ILC, they can be performed only
at $\sqrt s=1$ TeV. Most of the SUSY spectrum (except for  $\tilde \tau_1$ and
$\chi_1^0$) is anyway heavy and can be produced only at a 1 TeV 
machine.\vspace*{-3mm} 

\end{description}

\begin{table}[!h]
\vspace*{-.5cm}
\caption[Parameter sets for scenarios with dark matter in the constrained MSSM.]
{cMSSM parameter sets for four illustrative scenarios of $\chi_1^0$
DM (with sign($\mu)>0$ and $A_0=0$ except for LCC1 where $A_0=-100$ GeV). 
The accuracy in the determination of the LSP mass and the relic density at 
the ILC are also shown (and compared to that obtained from LHC
measurements only).}
\label{tab:LCCpoints}
\centering
\begin{tabular}{|l|ccc|ll|lll|}\hline
   Point &  $m_0$ & $m_{1/2}$ & $\tan\beta$ & 
   $m_{\chi_1^0}$ & $\Delta$ILC  & $\Omega_\chi h^2$ &  $\Delta$\,ILC &
   ($\Delta$\,LHC) \\ \hline
   LCC1 &   100 &  250 &  10   & $96.1$ & $\pm0.05$ &
                      0.192 &             $\pm$  $0.24$\% & (7.2\%)\\
   LCC2 & 3280 & 300 & 10   & $107.9$ &$\pm 1.0$ &
                        0.109 &     $\pm$ 7.6\%   & (82\%) \\
   LCC3 &  213 & 360 &  40   & $142.6$&$ \pm 0.1$ &
                         0.101 &   $\pm$ 18\%   & (167\%) \\
   LCC4 &  380 & 420 &  53     & 169.1 & $\pm 1.4$ &
                0.114 &   $\pm$ 19\% & (405\%) \\ \hline
\end{tabular}
\vspace*{-2mm}
\end{table}

\begin{figure}[!h]
\vspace*{-3mm}
\centering
\includegraphics[width=3.7cm,height=4.8cm]{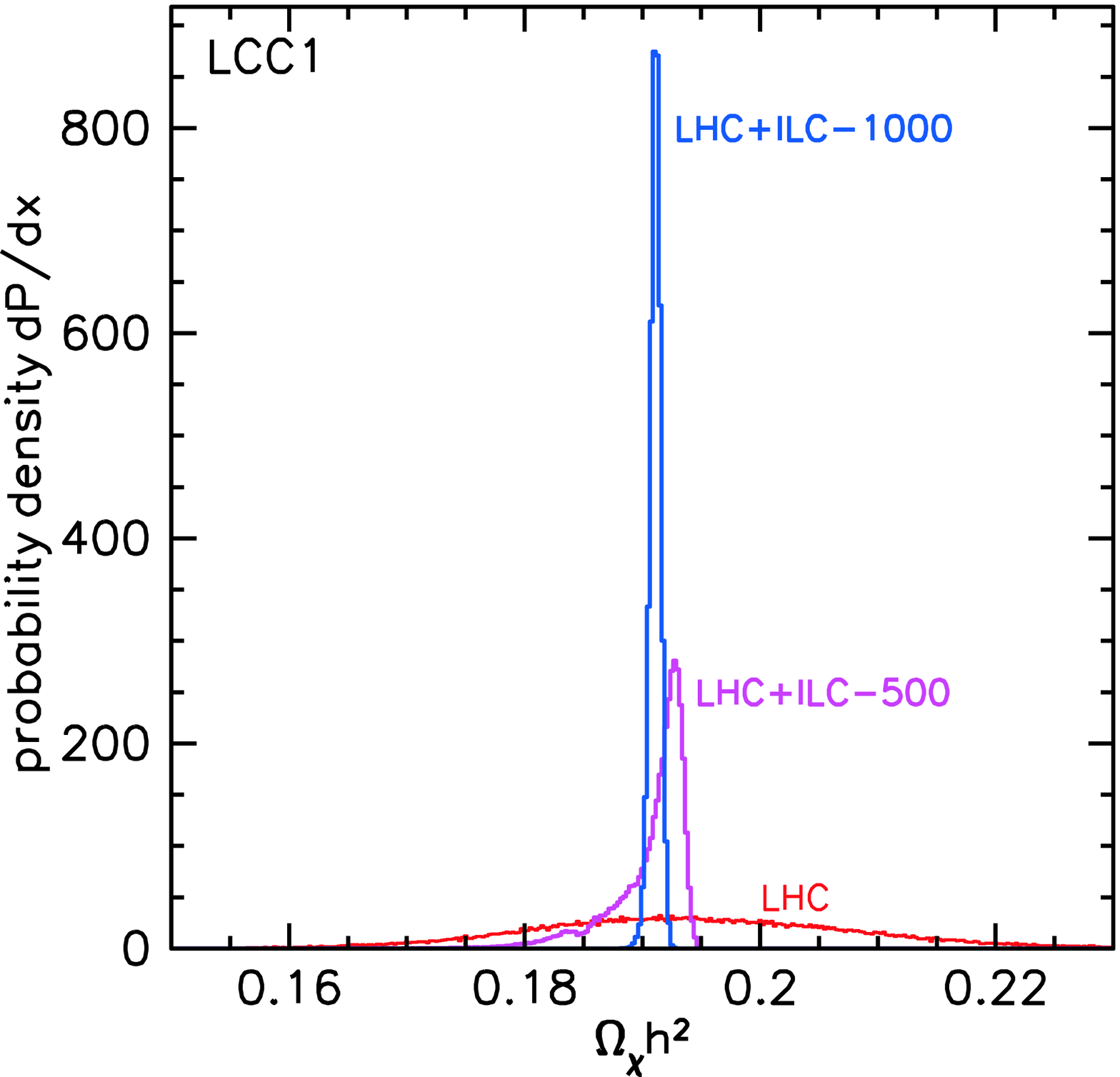} \hskip 0.1cm
\includegraphics[width=3.7cm,height=4.8cm]{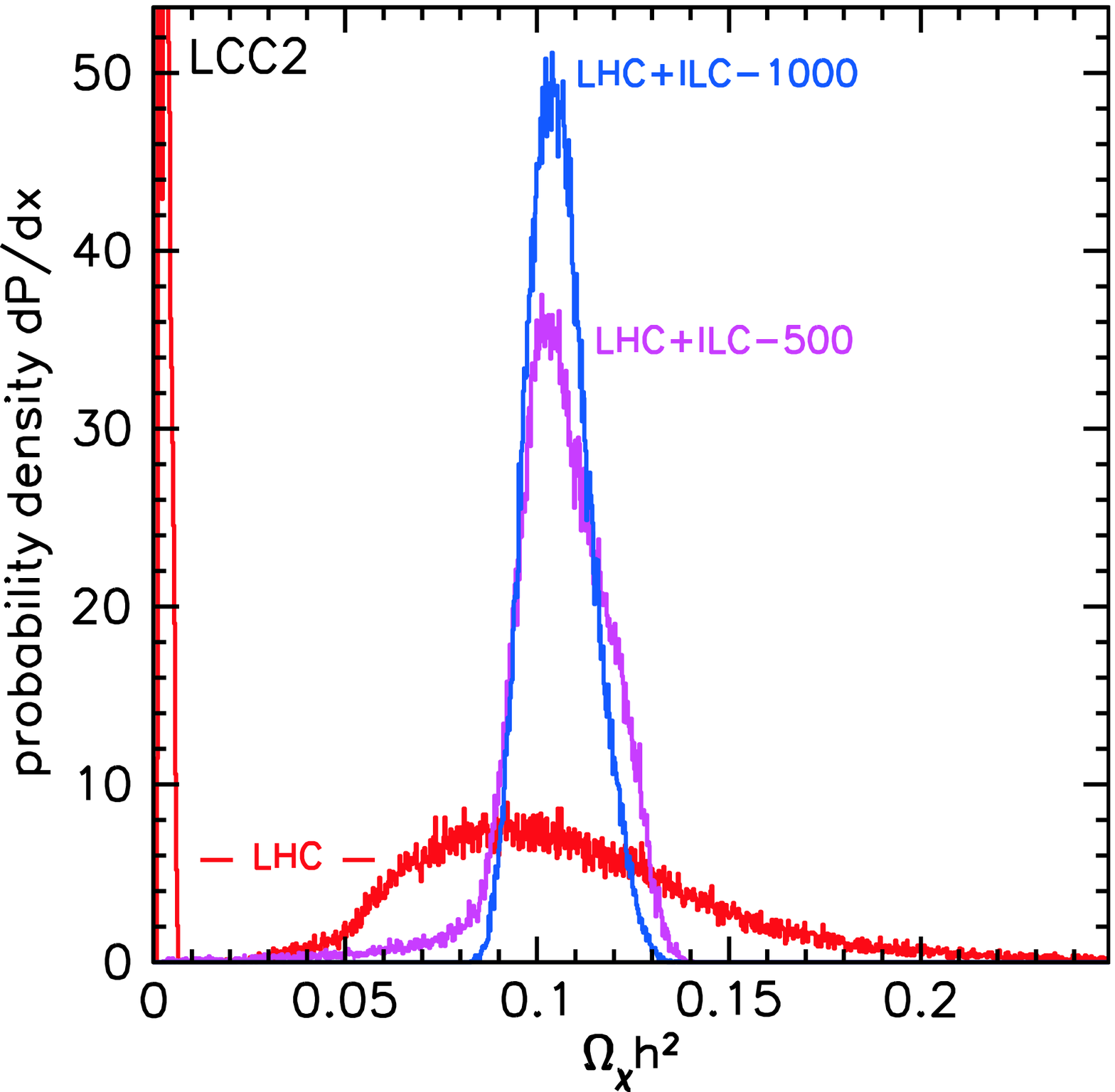} \hskip 0.1cm
\includegraphics[width=3.7cm,height=4.8cm]{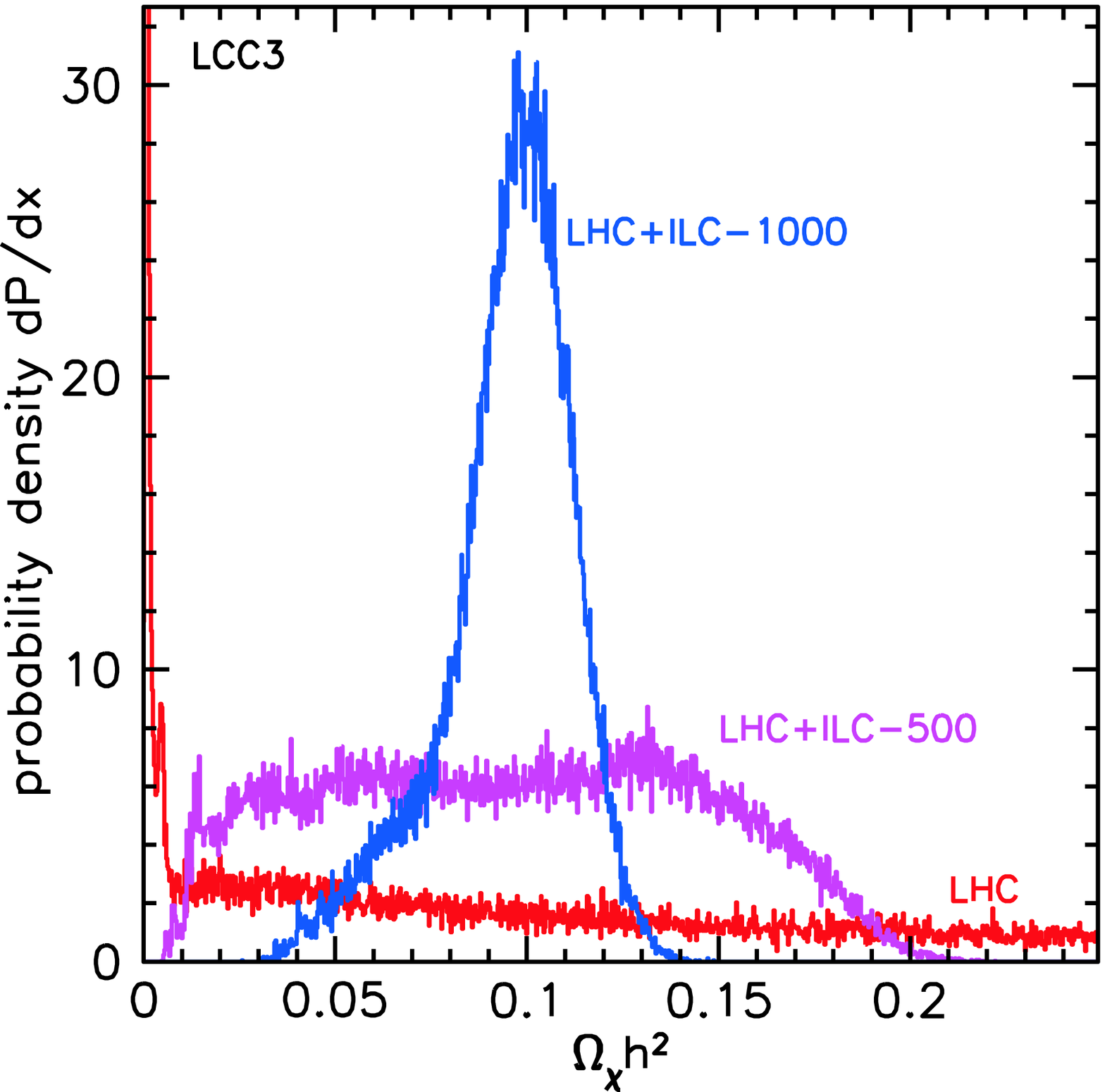} \hskip 0.1cm
\includegraphics[width=3.7cm,height=4.8cm]{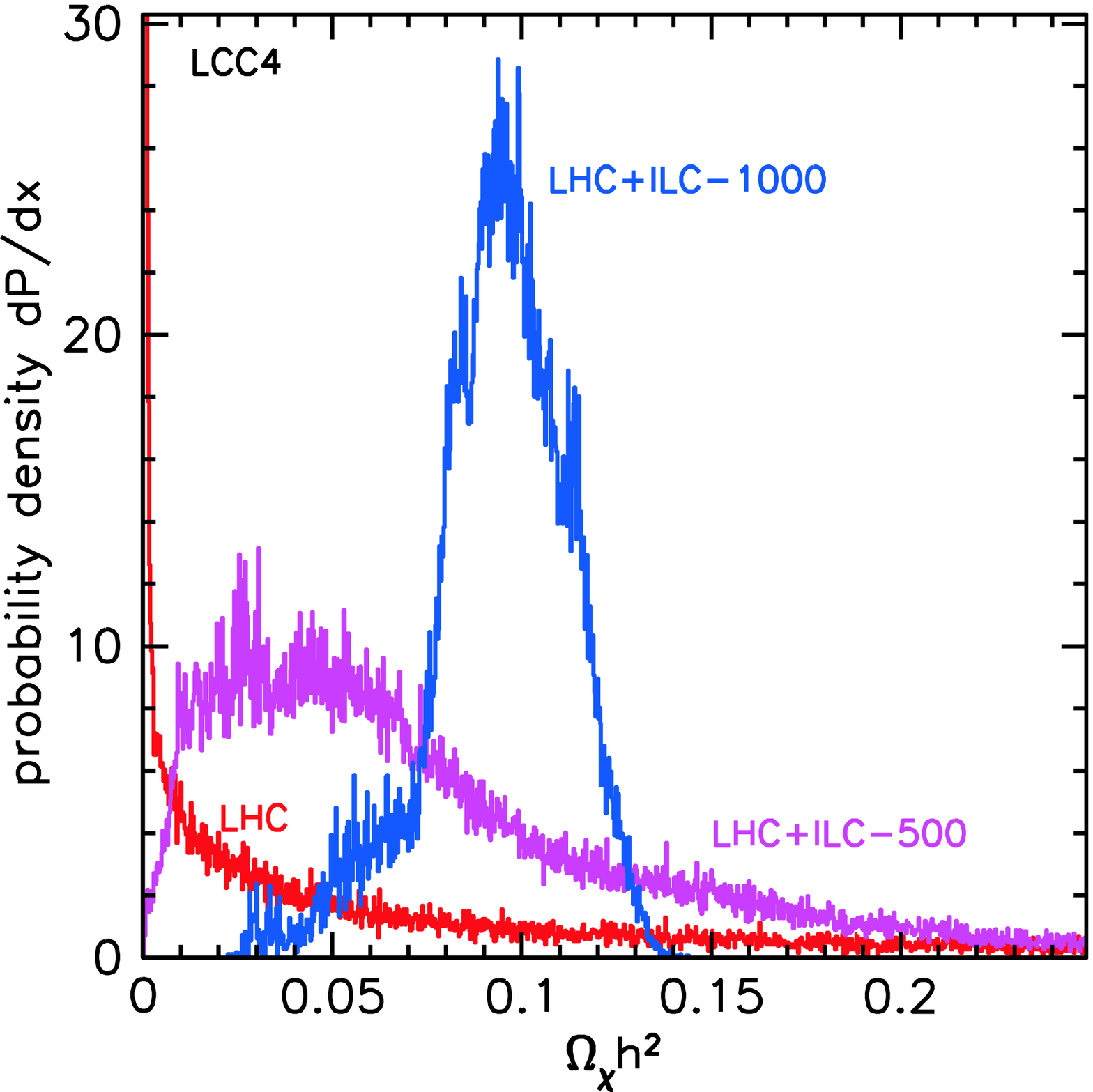}\\
\vspace*{-6mm}
\caption[Determination of the neutralino relic density at the ILC and LHC]
{Probability distribution  of predictions for $\Omega_\chi h^2$ from
measurements at the ILC with $\sqrt s\!=\!0.5$ and 1\,TeV, and LHC (after 
qualitative identification of the model); from Ref.~\cite{DM-Peskin}.} 
\label{fig:LCCcomp}
\vspace*{-3mm}
\end{figure}

Many detailed studies of the determination of the DM density from collider
measurements in scenarios close to the LCC ones have been performed
\cite{LHC-benchmark,DM-LHC,DM-ILC}. A particular focus has  been put
recently on the LCC3 $\tilde \tau_1$--$\chi_1^0$ co--annihilation point
\cite{DM-ILC} which is known to be difficult and very demanding for ILC
detectors as an optimal detection of energetic electrons in the very forward
region and a very efficient rejection of the $\gamma \gamma$ background is
required.  Here, we will rely on a recent comprehensive analysis performed in
Ref.~\cite{DM-Peskin} to summarize the main results. In this study, the four LCC
points have been described in terms of 24 effective MSSM parameters to be as
model independent as possible, over which full scans [using a Markov Chain Monte
Carlo algorithm] are performed to determine the MSSM models that are compatible
with the experimental measurements. The neutralino relic density calculated
using {\tt microMRGAS} \cite{micromegas} and the  precision from the ILC
measurements are summarized  for these points in the right--handed column of
Tab.~\ref{tab:LCCpoints}. The accuracies range from less than 1\% in the
LLC1/SPS1a scenario to 20\% in the difficult LCC3 co--annihilation  and LCC4
``$A$--pole" scenarios; a few percent accuracy is reached in the LCC2
``focus--point" scenario.  The analysis also leads to the probability
distributions of predictions for $\Omega_\chi h^2$, using the various expected
measurements, which are shown in Fig.~\ref{fig:LCCcomp}. The ILC measurements at
$\sqrt s=500$ GeV and $1$ TeV for various sparticle masses and mixings, taking
into account LHC data, are compared to those which can be obtained using LHC
data alone (after a qualitative identification of the model), which in most
cases needs ILC data. As can be seen, the gain in sensitivity by combining LHC
and ILC data is spectacular.

Once the DM relic density is precisely obtained, one can turn to the
prediction (or the verification, if they have already been measured
in astroparticle experiments) of the cross sections in direct and
indirect detection of the DM. For both techniques, the detection
rates are convolutions of microscopic cross sections that can be
``determined" in particle physics experiments with densities that
can be measured in astrophysical experiments.  In indirect
detection, one looks for, e.g., high energy neutrinos or photons
originating from the annihilation of neutralinos in our galaxy and
the rate is directly proportional to the annihilation cross sections
which enter in the determination of the DM  relic density; however,
the distribution of DM has several orders of magnitude uncertainty.
In direct detection, i.e. in the search of the elastic scattering of
ambient neutralinos off a nucleus in a laboratory detector, the
astrophysical uncertainty is only a factor of two while the
LSP--nucleon scattering cross section has inherent  uncertainties
from strong interactions that are larger.

Nevertheless, if the modeling of the DM distribution and of the
$\pi$--nucleon interaction can be improved, a precise determination
of the detection rates can be performed by reconstructing the
microscopic cross sections using precision SUSY parameter
measurements at the ILC and at the LHC for the squark sector.  This
is clearly the case for the LSP annihilation cross section which is
similar to that giving $\Omega_\chi h^2$ but also for the
LSP--nucleon cross section when it is dominated by Higgs exchange
diagrams. In turn, the determination of the microscopic LSP cross
sections from ILC data could allow to significantly constrain in a
general way the distribution of DM in the galaxy; see
Ref.~\cite{DM-Peskin} for a discussion and a detailed study.

\subsubsection{Gravitino DM at the ILC}

SUSY particles other than the lightest neutralinos can also form the DM in the
Universe. While LSP sneutrinos have been ruled out by direct WIMP searches
\cite{DM-review}, the possibility of the axino \cite{DM-axinos} or the gravitino
\cite{DM-grav0} DM is still open. In many scenarios, one can arrange so that
these WIMPs have the required relic density by choosing appropriate values of
the masses and the reheat temperature after the phase of inflation, for
instance. These particles have extremely weak couplings to ordinary matter and
cannot be observed directly in astrophysical experiments; in contrast, they can
be studied at the ILC. Here, we briefly discuss the scenario of a gravitino LSP
and its implication for the ILC.

In mSUGRA--type models, the mass of the gravitino and those of the SM
superpartners $\tilde P$ are given by $m_{\tilde G, \tilde P}= \kappa_{\tilde G,
\tilde P} \cdot F/M_P$ where $M_P \simeq 2.4\cdot 10^{18}$ GeV is the reduced
Planck mass, $F \sim (10^{11}$ GeV)$^2$ is the square of the SUSY breaking
scale; $\kappa_{\tilde G}= \frac{1}{\sqrt 3}$ while $\kappa_{\tilde P}$ is
model--dependent and is expected to be ${\cal O}(1)$.  The gravitino can be
therefore the LSP with a mass in the range $m_{\tilde G}  \propto 10$-- 100 GeV.
However, its couplings to matter are very strongly suppressed by a factor
$1/M_P$ and, thus, the gravitino is a super--WIMP that cannot be directly
observed in astrophysical experiments.

In the early universe, gravitinos are generated via thermal production  through
processes involving SM and SUSY particles in the thermal bath and also in
non--thermal decay processes  of superparticles which are out of equilibrium.
These superparticles will first decay into the NLSP, which can be either a
neutralino, a charged slepton (generally a $\tilde \tau$) or a sneutrino, that 
first freezes out and then decays into the gravitino whose relic density is
given by $\Omega_{\tilde G} h^2 = m_{\tilde G}/m_{\rm NLSP} \cdot \Omega_{\rm
NLSP} h^2$.  Since the next--to LSP  decays gravitationally, NLSP $\to \tilde
G+X$, its lifetime is in principle of order  $\tau_{\rm NLSP} \propto
M_P^2/M_{\rm EWSB}^3 =  10^2$--$10^8\;$s and thus very long. It is therefore
constrained by cosmology, in particular by primordial nucleosynthesis (BBN) and
cosmic microwave background (CMB) data, and can eventually be tested at
colliders by the measurement of the NLSP mass and lifetime.

\begin{figure}[h!]
\begin{tabular}{cc}
\hspace*{-4mm}
\epsfig{file=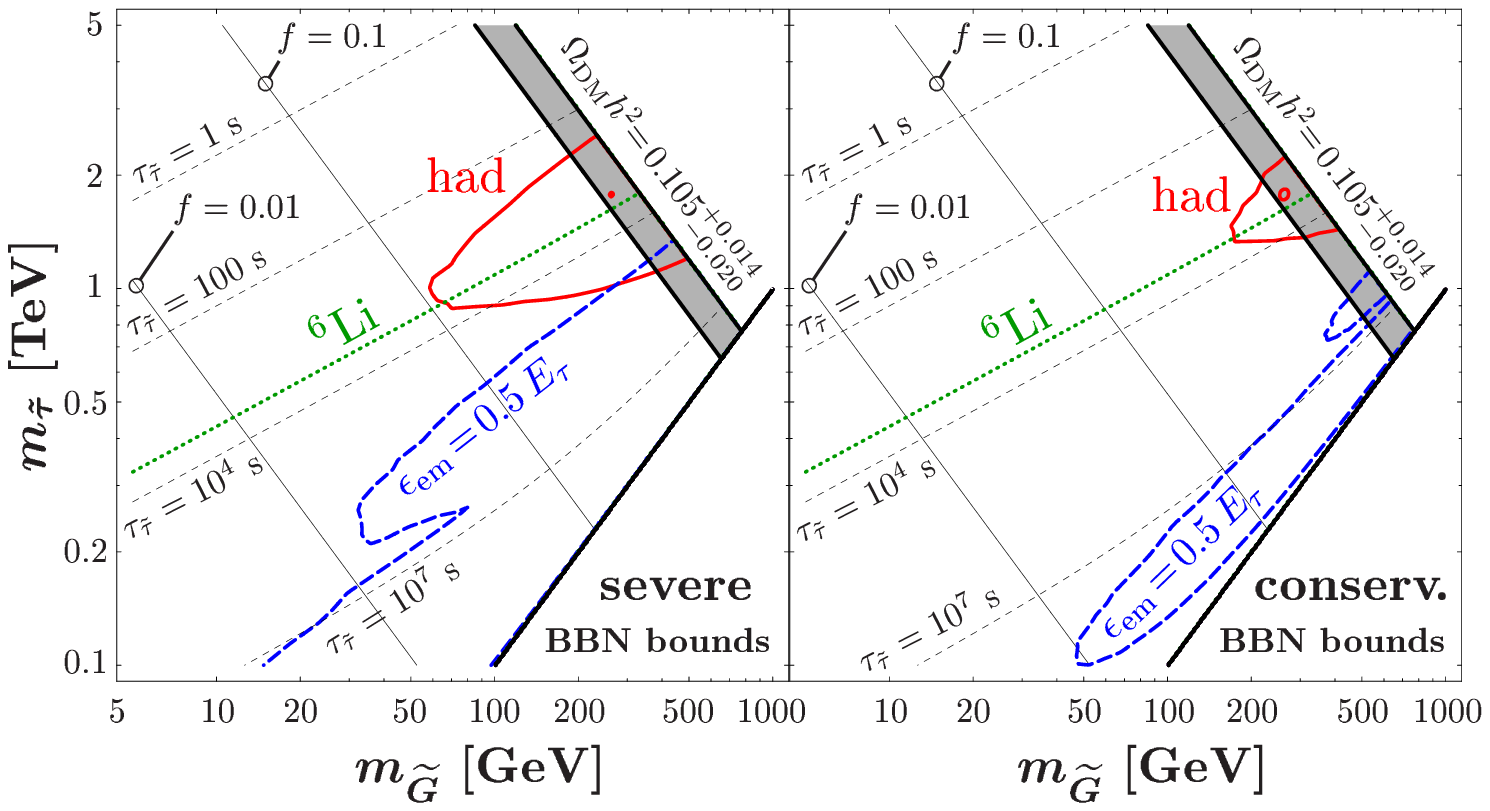,width=8.3cm,height=6cm}
    & \hspace*{-.5cm}
\epsfig{file=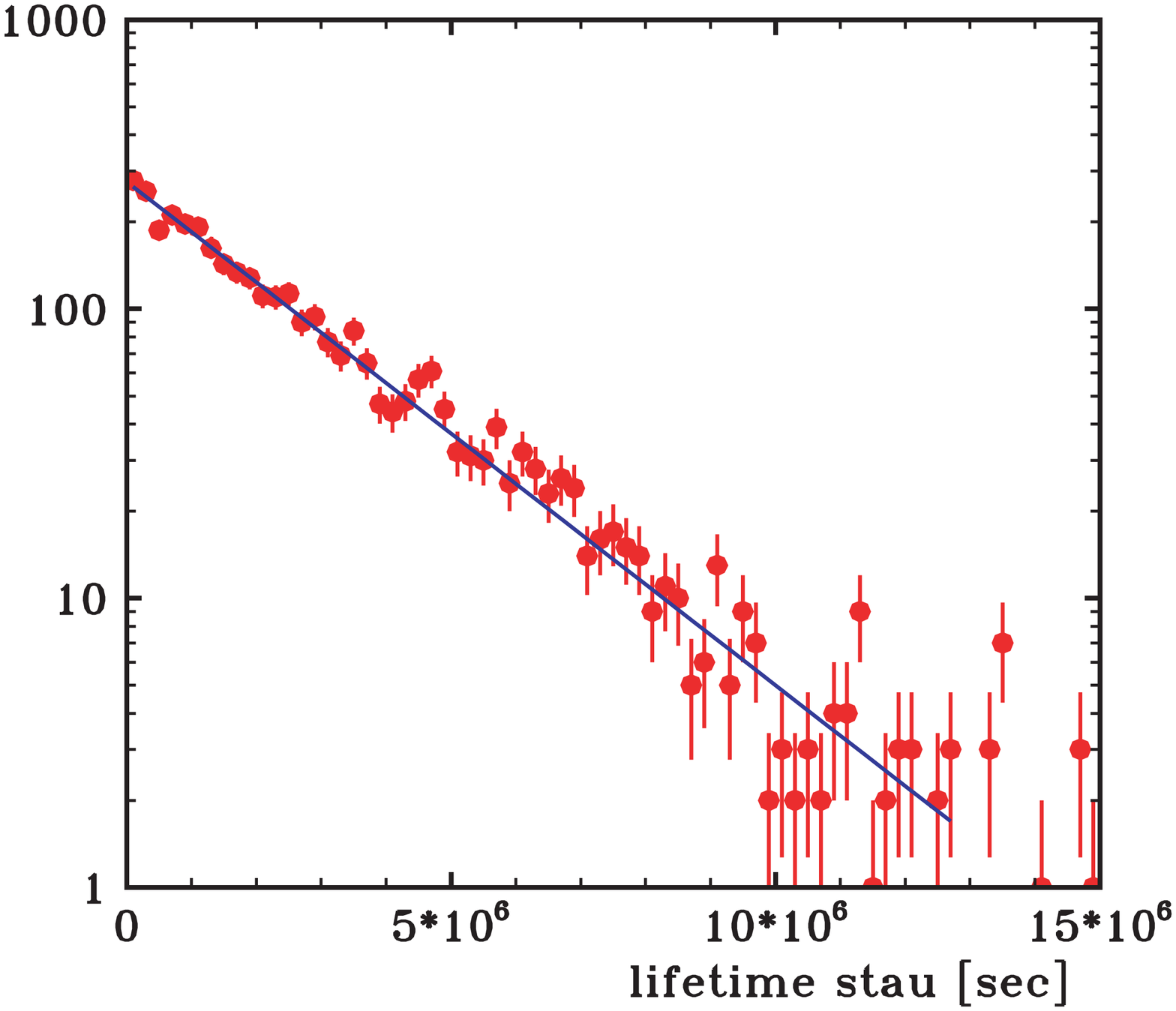,width=7cm,height=6.cm}
\end{tabular}
\vspace*{-9mm}
\caption[Constraints on gravitino and stau masses and ILC stau lifetime 
measurement] 
{Left:Cosmological constraints on the masses of the gravitino LSP and
the stau NLSP from severe and conservative BBN constraints; the thick solid
(red) and thick dashed (blue) curves are for the BBN bounds from late hadronic
and electromagnetic energy injection,  respectively, and the regions inside or
to the right of the corresponding  curves are excluded \cite{DM-grav1}. The
$\tilde \tau$ lifetime distribution in the decay $\tilde \tau_1 \to\tau \tilde
G$ at the ILC with $\sqrt{s}=500$ GeV and ${\cal L}=100$ fb$^{-1}$ (right);
from  Ref.~\cite{DM-Martyn}.} \label{fig:DM-grav} 
\vspace*{-4mm}
\end{figure}

Gravitinos with masses in the range $m_{\tilde G}  \propto 10$--100 GeV are also
good DM candidates.  However, strong  constraints from  BBN and in particular
recent data from the abundance of primordial light elements such as Lithium,
impose that the mass difference between the NLSP and the gravitino should be
relatively large. In the case where the NLSP is the $\tilde \tau$ slepton, the
constraints are shown in the left--hand side of  Fig.~\ref{fig:DM-grav}
\cite{DM-grav1}.  For stau leptons with masses below $m_{\tilde \tau} \lsim 400$
GeV,  a gravitino mass of $m_{\tilde G} \lsim 10$ GeV is required; the $\tilde
\tau$ lifetime  is also restricted to be in the $10^3$--$10^5\;$s range. Note,
however, that these bounds might be somewhat relaxed  with a better theoretical 
understanding of the bound state effects of Li production  and/or by possible
entropy production after $\tilde \tau$ decoupling.   Furthermore, all problems
from BBN constraints can be easily solved if one allows for a tiny amount of
R--parity violation; in this case there is no constraint on the $\tilde \tau$
mass and, for a successful thermal leptogenesis, one needs  $m_{\tilde G} \gsim
5$ GeV for the gravitino \cite{DM-grav2}.

At the ILC, a detailed study \cite{DM-Martyn} has been performed in an
mSUGRA--like scenario \cite{DM-grav3} in which $m_{3/2}=m_0=\frac{1}{22} m_{1/2}
\sim A_0= 20$ GeV, $\tb=15$ and $\mu>0$, leading to stau and gravitino masses of
$m_{\tilde \tau_1}=157.6$ GeV and   $m_{\tilde G}=20$ GeV; the stau lepton  has
a lifetime $\tau_{\tilde \tau_1}= 2.6 \cdot 10^6\;s$, i.e. approximately one
month,  and is stopped in the detector\footnote{Again, this scenario cannot be
considered to be realistic in view of the BBN  bounds discussed above. However,
most of the obtained results may be readily  taken up for a more viable
scenario.}.  Assuming a c.m. energy $\sqrt s=500$ GeV and a luminosity ${\cal
L}=100$ fb$^{-1}$ and, thanks to the relatively large cross section $\sigma(\ee
\to \tilde \tau_1 \tilde \tau_1+X) \sim 300$ fb, a very clean environment and good
detector (tracking, momentum and energy resolution, etc.) performance, one can
achieve very precise measurements. The stau mass can be determined from the mean
value of the  $\tilde \tau$ momentum with an accuracy of $\Delta m_{\tilde
\tau_1}\simeq 200$ MeV. The lifetime can be determined from  a fit to the decay
time distribution shown in the right--hand side of Fig.~\ref{fig:DM-grav} and
one obtains  $\tau_{\tilde \tau_1}= (2.6 \pm 0.05) \cdot 10^6\;s$. Assuming the
usual gravitational coupling, one then obtains the gravitino mass from the
$\tilde \tau$ mass and lifetime with a very good accuracy, $\Delta m_{\tilde
G}=\pm 200$ MeV. In fact, one can also measure directly the gravitino mass from
the recoil of the tau lepton in the decay $\tilde \tau_1 \to \tau \tilde G$ and
an accuracy of $\pm 4$ GeV can be achieved. This allows the unique opportunity
to have an independent access in a microscopic experiment  to the value of the
reduced Planck scale, $M_P \simeq (2.4 \pm 0.5) \cdot 10^{18}$ GeV and, hence,
to Newton's constant, $G_N=1/(8\pi M_P^2)$. Therefore, also in this scenario,
precision measurements at the ILC would allow to derive very important
informations on cosmological phenomena.

Note that in scenarios in which a small amount of R--parity violation is 
introduced in order to avoid  BBN constraints, the $\tilde \tau$ state will have
two--body $\not \hspace*{-1.5mm} R_p$ decays, yielding visible tracks  in the
detector macroscopic times later; however, in this case, one cannot determine
the Planck mass anymore \cite{DM-grav2}. 

\subsection{DM in extra dimensional scenarios}

An interesting feature in the simplest version of universal extra dimension
(UED) models discussed in chapter \ref{sec:alternatives}, is the presence  of  a
discrete conserved quantity, the so called KK--parity $(-1)^n$ where $n$ is the
KK level. KK parity ensures the presence of a stable massive particle,  the LKP,
which can be a cold DM candidate \cite{LKP-heavy}. Several possible LKP
candidates  are the first KK excitations of Higgs or gauge bosons, such  as the
particle corresponding to the hypercharge gauge boson $B_1$ which is naturally
obtained in minimal UED  (MUED) models, and the KK  excitation of a neutrino. In
warped extra dimensional models embedded in a GUT, the $Z_3$ symmetry introduced
to prevent rapid proton decay also guarantees the stability of the lightest KK
fermion, a  right--handed neutrino \cite{DM-LZP}. This particle is called the
LZP and  can be also a good cold DM candidate. In the following, we briefly
discuss the two options of a $B_1$ LKP and a $\nu_R$ LZP, and their implications
at the ILC. 

\subsubsection{DM in universal extra dimensions}

In MUED models, the LKP naturally turns out to be the KK partner  of  the
hypercharge gauge boson and, if only annihilation processes are considered, its
cosmological relic density is typical of a WIMP candidate. In order to  explain
all of the DM, the $B_1$ mass should be in the range $M_{\rm B_1} = 600$--800
GeV, depending on the rest of the KK  spectrum. The mass is clearly too large
for this particle to be produced at the ILC. However, it has been  realized that
one needs to include co--annihilation processes with the ${\rm SU(2)}$ singlet
KK leptons, which in MUED are the lightest among the  remaining $n=1$ KK
particles, as well as  the influence  of gravitons on the final relic density
results. 

\begin{figure}[!h]
\vspace*{-.2cm}
\begin{center}
\includegraphics[width=7cm]{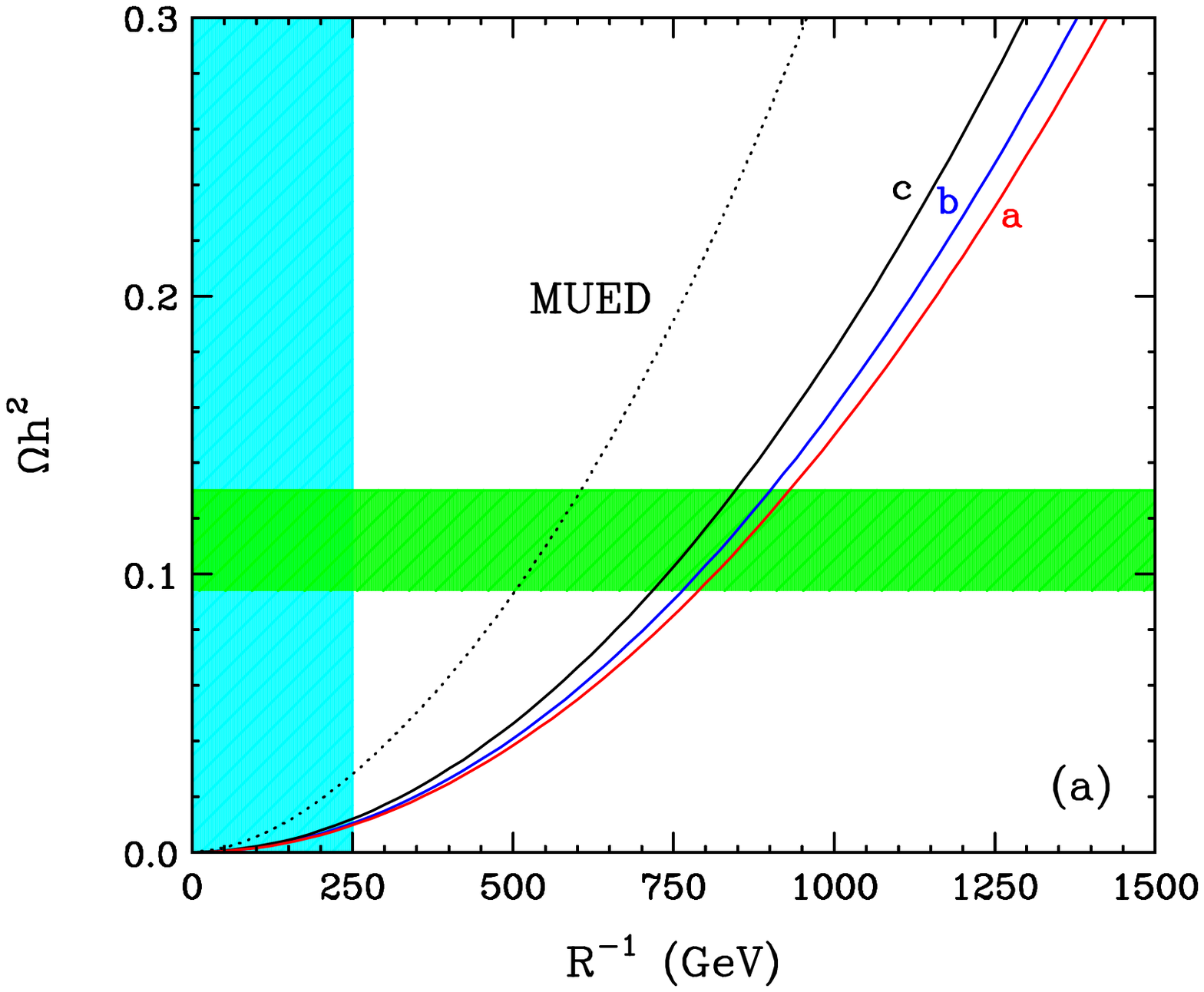}\hspace*{.7cm}
\includegraphics[width=7cm]{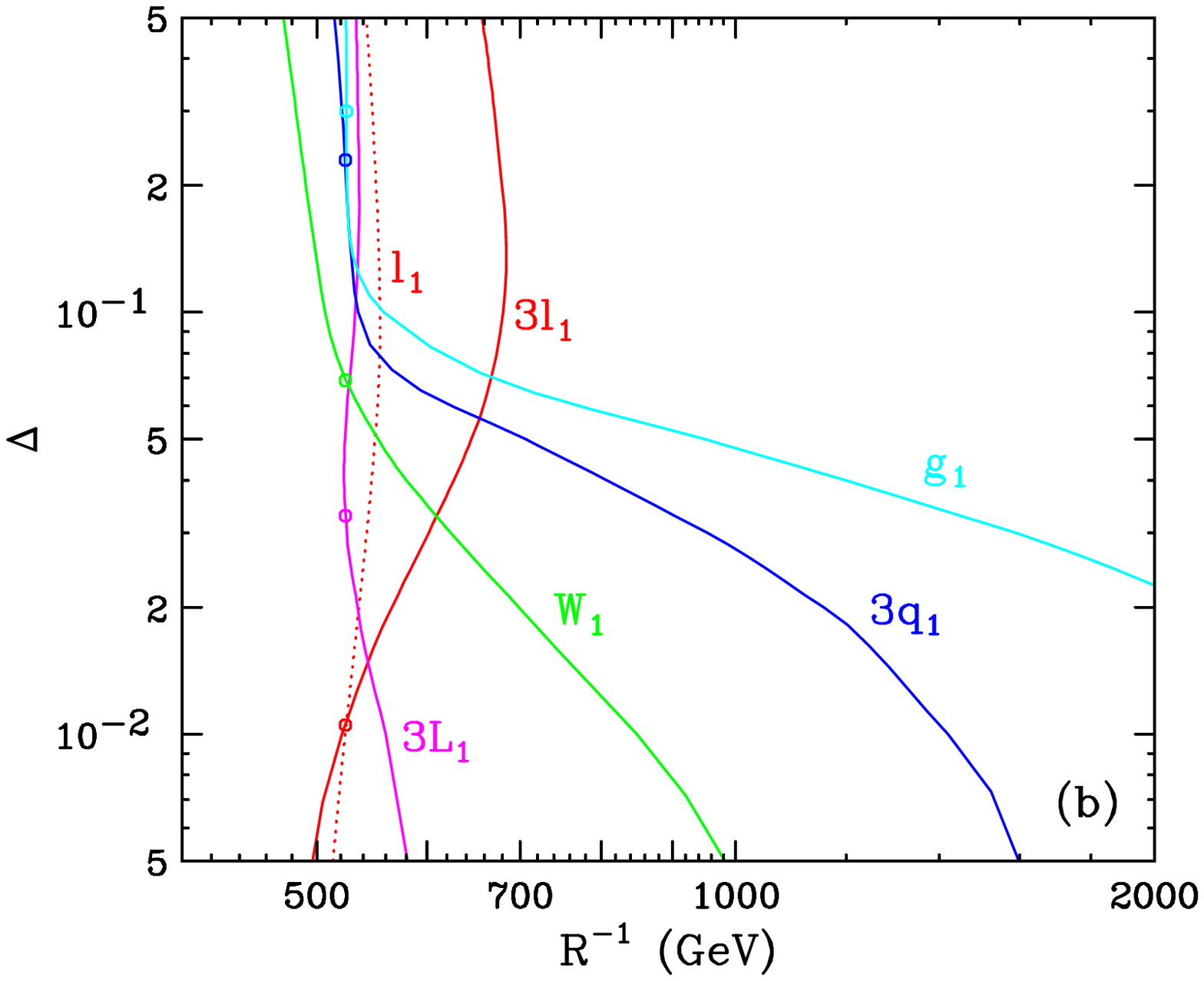}
\end{center}
\vspace*{-9mm}
\caption[Relic density of the lightest KK particle in universal extra 
dimensions models]
{Left: relic density of the LKP as a function of $R^{-1}$ in the  MUED
model with and without co--annihilation. Right: the change in the cosmologically
preferred value for $R^{-1}$ as a result of varying the different KK masses away
from their nominal  MUED values. From Ref.~\cite{LKP-Kong}.}
\label{fig:DM-KK}
\vspace*{-4mm}
\end{figure}

The left--hand of Fig.~\ref{fig:DM-KK} shows the relic density of the LKP as a
function of the inverse of the size of the extra dimension $R^{-1}$, in the MUED
model \cite{LKP-Kong}. The lines marked ``a,b,c'' are for the results obtained
when  considering only their annihilation with various assumptions
on the KK mass spectrum, while  the dotted line is the result from the full
calculation in MUED, including all co-annihilation  processes and with the
proper choice of masses. The green horizontal and the blue vertical bands are,
respectively,  for the WMAP preferred range and the $R^{-1}$ regions disfavored
by precision data. As can be seen, LKP particles in the mass range close to 500
GeV are compatible with DM. In the right--hand side of Fig.~\ref{fig:DM-KK},
shown is the change in the cosmologically preferred value for $R^{-1}$ as a
result of varying away from their nominal  MUED values the KK masses of the
different particles: three generations of SU(2) singlet and doublet KK
leptons and quarks as well as KK gluons and gauge bosons. As can also be seen,
visible KK states in the vicinity of $R^{-1}=500 $ GeV are also possible.

Thus, if the energy of the ILC is slightly raised or the KK masses compatible
with DM are lowered by some mechanism, the new particles can be produced at the
ILC. At least the lighter KK states are accessible as the mass difference with
the LKP can be small to allow for co--annihilation. In many cases, the signals
will mimic those of SUSY particles, in particular the presence of missing
transverse energy. The determination of the mass and mixing of these particles,
as well well as their spin and CP--quantum numbers [which are important in this
context as the LKP is a spin--one boson while the LSP neutralino in SUSY models
is a Majorana fermion], will allow to discriminate between the two scenarios
\cite{SUSY-spin-chi+,Battaglia:2005zf}. 

\subsubsection{DM in warped extra dimensions} 

As discussed in chapter \ref{sec:alternatives}, the most promising and realistic
warped extra-dimensional scenarios need the electroweak  gauge group to be
extended to ${\rm SU(2)_L\times SU(2)_R \times U(1)_X}$. In this context, KK
Dirac neutrinos charged under the ${\rm SU(2)_R}$ group are necessary parts of
the models. Implementing baryon number conservation in these warped GUT models
leads to a KK right--handed neutrino $\nu_R$ that is absolutely stable and thus,
a potential candidate for cold DM \cite{DM-LZP}. In fact, even in the absence of
this  additional symmetry, $\nu_R$ can be stable at cosmological scales if the
couplings involved in its decay are strongly suppressed, which can occur also if
it has a large annihilation cross section, providing the correct relic density.

\begin{figure}[!htb]
\vspace*{-1.8cm}
\hspace*{0.5cm}
  \centerline{\epsfig{file=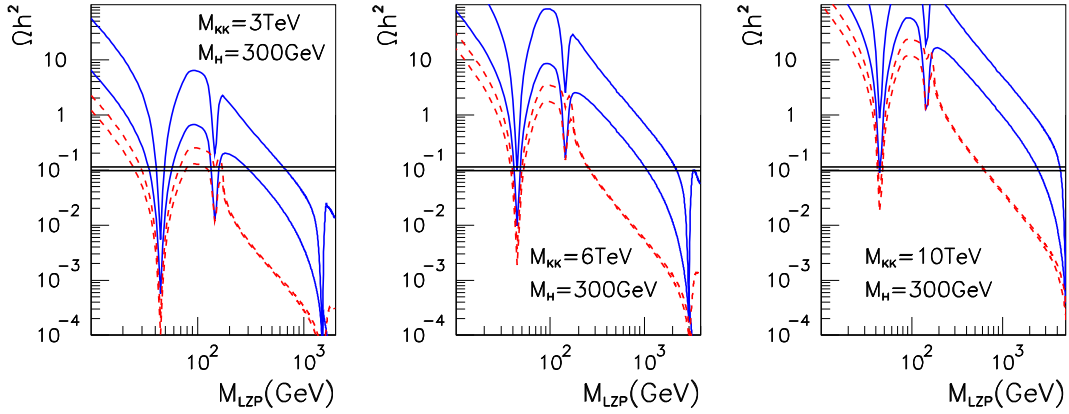, width=16cm}}
\vspace*{-1.9cm}
  \caption[Relic density  of the dark matter particle in warped extra
  dimensional models]
  {The relic density  of the LZP in annihilation for three $M_{\rm KK}$
  values, $g_{10}=0.3$ (dashed) and 1.2 (solid lines) and two values of 
  $c_{\nu_L}=0.9$ (lower) and $-0.1$ (upper curves); from Ref.~\cite{LKP-GGS}.}
  \label{fig:DM-lzp}
\vspace*{-3mm}
\end{figure}

In a RS scenario embedded in the SO(10)  GUT group, the $\nu_R$ has no direct
couplings to the $Z$ boson but a small $Z\bar \nu_R \nu_R$ coupling is induced
by the mixing between the  $Z$--$Z^\prime$ mixing. The $Z^\prime$ boson couples  with full
strength to the  $\nu_R$ LKP state but, as it must be heavier than $M_{\rm KK}
\sim 3$ TeV, the resulting interactions are rather weak.  These arguments make
that, although of the Dirac type, KK right--handed neutrinos  with masses in the
range of 1 GeV to 1 TeV can have the required  relic abundance without being in
conflict with the bounds from direct detection experiments \cite{LKP-GGS}. The
DM density is shown in Fig.~\ref{fig:DM-lzp} as a function of the LZP mass for
two values of the SO(10) coupling $g_{10}$  and two different localizations of
the left--handed neutrino $\nu_L$ (which also mixes with $\nu_R$); the masses of
the KK gauge bosons are assumed to be $M_{\rm KK}= 3,6$ and 12 TeV while the SM
Higgs mass is fixed to be $M_H=300$ GeV.  One notices the effect of the $Z$,
Higgs and $Z^\prime$ resonances which allow for the relic  density to be compatible
with the WMAP range. Since all KK fermions belonging to the multiplet containing
the right--handed top quark,  except for its KK mode, are expected to be light
compared to the KK gauge bosons and close in mass to  the LZP, co--annihilation
with the KK leptons for instance can play a non--negligible role
\cite{LKP-GGS}.  

If the LZP and the KK fermions which are in the same multiplet have not too
large masses, the ILC will be the ideal instrument to produce them  and to study
in great detail their properties. Again, threshold scans would allow for precise
mass measurements and the study of the cross sections as well as various
production and decay distributions would allow  for the determination of the
couplings and spins of the particles. These measurements could be then used to
predict the DM density and compare it with the experimental value.

\section{The baryon asymmetry}

\subsection{Electroweak baryogenesis in the MSSM}

Electroweak baryogenesis is an interesting possibility where the baryon
asymmetry of the universe is generated at the electroweak phase transition.
Since the relevant energy scale is the weak scale, this scenario has potential
impacts on the Terascale physics. As a strong first--order phase transition is a
necessary condition of successful electroweak baryogenesis, the Higgs sector
should be extended from the minimal one Higgs doublet SM in which, in view of
the current bound on the Higgs boson mass, it is not the case.  A strong
first--order phase transition is possible in various extensions of the Higgs
sector such as the SM supplemented with a scalar singlet field, the two Higgs
doublet model, the MSSM and the next-to-minimal supersymmetric Standard Model
(NMSSM).

The electroweak baryogenesis scenario in the MSSM has been studied in detail in
the literature; see Refs.~\cite{Baryo-studies} for reviews.  In order to account
for the observed amount of baryon asymmetry, a rather specific choice of SUSY
parameters is required. First, one of the top squarks, mostly  right--handed,
has to be lighter than the top quark in order that a strong first--order phase
transition is realized. The mass of the other stop, on the other hand, becomes
larger than 1 TeV. A new source of CP violation necessary for the generation of
the baryon asymmetry is provided by the CP phases of the chargino and neutralino
mass matrices. Since the new phases contribute to the electron and neutron
electric dipole moments, scalar fermions of the first and second generations
should be heaver than a few TeV, while charginos and neutralinos can be in the
few 100 GeV range. Finally, the lightest Higgs boson mass is predicted to be
close to the present experimental bound, $M_H \sim 114$ GeV.  If the lightest
neutralino is to account for the DM in this scenario, the mass difference
between the light stop and the LSP should not be large, and stop--neutralino
co--annihilation \cite{DM-stop} is the primary mechanism which generates an  LSP
relic abundance which matches the WMAP value.

These features are important to test this scenario at the  LHC and ILC 
\cite{Murayama:2002xk,Baryo-Carena}. The discovery of a light top squark and a
SM--like Higgs boson with a mass close to 120 GeV would be a strong indication
that electroweak baryogenesis is the mechanism for the generation of the baryon
asymmetry. In order to confirm this picture, one needs to determine that $\tilde
t_1$  is mainly right-handed and check that the masses and compositions of the
charginos and neutralinos  are compatible with the required  values  and
finally, compute the DM  the relic abundance so as to compare with cosmological
observations. If $\tilde t$--$\chi_1^0$ co--annihilation is relevant, it is
important to determine the stop--neutralino mass difference very precisely. A
detailed  analysis of the stop, chargino  and neutralino sectors at the ILC has
been  performed for this scenario in Ref.~\cite{Baryo-Carena}. It is found that
the experimental accuracies in the measurements of the stop and ino parameters,
as discussed in chapter \ref{sec:susy}, allow to determine the strength of the
phase transition  with  a reasonable precision, $\Delta_{\rm exp}\left[v(T_{\rm
c})/T_{\rm c}\right] \lsim 10\%$, if the theoretical error is ignored. The
second crucial ingredient for electroweak baryogenesis,  the CP--violating
source responsible for the baryon asymmetry, remains however unconstrained as
only an upper bound on the phase of the $\mu$ parameter,   $|\phi_\mu| \lsim
0.7$,  can be derived.

\begin{figure}[!h]
\begin{center}
\epsfig{file=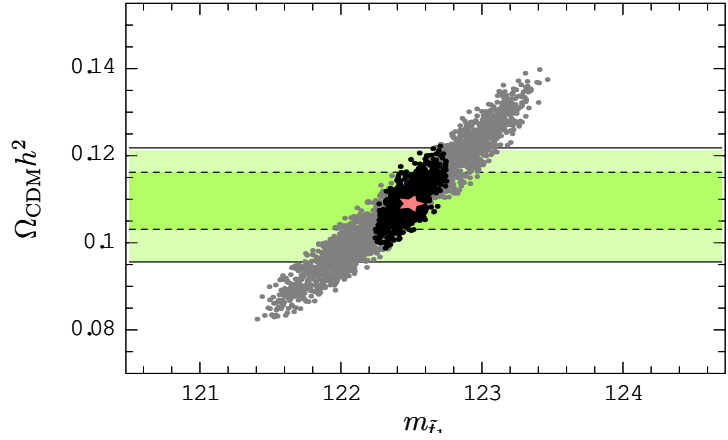,width=10cm}
\end{center}
\vspace*{-7mm}
\caption[Dark matter abundance in an MSSM scenario with electroweak
baryogenesis] 
{The DM abundance $\Omega_{\rm \chi} h^2$ as a function of the stop
mass for the electroweak baryogenesis scenario, taking into account 
experimental errors for stop, ino and Higgs measurements at the ILC; the dots 
correspond to a scan over the 1$\sigma$ region allowed by these errors; from 
Ref.~\cite{Baryo-Carena}.} 
\label{fig:DM-stop}
\vspace*{-3mm}
\end{figure}

In addition, the collider measurements  can be used to predict rather precisely
the DM relic density. By determining the stop and lightest neutralino masses and
the stop mixing parameters, the stop--neutralino co--annihilation cross section
can be strongly constrained and the DM relic density predicted with a precision
of the same order as current astrophysical results. This is exemplified in 
Fig.~\ref{fig:DM-stop} which shows the accuracy in the determination of the DM
abundance as a function of the stop mass in the electroweak baryogenesis
scenario of Ref.~\cite{Baryo-Carena}. While an experimental error $\Delta 
m_{\tilde t_1}=1.2$ GeV (grey dots) leads to a relatively loose constraint, a
precision $\Delta  m_{\tilde t_1}=0.3$ GeV (dark dots) matches the original
scenario used as input (the red star) and the  1$\sigma$ and 2$\sigma$ WMAP
constraints (horizontal shaded bands). Refinements in the determination of the
stop mass can thus improve this result significantly. 

In non--SUSY scenarios, a  strong first--order electroweak phase transition 
needed to generate the baryon asymmetry can also be made possible. For instance,
this phase transition can be induced if the SM effective theory with one Higgs
doublet $\Phi$ is augmented with a dimension--six Higgs operator
\cite{Grojean:2004xa}, leading to a scalar Higgs potential of the form 

\centerline{$V= \lambda (\Phi^\dagger \Phi -\frac12 v^2)^2 + \frac{1}
{\Lambda^2}  (\Phi^\dagger \Phi-\frac12 v^2)^3$.} 

This additional term can be generated by strong dynamics at the TeV scale or by
integrating out heavy particles such as an additional singlet scalar field
\cite{Baryo-Quiros} or the heavier Higgs particles of a general two--Higgs
doublet model \cite{Kanemura:2004ch}. 

At zero--temperature, the CP--even Higgs state can be expanded in terms of its
usual vev, $\langle \varphi \rangle = v_0 \simeq 246$ GeV and the physical Higgs
boson field $\Phi=\varphi/{\sqrt 2}= (H+v_0)/{\sqrt 2}$. From the requirement
that the phase transition is first order and that the minimum at
zero--temperature is a global minimum, one  obtains,  respectively, an upper and
a lower bound on the cut--off $\Lambda$ for a given Higgs mass. For  a low
cut--off scale, $\Lambda \lsim 1$ TeV, the required electroweak phase 
transition can be achieved for Higgs masses  $M_H \gsim 114$ GeV
\cite{Grojean:2004xa}. 

\begin{figure}[!h]
\vspace*{-3mm}
\includegraphics[width=7.3cm,height=5.5cm]{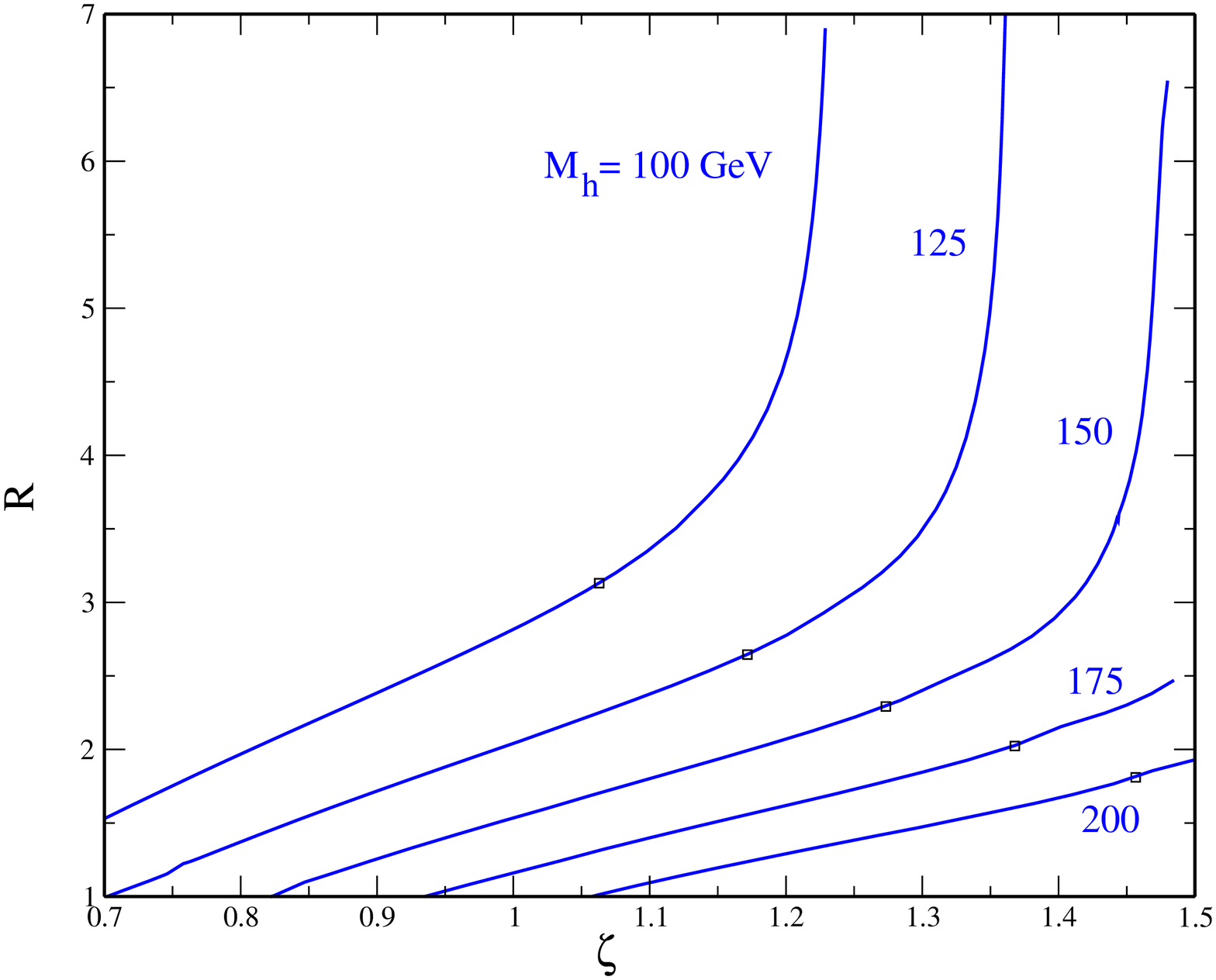}\hspace*{0.7cm}
\includegraphics[width=7.3cm,height=5.5cm]{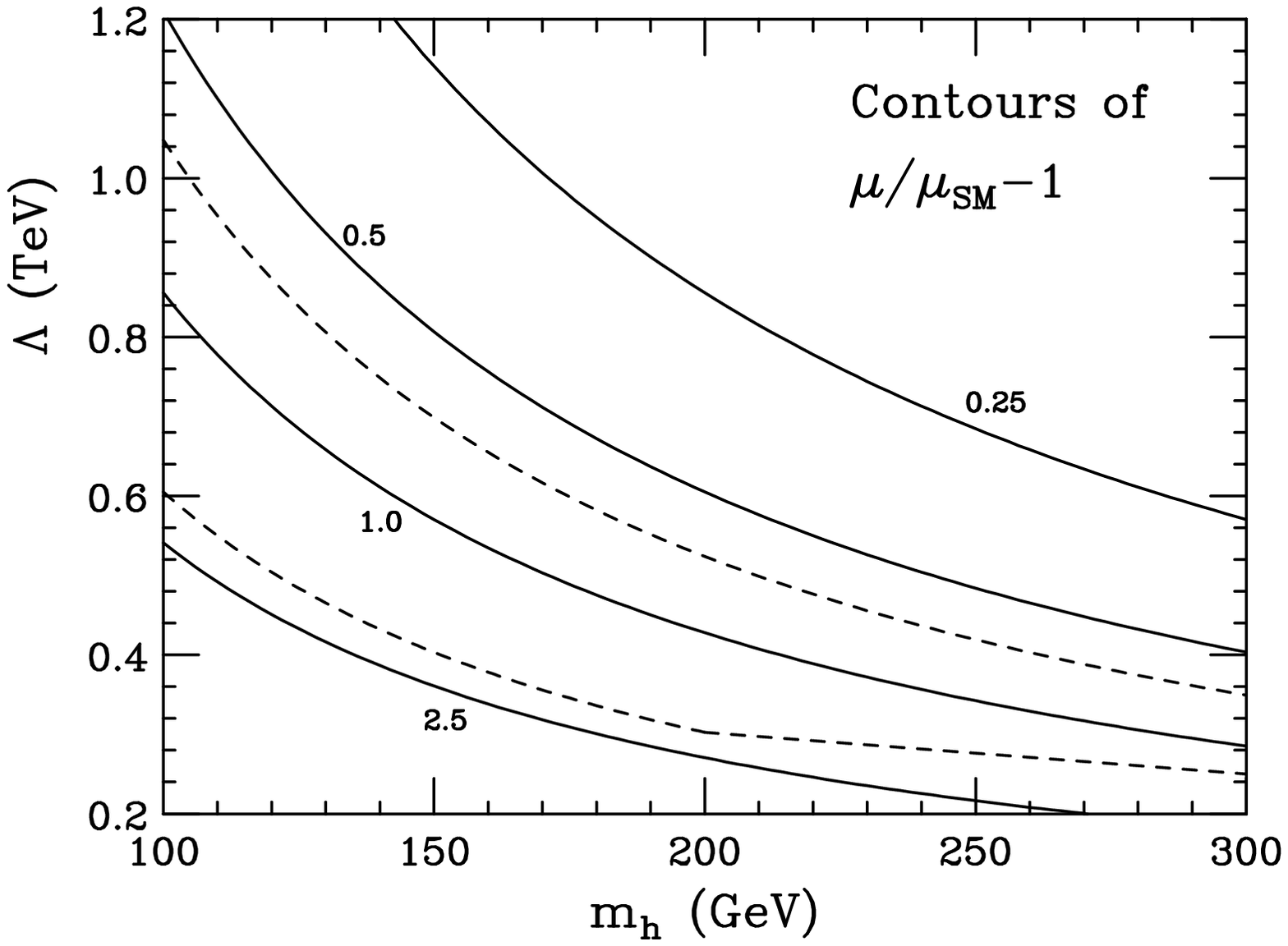}
\vspace*{-5mm}
\caption[Constraints and new effects in scenarios with electroweak 
baryogenesis]
{Left: the ratio $R\equiv \langle v(T_c)\rangle/T_c$ as a function of 
the parameter $\zeta$ for several $M_H$ values \cite{Baryo-Quiros}. 
Right: contours of constant $\mu/\mu_{\rm SM}-1$  in the $\Lambda$ vs.\ $M_H$ 
plane; the dashed lines delimit the region in which electroweak baryogenesis
can take place \cite{Grojean:2004xa}.}
\label{fig:DM-Hself} 
\vspace*{-3mm}
\end{figure}

As a concrete example of a possible origin of the dimension--six operator, one
can have a scalar singlet $N$ coupled to the Higgs field via an interaction of 
the form $\zeta^2 \Phi^\dagger \Phi N^2$.   If the singlet field has a mass
$m_N$ that is larger than the weak scale, it can be  integrated out and gives
rise to the additional Higgs interactions, $ \Delta V \propto \zeta^2 /m_N^2
\cdot |\Phi|^6$.  The baryogenesis condition of the non--erasure of the
generated baryon asymmetry is $R= \langle v T_c \rangle /T_c \gsim 1$ where
$T_c$ is the critical temperature at which the origin and the non-trivial
minimum at $\langle v(T_c)\rangle$ become degenerate. The dependence of this
ratio on the parameter $\zeta$ in the $\Phi^\dagger \Phi N^2$ interaction is
displayed  in Fig.~\ref{fig:DM-Hself}  for several values of the Higgs mass
$M_H$. As can be seen, $R$ values larger than unity can be obtained for Higgs
masses as large as $M_H \sim 200$ GeV.

Since the Higgs potential is altered by the dimension--six  operator with a
low--scale cutoff, large shifts in the Higgs boson self--couplings from their SM
values are generated. For instance, the trilinear Higgs coupling becomes
$\lambda_{HHH} \equiv \mu =3M_H^2/v_0+6v_0^3/\Lambda^2$ and the SM value 
$\mu_{\rm SM}$ is recovered only for $\Lambda \to \infty$.  In
Fig.~\ref{fig:DM-Hself}, the deviation of the trilinear Higgs coupling
normalized to its SM value, $\mu/\mu_{\rm SM}-1$, is displayed in the $[M_H,
\Lambda]$ plane and one sees that shifts of order unity can be obtained. This is
particularly true in the allowed regions (delimited by the dashed lines) for the
cut--off scale and the Higgs mass. 

Thus, if the electroweak phase transition plays an important role for the
generation of the baryon asymmetry of the universe, there is a  possibility to
test this mechanism  in collider experiments and, in particular, at the ILC. A
first hint may be obtained in Higgs physics as the nature of the electroweak
phase transition is closely related to the structure of the Higgs potential
and,  as  illustrated above, large deviations of the Higgs self--couplings from
their SM values are expected in this case.  Another important ingredient is the
new source of CP violation that triggers the separation of particles and
anti-particles during the first--order phase transition. Since the new CP phases
are carried by states that are  present at the phase transition temperature,
that is in the range the electroweak symmetry breaking scale, some of these
particles are very likely to be within the kinematical reach of the ILC. Precise
determination of particle masses, couplings and CP phases at the ILC will be
thus essential to confirm or disprove the electroweak baryogenesis scenario.

\subsection{Leptogenesis and right--handed neutrinos}

If leptogenesis \cite{DM-leptogen} is the origin of the observed baryon asymmetry in
the universe,  the roots of this phenomenon are located near the GUT or the
Planck scale. CP--violating decays of heavy right--handed Majorana neutrinos
generate a lepton asymmetry which is transferred to the quark/baryon sector by
sphaleron processes. Heavy neutrino mass scales as introduced in the seesaw
mechanism \cite{SUSY-seesaw} for generating light neutrino masses and the size
of the light neutrino masses needed for leptogenesis define a self--consistent
frame which is compatible with all experimental observations
\cite{DM-neutrino}.

As discussed in chapter \ref{sec:susy}, in some supersymmetric models, the size
of the heavy seesaw scales can be related to the values of the charged and
neutral slepton masses \cite{SUSY-FPZ}.  Of particular interest is the
comparison of scalar masses in the tau and the electron sector. If the scalar
mass parameters are universal at the GUT scale, as in minimal supergravity for
instance, this regularity can be unraveled in the first and second generation of
the scalar masses at the electroweak scale. However,  slepton masses of the
third generation will be different from the first two in theories incorporating
the seesaw mechanism. The running of the slepton masses from the GUT to the
electroweak scale will be affected by loops involving the heavy right--handed
neutrino, with masses in the range  $10^{10}$--$10^{15}$ GeV, which have large
Yukawa couplings in the third generation. Sum rules for mass differences of
sneutrinos and selectrons between the first and third generation can be
constructed that project out this contribution. 

\begin{figure}[h!]
\begin{center}
\psfig{figure=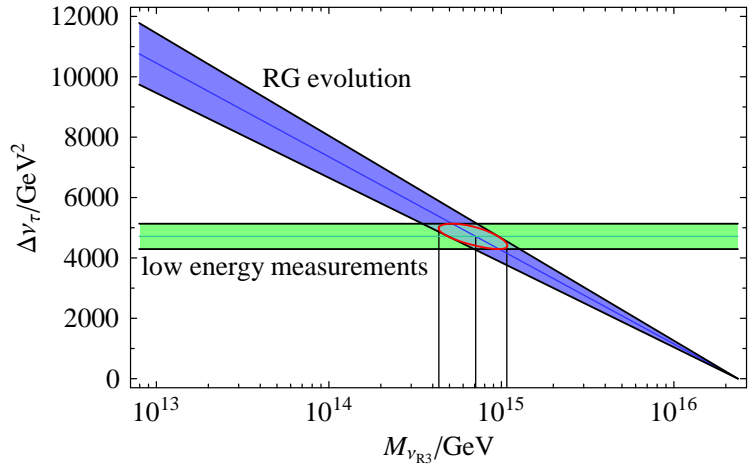,width=10cm}
\end{center}
\vspace*{-5mm}
\caption[ILC resolution on the right--handed neutrino mass 
in leptogenesis scenarios]
{ILC resolution in the estimate of the mass of the heaviest 
right--handed neutrino from the RGE evolution of slepton mass \cite{SUSY-FPZ}.}
\label{fig:sneutrino}
\end{figure}

Being approximately linear in the seesaw scale, the scale can be estimated from
the sneutrino and slepton masses with a rather good accuracy. In this way a
method has been found by which the large right--handed neutrino mass can, at
least indirectly, be measured  \cite{SUSY-FPZ}. The excellent resolution of ILC
can be exploited in this way to estimate the mass of the heaviest right-handed
neutrino within a factor of two as illustrated in Fig.~\ref{fig:sneutrino}.

Thus, by means of extrapolations governed by the renormalization group, the high
accuracy that can be achieved at the ILC in the slepton and sneutrino mass 
measurements, as discussed in chapter \ref{sec:susy}, can be exploited to
determine high-scale parameters that cannot be accessed directly. ILC
high--precision measurements in the SUSY sector may shed light on the heavy
neutrino sector and on the baryon asymmetry in the universe when realized via
leptogenesis, even at scales close to the GUT scale, as it might provide a very
valuable input which is the scale of the heavy right--handed neutrinos. 

%
%

\clearpage

%
\listoffigures %
\listoftables

\typeout{Lastpage = \arabic{page}}
\end{document}